%% file: main.tex
\documentclass[twoside,a4paper,10pt]{book}

\usepackage{amsmath}
\usepackage{amssymb}
\usepackage{graphicx}
\usepackage{fancyhdr}              	% Fancy Header and Footer
\font\msytw=msbm9 scaled\magstep1

\let\a=\alpha \let\b=\beta  \let\g=\gamma  \let\d=\delta \let\e=\varepsilon
\let\z=\zeta  \let\h=\eta   \let\th=\theta \let\k=\kappa \let\l=\lambda
\let\m=\mu    \let\n=\nu    \let\x=\xi     \let\p=\pi    \let\r=\rho
\let\s=\sigma \let\t=\tau   \let\f=\varphi \let\c=\chi
  \let\y=\upsilon \let\o=\omega
\let\G=\Gamma \let\D=\Delta  \let\Th=\Theta\let\L=\Lambda 
\let\P=\Pi    \let\Si=\Sigma     \let\Ps=\Psi
\let\O=\Omega \let\Y=\Upsilon
\let\ee=\epsilon

\let\==\equiv
\let\io=\infty 
\def\ie{{i.e. }}
\def\eg{{e.g. }}
\let\dpr=\partial

\let\T=\tau
\newcommand{\su}{\sigma^{\Th}}
\newcommand{\sd}{\sigma^{baths}}
\newcommand{\st}{\sigma^{eff}}
\newcommand{\Iint}{\int_{-\infty}^\infty}

\def\EE{{\cal E}} 
\def\CC{{\cal C}}\def\FF{{\cal F}} 
\def\TT{{\cal T}}\def\NN{{\cal N}} 
  
\def\DD{{\cal D}} 
\def\KK{{\cal K}} \def\QQ{{\cal Q}}

\def\re{{\rm Re}\,}
\def\im{{\rm Im}\,}

\def\ul{\underline}
\def\erf{\text{erf}}

\def\ol#1{{\overline #1}}

\def\V#1{{\underline#1}}
\def\Prob{{\rm Prob}}
\def\qed{\raise1pt\hbox{\vrule height5pt width5pt depth0pt}}

\def\to{\rightarrow}
\def\la{\left\langle}
\def\ra{\right\rangle}
\def\wh{\widehat}
\def\wt{\widetilde}
\def\Tr{{\rm Tr}\,}

\mathchardef\aa   = "050B
\mathchardef\bb   = "050C
\mathchardef\ggg  = "050D
\mathchardef\xxx  = "0518
\mathchardef\zzzzz= "0510
\mathchardef\oo   = "0521
\mathchardef\lll  = "0515
\mathchardef\mm   = "0516
\mathchardef\Dp   = "0540
\mathchardef\H    = "0548
\mathchardef\FFF  = "0546
\mathchardef\ppp  = "0570
\mathchardef\nn   = "0517
\mathchardef\ff   = "0527
\mathchardef\pps  = "0520
\mathchardef\FFF  = "0508
\mathchardef\nnnnn= "056E

\def\RRR{\hbox{\msytw R}}

\def\NNN{\hbox{\msytw N}} 
\def\ZZZ{\hbox{\msytw Z}}

\def\ul#1{{\underline#1}}

\newcommand{\beq}{\begin{equation}}
\newcommand{\eeq}{\end{equation}}
\newcommand{\bea}{\begin{eqnarray}}
\newcommand{\eea}{\end{eqnarray}}

\begin{document}

\frontmatter

\include{prima}

\pagestyle{empty}
\pagenumbering{roman}
\mbox{}
\newpage

%\mbox{}
%\newpage
\setcounter{page}{1}
%\include{cita}
%\newpage
\tableofcontents
\newpage
\pagestyle{fancy}

\addcontentsline{toc}{chapter}{Introduction}

\include{intro}

\mainmatter

\pagenumbering{arabic}
\setcounter{page}{1}

\part{Equilibrium}

\include{cap1}

\include{cap2}
\include{cap3}

\part{Nonequilibrium}

\include{cap4}
\include{cap5}

\include{cap6}

\include{cap7}

%\include{conclu}

\include{ringra}

\newpage
\pagestyle{fancy}
\addcontentsline{toc}{chapter}{List of figures}
\listoffigures

\backmatter

\addcontentsline{toc}{chapter}{Bibliography}

\include{biblio}
\end{document}

%% file: prima.tex
\begin{titlepage}

\thispagestyle{empty}
\begin{center}
{\large \textsc{UNIVERSIT\`A DEGLI STUDI DI ROMA ``LA SAPIENZA''}} \\
%\vspace{0.5cm}
%{\large \textsc{``LA SAPIENZA''}}

\vspace{1cm}

\includegraphics[width=3cm]{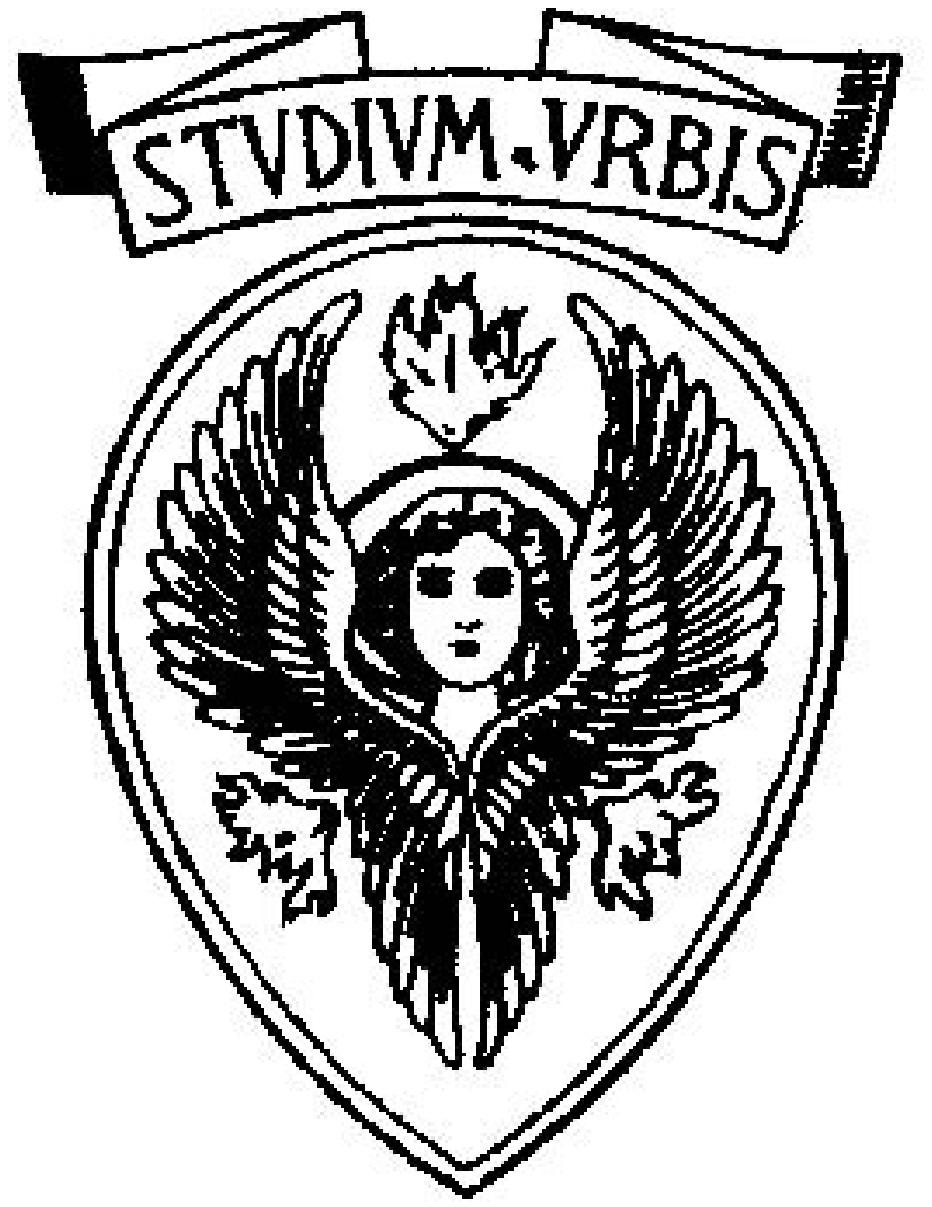}

\vspace{1cm}

{\normalsize \textbf{FACOLT\`A DI SCIENZE MATEMATICHE, 
FISICHE E NATURALI}} \\[.5cm]
{\large \textbf{Dottorato in Fisica - Ph.~D. in Physics}}\\[.25cm]
{\large \textbf{XVIII ciclo}}

\vspace{2.5cm}

{\huge

\textbf{Some applications of recent theories} \\
[.3cm]
\textbf{of disordered systems}\\
%[.3cm]
%\textbf{titolo}

} 
\end{center}

\vspace{2.8cm}
{\large
%\noindent Coordinator:

%\vspace{.1cm}

%\noindent Prof.~\textsl{Valeria Ferrari}
%\\
%\mbox{} \\
\noindent Supervisors: \hfill Candidate:

\vspace{.2cm}

\noindent Prof.~\textsl{Giorgio Parisi} 
\hfill Dr.~\textsl{Francesco Zamponi}\\[1mm]
\noindent Prof.~\textsl{Giancarlo Ruocco}
}

\vspace{1.2cm}

\begin{center}
{\large {October, 2005}}
\end{center}

\end{titlepage}

%% file: intro.tex
\chapter*{Introduction}

This Ph.D. thesis is divided in two parts.
The first one concerns the equilibrium properties of glassy
systems, \ie properties that can be derived from the Gibbs
distribution\footnote{The glassy state of matter is often a 
{\it metastable} state, due to the presence of a crystalline state
with lower free energy. The properties of the ``equilibrium'' glass
can be studied if one assumes that in some way the nucleation of the
crystal can be avoided. It is not obvious that this is possible, and
this point has always been matter of debate. If the existence of the
crystal can be neglected, one
can study the ``equilibrium'' properties of the glass by restricting
the Gibbs distribution to the amorphous configurations.}.
Non--glassy equilibrium systems are very well understood. Many different
thermodynamic phases of classical many-body systems are known and their
properties can be computed starting from the Gibbs distributions
or its decomposition in {\it pure states}.
Quantum effects can be taken into account leading to new thermodynamic
phases (superfluids, or superconductors) whose statistical properties
are also well understood. The {\it phase transitions} between different
phases have been extensively studied in the last century and their
current understanding is very satisfactory.

The theoretical understanding of the glass phase and the related 
glass transition, on the contrary, is still poor, even if many 
important progresses have been recently achieved.
Despite the existence of a number of mean field models which
reproduce the basic phenomenology of glassy systems, and for which
the glass transition can be fully characterized, the existence
of a thermodynamic glass transition in finite dimension is still a
matter of debate. Many authors believe that the glass transition
in finite dimension is a purely {\it dynamical} phenomenon that
cannot be derived from the Gibbs distribution.
The situation is complicated by the absence of a simple finite
dimensional glassy model which could play, in the context of glassy
systems, the role that the Ising model played in the context of
second order phase transitions.
Experiments and numerical simulations
can only investigate the {\it nonequilibrium}
counterpart of the (eventual) thermodynamical glass transition,
so experimental data on pure thermodynamical glassy states are
not available.
Thus, the problem of the existence of a thermodynamic glass
transition in finite dimension, and many related problems,
such as the existence of a diverging correlation length,
can only be addressed by analytical 
solution, either exact or approximate, of ``glassy'' models.

Mean field models are - up to now - the 
only solvable models of glassy systems:
they provide an useful framework to describe
the basic phenomenology observed in experiments.
Their detailed investigation revealed that
the glass transition is connected with the existence
of an exponential number (in the size of the system) of 
{\it metastable states}. The characterization of these metastable
states allowed to understand their relevance for the dynamics
of the system: it emerged that they play a key role in the
{\it nonequilibrium} dynamics of glasses and are responsible
for the existence of a {\it nonequilibrium glass transition} which
closely reflects the one that is observed in real glassy materials.

Some aspects of the phenomenology of glasses and a theory
attempting to describe them are presented in chapter~\ref{chap1}.
As an example of simple model for the glass transition in finite
dimension, I studied the Hard Sphere liquid (in collaboration with
G.~Parisi). This study is presented
in chapter~\ref{chap2}. Obviously the model
is not exactly solvable, but it was possible to solve it 
{\it approximately} by means of a replica trick and of the HNC
approximation - a standard approximation in the theory of simple
liquids. 
This strategy was already successfully applied by M.~M\'ezard and
G.~Parisi to analytic potentials (\eg Lennard-Jones), 
but the application to Hard
Spheres required some additional work due to the singularity
of the interaction potential.
In this approximation, a thermodynamic glass transition
is found. The equation of state of the glass and its pair correlation
function $g(r)$ can be computed. This allows also to obtain an estimate
of the {\it random close packing} density and of the mean coordination
number in the amorphous packings. The results agree well with the
available numerical data and with other theories. This is encouraging
but does not solve the problem of the existence of the glass transition
in finite dimension because of the approximations involved.

As discussed above mean field models reproduce many aspects of 
the phenomenology of glasses. 
In chapter~\ref{chap3} it
is shown (in collaboration with G.~Parisi and G.~Ruocco) 
that these models also reproduce a correlation between the fragility
of a liquid - to be defined in chapter~\ref{chap1} - 
and the vibrational properties of its glass that has been recently
found by T.~Scopigno {\it et al.} analyzing experimental data on a 
wide class of glassy materials. This result is - in our opinion -
an interesting confirmation of the relevance of mean field models
in the description of the phenomenology of real glasses.
An outcome of this study is that the number of metastable 
states is a decreasing function of fragility; 
this prediction differs from the one that has
been obtained by other authors and can be tested, in principle, on
real materials.

The second part of the thesis
concerns some recent attempts - discussed in chapter \ref{chap4} -
to build a statistical
theory of nonequilibrium stationary states induced by the application
of an external driving force on a thermostatted system.
From the {\it chaotic hypothesis}, an extension of the ergodic hypothesis
to nonequilibrium systems proposed by E.~G.~D.~Cohen and G.~Gallavotti,
an explicit expression for the measure describing the system in
stationary state can be derived. For time-reversible systems, an
interesting prediction of this theory is the validity
of the {\it fluctuation relation}:
a relation between the probability of positive and negative 
large fluctuations of the {\it phase
space contraction rate} $\s$, 
often identified with the {\it entropy production rate}. 
What is remarkable is that the fluctuation relation is {\it universal},
in the sense that it contains no model-dependent parameters.

A test of the fluctuation relation
is then a rather stringent test of the
theory, and indeed it has been performed in a number of cases, in the last
decade, with positive result. 
In chapter \ref{chap5}
(in collaboration with A.~Giuliani and G.~Gallavotti)
the fluctuation relation is tested
in a numerical simulation of a system of particles interacting via
a Lennard-Jones--like potential and subjected to an external driving
force and to a thermostatting force (isokinetic constraint).
With respect to previous studies of similar systems, an important
progress has been obtained: the observation of non-Gaussian tails
in the probability distribution of $\s$. 
This is important because the fluctuation relation is 
related to the Green-Kubo relations at the Gaussian level,
so a test that is really independent from linear response theory
requires the observation of non-Gaussian tails.
This progress was possible thanks to the
increase of computational power in the last years.

In chapter \ref{chap6} some aspects of the driven nonequilibrium 
dynamics of glassy systems are discussed. In the limit
of small driving force (small entropy production), it has been shown
by L.~Cugliandolo and J.~Kurchan that a nonequilibrium 
{\it effective temperature} can be introduced, which has the property
of being a temperature in the thermodynamic sense: it controls heat
flows and enters the relation between spontaneous fluctuations and
response to external perturbations as in equilibrium.
The systems reaches a stationary state and it is possible to
decompose the dynamics in different time scales. On each time scale,
a single effective temperature is defined. The system behaves as if
composed by many non-interacting subsystems, evolving on well separated
time scales, each one characterized by the corresponding 
effective temperature. 

The fluctuation relation is related to the definition of temperature
out of equilibrium. For driven systems evolving on a single time scale and
in contact with an equilibrated bath at temperature $T$, the temperature
of the bath controls the fluctuations of the entropy production rate.
Thus, one can ask if, for driven glasses, a modified fluctuation relation
can be introduced, in which the effective temperature enters instead of
the temperature of the bath.
This idea was first investigated by M.~Sellitto, and many proposals in
this direction subsequently appeared.
I investigated (in collaboration with F.~Bonetto, L.~Cugliandolo
and J.~Kurchan) a very simple model for glassy dynamics: a Brownian
particle in contact with a bath whose correlation and response function
do not satisfy the fluctuation--dissipation relation.
An effective temperature can be defined, and we showed that a modified
fluctuation relation holds, in which the temperature of the bath is
replaced by the effective temperature. These results are presented
in chapter \ref{chap:sette} where they are also compared with similar results
that recently appeared in the literature.
Some numerical data, obtained on a sheared
Lennard-Jones--like system in the
glassy regime (in collaboration with L.~Angelani and G.~Ruocco), are also
presented. They partially confirm the results obtained analytically.
Unfortunately, a numerical check of all the predictions of the model
is impossible because the time scales involved are beyond the ones accessible
to the numerical simulation.

Chapters \ref{chap1}, \ref{chap4} and \ref{chap6} are introductory
chapters, while in chapters \ref{chap2}, \ref{chap3}, \ref{chap5} and
\ref{chap:sette} original results are presented.
It is important to remark that this is
{\it not a review article}. In the introductory chapters, I made
no attempt to quote all the theories, numerical data, experimental results
avalaible on the subject.
For example, in chapter \ref{chap1} the inherent structures approach is
missing, and in chapter \ref{chap6} only the dynamics of mean field models
is discussed,
without any attempt to review the rich dynamical phenomenology of real
materials and the theories attempting to describe it (\eg Mode-Coupling
theories).
{\it Only the notions that were needed to present the original
results have been included in the introductory chapters.}
This does not necessarily 
mean that I prefer the theories presented in these chapters
to other ones.

The results collected here have been published in:
\begin{itemize}
\item Chapter \ref{chap2}: G.~Parisi and F.~Zamponi, 
J.~Chem.~Phys. 123, 144501 (2005).
\item Chapter \ref{chap3}: G.~Parisi, G.~Ruocco and F.~Zamponi, 
Phys.~Rev.~E {\bf 69}, 061505 (2004).
\item Chapter \ref{chap5}: A.~Giuliani, F.~Zamponi and G.~Gallavotti,
J.~Stat.~Phys. {\bf 119}, 909 (2005).
\item Chapter \ref{chap:sette}: F.~Zamponi, G.~Ruocco and L.~Angelani, 
Phys.~Rev.~E {\bf 71}, 020101(R) (2005); \\ \mbox{} \hskip0.1cm 
F.~Zamponi, F.~Bonetto, L.~Cugliandolo and J.~Kurchan,
J.~Stat.~Mech.~(2005)~P09013.

\end{itemize}

%% file: cap1.tex
\chapter{The glass transition}
\label{chap1}

%%%%%%%%%%%%%%%%%%%%%%%%%%%%%%%%%%%%%%%%%%%%%%%%%%%%%%%%%%%%%%%%%%%%%%
%%%%%%%%%%%%%%%%%%%%%%%%%%%%%%%%%%%%%%%%%%%%%%%%%%%%%%%%%%%%%%%%%%%%%%

\section{Basic phenomenology}

Although liquids normally crystallize on cooling, there
are members of all liquids types (including molecular, ionic and
metallic) that can be supercooled below the melting temperature
$T_m$ and then solidify at some temperature $T_g$, the 
{\it glass transition temperature}. 
The viscosity $\h(T)$ of the liquid increases continuously
but very fast below $T_m$ and at some point reaches values so high that
the liquid does not flow anymore and can be considered a solid for
all practical purposes: at low temperatures, an amorphous solid
phase is observed.
The temperature $T_g$ marking the transition between the liquid and
the glass is often defined by the condition $\h(T_g)=10^{13}$ Poise, 
but many other definitions are possible.

\begin{figure}[t]
\includegraphics[height=230pt]{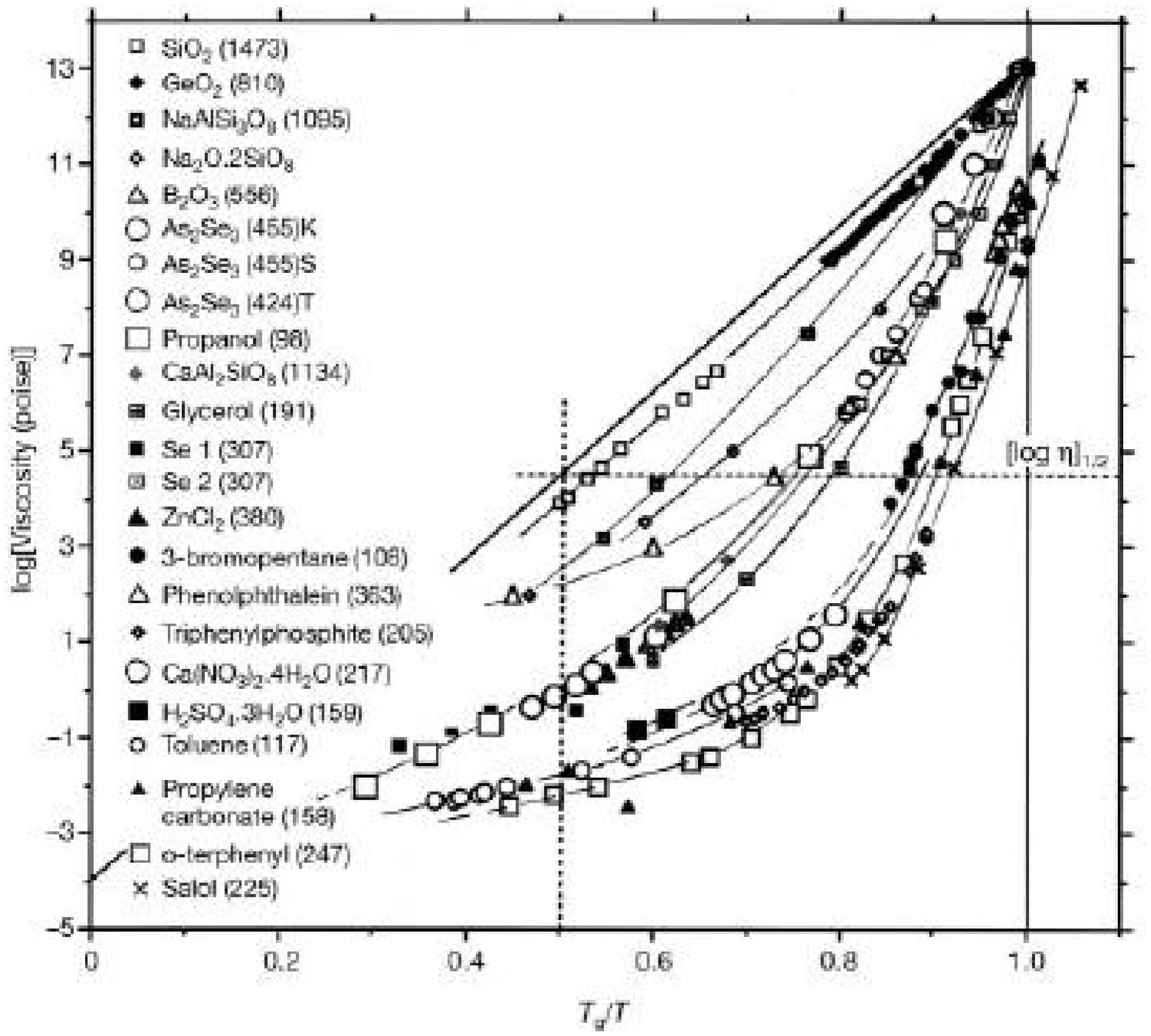}
\includegraphics[height=230pt]{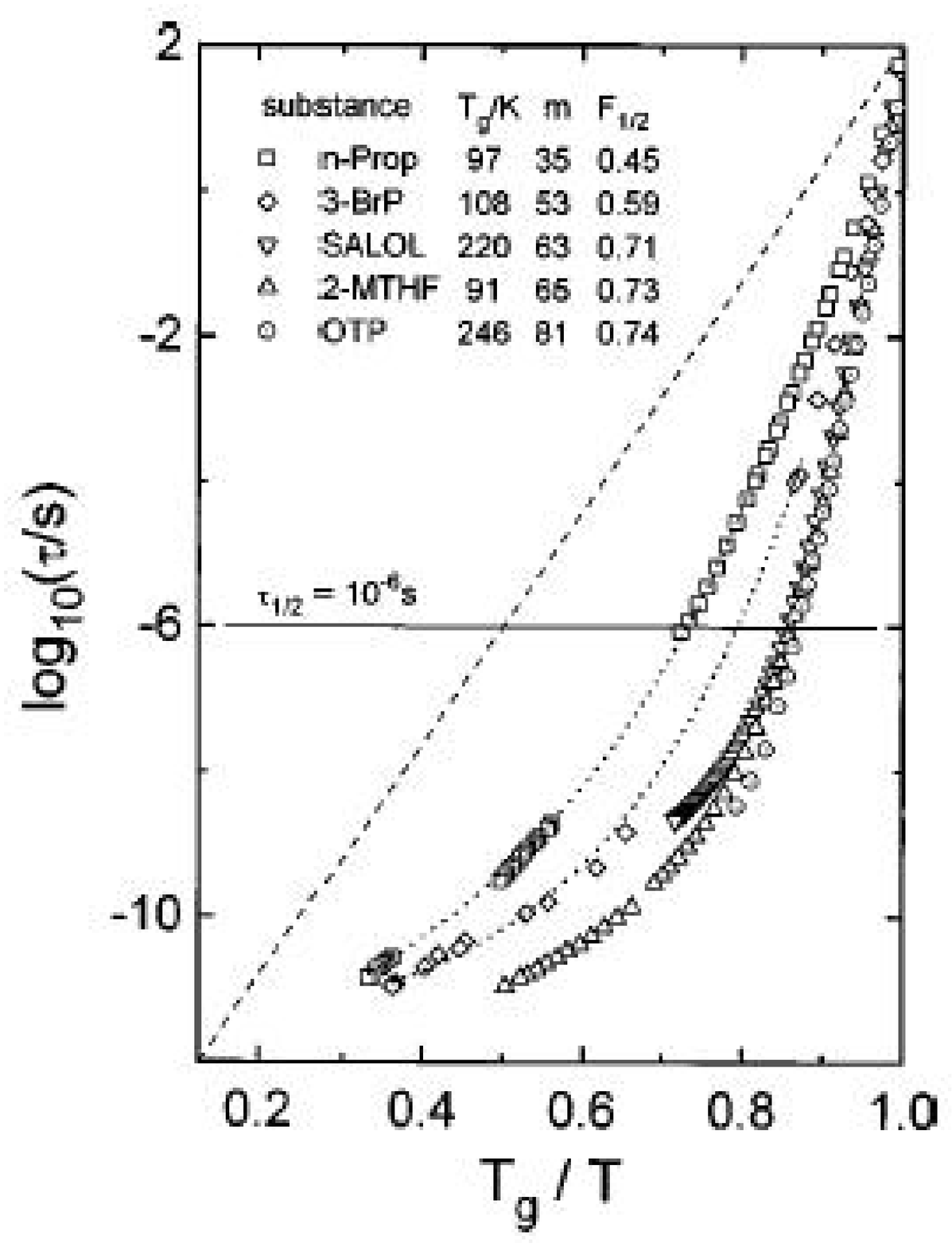}
\caption[Viscosity and relaxation time of many glass formers]
{(Left, from~\cite{MA01}) Viscosity data for many glass forming liquids.
The logarithm of the viscosity measured in Poise is reported as a function
of $T_g/T$. The (calorimetric) $T_g$ is defined as the temperature at which 
the enthalpy relaxation time is $\sim 200$s, and its value is 
reported in parenthesis in the key of the figure. 
Note that for some systems the value of the calorimetric
$T_g$ does not satisfy exactly the condition $\h(T_g)=10^{13}$ Poise. 
Fragility is the slope of the curves in $T_g/T=1$.
(Right, from~\cite{RA98}) Structural relaxation time obtained from 
dielectric relaxation measurements.
The dashed line indicates Arrhenius behavior. 
The value of $T_g$, obtained from $\t_\a(T_g)=100$s,
and of fragility are reported in the key.}
\label{fig1:rel}
\end{figure}

As an example of this phenomenon, in Fig.~\ref{fig1:rel} the 
viscosity of many glass forming liquids is reported as a function 
of the temperature. Following Angell~\cite{An95,An97,MA01}, the quantity
$\log_{10} \left[\frac{\h(T)}{\rm Poise}\right]$ is reported as a 
function of $T_g/T$.
The viscosity increases of about $17$ orders of magnitude on decreasing the
temperature by a factor $2$. Note that as the increase of viscosity
is so fast, the dependence of $T_g$ on the particular value of viscosity
($10^{13}$ Poise) which is chosen to define it is very weak.

It is often found that the viscosity around $T_g$ follows the 
Vogel--Fulcher--Tamman (VFT) law \cite{VFT},
\beq
\label{VFT}
\h(T) = \h_\io e^{\frac{\D}{T-T_0}} \ ,
\eeq
where $\h_\io$, $\D$ and $T_0$ are system--dependent parameters.
If $T_0 = 0$ this relation reduces to the Arrhenius law; otherwise,
the extrapolation of the viscosity below $T_g$ leads to a divergence
at $T=T_0$.

\subsection{Fragility}
\label{sec1:fragility}

The {\it fragility} concept has been introduced by Angell~\cite{An85}.
It describes how fast the viscosity increases with decreasing temperature
on approaching $T_g$. ``Strong'' glasses (low values of fragility)
show a ``weak'' $T$ dependence of $\h(T)$, which is often described
by the Arrhenius law (see \eg the curve for ${\rm SiO}_2$ in 
Fig.~\ref{fig1:rel}), while ``fragile'' glasses show a 
much faster $T$ dependence of the
relaxation time, often described by the VFT law with $T_0 \neq 0$.
A common example of fragile glass former is the 
{\it o-terphenyl} (OTP), see Fig.~\ref{fig1:rel}.

If the VFT law holds,
the ratio $\frac{T_g}{T_g - T_0}$ can be taken as a {\it fragility index}:
it ranges from $1$ for strong glasses to $\sim 10$ for the most fragile 
glasses~\cite{An97}. However, the common definition of the fragility index, which is
also independent of the VFT law, is
\beq
\label{mAdef}
m_A \equiv \left. \frac{d \log_{10} \left[ \frac{\h(T)}{\rm Poise} \right]}{d(T_g/T)}
\right|_{T=T_g} \ ,
\eeq
\ie it is given by the slope of the curves in Fig.~\ref{fig1:rel} at $T_g/T=1$.
This definition involves only the derivative of the viscosity at $T_g$, without
any assumption on the global behavior of $\h(T)$. According to this definition,
a strong glass (strictly Arrhenius behavior) would have $m_A \sim 17$
(being $\h_\io=10^{-4}$ Poise and $\h(T_g)=10^{13}$ Poise), while the most fragile
systems reach $m_A \sim 160$~\cite{An97}. If the VFT law holds it is easy to show
that $m_A \sim \frac{17 T_g}{T_g-T_0}$.

\subsection{Structural relaxation time}

The viscosity is related to the {\it structural relaxation time} $\t_\a$
by the Maxwell relation, $\h = G_\io \t_a$, where $G_\io$ is the 
infinite--frequency shear modulus of the liquid.
The structural relaxation time is related to the decorrelation of
density fluctuations. In glass forming liquids, for $T_m \gg T \geq T_g$,
the decorrelation of density fluctuations happens on two well separated
time scales: a ``fast'' time scale ($\sim 10^{-12} s$), which is related to 
vibrations of the particles around the disordered instantaneous positions,
and a ``slow'' time scale $\t_\a$, which is related to cooperative 
rearrangements of the disordered structure around which the fast vibrations
take place. Through the Maxwell relation, the fast increase of viscosity
around $T_g$ is then related to a marked slowing down of the structural
dynamics; usually, at $T_g$ one has $\t_\a \sim 100s$, while in the liquid
phase $\t_\a \sim \m s$.

The structural relaxation time, obtained from dielectric relaxation data,
of some fragile glass forming liquids is reported in the right panel
of Fig.~\ref{fig1:rel}. The behavior of $\t_\a(T)$ is also described
by a VFT law with an apparent divergence at $T=T_0$. This leads to the
interpretation of $T_0$ as a temperature at which a structural arrest
takes place.

A common pictorial interpretation of the dynamics of glass forming liquids
above $T_g$ is the following: for short times the particles are ``caged'' by
their neighbors and vibrate around a local structure on a nanometric scale;
the structural relaxation is then interpreted as a slow cooperative rearrangement
of the cages. Note that on the time scale of the structural relaxation time $\t_\a$,
the mean square displacement of the particles is smaller than the particle radius,
so one cannot think to the structural relaxation as a process of single--particle 
``jumps'' between adjacent cages.

\subsection{Configurational (or excess) entropy}

The idea that the dynamics in the supercooled phase is separated in a fast
intra--cage motion and in a slow cooperative rearrangement of the structure
suggests to split the total entropy of the liquid in a ``vibrational''
contribution, related to the volume of the cages,
and a ``configurational'' contribution, that counts the
number of different disordered structures that the liquid can assume~\cite{Ka48}:
\beq
S_{liq}(T) \sim S_{vib}(T) + S_c(T) \ .
\eeq
To estimate the vibrational contribution to the entropy of the liquid, one
can assume that it is roughly of the order of the entropy of the corresponding
crystal. It is then possible to estimate $S_c(T)$ as
\beq
\label{ScDEF}
S_c(T) = S_{liq}(T)-S_{cryst}(T)=\D S_m - \int_T^{T_m} d \log T' \,
\big[ C_{liq}(T') - C_{cryst}(T') \big] \ ,
\eeq
where $\D S_m \equiv S_{liq}(T_m)-S_{cryst}(T_m)$ 
is the entropy difference between the liquid and the crystal at
the melting temperature $T_m$, and 
$C(T) = T \frac{\dpr S}{\dpr T}$ is the specific heat.
Note that in experiments one usually works at constant pressure, $C=C_p$, while
in numerical simulations and in theoretical computations one usually works 
at constant volume, $C=C_v$. The configurational entropy $S_c$ is sometimes called
``excess entropy''.

\begin{figure}[t]
\centering
\includegraphics[width=450pt]{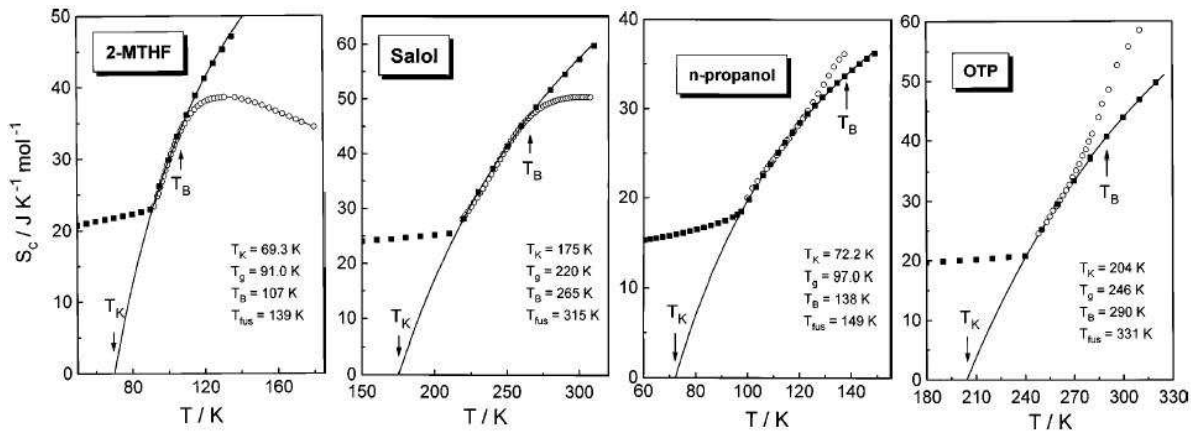}
\caption[Configurational entropy of four different fragile glass formers]
{(From~\cite{RA98} and references therein) Configurational entropy $S_c(T)$ of four 
fragile glass formers. The black squares are obtained from calorimetric measurements
of the specific heat of the liquid and of the crystal, see Eq.~(\ref{ScDEF}). Below
$T_g$ (reported in the key) the liquid falls out of equilibrium. The black line
is the extrapolation according to Eq.~(\ref{ScFIT}) of the equilibrium data for 
$T \geq T_g$ below $T_g$, that goes to zero at $T=T_K$. 
The open white circles are derived from the dielectric relaxation data of 
Fig.~\ref{fig1:rel} using the Adam--Gibbs relation, Eq.~(\ref{AGrel}).
The coincidence of the two estimates of $S_c(T)$ proves the validity of the
Adam--Gibbs relation for $T_B \geq T \geq T_g$.}
\label{fig1:Sconf}
\end{figure}

In Fig.~\ref{fig1:Sconf} the estimate of $S_c$, obtained from calorimetric measurements
of the specific heat and using Eq.~(\ref{ScDEF}), is reported
for four different fragile glass formers. Below $T_g$ the liquid falls out of equilibrium
as the structural relaxation time becomes of the order of the 
experimental time scale ($\sim 100$s).
This means that the structural rearrangements are ``frozen'' on the 
experimental time scale and the only contribution to the specific heat comes
from the intra--cage vibrational motion; in this situation the specific heat of
the liquid becomes of the order of the one of the crystal and $S_c(T)$ approaches
a constant value. However, one can ask what would happen if the time scale of the
experiment were much bigger, say $10^6$s. In this case, the glass transition temperature
$T_g$ would be lower and the {\it plateau} would be reached at smaller values
of $S_c$. If one assumes to be able to perform an {\it infinitely slow} experiment,
one can imagine to follow the extrapolation of the data collected above $T_g$ to
lower temperatures. For fragile liquids, it is found that a good extrapolation
is
\beq
\label{ScFIT}
S_c(T) = S_\io \left( 1 - \frac{T_K}{T} \right) \ ,
\eeq
where the parameters $S_\io$ and $T_K$ are fitted from the data above $T_g$.
This extrapolation is reported as a full line in Fig.~\ref{fig1:Sconf}.

The outcome of this procedure is that the configurational entropy seems to vanish
at a finite temperature $T_K$. As $S_c$ counts the number of different structures
that the liquid can access, it is not expected to become negative; also, negative
values of $S_c$ imply that the entropy of the liquid becomes smaller than the
entropy of the crystal, which is a very counterintuitive phenomenon.
A possible explanation of this paradoxical behavior was proposed by 
Kauzmann~\cite{Ka48}, who argued that at some temperature between $T_g$ and $T_K$
the free energy barrier for crystal nucleation becomes of the order of the free
energy barrier between different structures of the liquid. This means that
the time scale for crystal nucleation becomes of the order of the structural
relaxation time $\t_\a$ of the liquid, and one cannot think anymore to 
an ``equilibrium'' liquid as crystallization will occur on the same time scale
needed to equilibrate the liquid. The extrapolation of $S_c(T)$ down to $T_K$ is
then meaningless, and the paradox is solved. This argument has been recently 
reconsidered, see \eg~\cite{CGG03}, and its implications are still under 
investigation.

\subsection{The ideal glass transition}

Alternatively, one can assume that the existence of the crystal is irrelevant,
because crystallization can be in some way strongly inhibited: for instance, by
considering binary mixtures, or --in numerical simulations-- by adding
a potential term to the Hamiltonian that forbids nucleation.
If crystallization is neglected, the extrapolation of $S_c$
suggests that at $T_K$ a phase transition happens: at $T_K$, the number
of structures available to the liquid is no more exponential, as $S_c=0$,
and the system is frozen in one amorphous structure which can be called
an {\it ideal glass}. Below $T_K$, the only contribution to the entropy
of the ideal glass is the vibrational one, so the specific heat has a jump
{\it downward} at $T_K$. The transition is expected to be of second order
from a thermodynamical point of view.

An evidence that support this picture is the fact that in almost all the fragile
glass formers it is found that $T_K \sim T_0$. For instance, in~\cite{An97} some
30 cases where $T_0 = T_K$ with an error of order $3\%$ are reported.
This means that both the structural relaxation time and the viscosity diverge at
$T_K$, so that the structures that are reached at $T_K$ are thermodynamically
stable, being associated to an infinite structural relaxation time.
The exact solution of a class of mean field disordered models which share many aspects
of the phenomenology with fragile glass formers also supports the picture that
a thermodynamic transition happens at $T_K$, as will be discussed later.

Of course, the ideal glass transition that occurs {\it in equilibrium} is not
observable: at some temperature $T_g > T_K$ where $\t_\a(T_g) = \t_{exp}$
a {\it real} glass transition, freezing the system in a nonequilibrium 
amorphous state (a {\it real} glass), happens. The value of $T_g$, as well as
the properties of the nonequilibrium glass (density, structure, etc.) depend
on the value of $\t_{exp}$, which is usually $\sim 100$s as already discussed.

\subsection{Adam--Gibbs relation}

\begin{figure}[t]
\centering
\includegraphics[width=8cm]{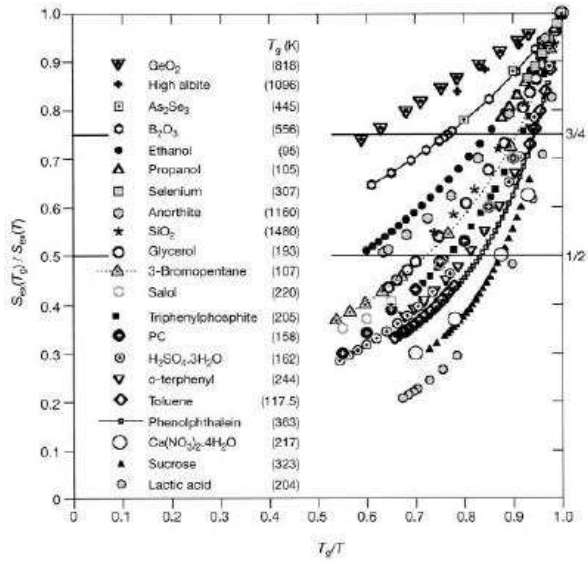}
\caption[Angell plot of the configurational entropy]
{(From~\cite{MA01}) Scaled configurational entropy $S_c(T_g)/S_c(T)$
(obtained from calorimetric data) for some of
the substances of Fig.~\ref{fig1:Sconf} as a function of $T_g/T$.
The slope of the curves in $T=T_g$ is related to the fragility by
Eq.~(\ref{frag-DCp}).
}
\label{fig1:Sconf_scaled}
\end{figure}

The identity of $T_0$ and $T_K$ suggests that the divergence of $\t_\a$ is
related to the vanishing of $S_c$. Indeed, Adam and Gibbs~\cite{AG65} proposed
that the following relation holds for $T$ close to $T_g$:
\beq
\label{AGrel}
\t_\a(T)=\t_\io \exp \left(\frac{\EE}{T S_c(T)}\right) \ ,
\hskip20pt
S_c(T)= \frac{\EE}{T \log[\t_\a(T)/\t_\io]} \ ,
\eeq
where $\EE$ is a system dependent parameter with the dimension of an energy
that is somehow related to the energy barrier for activated processes of 
transition between different liquid structures. A similar relation for
the viscosity is obtained by the Maxwell relation $\h=G_\io \t_\a$.
Eq.~(\ref{AGrel}) has been successfully tested in a wide number of experiments
and numerical simulations. As an example, in Fig.~\ref{fig1:Sconf} the
configurational entropy obtained from dielectric relaxation measurements
of $\t_\a$ via Eq.~(\ref{AGrel}) is compared with the calorimetric
measurement of $S_c$. The results show that Eq.~(\ref{AGrel}) is very well
satisfied in a range of temperatures above $T_g$.

The original Adam--Gibbs theory leading to Eq.~(\ref{AGrel}) was reconsidered
and improved in recent works~\cite{KTW87a,Wo97,XW01,LW03,BB04}, which will be
discussed later.

Eq.~(\ref{AGrel}) allows to rewrite the fragility defined by Angell as
\beq
\label{frag-DCp}
\frac{m_A}{17} = 1 + \frac{T_g}{S_c(T_g)} \frac{dS_c}{dT}(T_g) =
1 + \left.\frac{d (S_c(T_g)/S_c(T))}{d(T_g/T)}\right|_{T=T_g} \ .
\eeq
As $T_g \frac{dS_c}{dT}(T_g)$ is the specific heat jump at 
$T_g$\footnote{Because the entropy of the liquid slightly above 
$T_g$ is $S_{vib}(T_g)+S_c(T_g)$ while, slightly below $T_g$, 
the structure is frozen and the entropy is simply $S_{vib}(T_g)$.},
the Adam--Gibbs relation implies that fragility is linearly related
to $\D C(T_g) / S_c(T_g)$.
In Fig.~\ref{fig1:Sconf_scaled} $S_c(T_g)/S_c(T)$ is reported as a function of $T_g/T$
for many of the substances whose viscosity is reported in Fig.~\ref{fig1:rel}.
The close similarity between the two plots is another indication of the validity
of the Adam--Gibbs relation.

\subsection{An order parameter for the glass transition}
\label{sec1:orderparameter}

To better investigate the possibility that a thermodynamic transition happens
at $T_K$, one should define an order parameter to discriminate between the
liquid and the (ideal) glass phase~\cite{MP99}.
Before going to a purely static description of the order parameter, it is
easier to discuss a dynamical one. Around $T_g$, the dynamics of the
particles happens on two time scales, the fast one related to the intra--cage
motion, the slow one related to cooperative structural rearrangements.
The latter are frozen at $T_K$: at an atomic level, one tends to
associate the glass transition with the divergence of the time scale on
which a given particle can get out of the cage made by its neighbors.
While this is an intuitive picture, it is not possible to translate it 
into a good definition of an amorphous solid phase: because of the
excitation and movements of vacancies and other defects, this individual
trapping time scale is always finite, although it will increase exponentially
when the temperature gets small. What is really divergent is the time scale
needed for a {\it large scale} rearrangement of the structure. This means
that, even if single particles can always escape their traps in finite time,
in the thermodynamic limit density fluctuations remain partially correlated
also for $t \to \io$. 
Considering a system of $N$ particles, a proper dynamical definition of 
the order parameter is, for example, the so-called
{\it nonergodicity factor}~\cite{MP99,Ge83}
\beq
\label{nonergdyn}
f_{dyn}(\ul k) = \lim_{t\to\io} \lim_{N\to\io} \frac{1}{N}\sum_{jl}
\langle e^{i \ul k \, [ \, \ul x_j(t)-\ul x_l(0) \, ]} \rangle = 
\lim_{t\to\io} \lim_{N\to\io} \langle \r_{\ul k}(t)\r_{-\ul k}(0) \rangle_{dyn} \ ,
\eeq
where $x_j(t)$ is the position of particle $j$ at time $t$, $\ul k$ is
an arbitrary wave vector of the order of magnitude of the inverse 
interparticle distance, $\r_{\ul k} = N^{-1/2} \sum_j e^{i \ul k \, \ul x_j}$
is a Fourier component of the density fluctuations. The
thermodynamic limit has to be taken first, because a finite
number of particles always has a finite relaxation time.
The average $\langle \bullet \rangle_{dyn}$ is on the dynamical history
of a single system.
$f_{dyn}(\ul k)$ is expected to vanish in the liquid phase and to be different
from $0$ in the glass phase.

In order to construct a static order parameter, one needs to identify
a macroscopic quantity that discriminates between the different equilibrium
states that the system can access. Unfortunately, for amorphous states 
it is impossible to construct such a quantity: in the glass case, in
order to choose a state, one should first know the average position
of each particle in the solid, which requires an infinite amount of
information. This situation is very different from the one that
characterizes an ordered solid in which the Fourier components of
the density $\r_{\ul k}$ develop strong Bragg peaks in the solid
phase. However, a simple method to deal with amorphous states has
been developed in the context of spin glasses: the idea is to
consider two identical copies of the original system coupled by a small
extensive attraction of amplitude $\epsilon$. One takes first the
thermodynamic limit, and then the limit $\epsilon \to 0$. In the
liquid phase, the two copies are able to decorrelate also in the
thermodynamic limit, while in the glass phase an infinitesimal
attraction is enough to keep the copies close
to each other. The order parameter is then defined as
\beq
\label{nonergeq}
f_{eq}(\ul k) = \lim_{\epsilon \to 0} 
\lim_{N\to \io} \frac{1}{N} \sum_{jl}
\langle e^{i \ul k \, [ \, \ul x_j-\ul y_l \, ]} \rangle_{\epsilon} \ ,
\eeq
which is the static analogue of $f_{dyn}$. The average 
$\langle \bullet \rangle_\epsilon$ is now on
the {\it equilibrium} distribution of the two coupled copies.

It is observed that $f(\ul k)$ jumps
{\it discontinuously} to a finite value when crossing the glass
transition temperature $T_g$. Thus, the glass transition is
a second order transition from a thermodynamical point of view
but it is of first order if one looks to the order parameter.

%%%%%%%%%%%%%%%%%%%%%%%%%%%%%%%%%%%%%%%%%%%%%%%%%%%%%%%%%%%%%%%%%%%%%%
%%%%%%%%%%%%%%%%%%%%%%%%%%%%%%%%%%%%%%%%%%%%%%%%%%%%%%%%%%%%%%%%%%%%%%

\section{A mean field scenario}

So far, the only systems for which the phenomenology described above
could be {\it analytically} derived are some type of mean field 
spin glasses~\cite{KTW87a,KTW87b,GM84,Ri92,KPV93,CS92,CS95,CGP97,CC05}:
the so-called $p$-spin glasses.
These systems show an {\it equilibrium} Kauzmann
transition at a finite temperature $T_K$, where the configurational
entropy vanishes, the specific heat jumps downward and the order parameter 
discontinuously jumps to a finite value. Their dynamics is very similar
to the one of glass forming liquids in the region of temperature 
$T_m > T \gg T_g$, but the VFT behavior of the relaxation time is
not reproduced by these models: instead, a power law divergence of the
relaxation time is found at a temperature $T_d>T_K$. Although
this phenomenon is due to the mean field nature of these models, it
is not completely unrelated to what is observed in glass forming liquids, where a
power law behavior of $\t_\a$ is found at temperature $T$ not too close to $T_g$.
Indeed, the equations that describe the dynamics of the $p$-spin glass models
are formally very similar to the {\it Mode--Coupling} equations~\cite{GS91,vMU93}
that describe well the dynamics of supercooled liquids in a range of temperature
below $T_m$ but not too close to $T_g$~\cite{Cu02}.
Moreover, many properties of the free energy landscape of these models
(pure states, metastable states, barriers, etc.) could be investigated,
allowing for a deep understanding of the mechanisms leading to the Kauzmann
transition and to the slowing down of the dynamics close to $T_d$.

Excellent reviews on the properties of the $p$-spin models have been recently
published~\cite{CC05,BCKM98,Cu02}; in the following only the main results will be reviewed,
referring to~\cite{CC05,BCKM98,Cu02} and to the original
papers~\cite{KTW87a,KTW87b,GM84,Ri92,KPV93,CS92,CS95,CGP97} for all the details.

\subsection{Mean field $p$-spin models: the replica solution and the dynamics}
\label{sec1:meanfieldgeneral}

The model is defined by the Hamiltonian
\begin{equation}
\label{Hpspin}
H_p(\s) = -\sum_{i_1 < \cdots < i_p} J_{i_1,\cdots,i_p} \sigma_{i_1} \cdots \sigma_{i_p} \ ,
\end{equation}
where $\sigma_i$ are either real variables subject to a spherical constraint 
$\sum_i \sigma^2_i =N$, or Ising variables, $\s_i=\pm 1$,
and $ J_{i_1,\cdots,i_p}$ are independent quenched random Gaussian variables with zero 
mean and variance $p! J^2/(2 N^{p-1})$. The sum is over all the 
{\it ordered} $p$-uples of indices $i_1 < \cdots < i_p$.
It is a mean field model because each degree of freedom interact with all the 
others with a strength that vanishes in the thermodynamic limit, in order to have
an extensive average energy.

The replica trick~\cite{MPV87} allows to solve the model at all temperatures.
A thermodynamic transition is found at $T_K$ corresponding to a $1$-step breaking
of the replica symmetry ({\sc 1rsb}). At the transition, the specific heat jumps downward.
The order parameter of the transition is the self-overlap between two different
replicas $a$ and $b$:
\beq
\label{overlapeq}
q_{ab} \equiv \frac{1}{N} \sum_{i=1}^N \s_i^a \s_i^b \ ,
\eeq
which plays the role, in the context of spin glass theory, of the nonergodicity
factor (\ref{nonergeq}). The average value of $q_{ab}$ jumps from $0$ to a finite
value $q_s$ at $T_K$.

The Langevin dynamics of the model can also be solved exactly~\cite{Cu02}. 
A dynamical transition
is found at a temperature $T_d > T_K$; the relaxation time of the spin-spin
autocorrelation function $C(t)=N^{-1} \sum_i \langle \s_i(t) \s_i(0) \rangle$
shows a power-law divergence for $T \to T_d$. A dynamical order parameter
can be defined as
\beq
\label{overlapdyn}
q_d = \lim_{t \to \io} \lim_{N\to \io} \frac{1}{N} \sum_i \langle \s_i(t) \s_i(0) \rangle
=  \lim_{t \to \io} \lim_{N\to \io} C(t) \ ;
\eeq
it is the analogue of the dynamical nonergodicity factor defined in (\ref{nonergdyn})
and jumps to a finite value at $T_d$. Below $T_d$ the system is no more able to
equilibrate with the thermal bath and enters a {\it nonequilibrium} regime.
This result gives a strong indication that {\it metastable} states, which do not
appear in the equilibrium calculation, are responsible for the slowing down of
the dynamics and for the dynamical transition at $T_d$.

\subsection{The TAP free energy}
\label{sec1:TAP}

To better understand what is going on in the model one has to investigate
the structure of its phase space. In particular, one wishes to characterize
the {\it equilibrium states} in order to understand the nature of the 
thermodynamical transition at $T_K$, as well as the structure of the
{\it metastable states} which seems to trap the system at $T_d$ and to be
responsible for the existence of a dynamical transition.
It turns out that a complete characterization of the structure of the states
is possible by mean of the {\it TAP free energy}.

A general result of statistical mechachics (see \eg \cite{MPV87,Ga99}) states 
that is always possible to decompose the equilibrium 
probability distribution as a sum 
over {\it pure} states\footnote{In a fully-connected system there is
no space notion: thus no boundary conditions can be applied to the system and
the pure states can be selected only using an external field.}:
\beq
P(\s_1,\cdots,\s_N) = \sum_\a w_\a P^\a(\s_1,\cdots,\s_N) \ ,
\eeq
where $\a$ is an index labelling the states and $w_\a$ is the weight of
each state, $\sum_\a w_\a = 1$. The probability distributions of the pure
states are characterized by the {\it clustering} property, that in mean
field reads
\beq
P^\a(\s_1,\cdots,\s_N) = \prod_{i=1}^N P^\a_i(\s_i) \ .
\eeq
The single-spin probability distribution is in turn specified by the average
magnetization of the spin $\s_i$, $m_i^\a=\sum_\s \s P^\a_i(\s)$. 
Thus, a {\it pure state} $\a$ is completely determined by the set of local
magnetizations $m^\a_i$, $i=1,\cdots,N$. Moreover, a variational principle
exists, stating that the local magnetizations of pure states must be minima of
some free energy function $F(m_i)$. This function, in the context
of spin glasses, has the name of Thouless--Anderson--Palmer (TAP) 
free energy~\cite{MPV87,TAP77}.

The weight $w_\a$ of state $\a$ is proportional to $\exp [-\b N f_\a]$, where
$f_\a = F(m_i^\a)/N$. Thus, in the thermodynamic limit only the lowest
free energy states are relevant. Local minima of $F$ having a free energy
density $f > f_{min}$ for $N\to \io$ are {\it metastable} states.
The TAP free energy $F(m_i)$ depend, in general, explicitly
on the temperature, so the whole structure of the states may depend strongly
on temperature.

\begin{figure}[t]
\centering
\includegraphics[width=8cm,angle=-90]{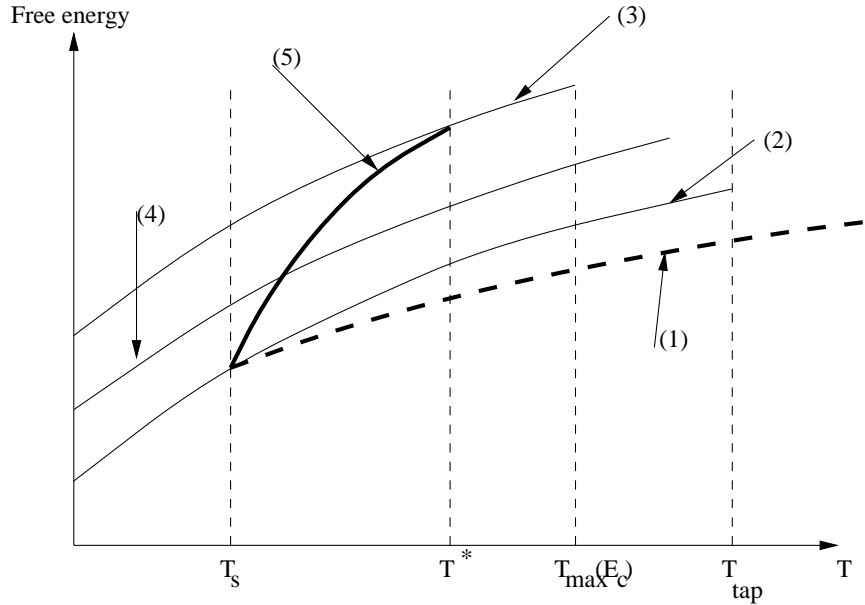}
\caption[Evolution of the TAP states of the spherical $p$-spin model]
{(From~\cite{BCKM98}; $T_s$ corresponds to $T_K$, $T^*$ to $T_d$) 
Sketch of the evolution in temperature of the TAP
states for the spherical $p$-spin model. Each TAP state like (4) can be
followed in temperature until it becomes unstable and disappears. The
complexity vanished continuously at the ground state $f_{min}$ (2) 
and goes abruptly to 0 above the maximum free energy $f_{max}$ (3). 
(5) is the free energy $f^*$ of the states that dominate the partition function. 
(1) is the equilibrium free energy $f^*-T\Si(f^*)$ that takes into account 
the entropic contribution of the degeneracy of the states.
}
\label{fig1:TAP}
\end{figure}

In mean field $p$-spin models, the expression of the TAP free energy can be
explicitly derived~\cite{Ri92,KPV93}, and the distribution of the states
can be computed. A peculiar property of the spherical $p$-spin model, which
simplify a lot the description of the results of the TAP computation, is that
the dependence of the free energy functional on $T$ is very simple.
Indeed, the states are labeled by their energy $E$ at $T=0$. The number of
states of energy $E$ is $\O(E)=\exp N\Si(E)$; the function $\Si(E)$ is called 
{\it complexity}: it is a concave function that vanishes continuously at the 
ground state energy $E_{min}$ and goes discontinuously to $0$ above some value 
$E_{max}$. At finite temperature, the minima get ``dressed'' by thermal 
fluctuations but they maintain their identity and one can follow their
evolution at $T>0$. At some temperature $T_{max}(E)$, thermal fluctuations
are so large that the states with energy $E$ become unstable and disappear,
until, at high enough temperature $T>T_{TAP}$, 
only the paramagnetic minimum, $m_i\equiv 0$, survives. 

At finite temperature, the number of states of given free energy
density $f$ is $\O(f)=\exp N \Si(f)$, where $\Si(f)=\Si(E(f))$ and $E(f)$ 
is the $T=0$ energy of the states of free energy $f$. The function $\Si(f)$
vanishes continuously at $f=f_{min}$ and drops to zero above $f=f_{max}$;
a qualitative plot of $\Si(f)$ is reported in Fig.~\ref{fig1:Scqualit}.
A similar behavior is found in all $p$-spins model like the Ising $p$-spin
glass\footnote{In Ising models as well as in perturbations of the spherical
model the picture is complicated by the presence of full RSB
metastable states~\cite{MR03}.}. The main peculiarity of $p$-spin models is that an 
{\it exponential number} of metastable states is present at low 
enough temperature.

One can write the partition function $Z$, at low enough temperature and for $N \to \io$, 
in the following way:
\beq
\label{Zm1}
\begin{split}
Z = e^{-\b N F(T)} \sim \sum_\a e^{-\b N f_\a}
= \int_{f_{min}}^{f_{max}}df \, e^{N [\Si(f)-\b f]}
\sim e^{N [\Si(f^*)-\b f^*]} \ ,
\end{split}
\eeq
where $f^* \in [f_{min},f_{max}]$ is such that $\Phi(f)=f - T \Si(f)$ 
is minimum, \ie it is the solution of
\beq
\frac{d\Si}{df} = \frac{1}{T} \ ,
\eeq
provided that it belongs to the interval $[f_{min},f_{max}]$.
%Depending on temperature, the first or the second contribution in
%the last line of Eq.~(\ref{Zm1}) will dominate.
Starting from high temperature, one encounters three temperature regions:
\begin{itemize}
\item For $T > T_d$, the free energy density of the paramagnetic state is
smaller than $f - T\Si(f)$ for any $f\in [f_{min},f_{max}]$, so the paramagnetic
state dominates and coincides with the Gibbs state (in this region the decomposition
(\ref{Zm1}) is meaningless).
\item For $T_d\geq T \geq T_K$, a value $f^* \in [f_{min},f_{max}]$ is found, such that
$f^* - T \Si(f^*)$ is equal to $f_{para}$. This means that the paramagnetic state
is obtained from the superposition of an
{\it exponential number} of pure states of {\it higher} individual free energy
density $f^*$. The Gibbs measure is splitted on this exponential number of
contributions: however, no phase transition happens at $T_d$ because of the
equality $f^* - T \Si(f^*)=f_{para}$ which guarantees that the free energy is
analytic on crossing~$T_d$. 
\item For $T < T_K$, the partition function is dominated by the lowest free 
energy states, $f^* = f_{min}$, with $\Si(f_{min})=0$ and 
$F(T)=f_{min} - T \Si(f_{min}) = f_{min}$. At $T_K$ a phase transition occurs,
corresponding to the 1-step replica symmetry breaking transition found in the
replica computation.
\end{itemize}
In the range of temperatures $T_d > T > T_K$, the phase space of the model
is disconnected in an exponentially large number of states, giving a contribution
$\Si(T) \equiv \Si(f^*(T))$ to the total entropy of the system.
This means that the entropy $S(T)$ for $T_d > T > T_K$ can be written as
\beq
S(T) = \Si(T) + S_{vib}(T) \ ,
\eeq
$S_{vib}(T)$ being the individual entropy of a state of free energy $f^*$.
From the latter relation it turns out that the complexity $\Si(T)$ is the $p$-spin
analogue of the configurational entropy $S_c(T)$ of supercooled 
liquids\footnote{In the interpretation of experimental data one should remember that
in experiments $S_{vib}$ can be estimated only by the entropy of the crystal. However,
the vibrational properties of the crystal can be different from the vibrational 
properties of an amorphous glass, see \cite{RS01} for a review. 
Corrections due to this fact must be taken into account: in many cases, 
the difference is reduced to a proportionality factor between 
$S_c$ and $\Si$~\cite{corezzi}.}.

The TAP approach provides also a pictorial explaination of the presence of
a dynamical transition at $T_d$. If the system is equilibrated at high 
temperature in the paramagnetic phase, and suddenly quenched below $T_d$,
the energy density start to decrease toward its equilibrium value.
This relaxation process can be represented as a descent in the free 
energy landscape at fixed temperature starting from high values of $f$.
What happens is that when the sistem reaches the value $f_{max}$ it becomes
trapped in the highest metastable state and is unable to relax to the
equilibrium states of free energy $f^*$, as the free energy barriers between
different states cannot be crossed in mean field~\cite{BCKM98,Cu02}. 
For this reason below $T_d$ the systems is unable to equilibrate.
What happens in real glasses is that activated processes of jump between
different metastable states allow the system to relax toward equilibrium
also below $T_d$. Activated processes give rise to the VFT behavior of
the relaxation time, as will be discussed in the following.

%%%%%%%%%%%%%%%%%%%%%%%%%%%%%%%%%%%%%%%%%%%%%%%%%%%%%%%%%%%%%%%%%%%%%%
%%%%%%%%%%%%%%%%%%%%%%%%%%%%%%%%%%%%%%%%%%%%%%%%%%%%%%%%%%%%%%%%%%%%%%

\section{Two methods to compute the complexity}

If a given system presents a structure of the free energy landscape similar
to $p$-spin glasses, two general methods to compute the complexity as a
function of the free energy of the states without directly solving the TAP
equations exist; they have been developed in 
\cite{MP99,FPV92,FP95,Mo95,FP97,BFP97,Me99,MP00}.
Both methods consider a number of copies of the system coupled
by a small field conjugated to the order parameter~(\ref{nonergeq}).

\subsection{Real replica method}
\label{sec1:realreplica}

\begin{figure}[t]
\centering
\includegraphics[width=7cm,angle=0]{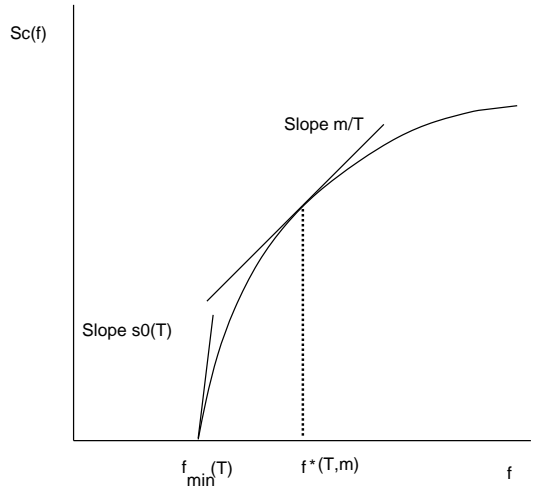}
\caption[Qualitative behavior of the complexity]
{(From~\cite{MP99}) A sketch of the complexity as a function of the
free energy density for system belonging to the $p$-spin class. The value
$f^*(m,T)$, solution of $\frac{d\Si}{df}=\frac{m}{T}$, is also reported.
}
\label{fig1:Scqualit}
\end{figure}

The idea of \cite{MP99,Mo95} is to consider $m$ copies of the original system, 
coupled by a small attractive term added to the Hamiltonian.
The coupling is then switched off after the thermodynamic limit has been taken.
For $T<T_d$, the small attractive coupling is enough to constrain
the $m$ copies to be in the same TAP state.
At low temperatures, the partition function of the replicated system is 
then
\beq
\label{Zm}
Z_m = e^{-\b N \Phi(m,T)} \sim \sum_\a e^{-\b N m f_\a}
= \int_{f_{min}}^{f_{max}}df \, 
e^{N [\Si(f)-\b m f]}
\sim  e^{N [\Si(f^*)-\b m f^*]} \ ,
\eeq
where now $f^*(m,T)$ is such that $\Phi(m,f)=m f - T \Si(f)$ is minimum and
satisfies the equation
\beq
\frac{d\Si}{df} = \frac{m}{T} \ .
\eeq
If $m$ is allowed to assume real values by an analytical continuation, 
the complexity can be computed from the knowledge
of the function $\Phi(m,T)=m f^*(m,T) - T \Si(f^*(m,T))$. 
Indeed, it is easy to show that
\beq
\label{mcomplexity}
\begin{split}
&f^*(m,T) = \frac{\partial \, \Phi(m,T)}{\partial m} \ , \\
&\Si(m,T) = \Si(f^*(m,T)) = m^2 \frac{\partial \,[ m^{-1} \b \Phi(m,T)]}{\partial m} = 
m \b f^*(m,T) - \b \Phi(m,T) \ .
\end{split}
\eeq
The function $\Si(f)$ can be reconstructed from the parametric plot of $f^*(m,T)$ 
and $\Si(m,T)$ by varying $m$ at fixed temperature.

The glass transition happens when $\b$ equals the slope $s_0(T)$ of $\Si(f)$ 
in $f=f_{min}$, so $T_K$ is defined by $T_K s_0(T_K)=1$. 
If $m < 1$, the value of $f^*(m,T)$ correspond to a {\it smaller}
slope with respect to $m=1$, so the glass transition is shifted towards
lower values of the temperature, see Fig.~\ref{fig1:Scqualit}. 
For any value of the temperature $T$ below $T_K$
it exists a value $m^*(T) < 1$ such that for $m < m^*$ the system is in the liquid phase.
The free energy for $T<T_K$ and $m < m^*(T)$ can be computed by analytic continuation of 
the free energy of the high temperature liquid. As the free energy is always continuous
and it is {\it independent} of $m$ in the glass phase (being simply the value $f_{min}(T)$
such that $\Si(f_{min})=0$), one can compute the free energy of the glass below $T_K$
simply as $F_{glass}(T)=f_{min}(T)=\Phi(m^*(T),T)/m^*(T)$.

This method allows to compute the complexity as well as the free energy of the glass,
\ie of the lowest free energy states, at any temperature, if one is able to compute
the free energy of $m$ copies of the original system constrained to be in the same
free energy state and to perform the analytical continuation to real $m$.
In~\cite{Me99} it was applied to the spherical $p$-spin 
system and it was shown that the method reproduces the results obtained from the
explicit TAP computation.

%%%%%%%%%%%%%%%%%%%%%%%%%%%%%%%%%%%%%%%%%%%%%%%%%%%%%%%%%%%%%%%%%%%%%%
%%%%%%%%%%%%%%%%%%%%%%%%%%%%%%%%%%%%%%%%%%%%%%%%%%%%%%%%%%%%%%%%%%%%%%

\subsection{Potential method}
\label{sec1:potentialmethod}

The second method~\cite{FPV92,FP95,FP97,BFP97} starts from a reference configuration
$\s$ of the original system and consider the partition function of an identical system 
$\t$ which Hamiltonian has been corrected by the addition of a coupling to the
configuration $\s$:
\begin{equation}
Z(\sigma,\ee,T)=\int d\tau \, e^{-\beta H(\tau)+\beta N \ee q(\s,\t)} \ ,
\end{equation}
where $q(\s,\t) = N^{-1} \sum_i \s_i \t_i$ as in (\ref{overlapeq}).
If the reference configuration $\s$ is extracted from the equilibrium distribution
at temperature $T$, the free energy $F(\s,\ee,T)=-TN^{-1} \log Z(\sigma,\ee,T)$ 
should not depend on the particular choice of $\s$ for $N\to\io$. 
Thus one averages over the equilibrium distribution of $\s$
at temperature $T$ and defines
\beq
F(\ee,T)=-\frac{T}{N} 
\int d\sigma \frac{e^{-\beta H(\sigma)}}{Z(\beta)} \log Z(\sigma,\ee,T) \ .
\eeq
If, in the limit $\ee\to 0$, the correlation between $\s$ and $\t$ is lost, 
one has $F(\ee=0,T)=F(T)$. Otherwise, one can study the effect of the correlation
in the limit of vanishing coupling between the replicas.

Being interested in the behavior at $\ee=0$, one considers the Legendre transform
of $F(\ee,T)$,
\beq
V(q,T) = \max_\ee \big[ F(\ee,T) + \ee q \big] \ ,
\hskip20pt
q(\ee) = -\frac{\dpr F(\ee,T)}{\dpr \ee} = \langle q(\s,\t) \rangle_\ee \ .
\eeq
The thermodynamic potential $V(q,T)$ is the free-energy of the system $\t$ constrained
to be at a fixed overlap~$q$ with $\s$:
\begin{equation}
\label{Vqdef}
\begin{split}
&V(q,T)=-\frac{T}{N} 
\int d\sigma \frac{e^{-\beta H(\sigma)}}{Z(\beta)} \log Z(\sigma,q,T) \ , \\
&Z(\sigma,q,T)=\int d\tau \, e^{-\beta H(\tau)} \delta(q-q(\sigma,\tau)) \ .
\end{split}
\end{equation}
As $\frac{d V}{dq} = \ee(q)$,
the average value of the order parameter in the limit $\ee\to 0$ is the value of
$q$ that solves $\frac{d V}{d q} = 0$; the minima of $V(q)$ correspond to the
possible phases in the limit of zero coupling.

\begin{figure}[t]
\centering
%\vspace{.05cm}
\includegraphics[width=.6\textwidth,angle=0]{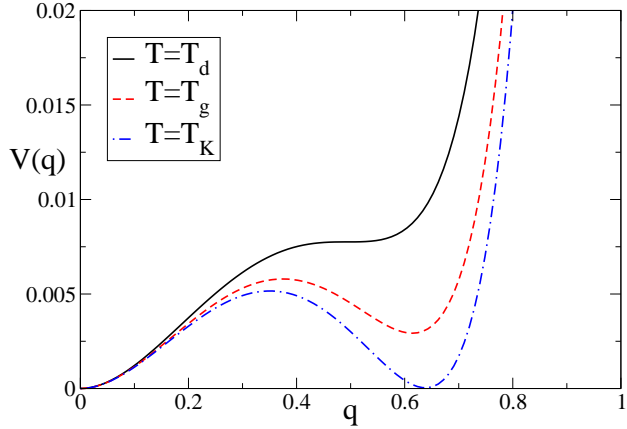}
\caption[The two-replica potential in the spherical $p$-spin model]
{The two-replica potential, $V(q,T)-F(T)$, 
for $T$$\in$$[T_K,T_d]$ in the spherical $p$-spin model.}
\label{fig_1}
\end{figure}

The qualitative behavior of $V(q,T)-F(T)$ is shown in Fig.~\ref{fig_1} for the 
spherical mean field $p$-spin model:
for $T>T_d$ it is a convex function of $q$ with only one minimum at $q=0$.
At the dynamical transition temperature $T_d$ a secondary minimum starts to develop at finite $q$.
On lowering the temperature below $T_d$, the value of $V$ at the minimum
decreases and vanishes at the thermodynamical transition temperature $T_K$.
Indeed, for $T>T_d$ there is only one phase in which the two copies $\s$ and $\t$ are
uncorrelated and the average overlap vanishes. Below $T_d$, a new phase in which the
two copies are in the same TAP state appears; this phase is metastable because
there is an exponential number of TAP states so the probability of finding the two
copies in the same state is exponentially small in absence of coupling. 
The value of $q$, $q_{min}(T)$, corresponding to this secondary minimum is the 
self-overlap of the equilibrium TAP states at temperature $T$.
For $T<T_K$, the value of $V(q_{min})$ becomes equal to $V(0)$, as the number of
states is no more exponential and a vanishing coupling is enough to constrain the two
copies to be in the same state. This correspond to the {\sc 1rsb} phase transition.
This approach underlines the first order nature of the transition from the point
of view of the order parameter.

The value of $V(q,T)$ at the secondary minimum for $T$$\in$$[T_K,T_d]$, \ie
the average free energy of the configuration $\t$ at $q=q_{min}=q(T)$,
is the free energy $f^*(T)$ of the equilibrium TAP states. 
From Eq.~(\ref{Vqdef}), recalling that $F(T)=f^*(T)-T\Si(f^*(T))$, one has
\beq
V(q_{min}(T),T)-V(0,T)=f^*(T)-F(T)=T\Si(f^*(T)) \ ,
\eeq
where $\Si(f^*(T))$ is the equilibrium complexity. The vanishing of 
$V(q_{min}(T_K),T_K)-F(T_K)$ corresponds to the vanishing of the complexity at $T_K$.

From the potential $V(q,T)$ one can extract the values of the dynamical and thermodynamical
transitions as well as the free energy of the equilibrium states $f^*(T)$ and their 
complexity $\Si(f^*(T))$.
To obtain informations about the {\it metastable} TAP states one needs to consider
a reference configuration equilibrated at a different temperature $T'$:
\begin{equation}
\label{Vqdef2T}
\begin{split}
&V(q,T,T')=-\frac{T}{N} 
\int d\sigma \frac{e^{-\beta' H(\sigma)}}{Z(\beta')} \log Z(\sigma,q,T) \ , \\
&Z(\sigma,q,T)=\int d\tau \, e^{-\beta H(\tau)} \delta(q-q(\sigma,\tau)) \ .
\end{split}
\end{equation}
If $T'$$\in$$[T_K,T_d]$ and if the evolution of the TAP states in temperature is described
by the curves in Fig.~\ref{fig1:TAP}, the configuration $\s$ is in one of the
{\it equilibrium} TAP states at temperature $T'$, while the configuration $\t$ is
constrained to be close to it (\ie in the same TAP state) but at temperature $T$.
The free energy of $\t$ for $q=q_{min}(T,T')$ is the free energy of an equilibrium TAP state 
at temperature $T'$ when followed at temperature $T$. The TAP states are labeled 
by their zero-temperature energy $E$; their free energy is $f_{TAP}(E,T)$.
Thus one has\footnote{This equation is slightly different from the one reported
in~\cite{FP95} because the equilibrium free energy $F(T)$ has not been subtracted
in the definition of $V(q,T,T')$.}
\beq
\label{FTAPVq}
V(q_{min}(T,T'),T,T') = f_{TAP}(E(T'),T) \ ,
\eeq
where $E(T')$ is the $T=0$ energy of the equilibrium TAP states at temperature $T'$
and $f^*(T')=f_{TAP}(E(T'),T')$.

The procedure to compute the properties of all the TAP states using the potential
method is the following:
\begin{itemize}
\item First one consider the potential for $T=T'$ and computes the free energy
$f^*(T)$ and the complexity $\Si(f^*(T))$ for $T$$\in$$[T_K,T_d]$. This give access
to the complexity $\Si(f)$.
\item Then one fixes $T'$$\in$$[T_K,T_d]$ and computes, using Eq.~(\ref{FTAPVq}) 
the free energy $f_{TAP}(E(T'),T)$ as a function of $T$ down to $T=0$.
\item In particular, the free energy of the glass (\ie of the lowest TAP states)
is obtained considering the limit $T'\to T_K$ (from above) in 
Eq.~(\ref{FTAPVq}).
\end{itemize}
It was shown in \cite{FP95,FP97} that the result is consistent with the direct
computation using the TAP equations.

\subsection{Connection with the standard replica method}
\label{sec1:compreplica}

It is interesting to consider the relation between the two methods described
above and the {\sc 1rsb} free energy, also because some of the formulae will
be useful in the applications of the next chapters.

In the spherical $p$-spin model, the average over the distribution of the 
couplings $J$ (indicated by an overbar)
of the $n$ times replicated partition function can be rewritten as~\cite{CS92}
\begin{equation}\label{reppart}
\begin{split}
&\overline{Z^n(J)}=\overline{\left( \int d\sigma e^{-\beta H(\sigma)} \right)^n} =
\overline{\int d\sigma_a e^{-\beta \sum_a H(\sigma_a)}} =
\int d Q_{ab} \ e^{N f(Q)} \ , \\
&f(Q)=\frac{\beta^2}{4} \sum_{ab} Q_{ab}^p + \frac{1}{2} \log \det Q \ ,
\end{split}
\end{equation}
where $a,b=1,\cdots,n$ and $Q_{ab}$ in the $n\times n$ overlap matrix~\cite{MPV87}.
The substitution of the {\sc 1rsb} {\it ansatz} for $Q$ (in the example, $n=6$ and
$m=3$):
\begin{equation}
\nonumber
\QQ=\left(
\begin{array}{cc}
\left(
\begin{array}{ccc}
1 & q & q \\
q & 1 & q \\
q & q & 1 \\
\end{array}
\right) & 0 \\
0 & 
\left(
\begin{array}{ccc}
1 & q & q \\
q & 1 & q \\
q & q & 1 \\
\end{array}
\right) \\
\end{array}
\right)
\end{equation}
in Eq.~(\ref{reppart}) gives, for $N\to\io$,
\beq
\begin{split}
&\overline{Z^n} \sim \exp \big[ -\b n N \phi_{1RSB}(m^*,q^*,T) \big] \ , \\
&F(T) = -\frac{T}{N} \lim_{n\to 0} \partial_n \overline{Z^n} =
\phi_{1RSB}(m^*,q^*,T) \ ,
\end{split}
\eeq
where the {\sc 1rsb} free energy is
\beq
\phi_{1RSB}(m,q,T) = -\frac{1}{2\b}\left\{ \frac{\b^2}{2}\big[1+(m-1)q^p\big]
+\frac{m-1}{m} \log(1-q) +\frac{1}{m}\log\big[1+(m-1)q\big] \right\} \ ,
\eeq
and $m^*$, $q^*$ are solutions of $\partial_m \phi_{1RSB}=0$ and $\partial_q \phi_{1RSB}=0$.
For $T>T_K$ the solution $q^*=0$, $m^*=1$ is the stable one, even if a solution with
$q^* \neq 0$ appears for $T<T_d$. Below $T_K$ this solution, with $q^* \neq 0$ and $m^* < 1$, 
is the free energy of the glass.

\subsubsection{Real replica method}

In the real replica method the partition function of $m$ copies of the system
is considered. Using the replica trick to compute the free energy,
\beq
\Phi(m,T)=-\frac{T}{N}\overline{\log Z_m} = -\frac{T}{N}
\lim_{n\to 0} \partial_n \overline{ ( Z_m )^n } = -\frac{T}{N}
\lim_{n\to 0} \partial_n \overline{ Z^{mn} }  \ ,
\eeq
one obtains the partition function of $nm$ copies of the system, with the constraint
that each block of $m$ replicas has to be in the same state. This leads naturally
to the {\sc 1rsb} structure for the overlap matrix (with $m$ fixed) and
\beq
\Phi(m,T) =  -\frac{T}{N}
\lim_{n\to 0} \partial_n\exp \big[ -\b n m N \phi_{1RSB}(m,q^*,T) \big] = 
m \phi_{1RSB}(m,q^*,T) \ .
\eeq
Note that the hypothesis that the $m$ replicas are in the same state implies that for
any value of $(m,T)$ the correct solution is the one with $q^* \neq 0$. Above $T_d$ this
solution disappears as a vanishing coupling cannot constrain the replicas to stay close
to each other.

The free energy of the real replica method is the {\sc 1rsb} free energy
as a function of $m$ at the value $q^* \neq 0$ that solves $\dpr_q \phi_{1RSB}=0$. 
Using Eq.~(\ref{mcomplexity}) the complexity as a function of $m$ is
\beq
T \Si(m,T) = m^2 \partial_m \big[ m^{-1} \Phi(m,T) \big] = 
m^2 \partial_m \phi_{1RSB}(m,q^*,T) \ ,
\eeq
and the equilibrium complexity is
\beq
\Si(T)=\Si(1,T)=-\frac{1}{2} \left[ \frac{\b^2}{2} ( q^* )^p + \log (1-q^*) + q^* \right] \ .
\eeq
As $\Si(m,T) \propto \dpr_m \phi_{1RSB}$, the value $m^*$ that optimizes the 
{\sc 1rsb} free energy below $T_K$ coincides
with the value $m^*$ defined by $\Si(m^*,T)=0$ of the real replica method.

\subsubsection{Potential method}

Using the replica trick~\cite{FP95} the following expression for $V(q,T,T')$ is 
derived\footnote{In \cite{FP95}, Eq.~(15), the factor $n^{-1}$ is missing
probably due to a misprint.}:
\begin{equation}\label{Vqrepliche}
V(q,T,T') = - \lim_{n\rightarrow0} \lim_{m\rightarrow 1} \frac{T}{Nn}
\frac{\partial}{\partial m} 
\overline{ \left( \int d\sigma \, e^{-\beta' H(\sigma)} Z(\sigma,q,T)^{m-1} \right)^n } \ .
\end{equation}
The last integral can be rewritten as
\begin{equation}
\left( \int d\sigma \, e^{-\beta' H(\sigma)} Z(\sigma,q,T)^{m-1} \right)^n =
\int d\sigma_{a \alpha} e^{- \sum_{a \alpha} \b_\a H(\sigma_{a\alpha})}
\prod_{a=1}^n \prod_{\alpha=2}^m \delta(q-q(\sigma_{a1},\sigma_{a\alpha})) \ ,
\end{equation}
where $a=1,\cdots,n$, $\alpha=1,\cdots,m$, 
and $\b_\a=\b' \d_{1\a} + \b (1-\d_{1\a})$.
This is again the expression
of the $nm$ times replicated equilibrium partition function, with the additional
constraint given by the $\delta$-functions.
The average over the disorder gives
\begin{equation}\label{fMF}
\begin{split}
&\int d Q_{a\alpha,b\beta} \ e^{N f(Q)} \ \prod_{a=1}^n \prod_{\alpha=2}^m \delta(q-Q_{a1,a\alpha}) \ , \\
&f(Q)=\frac{1}{4} \sum_{a\alpha,b \beta} \b_\a \b_\b Q_{a\alpha, b \beta}^p + \frac{1}{2} \log \det Q \ .
\end{split}
\end{equation}
Evaluating the integral at the saddle point, one has
\begin{equation}
\label{Vq1}
V(q,T,T')=- \lim_{n\rightarrow0} \lim_{m\rightarrow 1} \frac{T}{n}
\frac{\partial}{\partial m} f(\overline{Q}) \ .
\end{equation}
The matrix $\overline{Q}$ is defined by the following conditions: \\
{\it i)} the elements on the diagonal are equal to 1; \\
{\it ii)} the elements $\overline{Q}_{a1a\alpha}$, $\alpha>1$, are equal to $q$; \\
{\it iii)} all the other elements are determined by the maximization of $f(Q)$. \\
As usual, one needs a parametrization of the matrix $Q$ in order to perform the analytic
continuation to non-integer $n$ and $m$. A possible {\it ansatz} is \cite{FP95} (in the example,
$n=2$, $m=4$):
\begin{equation}
\label{1RSBpotenziale}
\overline{Q}=\left(
\begin{array}{cc}
\left(
\begin{array}{cccc}
1 & q & q & q \\
q & 1 & r & r \\
q & r & 1 & r \\
q & r & r & 1 \\
\end{array}
\right) & 0 \\
0 & \left(
\begin{array}{cccc}
1 & q & q & q \\
q & 1 & r & r \\
q & r & 1 & r \\
q & r & r & 1 \\
\end{array}
\right) \\
\end{array}
\right) \ .
\end{equation}
and corresponds to the following structure: each replica of $\s$ 
is independent from the others, and for each $\s$ there are $m-1$
copies of $\t$ which have overlap $q$ with $\s$ and overlap
$r$ within themselves.
Within this {\it ansatz}, and using the relation
\begin{equation}
\det  \left(
\begin{array}{cccc}
1 & q & q & q \\
q & 1 & r & r \\
q & r & 1 & r \\
q & r & r & 1 \\
\end{array}
\right) = (1-r)^{m-2} [ 1-2r+rm-(m-1)q^2 ] \ ,
\end{equation}
one gets
\begin{equation}
V(q,T,T')=-\frac{1}{2\b} \left[ \beta \beta' q^p - \frac{\beta^2 r^p}{2} + \frac{\b^2}{2}
 + \log(1-r) + \frac{r-q^2}{1-r} \right] \ ,
\end{equation}
where $r(q)$ is determined by $\partial_r V=0$.

For $T=T' \geq T_K$ it is easy to check that the condition $\frac{dV}{dq}=\partial_q V=0$, together 
with $\partial_r V=0$, is satisfied if $q=r$. Thus, when $V(q)$ is stationary, $r(q)=q$ and
the potential $V(q)$ reduces to
\beq
V(q,T)=-\frac{\b}{4}-\frac{\beta q^p}{4}
 - \frac{T}{2} \big[ \log(1-q) + q \big] = F(T) + \lim_{m\to1} \dpr_m \phi_{1RSB}(m,q,T) \ .
\eeq
The latter relation can be proven in general and follows from the observation that
when $\frac{dV}{dq}=0$ the matrix $\ol{Q}$ reduces to the usual {\sc 1rsb} overlap matrix
${\cal Q}$. This is because the condition $\frac{dV}{dq}=\dpr_q V=0$ together with $\dpr_r V=0$
is equivalent, from Eq.~(\ref{Vq1}), to
\begin{equation}
\frac{df(Q)}{dQ}=0 \ .
\end{equation}
This means that the function $f(Q)$ must be stationary with respect to all the elements of $Q$,
and the {\sc 1rsb} matrix ${\cal Q}$ provides a solution to this condition.
If $\ol{Q}={\cal Q}$, one has
\begin{equation}
\frac{f(\ol{Q})}{nm} = -\beta \phi_{1RSB}(m,q,T) \ .
\end{equation}
Substituting this expression in Eq.~(\ref{Vq1}), one obtains
\begin{equation}
V(q,T)=\lim_{m\rightarrow1} \partial_m \Big( m \phi_{1RSB}(m,q,T) \Big)
= F(T) + \lim_{m\rightarrow1} \partial_m \phi_{1RSB}(m,q,T) \ ;
\end{equation}
using the relation $\phi_{1RSB}(m=1)=F(T)$ that holds above $T_K$.
Therefore, {\it on its stationary points}, $V(q,T)$ is given (at the {\sc 1rsb} level)
by this simple expression, that can be easily calculated in several models.
Note that, as discussed in \cite{BFP97}, full RSB effects can be important for the computation
of $V(q,T)$ even in {\sc 1rsb} models such as the $p$-spin spherical model.

If $T'=T_K$ and $T < T_K$, the value of $V$ in the secondary minimum can still
be computed using the simple {\it ansatz} (\ref{1RSBpotenziale}). It can be seen,
using the relation
\beq
\frac{T}{T_K} = m^* \left[ \frac{1+(m^*-1)q^*}{m^*} \right]^{p/2} \ ,
\eeq
that follows from the equations that define $m^*$ and $q^*$, that the solutions to
$\dpr_r V=0$, $\dpr_q V=0$, is
\beq
\begin{cases} &r = q^* \ , \\
&q=q^* \sqrt{\frac{m^*}{1+(m^*-1)q^*}} \ .\\
\end{cases}
\eeq
Indeed $r$ is the self--overlap of the replicas inside the equilibrium states at
$T_K$, so it is equal to the self--overlap $q^*$ of the glass. Substituting these espressions
in $V(q,T,T')$ one obtains
\beq
V(q_{min},T,T_K) = \phi_{1RSB}(m^*,q^*,T) = F_{glass}(T) \ ,
\eeq
as expected from Eq.~(\ref{FTAPVq}).

\subsubsection{Discussion}

The explicit relation between the free energies $\Phi(m,T)$ and $V(q,T,T')$ and the 
{\sc 1rsb} free energy $\phi_{1RSB}(m,q,T)$ derived for the spherical
$p$-spin model confirms that:
\begin{itemize}
\item
the real replica potential $\Phi(m,T)$ is related to the {\sc 1rsb} free energy as 
a function of $m$ for $q=q^*(m,T) \neq 0$. For this reason it allows to study the
properties in the glass phase at $m=m^*<1$. Remarkably, it also allows to compute
the free energy of the metastable states and their complexity for $T$$\in$$[T_K,T_d]$.
\item
the potential $V(q,T,T')$ for $T=T'$ is related to the (derivative of the) {\sc 1rsb} 
free energy at $m=1$, as a function of $q$. 
Thus it is not suitable to study the region below $T_K$ where
$m \neq 1$, but it allows to study in detail the properties of the intermediate
phase $T$$\in$$[T_K,T_d]$, in particular the metastability of
the $q\neq 0$ phase for $T$$\in$$[T_K,T_d]$ and to estimate the barrier between the
metastable and stable regions \cite{PRZ04,Fr05}.
\item
to compute the free energy of the metastable states and, as a particular case,
the free energy of the glass, one needs to consider an extended definition of the
potential, $V(q,T,T')$, see Eq.~(\ref{Vqdef2T}).
The relation between this potential and $\phi_{1RSB}$ is more involved, but at
least for $T'=T_K$ one has $V(q_{min},T,T_K)=\phi_{1RSB}(m^*,q^*,T)=F_{glass}(T)$.
\end{itemize}

%%%%%%%%%%%%%%%%%%%%%%%%%%%%%%%%%%%%%%%%%%%%%%%%%%%%%%%%%%%%%%%%%%%%%%
%%%%%%%%%%%%%%%%%%%%%%%%%%%%%%%%%%%%%%%%%%%%%%%%%%%%%%%%%%%%%%%%%%%%%%

\section{Beyond mean field}
\label{sec1:beyondmeanfield}

The {\it random first order} scenario that emerges from the analytical solution of
$p$-spin disordered models explains most of the phenomenology of the glass
transition. However, some big issues remain unexplained.
The main problem of the mean field approach --as usual-- is the existence of 
metastable states
with {\it intensive} free energy higher than the free energy of the ground states,
$f > f_{min}$. These states are responsible for the existence of a finite
complexity. Their lifetime is infinite, so they are able to trap the system 
below $T_d$. This is the reason why the dynamical transition, \ie the divergence
of the structural relaxation time, happens at a temperature $T_d > T_K$.

In a model with short range interactions, metastable states have a finite lifetime
due to the nucleation of bubbles of the stable phase inside the metastable one,
so they are not thermodynamically stable.
One should expect the existence of well defined states with $f>f_{min}$
to be impossible; but the analogy between mean field models and real glasses
is mainly based on the analogy between the complexity $\Si(T)$ and the
configurational entropy $S_c(T)$. How can one explain the existence of a finite
configurational entropy, related to well defined metastable states, in a short
range system?

Moreover, the observed crossover of the relaxation time from a power--law behavior
to a Vogel--Fulcher--Tamman law (\ref{VFT}) as well as the Adam--Gibbs relation
(\ref{AGrel}) are not explained by the mean field theory, which predicts a strict
power--law divergence of $\t_\a$ for $T \to T_d^+$. 
The observation of a finite relaxation time below $T_d$ is again related to the finite
lifetime of metastable states. The system, instead of being trapped forever into a
state, is able to escape, due to nucleation processes; it is then trapped by
another state, and so on. These processes of jump between metastable states are
{\it activated} processes: the system has to cross some free energy barrier in order
to jump from one state to another. The relaxation time is then expected to scale
as
\beq
\t_\a(T) \sim \t_\io \exp \big[ \b \D F(T) \big] \ ,
\eeq
$\D F(T)$ being the typical free energy barrier that the system has to cross at
temperature $T$.
The VFT law and the observation that $T_0 \sim T_K$ suggest that the barrier should
diverge at $T_K$, $\D F(T) \sim (T-T_K)^{-1}$; more generally, the Adam--Gibbs formula relates
this divergence to the vanishing of the configurational entropy, $\D F(T) \sim S_c(T)^{-1}$.
It is then essential to understand what is really the meaning of $S_c(T)$ in finite dimension
and why it is related to the free energy barrier for nucleation.

A crucial observation is that the divergence of the relaxation time at $T_K$, in short range
systems, is possible {\it only} if the cooperative processes of structural rearrangement involve
atoms that are correlated on a typical length scale $\xi$, which diverges at $T_K$.
If no divergent length scale is present in the system, it is always possible to divide it
in {\it finite} subsystems, each one relaxing {\it independently} of the others: and the relaxation
of a finite system usually happens in finite time, if the interactions are finite and have short range.

A simple idea that follows from the above observation and can explain how the mean field picture 
is modified in short range systems is the following 
\cite{KTW87a,Wo97,XW01,LW03,BB04,FPV94,Pa94,FT04,Fr05,DSW05}.
It exists a typical lenght scale $\xi(T)$ over which structural relaxation processes
happens. If one looks at smaller length scales, the system behaves as if it were mean field:
metastable states are stable for $l < \xi$, yielding a finite {\it local} complexity.
However, for large scales $l > \xi$, metastability is destroyed and only the lowest free energy
states are stable. For $T \to T_K^+$, $\xi \to \io$, so below $T_K$ a stable {\it ideal} glass
phase is possible. This idea leads naturally to the identification of the configurational 
entropy $S_c(T)$ with the {\it local} complexity $\Si(T)$, and to a derivation of an
Adam-Gibbs--like relation between the relaxation time $\t_\a$ and $\Si(T)$.

\subsection{Dynamical heterogeneities: a derivation of the Adam-Gibbs relation}
\label{sec1:BBargument}

The above considerations can be formalized as follows~\cite{BB04}. 
An homogeneous {\it equilibrium state} in a finite dimensional system is defined 
as the probability distribution that is reached in each finite volume inside the container
when the thermodynamic limit is taken with a given sequence of boundary conditions~\cite{Ga99}:
\eg for a ferromagnet at low temperatures the two states $+$ and $-$ can be obtained
taking the thermodynamic limit with the spins on the boundary fixed to $+$ or $-$,
respectively.

For glassy systems this simple procedure does not work because the order parameter
(\ref{nonergdyn}) is the self--overlap of the configurations of the same system for
$t\to \io$ or, equivalently, the overlap (\ref{nonergeq}) between two coupled copies
of the system, and it is not clear how to fix it using boundary conditions.

To overcome this problem, {\it assume} that an equilibrium state $\a$ of free energy
density $f_\a$ exists. Assume also that a whole distribution of states of complexity $\Si(f)$
(per unit volume) exists for $f\in [f_{min},f_{max}]$. 
Then, consider a configuration belonging to the
state $\a$ and a bubble of radius $R$ inside the system; all the particles outside 
the bubble are frozen in their position and act as boundary terms, 
and one consider the partition function of the bubble in presence of these
boundary conditions. The idea is to find a self--consistency condition for the radius
of the bubble $R$ requesting that the particles inside the bubble remain in the
state $\a$ due to the boundary conditions.

The partition function of the bubble is\footnote{\label{nota1}To simplify the equations,
in the following $O(1)$ constants related to the shape
of the bubble will be neglected, \eg one should write $v_d R^d$ instead of $R^d$, $v_d$ being the
volume of a sphere of radius $R=1$, and $s_\th R^\th$ instead of $R^\th$, $s_\th$
taking into account the shape of the interface. These constants do not change the
qualitative results of this section, and will eventually be included later.}
\beq
Z_R \sim e^{-\b f_\a R^d} + \sum_{\g\neq \a} e^{-\b f_\g R^d - \b \Y R^\th}  =
e^{-\b f_\a R^d} +
\int_{f_{min}}^{f_{max}} df \, e^{\Si(f) R^d-\b f R^d- \b \Y R^\th} \ .
\eeq 
The first term represents the bubble in the same state $\a$ of the 
particles outside the boundary,
while the second term represents the situation where the bubble is in a different state.
In this case, the term $\b \Y R^\th$ represents the free energy cost of the
interface between the states $\a$ and $\g$ at the boundary of the bubble, which
should scale as $R^\th$ with $\th \leq d-1$ if the interactions have short range.
If the state $\a$ is chosen to be
an equilibrium state, of energy $f_\a=f^*$ such that $\frac{d\Si}{df}(f^*)=\b$, the partition
function becomes
\beq
Z_R \sim e^{-\b f^* R^d} + e^{\Si(f^*) R^d-\b f^* R^d- \b \Y R^\th} \ ,
\eeq
where $\Si(f^*(T))=\Si(T)$ is the equilibrium complexity as usual. It is clear that
if $\Si(f^*) R^d- \b \Y R^\th > 0$, the second term dominates and the bubble
is in a different state, otherwise the first term dominates and the boundary
conditions are able to keep the particles inside the bubble in the state $\a$.
If the bubble is in the state $\a$ it gains the term due to the interface,
$\b \Y R^\th$; however the probability of changing state is very large due
to the exponential degeneracy of the states, as expressed by the term $\Si(T)R^d$.
In this sense, one can think to $\D F_v(T) = -T\Si(T)$ as the bulk free energy gain that drives
the escape from the state $\a$: it is not a free energy difference between the stable
and metastable phase, as in ordinary nucleation problems, rather it is the contribution
coming from the large number of possibilities that one has to choose a different state 
{\it with the same free energy density}.

As $\th \leq d-1$ in short range systems, it is clear that for $R\to\io$ the second
term is always dominant and the bubble always escapes from the state $\a$. This implies
that the initial assumption on the existence of the state $\a$ is not consistent as long
as $\Si(f)>0$. This is a formalization of the statement that {\it an exponential number of
states cannot exists in short range systems}: in other words, 
there are no boundary conditions trough which one can select an exponential number 
of different states.

However, if $R$ is small enough, one has $\Si(f^*) R^d- \b \Y R^\th < 0$ and the
bubble remains in the state $\a$. This happens for 
\beq
R < \xi(T) = \left(\frac{\Y(T)}{T\Si(T)}\right)^{\frac{1}{d-\th}} \ .
\eeq
The conclusion is that {\it it exists a temperature dependent length scale, $\xi$, such
that for $R < \xi$ there is an exponential number of stable states}. These states are
destroyed by relaxation processes that change the state inside the bubble if $R > \xi$.

The argument can be rephrased as follows: the free energy cost for creating a bubble of
radius $R$ of a state $\g \neq \a$ inside the state $\a$ is
$\D F \sim -T\Si(T) R^d + \Y(T) R^\th$. If one is able to create, by a fluctuation,
a bubble of radius $R>\xi(T)$, then the bubble will never go back into the state $\a$
and a (local) activated process of escaping from a (local) state has taken place.
To do that one has to cross the barrier given by the maximum of $\D F(R)$ in the interval
$[0,\xi]$. This maximum is at
\beq
R^*(T) = \left( \frac{\th \Y(T)}{d T\Si(T)}\right)^{\frac{1}{d-\th}} = 
\left(\frac{\th}{d}\right)^{\frac{1}{d-\th}} \xi(T) \ ,
\eeq
and the value of the free energy barrier is
\beq
\D F^* = \D F(R^*) =  A(d,\th)\frac{\Y(T)^{\frac{d}{d-\th}}}{[T\Si(T)]^{\frac{\th}{d-\th}}} \ ,
\hskip20pt
A(d,\th)=\left(\frac{\th}{d}\right)^{\frac{\th}{d-\th}} -
\left(\frac{\th}{d}\right)^{\frac{d}{d-\th}} \ .
\eeq
Then the relaxation time should scale as
\beq
\t_\a \sim \exp [\b \D F^*] \sim 
\exp\left\{ \b \frac{\Y(T)^{\frac{d}{d-\th}}}{[T\Si(T)]^{\frac{\th}{d-\th}}}\right\} \ ,
\eeq
which in $d=3$ for $\th = d-1 =2$ gives
\beq\label{AGBB}
\t_\a \sim \exp \left\{ \b \frac{\Y(T)^3}{[T\Si(T)]^2}\right\} \ .
\eeq
The latter relation is very similar to the Adam--Gibbs relation (\ref{AGrel}) even if it
differs from it in the exponents\footnote{It is worth to note that the extrapolations based 
on the avalaible experimental data cannot really discriminate between different exponents
in Eq.~(\ref{AGBB}).}.

The function $\Si(T)$ is interpreted in this way as the {\it local} complexity, \ie
the number of different states the system can visit on a scale $\xi(T)$. The interpretation
of $-T\Si(T)$ as a driving force for nucleation leads then to the Adam--Gibbs relation.
From $\Si(T) \sim T-T_K$ close to $T_K$ and assuming that $\Y(T_K)$ is finite it follows, 
for $\th=d-1$, that
\beq\label{VFTcol2}
\xi(T) \sim R^*(T) \sim (T-T_K)^{-1} \ , \hskip20pt  \t_\a \sim e^{\frac{1}{(T-T_K)^2}} \ ,
\eeq
so the correlation length diverges at $T_K$ as expected and 
a VFT like relation is derived for the relaxation time $\t_\a$, again with exponent $2$.
Note that the Adam--Gibbs relation and the VFT law are recovered if one assumes 
that $\th = d/2$; an argument
in favor of this scaling for the surface tension has been proposed recently 
in~\cite{DSW05}.

In \cite{BB04} the argument was extended also to the case in which the state $\a$ has a
free energy $f_\a < f^*$. In this case it is found that the typical decay length $\xi(f,T)$
is bigger than $\xi(T)$. The distribution of states then induces a distribution of lengths,
and in turn this gives a distribution of local relaxation times that can explain the
observed heterogeneity of the relaxation in glassy systems close to $T_g$, 
see \eg~\cite{Ed00}.

\subsection{The potential method beyond mean field}
\label{sec1:potbeyMF}

\begin{figure}[t]
\includegraphics[width=7.7cm,angle=0]{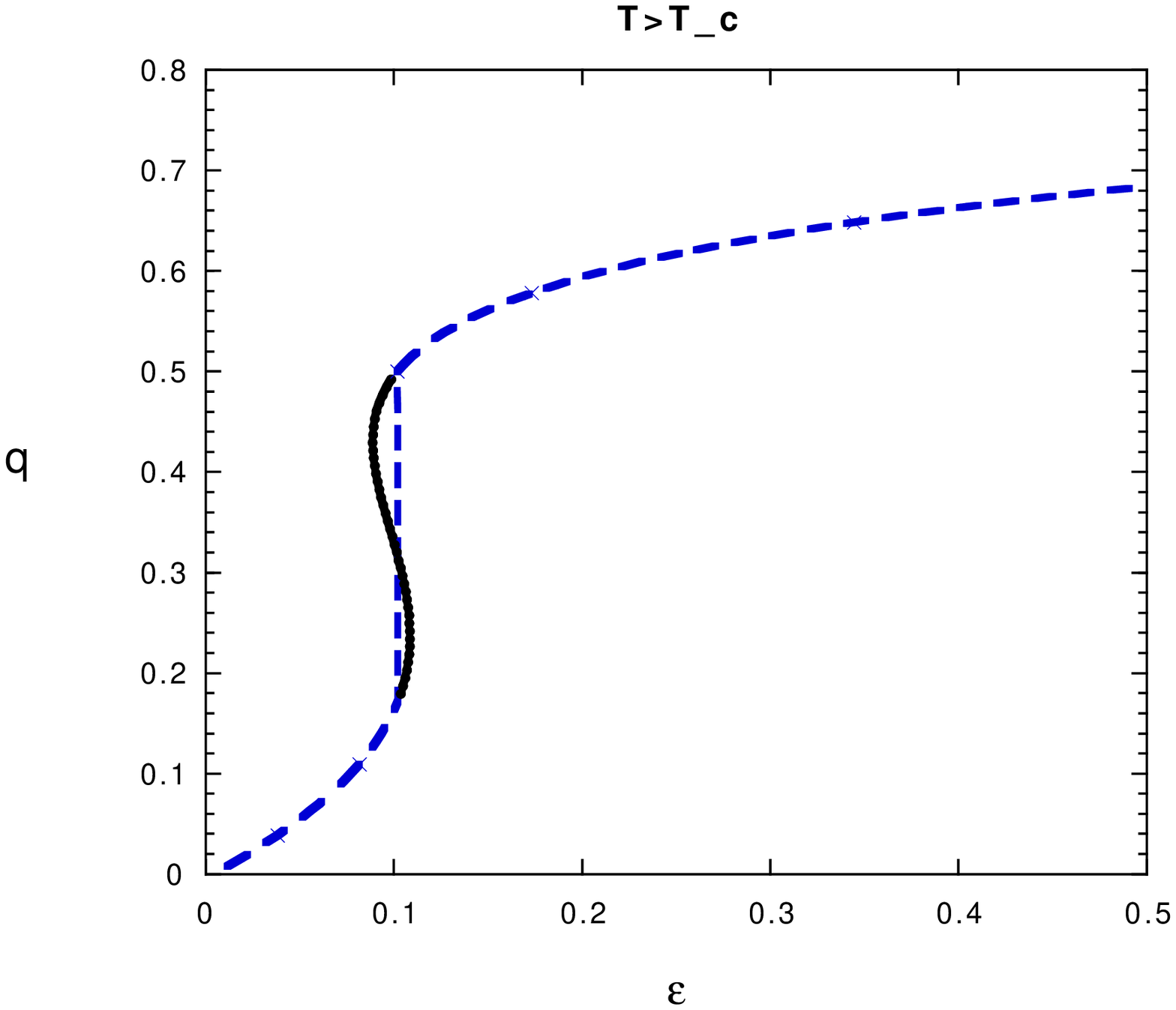}
\includegraphics[width=7.7cm,angle=0]{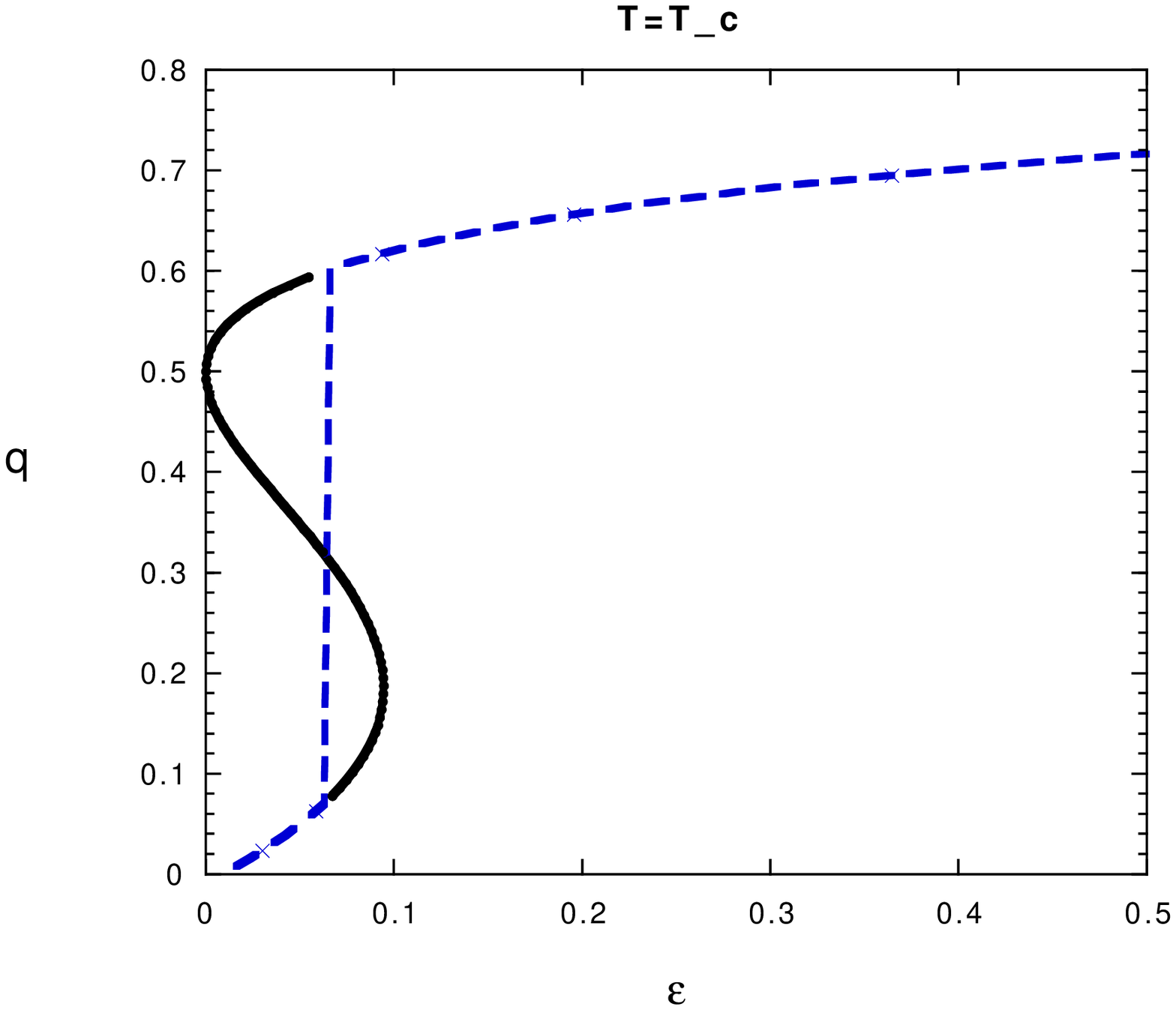}
\includegraphics[width=7.7cm,angle=0]{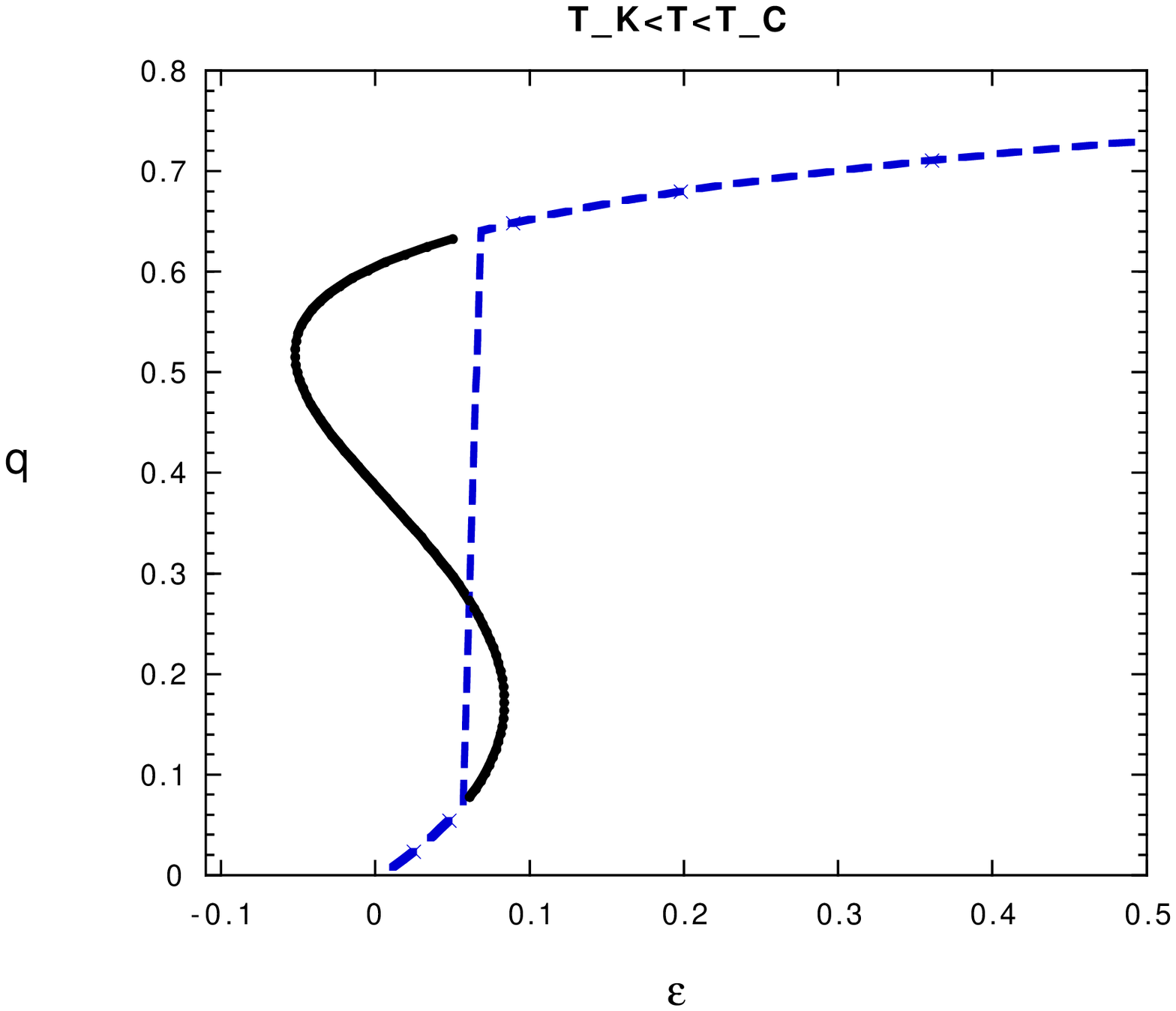}
\includegraphics[width=7.7cm,angle=0]{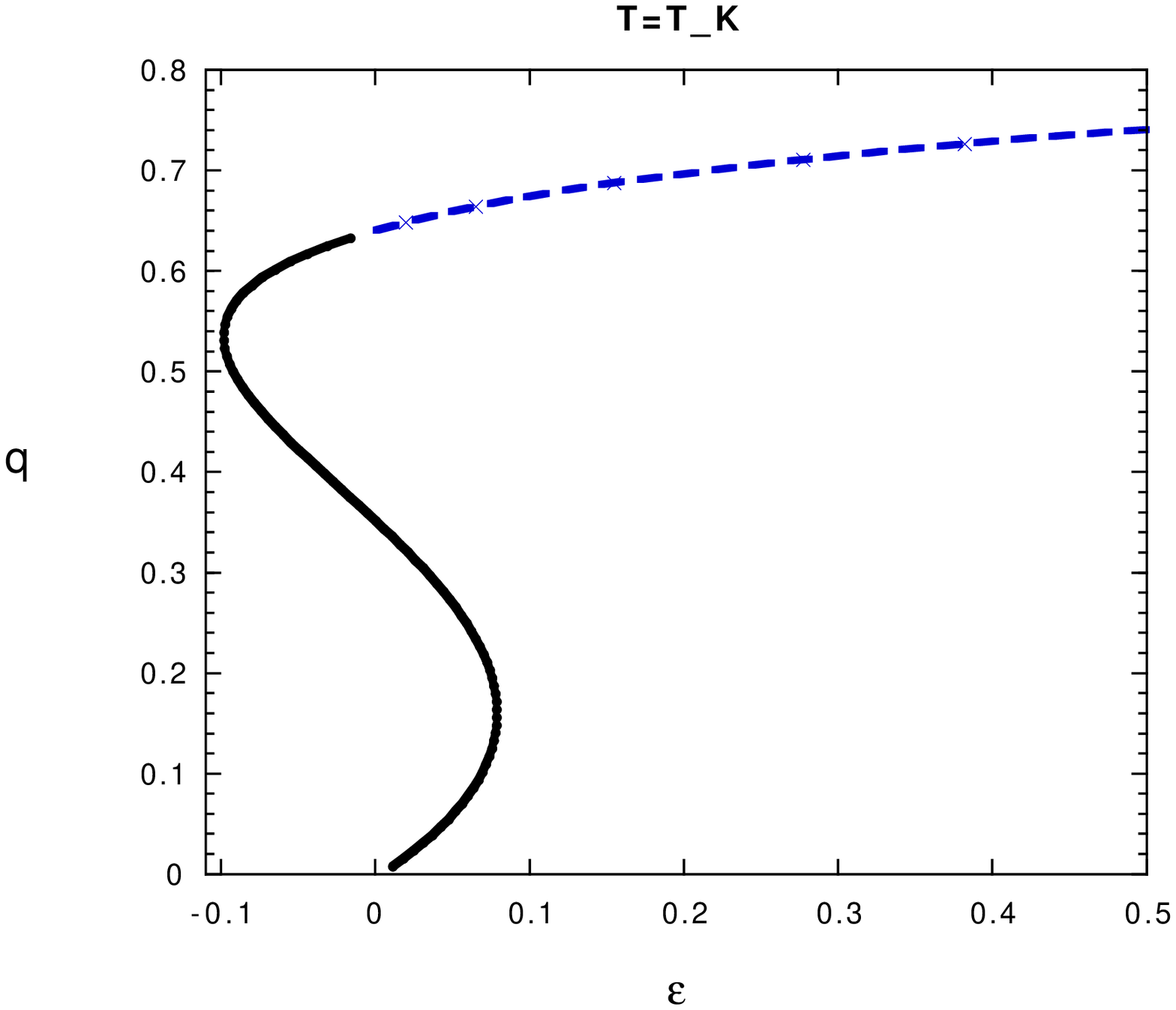}
\caption[The function $q(\ee)$ in short range systems]
{(From~\cite{MP00}; $T_c$ corresponds to $T_d$) 
Sketch of the function $q(\ee)$ in short range system (dashed line). 
A first order transition happens at $\ee_c(T) > 0$ for $T>T_K$; at $T_K$, $\ee_c(T_K)=0$.
The correlated phase is metastable for $\ee < \ee_c(T)$ and follows approximately
the mean field behavior (full line). For $T \sim T_d$ the metastability limit of
the correlated phase is around $\ee =0$ and the complexity cannot be defined.
}
\label{fig1:qeps}
\end{figure}

An interesting question is how one can estimate the (local) complexity in short 
range systems. A possible way is to consider again the two--replica potential $V(q,T)$,
Eq.~(\ref{Vqdef}), and its Legendre transform $F(\ee,T)$.
For mean field systems $V(q,T)$ is sketched in Fig.~\ref{fig_1} and
the difference between the secondary minimum and the primary one is $T\Si(T)$. 
The value of $q(T)$ at the secondary minimum is given by 
$\lim_{\ee \to 0} q(\ee,T)$, where $q(\ee,T)=-\frac{\dpr F}{\dpr \ee}$ is
the mean overlap of the two replicas in presence of a coupling proportional
to $\ee$. The function $q(\ee,T)$ is sketched in Fig.~\ref{fig1:qeps} in the
different regions of temperature. Below $T_d$ the extrapolation of $q(\ee,T)$
down to $\ee=0$ starting from high values of $\ee$ gives the value of $q(T)$.

In short range systems, as the metastable phase corresponding
to the secondary minimum has a finite lifetime, the true potential $V(q)$ is a
concave function of $q$ and has only one minimum in $q=0$ above $T_K$~\cite{MP00}.
For $\ee$ large enough, the phase in which the two replicas are highly correlated
is stable. However, for any $T > T_K$ it exists a value $\ee_c(T)$ where a first
order transition to the small $q$ phase happens
(dashed lines in Fig.~\ref{fig1:qeps}). One expects that $\ee_c(T) > 0$ for $T>T_K$
and that $\ee_c(T_K)=0$, so that the correlated phase becomes stable up to $\ee=0$ for
$T<T_K$. For $\ee < \ee_c(T)$ the correlated phase is metastable. This means that if
one prepares the system at $\ee$ large enough and slowly decreases the value of $\ee$ below
$\ee_c$, the system follows the metastable branch of the curve $q(\ee)$ until,
after some time, a bubble of the stable phase nucleates driving the transition to
$q \sim 0$. But, if the change of $\ee$ is fast enough, and if $T$ is close to $T_K$, one should
be able to ``supercool'' the correlated phase up to $\ee=0$ and to extrapolate
the value of $q(T)$ corresponding to the metastable minimum at $\ee=0$. 
The knowledge of the curve $q(\ee)$ up to $\ee=0$ in the metastable
branch allows to compute $V(q(T),T)$ and the complexity $\Si(T)$ as a function 
of~$T$~\cite{MP00,CMPV99,AFST05}.

An ambiguity in the definition of $\Si(T)$ is present because the function $q(\ee)$
below $\ee_c$ (slightly) depends on the time scale and in general on the history of the system.
However one can reasonably expect (relying on similar results obtained for Ising models, 
see \eg \cite{Ga99,Is84}) 
that the ambiguity is of the order of 
$\exp [-(\ee_c-\ee)^{-1}]$ for $\ee_c \gtrsim \ee$ so it becomes smaller and smaller on
approaching $T_K$. Close to $T_d$ the ambiguity becomes very large and $\Si(T)$
cannot be properly defined in short range systems.

\subsection{A first principles computation of the surface tension}
\label{sec1:instantons}

A way to compute the free energy barrier for nucleation $\D F(T)$ using again the two--replica
potential $V(q,T)$ has been recently proposed~\cite{Fr05,DSW05}. Indeed, the potential $V(q,T)$
allows to realize the situation considered in section~\ref{sec1:BBargument}
in a way that is suitable for analytical computations. 

The configuration $\s$ is a reference (frozen) configuration that belongs to an equilibrium
state at temperature $T$. The configuration $\t$ is constrained to have a fixed overlap $q$
with $\s$, so if $q = q(T)$ it belongs to the same state.
Consider now a system with long but finite range interactions, whose scale is $1/\g$, $\g \sim 0$,
enclosed in a volume $V=N \g^{-d}$, $N \to \io$, \ie the thermodynamic limit is taken at the
beginning of the calculation.
Consider an adimensional space variable $x$
obtained rescaling the space by $\g$, \ie $\int d^d x = N$.
We can define a local overlap $q(x)$
averaging the overlap over a large volume of linear dimension
smaller than $1/\g$, and consider the potential $V[q(x),T]$ as a functional of the local
overlap\footnote{This is the same procedure used in the study of nucleation problems, where 
the free energy is considered as a functional of the local corse--grained order parameter,
\eg the magnetization or the density.} $q(x)$. 
A configuration $\t$ such that 
\beq\label{boundaryist}
\begin{cases}
&q(x)=q(T) \hskip30pt \text{for} \, |x|\to \io \ , \\
&q(x) \sim 0 \hskip45pt \text{for} \, x=0 \ ,
\end{cases}
\eeq
is in the same state of $\s$ outside some
bubble of radius $R$ and is in another state inside the bubble. The quantity 
$\D F(T)=V[q(x),T]-V(q(T),T)$ represents exactly the free energy cost of this configuration with
respect to the configuration in which the two replicas are in the same state in all the points
of the space,
so it is exactly the free energy barrier defined in section~\ref{sec1:BBargument}.
The overlap profile $q(x)$ has to be determined by minimizing $V[q(x),T]$ with the boundary
conditions~(\ref{boundaryist}) in order to find the most probable transition state
for the nucleation~\cite{Fr05,DSW05}.

In systems with long but finite range\footnote{From now the discussion will be not technical;
for all the technical details as well as for a deep discussion of many important issues
the reader should refer to the original papers~\cite{FT04,Fr05,DSW05}.}
(Kac spin glasses) it has been shown~\cite{FT04} that the potential
$V[q(x),T]$ has a form very similar to the mean field one~(\ref{Vqrepliche})
\beq
V[q(x),T] = -\frac{T \g^d}{N} \lim_{n\to 0}\lim_{m\to 1} \frac{1}{n} \frac{\dpr}{\dpr m} 
\overline{Z_{mn}[q(x)]} \ ,
\eeq
where $Z_{mn}[q(x)]$ is the partition function of an $nm$-times replicated system such
that in each $m\times m$ subblock the first replica has overlap $q(x)$ with the other
$m-1$ replicas (as in mean field). The partition function has the form
\beq
Z_{mn}[q(x)]=\int {\cal D}Q(x) e^{\frac{1}{\g^d} S[Q(x)]} \ ,
\eeq 
where the matrix $Q(x)$ respects the constraint above, \ie it has a structure similar
to (\ref{1RSBpotenziale}), and the action $S[Q(x)]$ has the form
\beq
S[Q(x)] = \int d^dx \big\{ K[Q(x)] + f[Q(x)] \big\} \ ,
\eeq
with $K[Q(x)]$ a kinetic term\footnote{The coefficient of $\Tr[\nabla Q(x)]^2$ also depends
on $Q(x)$. This dependence is neglected here but does not affect the results.}, 
$K[Q(x)] \sim -\frac{\b}{2}\Tr[\nabla Q(x)]^2$, and $f[Q(x)]$ a potential
identical to the mean field one given in Eq.~(\ref{fMF}). The mean field potential then
plays the role of a local potential in each volume $\sim \g^{-d}$, while the contributions due
to space variations on a scale $1/\g$ are taken into account by the kinetic term.

\begin{figure}[t]
\centering
\includegraphics[width=7.7cm,angle=-90]{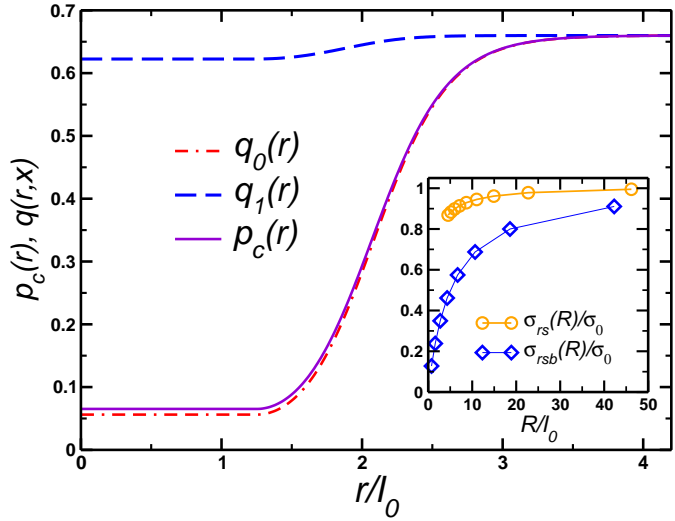}
\caption[Instanton profile close to $T_K$]
{(From~\cite{DSW05}) A typical instanton profile (full line) close to $T_K$. 
In the inset, the surface tension as a function of the droplet size is reported.
}
\label{fig1:instant}
\end{figure}

For $\g \to 0$, if one looks to homogeneous solutions $Q(x) \= Q$, all the results of
the mean field model are reproduced. To look for nonhomogeneous solution respecting the
boundary conditions~(\ref{boundaryist}), an {\it ansatz} of the form (\ref{1RSBpotenziale})
in each point $x$ has been proposed \cite{Fr05}; if the potential has to been minimized also
w.r.t. $q(x)$, one can assume that $r=q$ in each point $x$ as in the homogeneous case. This
is the simplest possible {\it ansatz} and one obtains
\beq\label{potinst}
V[q(x),T]= \frac{1}{N} \int d^d x \left\{ \frac{1}{2} [\nabla q(x)]^2 + V(q(x),T) \right\} \ ,
\eeq
where $V(q,T)$ is the mean field potential.
Then the equation for $q(x)$ has the form
\beq \label{istantone}
\nabla^2 q(x) = \frac{d V}{dq} \ ,
\eeq
and the boundary conditions~(\ref{boundaryist})
have to be imposed to the solution. If one is able to solve Eq.~(\ref{istantone}), 
substituting the solution $\ol q(x)$ into the potential
one can compute the barrier $\D F(T) = V[\ol q(x),T]-V(q(T),T)$. 
Subtracting from the barrier the bulk contribution $-T \Si(T) R^d$, one gets an estimate
of the surface tension. A typical profile of the solution and the corresponding surface
tension are reported in Fig.~\ref{fig1:instant}.

However, an approximate solution is possible if one looks for spherical solutions $q(r)$
with $r=|x|$ and, in the limit of large radius $R$, approximates the Laplacian,
close to the interface, with
\beq
\nabla^2 q \sim \frac{d^2 q}{dr^2} = \frac{dV}{dq} \ .
\eeq
In this case the problem becomes planar so the radius of the droplet remains undetermined.
For a given radius $R$ of the droplet, the bulk term$^{\ref{nota1}}$ is 
simply $[V(q(T),T)-V(q,0)] R^d=-T\Si(T) R^d$.
To estimate the surface tension, \ie the contribution to the integral (\ref{potinst}) due
to the interface, note that the quantity $E=\frac{1}{2} \left(\frac{d q}{dr}\right)^2 - V(q,T)$ 
is conserved and is equal, recalling that, for $r\to\io$, $\frac{dq}{dr} \to 0$, to $V(q(T),T)$, 
so one has, for $r \sim R$,
\beq
\frac{dq}{dr} = \sqrt{2 [V(q,T) - V(q(T),T)]} \ ,
\eeq
and the contribution coming from the region $r\sim R$ to the barrier $V[q(x),T]-V(q(T),T)$ is,
defining $r_-$ such\footnote{Note that for $r<r_-$ the approximation surely breaks down,
otherwise $\frac{dq}{dr}$ would become an imaginary number.} that $V(q(r_-),T)=V(q(T),T)$,
\beq\begin{split}
R^{d-1} \int_{r_-}^\io dr& \left\{\frac{1}{2} \left(\frac{dq}{dr} \right)^2 + 
V(q,T) - V(q(T),T) \right\} =
R^{d-1} \int_{r_-}^\io dr \left(\frac{dq}{dr} \right)^2 =\\
&R^{d-1} \int_{q_-}^{q(T)} dq \sqrt{2 [V(q,T) - V(q(T),T)]} \ ,
\end{split}\eeq
so we get the following expression for the surface tension~\cite{Fr05}:
\beq
\Y(T) =  \int_{q_-}^{q(T)} dq \sqrt{2 [V(q,T) - V(q(T),T)]} \ .
\eeq
As the difference $q(T)-q_-$ is always of order $1$, the surface tension scales
as 
\beq
\Y(T) \sim \sqrt{V(q_{max},T)-V(q(T),T)}
\eeq
and is finite at $T_K$. The outcome of
the simplest instanton calculation is that $\th=d-1$ and $\Y(T_K) \neq 0$.
This leads to the scalings (\ref{VFTcol2}).

It has been recently found that a more refined calculation that includes replica symmetry
breaking at the interface reduces the surface tension; from this observation an argument
that leads to $\th=d/2$ has been proposed. However, a detailed theory is still missing.

%% file: cap2.tex
\chapter{The ideal glass transition of Hard Spheres}
\label{chap2}

\section{Introduction}

The question whether a liquid of identical Hard Spheres undergoes
a glass transition upon densification is still open, see \eg
\cite{RT96,RLSB98,Sp98,TdCFNC04}.
It is interesting to apply the replica method to the Hard Spheres
liquid, following what has been done for 
Lennard--Jones systems~\cite{MP99,MP00,CMPV99} in order to
investigate the possibility of the existence of a Kauzmann transition.

In an Hard Sphere system, on increasing the density,
and if crystallization is avoided, one can access 
the metastable region of the phase diagram above the freezing
packing fraction $\f_f=0.494$, where $\f=\frac{N \p D^3}{6V}$, 
$D$ is the Hard Sphere diameter, $N$ is the number of particles
and $V$ is the volume of the container.
In this region the dynamics of the liquid becomes slower and slower
on increasing the density. The particles
are ``caged'' by their neighbors, and the dynamics separates into
a fast rattling inside the cage and slow rearrangements of the
cages. The typical time scale of these rearrangements increase
very fast around $\f_g \sim 0.56$ and many authors reported the
observation of a glass transition at these values of 
density~\cite{GS91,vMU93}.
Note that the Kauzmann density
is expected to be larger than the experimental glass transition density,
as at $\f_K$ the relaxation time is expected to diverge so that
the system freezes in a metastable state, on the experimental time
scale, for a density $\f_g$ smaller than~$\f_K$.

A related problem is the study of {\it dense amorphous packings}
of Hard Spheres. Dense amorphous packings are relevant in 
the study of colloidal suspensions, granular matter, powders,
etc. and have been widely studied in the 
literature~\cite{Be83,SK69,Fi70,Be72,Ma74,Po79,Al98,SEGHL02}.
The metastable states of the Hard Sphere liquid provide
examples of such packings: when the system freezes in one of these
states, if one is still able to increase the density in order to
reduce the size of the cages to zero (for example by shaking the
container \cite{SK69,Fi70} or using suitable computer 
algorithms \cite{Be72,Ma74,SEGHL02}), a {\it random close packed}
state is reached.
The problem of which is the maximum value of density $\f_c$ that
can be reached applying this kind of procedures has been tackled using
a lot of different techniques, usually finding values of $\f_c$ in the
range $0.62 \div 0.67$.
Another interesting problem is to estimate the mean
coordination number $z$, {\it i.e.} the mean number of contacts between
a sphere and its neighbors, in the random close packed states.
Many studies addressed this question usually finding values of 
$z \sim 6$.

Few estimates of the configurational entropy for Hard Spheres
are currently available \cite{Sp98,Luca05} and indicate
a value of $\f_K$ in the range $0.58 \div 0.62$. These estimates
were obtained following numerical procedures already succesfully
applied in Lennard--Jones systems \cite{CMPV99,SKT99} or the method
described in section~\ref{sec1:potbeyMF};
for the Lennard--Jones liquid
the results compare well with the theoretical predictions
of the replica theory~\cite{MP00,CMPV99}. A tentative replica study
of the Hard Spheres glass transition, based on the potential method
described in section~\ref{sec1:potentialmethod},
can be found in \cite{CFP98}, where a good estimate of $\f_K \sim 0.62$
was obtained. However, the configurational entropy computed in \cite{CFP98}
is two orders of magnitude smaller than the one found in numerical
simulations. This negative results is probably due to some technical
problem in the assumptions of \cite{CFP98}.

For technical reasons the real replica method 
(see section~\ref{sec1:realreplica}) 
of \cite{MP99,Mo95,MP00,CMPV99,MP99b}, that gives
very good results for Lennard--Jones systems,
cannot be extended straightforwardly to the
case of Hard Spheres; indeed at some stage is was assumed that the vibrations around the
equilibrium positions were harmonic in a first approximation. 
This approximation is not bad for soft 
potentials, but it clearly makes no sense for hard spheres.

In this chapter a way to adapt the replica method of 
\cite{MP99} to the case of the Hard Sphere liquid, and 
in general of potentials such that the pair distribution
function $g(r)$ shows discontinuities, will be developed.
This allows to compute from first principles the 
configurational entropy of the liquid as well as 
the thermodynamic properties of the glass and the random 
close packing density.
A very good estimate of the configurational 
entropy, that agrees well with the recent numerical 
simulations of \cite{Sp98,Luca05},
a Kauzmann density in the range $0.58 \div 0.62$
(depending on the equation of state we use to describe
the liquid state), and a random close packing density in the
range $0.64 \div 0.67$, are found. 
Moreover, the mean 
coordination number in the amorphous packed states is found to be
$z=6$ irrespective of the equation of state used for the liquid,
in very good agreement with the result of numerical 
simulations~\cite{Be72,Ma74,SEGHL02}.

\section{The molecular liquid}
\label{sec2:replica}

The starting point of the real replica method described in 
section~\ref{sec1:realreplica} is the free energy of a system of $m$
copies of the original liquid constrained to be in the same metastable
state. This means that each atom of a given replica is close to an atom of
each of the other $m-1$ replicas, \ie, the liquid is made of {\it molecules}
of $m$ atoms, each belonging to a different replica of the original system.
In other words the atoms of different replicas stay in the same cage.
The problem is then to compute the
free energy of a molecular liquid where each molecule is made of $m$ atoms.  The $m$ atoms are kept
close one to each other by a small inter-replica coupling that is switched off at the end of the
calculation, while each atom interacts with all the other atoms of the same replica via the original
pair potential.  This problem can be tackled by mean of the HNC integral equations~\cite{Hansen}.

\subsection{HNC free energy}

The traditional HNC approximation can be naturally extended to the case where particles have internal degrees of
freedom and also to the replica approach where one has molecules composed by $m$ atoms.

Let $x=\{\ul x_1,\cdots,\ul x_m\}$, $\ul x_a \in \RRR^d$ be the coordinate of a molecule
in dimension $d$.  The single-molecule density is
\beq
\r(x) = \langle \sum_{i=1}^N \prod_{a=1}^m \delta (\ul x_{ia} - \ul x_a) \rangle \ ,
\eeq
and the pair correlation is
\beq
\r(x) g(x,y) \r(y) = \langle \sum_{i,j}^{1,N} \prod_{a=1}^m \delta (\ul x_{ia} - \ul x_a)
\prod_{b=1}^m \delta (\ul x_{jb} - \ul y_b)  \rangle \ .
\eeq
Define also $h(x,y)=g(x,y)-1$. The interaction potential between two molecules
is $v(x,y)= \sum_a v(|\ul x_a-\ul y_a|)$.

The HNC free energy is given by~\cite{MP99,Hansen}
\beq
\label{HNCfree}
\begin{split}
\b \Ps[\r(x),g(x,y)] &= \frac{1}{2} \int dx dy \, \r(x) \r(y) 
\big[ g(x,y) \log g(x,y)
- g(x,y) + 1 + \b v(x,y) g(x,y) \big] \\
&+ \int dx \r(x) \big[ \log \r(x) -1 \big]
+ \frac{1}{2} \sum_{n\geq 3} \frac{(-1)^n}{n} \Tr [ h\r]^n \ ,
\end{split}
\eeq
where
\beq
\Tr [ h\r]^n = \int dx_1 \cdots dx_n h(x_1,x_2) \r(x_2) h(x_2,x_3) \r(x_3)
\cdots h(x_{n-1},x_n) \r(x_n) h(x_n,x_1) \r(x_1) \ .
\eeq
For Hard Spheres the potential term vanishes, 
$\int dx dy \, \r(x)\r(y) g(x,y)v(x,y) \equiv 0$, so the
reduced free energy $\b \Psi$ will not depend on the 
temperature in all the following equations.  Similarly, all the free energy functions that will be
consider below do not depend on the temperature once multiplied by $\b$.  In principle one could
stick to $\beta=1$ and slightly simplify the formulae. However, it is useful to keep explicitly
$\beta$, in order to conform to the standard notation for soft spheres (or for hard spheres with an
extra potential).

Differentiation w.r.t $g(x,y)$ leads to the HNC equation:
\beq
\log g(x,y) + \b v(x,y) = h(x,y)-c(x,y) \ ,
\eeq
having defined $c(x,y)$ from
\beq
h(x,y)=c(x,y)+\int dz \, c(x,z)\r(z)h(z,y) \ .
\eeq
The free energy (per particle) of the system is given by
\beq\begin{split}
&\phi(m,T)= \frac{1}{Nm} \min_{\r(x), g(x,y)} \Psi[\r(x),g(x,y)] \ , \\
&\Phi(m,T) = m \phi(m,T) \ ,
\end{split}\eeq
and once the latter is known one can compute the free energy of the states and the complexity
using Eq.s~(\ref{mcomplexity}).

\subsection{Single molecule density}

The solution of the previous equations for generic $m$ is a very complex problem (it is already
rather difficult for $m=2$). Some kind of {\it ansatz} is needed to simplify the computation, that may
become terribly complicated.

The single molecule density encodes the information about the inter-replica coupling
that keeps all the replicas in the same state. One can assume that this arbitrarily small coupling
has already been switched off, with the main effect of building molecules of $m$
atoms vibrating around the center of mass $\ul X \in \RRR^d$ of the molecule with a certain 
``cage radius'' $A$. The simplest {\it ansatz} for $\r(x)$ is then~\cite{MP99}
\beq
\label{rrho}
\r(x) = \wh\r \int d \ul X \prod_a \r(\ul x_a-\ul X) \ , \hspace{10pt} \int d \ul u \, \r(\ul u)=1 \ ,
\eeq
with
\beq
\label{rho}
\r(\ul u)=\frac{e^{-\frac{u^2}{2A}}}{(\sqrt{2\p A})^d} \ ,
\eeq
and $\wh\r=V^{-1} \int dx \, \r(x)$ the number density of molecules.
With this choice it is easy to show that
\beq
\frac{1}{N} \int  dx \, \r(x) \big[ \log \r(x) -1 \big] =
\log \wh\r -1 + \frac{d}{2} (1-m) \log ( 2\p A ) - \frac{d}{2} \log m + \frac{d}{2}(1-m) \ .
\eeq

\subsection{Pair correlation}

As the information about the inter-replica coupling is already encoded
in $\r(x)$, a reasonable {\it ansatz} for $g(x,y)$ is:
\beq
\label{gprod}
g(x,y) = \prod_a g(| \ul x_a-\ul y_a |) \ ,
\eeq
where $g(r)$ is rotationally invariant because so is the interaction
potential.
It is useful to define $G(r) \equiv [g(r)]^m$.
Using the {\it ansatz} above, it is easy to rewrite the free
energy (\ref{HNCfree}) as follows:
\beq
\label{HNCfree2}
\begin{split}
\b \Ps &= \frac{\wh\r N}{2} \int d \ul r \, 
\big\{m [F_0(r)]^{m-1} F_1(r) - [F_0(r)]^m + 1
+ m  [F_0(r)]^{m-1} F_v(r) \big\} \\
&+ \int dx \, \r(x) \big[ \log \r(x) -1 \big]
+ \frac{1}{2} \sum_{n\geq 3} \frac{(-1)^n}{n} \Tr [ h\r]^n \ ,
\end{split}
\eeq
where
\beq
\label{Fp}
\begin{split}
&F_p(|\ul r|) = \int d\ul u d\ul v \, \r(\ul u)\r(\ul v) \, 
g(|\ul r+\ul u-\ul v|) [\log g(|\ul r+\ul u-\ul v|)]^p \ , \\
&F_v(|\ul r|) = \int d\ul u d\ul v \, \r(\ul u)\r(\ul v) \, 
g(|\ul r+\ul u-\ul v|) \b v(|\ul r+\ul u-\ul v|) \ .
\end{split}
\eeq
Note that as $g(r)$ and $v(r)$ are rotationally invariant, so are $F_p(r)$ and $F_v(r)$. 
If $\r(\ul u)$ is given by Eq.~(\ref{rho}), one gets
\beq
F(|\ul r|) = \int d\ul u \, \frac{e^{-\frac{u^2}{4A}}}{(\sqrt{4\p A})^d} f(|\ul r+\ul u|) \ ,
\eeq
where $f(r) \in \{g(r), \, g(r)\log g(r), \, g(r) \b v(r)\}$.
For Hard Spheres $F_v \equiv 0$.

\section{Small cage expansion}
\label{sec2:smallcageexp}

The strategy of~\cite{MP99} was to expand the HNC free energy in a power series
of the cage radius $A$, assuming that the latter is small close to the glass
transition. The expansion is carried out easily if the pair potential $v(r)$ and
the pair correlation $g(r)$ are analytic functions of $r$. However this is not
the case for Hard Spheres, as $g(r)$ vanishes for $r < D$ and has a discontinuity
in $r=D$, so the formulae of~\cite{MP99} for the power
series expansion of $\Psi$ cannot be applied to our system. 

It is crucial to realize that, independently from any approximation, in the limit $A \to 0$, the
partition function becomes (neglecting a trivial factor) the partition function of a single atom
at an effective temperature given by $\b_{eff}=\b m$. In the case of Hard Spheres, where there is no
dependence on the temperature, the change in temperature is irrelevant.

In \cite{MP99} it was shown that the first term of the expansion is proportional to $A$ if $g(r)$ is
differentiable.  It will be shown in the following that, in the case of Hard Spheres, the presence of a
jump in $g(r)$ produces terms $O(\sqrt{A})$ in the expansion. At first order one can focus on these
terms neglecting all the contributions of higher order in $\sqrt{A}$.  This means that one can
neglect all the contributions coming from the regions where $g(r)$ is differentiable and concentrate
only on what happens around $r=D$.

\subsection{Expansion of $F_0(r)$}

The contribution one wants to estimate comes from the discontinuity of $g(r)$ in
$r=D$. Thus to compute this correction the form of $g(r)$ away from the singularity is
irrelevant and one can use the simplest possible form of $g(r)$.

It is convenient to discuss first the expansion of $F_0(r)$ in $d=1$. 
The simplest possible form of $g(r)$ is
\beq
g(r)=\theta(r-D) [1+ (y-1) e^{-\m (r-D)}] \ ;
\eeq
the amplitude of the jump of $g(r)$ in $r=D$ is given by $y$.
Remember that $\ul r \in \RRR$ and $r=|\ul r| \in \RRR^+$.
As the functions $F_0$ and $g$ are even in $\ul r$, one can write
\beq
\label{termine1}
\int_{-\io}^\io d\ul r [F_0(\ul r)^m - g(\ul r)^m] = 2 \int_0^\io dr [F_0(r)^m - g(r)^m] \ .
\eeq
Defining
\beq
\begin{split}
&\erf(t) \equiv \frac{2}{\sqrt{\p}} \int_0^t dx \, e^{-x^2} \ , \\
&\Th(t) = \frac{1}{2} [ 1 + \erf(t) ] \ ,
\end{split}
\eeq
these functions play the role of ``smoothed'' sign and $\th$-function
respectively; note also that the function $\Th(t)$ goes to $0$ as $e^{-t^2}$ for
$t\rightarrow -\io$. Then
\beq
\int_{-\io}^\io du \, \frac{e^{-\frac{u^2}{4A}}}{\sqrt{4\p A}} \,
\theta(r+u-D) = \frac{1}{2} 
\left[ 1 + \text{erf}\left(\frac{r-D}{\sqrt{4A}}\right) \right] 
\equiv \Th\left(\frac{r-D}{\sqrt{4A}}\right) \ ,
\eeq
and
\beq
\label{F0}
\begin{split}
F_0(r)&=\Th\left(\frac{r-D}{\sqrt{4A}}\right)
+ \Th\left(-\frac{r+D}{\sqrt{4A}}\right)
 + (y-1) e^{A \m^2}
\Big\{ e^{-\m(r-D)} \Th\left(\frac{r-D-2A\m}{\sqrt{4A}}\right)\\
& + e^{\m(r+D)}\Th\left(-\frac{r+D+2A\m}{\sqrt{4A}}\right)
\Big\} \ .
\end{split}
\eeq
As $r\geq 0$ one can neglect the terms proportional to $\Th\left(-\frac{r+D}{\sqrt{4A}}\right)$
in Eq.~(\ref{F0}), that give a contribution of order $\exp(-D^2/A)$ for $A \rightarrow 0$.
Defining the reduced variable $t= (r-D)/\sqrt{4A}$:
\beq
\begin{split}
g(t) &= \theta(t) [1+(y-1) e^{-\m 2 \sqrt{A} t}] \ , \\
F_0(t)&=\Th(t) + (y-1)  e^{-\m 2 \sqrt{A} t} \, e^{A \m^2}\Th(t+\m \sqrt{A}) \ ,
\end{split}
\eeq
and Eq.~(\ref{termine1}) becomes
\beq
\int_0^\io dr [F_0(r)^m - g(r)^m] =
2\sqrt{A} \int_{-\frac{D}{\sqrt{4A}}}^\io dt [ F_0(t)^m - g(t)^m ] \equiv 2 \sqrt{A} Q(A) \ .
\eeq
If the function $Q(A)$ has a finite limit $Q(0)$ for $A\rightarrow 0$ one has 
$Q(A) = Q(0) + o(1)$ and the leading correction to the free energy is 
$O(\sqrt{A} Q(0))$.
The limit for $A \rightarrow 0$ of $Q(A)$ is formally given by
\beq
Q(0) = y^m \int_{-\io}^\io dt \, [ \Th(t)^m - \th(t)^m ] \equiv y^m Q_m
\eeq
where $y^m \equiv Y$ is the jump of $G(r)\equiv g(r)^m$ in $r=D$ and 
$Q_m \equiv \int_{-\io}^\io dt \, [ \Th(t)^m - \th(t)^m ]$.
It is easy to show that $Q_m$ is a finite and smooth function of $m$ for
$m \neq 0$, that
\beq
\begin{split}
&Q_m = (1-m) Q_0 + O[(m-1)^2] \ , \\
&Q_0 = -\int_{-\io}^\io dt \, \Th(t) \log \Th(t) \sim 0.638 \ ,
\end{split}
\eeq
and that $Q_m$ diverges as $Q_m \sim \sqrt{\p/4m}$ for $m \rightarrow 0$.
Finally, recalling that $G(r)=[g(r)]^m$,
\beq
\label{Gterm}
\frac{1}{2}\int d\ul r \, F_0(r)^m = \frac{1}{2}\int d\ul r \, G(r) + 2 \sqrt{A} Y Q_m \ .
\eeq
In dimension $d>1$, recalling that $F_0(r)$ and $G(r)$ are both rotationally invariant, one has
\beq
\label{DDDD}
\int d\ul r \, [ F_0(r)^m - G(r)^m ] = \O_d \int_0^\io dr \, r^{d-1} \, [ F_0(r)^m - G(r)^m ] \ ,
\eeq
where $\O_d$ is the solid angle in $d$ dimension, $\O_d=2\pi^{d/2}/\G(d/2)$.
The function $F_0(r)$ can be written as
\beq
F_0(r) = \int d\ul u \, \frac{e^{-\frac{u^2}{4A}}}{(\sqrt{4\p A})^d} g(|r \widehat i +\ul u|) \ ,
\eeq 
where $\widehat i$ is the unit vector \eg of the first direction in $\RRR^d$. For small $\sqrt{A}$, 
the $u$ are small too. The function $g(|r \widehat i +\ul u|)$ is differentiable along the directions
orthogonal to $\widehat i$. Expanding in series of $u_\m$, $\m \neq 1$, at fixed $u_1$, one sees that
the integration over these variables gives a contribution $O(A)$, so
\beq
\label{5D}
F_0(r) = \int_{-\io}^\io du_1 \, \frac{e^{-\frac{u_1^2}{4A}}}{\sqrt{4\p A}} g(r + u_1) + O(A) \ ,
\eeq 
as in the one dimensional case. The function $F_0(r)^m - G(r)^m$ is large only for
$r - D \sim \sqrt{A}$ so at the lowest order one can replace $r^{d-1}$ with $D^{d-1}$
in Eq.~(\ref{DDDD}), and obtains
\beq
\int d\ul r \, [ F_0(r)^m - G(r)^m ] = \O_d D^{d-1} \int_0^\io dr \, [ F_0(r)^m - G(r)^m ] \ .
\eeq
The last integral, with $F_0(r)$ given by Eq.~(\ref{5D}) is the same as in $d=1$, so
\beq
\label{GtermD}
\frac{1}{2} \int d\ul r \, F_0(r)^m = \frac{1}{2} \int d\ul r \, G(r) + 
\sqrt{A} Y \Si_d(D) Q_m \ ,
\eeq
where $\Si_d(D)$ is the surface of a $d$-dimensional sphere of radius $D$,
$\Si_d(D) = \O_d D^{d-1}$.
This result can be formally written as
\beq
\label{corrQ}
F_0(r)^m \sim G(r) + 2\sqrt{A} Y Q_m \d(|r|-D)
\equiv G(r) + Q_0(r)
\eeq
as the correction comes only from the region close to the singularity of 
$g(r)$, $r-D \sim \sqrt{A}$.

\subsection{$G\log G$-term}

The correction coming from the term 
$\int dr \, m F_0(r)^{m-1} F_1(r)$ will now be estimated.
Using the same argument as in the previous subsection, one can
restrict to $d=1$.
Note first that $F_0(r)$, for $|r-D| \sim \sqrt{A}$, has the form
\beq
\label{F0sing}
F_0(r) = y \, \Th\left(\frac{r-D}{\sqrt{4A}}\right) + o(\sqrt{A}) \ ,
\eeq
where $y$ is the jump of the function $g(r)$ in $r=D$. Similarly,
$F_1(r)$ has the form
\beq 
F_1(r) =
\begin{cases} 
g(r) \log g(r) + O(A) \ , \hskip33pt  |r-D| \gg \sqrt{A} \ , \\
y\log y \, \Th\left(\frac{r-D}{\sqrt{4A}}\right) + o(\sqrt{A}) \ , \hskip10pt
|r-D| \sim \sqrt{A} \ .
\end{cases}
\eeq
The integral 
\beq
\int_0^\io dr [ m F_0(r)^{m-1} F_1(r) - m g(r)^m \log g(r) ]
\eeq
has then two contributions: the first comes from the region $|r-D| \gg \sqrt{A}$ and
is of order $A$ as if the function $g(r)$ were continuous. The other comes from the
region $|r-D| \sim \sqrt{A}$ and is of order $\sqrt{A}$ as in the previous case.
To estimate the latter one can use again the reduced variable $t$ and approximate
$F_1(t) \sim y \log y \, \Th(t)$, $F_0(t) \sim y \, \Th(t)$.
Then
\beq
\int_0^\io dr [ m F_0(r)^{m-1} F_1(r) - m g(r)^m \log g(r) ] =
Y \log Y \, 2\sqrt{A} Q_m + o(\sqrt{A}) \ ,
\eeq
in $d=1$ and finally, in any dimension $d$,
\beq
\label{GlogGtermD}
\frac{1}{2}\int d\ul r \, m F_0(r)^{m-1} F_1(r) =
\frac{1}{2}\int d\ul r \, G(r) \log G(r) + \sqrt{A} Y \log Y \, \Si_d(D) Q_m \ .
\eeq

\subsection{Interaction term}

Substituting Eq.~(\ref{rrho}) in the last term of the HNC free energy one obtains
\beq
%\begin{split}
\Tr[h\r]^n =\wh\r^n \int d\ul X_1 \cdots d\ul X_n \int du_1 \cdots du_n
\r(u_1) \cdots \r(u_n)
h(\ul X_1-\ul X_2,u_1-u_2) \cdots h(\ul X_n-\ul X_1,u_n-u_1) \ ,
%\end{split}
\eeq
where $h(X,u) = \prod_{a=1}^m g(X+u_a) - 1$ and 
$\r(u)=\prod_{a=1}^m \r(\ul u_a)$ with
$\r(\ul u)$ given by Eq.~(\ref{rho}).

The correction $O(\sqrt{A})$ to this integral comes from the regions where 
$|X_i-X_{i+1}| = D + O(\sqrt{A})$ for some $i=1,\cdots,n$. In these regions the functions
$h$ such that their arguments are not close to the singularity can be expanded in a power series
in $u$, the correction being $O(A)$~\cite{MP99}. Thus one can write,
defining $H(r)=G(r)-1$:
\beq
\begin{split}
\wh\r^{-n} \Tr[h\r]^n &= 
\int d\ul X_1 \cdots d\ul X_n H(\ul X_1-\ul X_2) \cdots H(\ul X_n-\ul X_1) +
n \int d\ul X_1 \cdots d\ul X_n \int du_1 du_2 \times\\ \times \r(u_1)& \r(u_2)
 \big[ h(\ul X_1-\ul X_2,u_1-u_2) - H(\ul X_1-\ul X_2) \big] 
 H(\ul X_2-\ul X_3) \cdots H(\ul X_n-\ul X_1)= \\
&\int d\ul X_1 \cdots d\ul X_n H(\ul X_1-\ul X_2) \cdots H(\ul X_n-\ul X_1) \\
&+ n \int d\ul X_1 \cdots d\ul X_n Q_0(\ul X_1-\ul X_2)
 H(\ul X_2-\ul X_3) \cdots H(\ul X_n-\ul X_1) \ ,
\end{split}
\eeq
where in the last step Eq.~(\ref{corrQ}) has been used:
\beq
\int du_1 du_2 \, \r(u_1) \r(u_2)
\big[ h(r,u_1-u_2) - H(r) \big] = F_0(r)^m - G(r) = Q_0(r) \ .
\eeq
Collecting all the terms with different $n$ one has
\beq
\begin{split}
\frac{1}{2} \sum_{n\geq 3} \frac{(-1)^n}{n} \Tr[h\r]^n &\sim
\frac{1}{2} \sum_{n\geq 3} \frac{(-1)^n}{n} \wh \r^n \Tr H^n 
+ \frac{\wh\r^3}{2} \int d\ul X_1 d\ul X_2 d\ul X_3 Q_0(\ul X_1-\ul X_2) H(\ul X_2-\ul X_3) 
\times\\ &\times
\sum_{n \geq 3} (-1)^n \wh\r^{n-3} \int d\ul X_4 \cdots d\ul X_n H(\ul X_3-\ul X_4)
 \cdots H(\ul X_n-\ul X_1) = \\
= \frac{1}{2} \sum_{n\geq 3} \frac{(-1)^n}{n}& \wh \r^n \Tr H^n
- \frac{\wh\r^3}{2} \int d\ul X_1 d\ul X_2 d\ul X_3 
Q_0(\ul X_1-\ul X_2) H(\ul X_2-\ul X_3) C(\ul X_3-\ul X_1) \ .
\end{split}
\eeq
Substituting the expression of $Q_0(r)$ and recalling that from the definition
of $C(\ul X)$ one has 
\beq
\wh \r \int d\ul Z H(\ul X-\ul Z) C(\ul Z-\ul Y) = H(\ul X-\ul Y) - C(\ul X-\ul Y) \ ,
\eeq
the following result is obtained (in any dimension $d$):
\beq
\label{inttermD}
\frac{1}{2} \sum_{n\geq 3} \frac{(-1)^n}{n} \Tr[h\r]^n
=\frac{1}{2} \sum_{n\geq 3} \frac{(-1)^n}{n} \wh \r^n \Tr H^n 
-N \wh\r Q_m \sqrt{A} y \Si_d(D) [ H(D)-C(D)] \ .
\eeq

\section{First order free energy}
\label{sec2:freeenergy}

\begin{figure} 
\centering 
\includegraphics[width=.55\textwidth,angle=0]{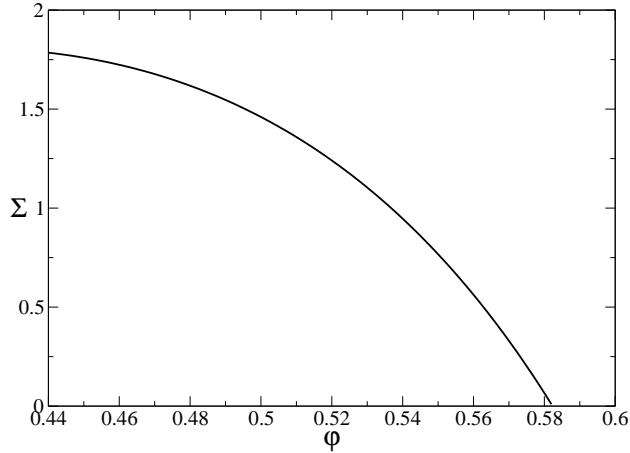}
\caption[The equilibrium complexity as a function of the packing fraction]
{The equilibrium complexity $\Si(\f)$ as a function of the packing fraction.}
\label{fig:Sieq} 
\end{figure} 

Substituting Eq.s~(\ref{GtermD}),~(\ref{GlogGtermD}) and (\ref{inttermD}) in Eq.~(\ref{HNCfree2})
one obtains the following expression for
the HNC free energy at first order in $\sqrt{A}$:
\beq
\begin{split}
\b F &= \frac{\b \Psi}{N} = \b F_0(A) + \b F_{eq}[G(r)] + \b \D F[A,G(r)] \ , \\
\b F_{eq} &= \frac{\wh\r}{2} \int d^d r \, \big[ G(r) \log G(r) - G(r) + 1 \big] + \log \wh\r -1 \\
&+ \frac{1}{2\wh\r} \int \frac{d^d k}{(2\p)^d} 
\left[ - \log[1+\wh H(k)] + \wh H(k) - \frac{1}{2} \wh H(k)^2 \right] \ , \\
\b F_0 &= \frac{d}{2} (1-m) \log (2\p A) + \frac{d}{2} (1-m) 
- \frac{d}{2} \log m \ , \\
\b \D F &= \wh\r Q_m \sqrt{A} \Si_d(D) \, G(D)
 \big[ \log G(D) - 1 - H(D) + C(D) ) \big] \ ,
\end{split}
\eeq
where $Q_m = Q_0 (1-m) + o((m-1)^2)$, $Q_0 \sim 0.638$ and the Fourier transform has been defined as
\beq
\label{Fdef}
\wh H(k) = \wh\r \int dr \, e^{i kr} H(r) \ .
\eeq

At the first order in $\sqrt{A}$ one only needs to know the function $G(r)$ determined by the
optimization of the free energy at the zeroth order in $\sqrt{A}$, \ie the usual free energy
$F_{eq}[G(r)]$: it satisfies the HNC equation $\log G(r) = H(r) - C(r)$.  Substituting this
relation in $\b \D F$ one simply obtains $\b \D F = -\wh\r Q_m \sqrt{A} \Si_d(D) \, G(D)$.

The derivative w.r.t. $A$ leads to the following expression for the cage radius:
\beq
\label{Amd}
\sqrt{A^*} = \frac{1-m}{Q_m} \frac{d}{\wh\r \Si_d(D) G(D)} \ ,
\eeq
which in $d=3$ becomes, defining again $Y=G(D)$:
\beq
\label{Am}
\frac{\sqrt{A^*}}{D} = \frac{1-m}{Q_m} \frac{1}{8 \, \f \, Y(\f)} \ ,
\eeq
where $\f = \frac{\pi D^3 \wh \r}{6}$ is the {\it packing fraction}. Substituting this result
in $\b \D F$ one has $\b \D F(A^*) =d(m-1)$.

Finally, the expression for the replicated free energy in $d=3$ is
\beq
\label{Phim}
\b \Phi(m,\f) = \b F_{eq}(\f) + \frac{3}{2} (1-m) \log [ 2\p A^*(m)] 
+ \frac{3}{2}(m-1) - \frac{3}{2} \log m \ .
\eeq
Note that for Hard Spheres one has $\b F_{eq}(\f) = -S(\f)$, $S$ being the total entropy 
of the liquid. Then
\beq
\label{Sm}
\begin{split}
\b f^*(&m,\f) = \frac{\partial \b \Phi}{\partial m} = 
-\frac{3}{2} \log [2\p A^*(m)]
+ \frac{3}{2} (1-m) \frac{d \log A^*(m)}{dm} + \frac{3}{2} \frac{m-1}{m} \ , \\
\Si(m&,\f) = m \b f^* - \b\Phi = S(\f) - \frac{3}{2} \log [ 2\p A^*(m)]
+\frac{3m}{2} (1-m) \frac{d \log A^*(m)}{dm}+ \frac{3}{2} \log m \ .
\end{split}
\eeq
For small enough density the system is in the liquid phase and  $m$ is equal to 1.
For $m=1$ one has:
\beq
\label{Sm1}
\begin{split}
&\frac{\sqrt{A^*(1)}}{D} = \frac{1}{8 Q_0 \, \f \, Y(\f)} \ , \\
&S_{vib}(\f) \equiv -\b f^*(1,\f) = \frac{3}{2} \log [2\p A^*(1)] \ , \\
&\Si(\f) = S(\f) - S_{vib}(\f) \ .
\end{split}
\eeq
This allows for a computation of $\Si(\f)$ once $S(\f)$ and $Y(\f)$ are known.
Note that $1+4\f Y(\f)=\b P/\r=-\f \frac{\partial S}{\partial \f}$, so a model for $S(\f)$
(or $Y(\f)$) is enough to determine all the quantities of interest.

\begin{figure} 
\centering 
\includegraphics[width=.55\textwidth,angle=0]{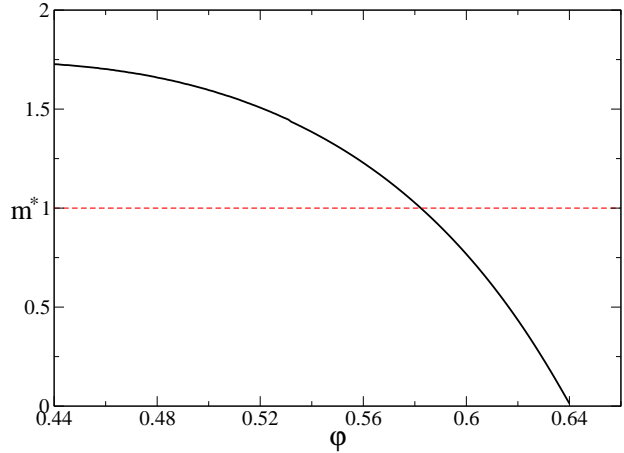}
\caption[Phase diagram of the molecular liquid]
{Phase diagram of the molecular liquid. For $m<m^*$ (full line)
the system is in the liquid phase, for $m>m^*$ it is in the glass phase.
}
\label{fig:mstar} 
\end{figure} 

\subsection{Results from the HNC free energy}
\label{sec2:results}

The functions $S(\f)$ and $Y(\f)$ have been computed solving numerically
the classical HNC equation for the 
Hard Sphere liquid up to $\f=0.65$.
This allows to compute $\b\Phi(\f,m)$ and gives access to all the thermodynamic quantities
using Eq.s~(\ref{Sm}) and (\ref{Sm1}).
In this section we discuss the results of this computation. The sphere
diameter will be set to $D=1$ in the following.

\subsubsection{Equilibrium complexity}

The equilibrium complexity $\Si(\f)$ is given by Eq.~(\ref{Sm1}). It is reported
in Fig.~\ref{fig:Sieq}. A complexity $\Si \sim 1$ is obtained as found in previous
calculations in Lennard-Jones systems~\cite{MP99,MP00,CMPV99,MP99b}, as well as in
the numerical simulations~\cite{CMPV99,SKT99}. The complexity vanishes at $\f_K = 0.582$, that
is the ideal glass transition density --or Kauzmann density-- predicted by
the HNC equations.

\subsubsection{Phase diagram in the $(\f,m)$ plane}

As discussed above, it exists a value of $m$, $m^*(\f)$, such that
for $m < m^*(\f)$ the system is in the liquid phase, and for $m>m^*(\f)$ is
is in the glassy phase. It is the
solution of $\Si(m,\f)=0$, where $\Si(m,\f)$ is given by
Eq.~(\ref{Sm}).
In Fig.~\ref{fig:mstar} $m^*$ is reported as a function of $\f$.
Clearly, $m^*=1$ at $\f=\f_K$ and $m^* < 1$ for $\f > \f_K$.
$m^*$ vanishes linearly at $\f_c=0.640$. As will be shown in the
following, above this value of $\f$ the glassy state does not
exist anymore.

\subsubsection{Thermodynamic properties of the glass}

\begin{figure} 
\centering 
\includegraphics[width=.55\textwidth,angle=0]{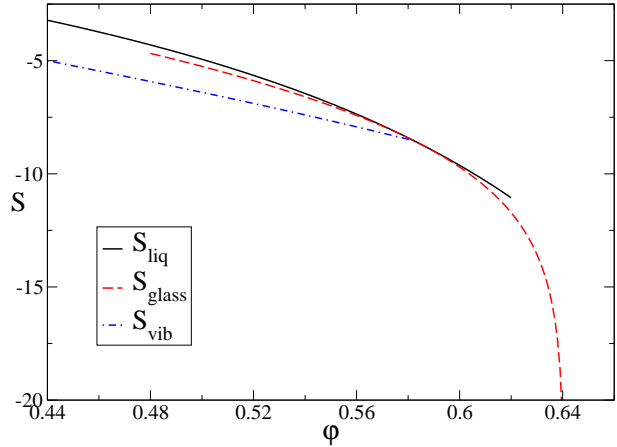}
\caption[Entropy of the liquid and of the glass]
{Entropy of the liquid (full line) and of the glass
(dashed line). The two curves intersect at $\f_K=0.582$ where
they are tangent and consequently the pressure is continuous
at the glass transition. The entropy of the glass goes to 
$-\io$ at $\f=\f_c=0.640$, so the glassy phase does not exist above
$\f_c$. The dot--dashed line is the entropy of the equilibrium
states of the liquid, $S_{vib}(\f)=S(\f)-\Si(\f)$.
}
\label{fig:entropy} 
\end{figure}
The knowledge of the function $m^*(\f)$ allows to compute
the entropy of the glass. Indeed, the free energy does not
depend on $m$ in the whole glassy phase, and it is continuous
along the line $m=m^*(\f)$, so the entropy of the glass is given by
\beq
\label{Sglass}
S_{glass}(\f)=-\b F_{glass}(\f)=-\frac{\b \Phi(m^*(\f),\f)}{m^*(\f)}
\eeq
This relation is true for $m^* < 1$. Below $\f_K$ one has $m^* > 1$
and the liquid phase is the stable one. Eq.~(\ref{Sglass}) for $m^*>1$ 
gives the entropy of the lowest states in the free energy landscape 
(see below) and can be regarded as the analytic
continuation of the glass entropy below $\f_K$. The reader should notice that
the glass phase for $m^*>1$ does not have a simple physical meaning and the interesting part of
the curves for the glass is in the region $\f>\f_K$.

In Fig.~\ref{fig:entropy} the entropies of the liquid and the
glass are reported as functions of the packing fraction. The glass phase becomes
stable above $\f_K = 0.582$; note that the entropy of the glass is
{\it smaller} than the entropy of the liquid, {\it i.e.} its free
energy is {\it bigger} than the free energy of the liquid.
The same happens also in Lennard-Jones systems and in mean-field
spin glass systems. However the physically relevant parts of the curves are the liquid one for
$\f<\f_K$ and the glassy one for $\f>\f_K$.

\begin{figure} 
\centering 
\includegraphics[width=.55\textwidth,angle=0]{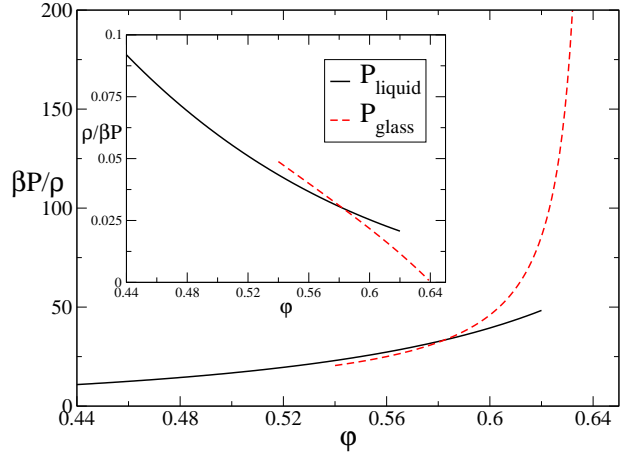}
\caption[Reduced pressure of the liquid and the glass]
{Reduced pressure $\b P/\r$ of the liquid and the glass as functions
of the packing fraction. The pressure is continuous at $\f_K$.
In the inset, the inverse reduced pressure is plotted; in the glass
phase it is proportional to $\f_c-\f$.
}
\label{fig:pressure} 
\end{figure}

The reduced pressure,
\beq
\frac{\b P}{\r} = -\f\frac{\partial S}{\partial \f} \ ,
\eeq
is reported in Fig.~\ref{fig:pressure}.
It is continuous at $\f_K$ and the glass transition
is a second order transition from the thermodynamical point of view.
Note that the pressure in the glass phase is well described by a
power law and it has a simple pole at $\f_c$:
\beq
\frac{\b P_{glass}}{\r} \propto \frac{1}{\f_c-\f} \ ,
\eeq
as one can see from the inset of Fig.~\ref{fig:pressure} where the inverse
reduced pressure is plotted as a function of~$\f$.

For $\f \rightarrow \f_c$ the pressure of the glass diverges and its
compressibility 
$\chi=\frac{1}{\f}\frac{\partial \f}{\partial P}$ vanishes and consequently
 $\f_c$ is the maximum density allowed for a disordered
state, {\it i.e.} it can be identified as the 
{\it random close packing density}.
The value $\f_c=0.640$ is in very good agreement with the values
reported in the literature.
Note that the compressibility jumps downward on increasing $\f$ across
$\f_K$, {\it i.e.} the compressibility of the glass is smaller than the
compressibility of the liquid.

\subsubsection{Cage radius}

\begin{figure} 
\centering 
\includegraphics[width=.55\textwidth,angle=0]{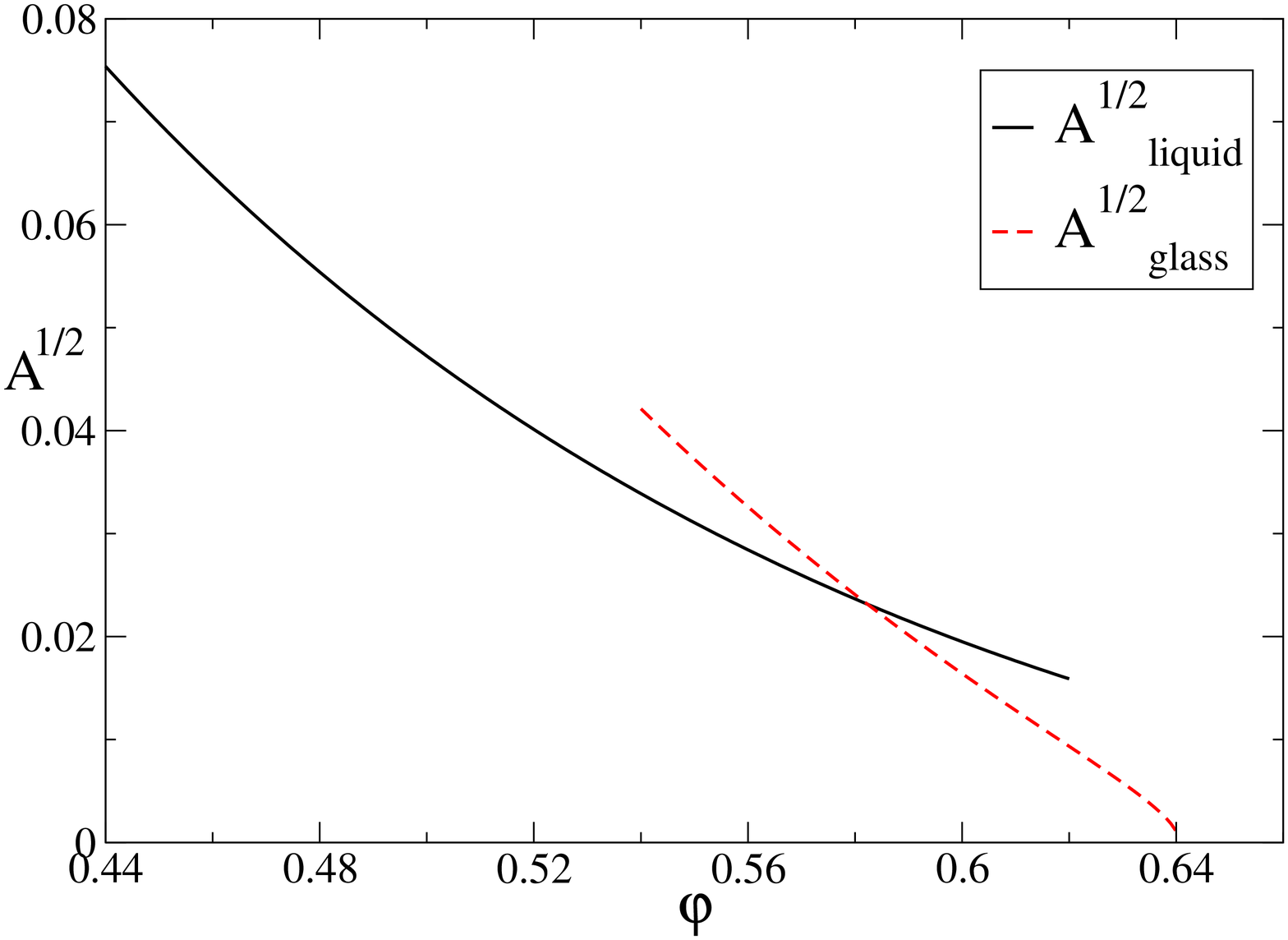}
\caption[Cage radius of the liquid and the glass]
{Cage radius $\sqrt{A}$ (in units of $D$) 
in the liquid and in the glass phase as function of $\f$.}
\label{fig:A} 
\end{figure}

The cage radius is given as a function of $m$ in Eq.~(\ref{Am}). 
In Fig.~\ref{fig:A} the cage radius in the liquid phase,
$\sqrt{A^*(1)}$, see Eq.~(\ref{Sm1}), and the cage radius in the
glass phase, defined as $\sqrt{A^*(m^*)}$, are reported. As $Q_m \sim \sqrt{\p/4m}$
for $m\sim 0$, the cage radius vanishes as $\sqrt{m^*}$ for $m^* \sim 0$,
{\it i.e.} it is proportional to $\sqrt{\f_c-\f}$.
The vanishing of the cage radius for $\f \rightarrow \f_c$ means that
at $\f_c$ each sphere is in contact with its neighbors, that is consistent
with the interpretation of $\f_c$ as the random close packing density.

\subsubsection{Complexity of the metastable states}

\begin{figure} 
\centering 
\includegraphics[width=.55\textwidth,angle=0]{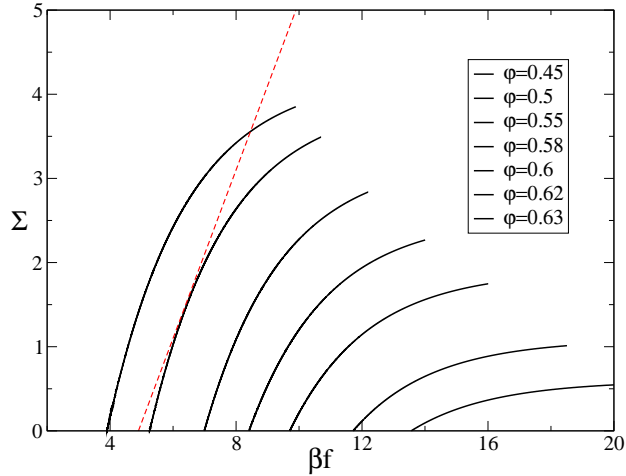}
\caption[Complexity of the metastable states]{
Complexity of the metastable states as a function of their free energy
$\b f$ for different values of $\f$.
From left to right, $\f=0.45, 0.5, 0.55, 0.58, 0.6, 0.62, 0.63$.
The curves are truncated arbitrarily at high $\b f$.
The dashed line has slope~$1$.
}
\label{fig:Sif} 
\end{figure}

From the parametric plot of $\b f^*(m,\f)$ and $\Si(m,\f)$ given in
Eq.~(\ref{Sm}) by varying $m$, one can reconstruct the function 
$\Si(\b f)$ for each value of the packing fraction. 
This function is reported in
Fig.~\ref{fig:Sif} for some values of $\f$ below and above $\f_K$.
The function $\Si(\b f)$ vanishes at a certain value $\b f_{min}$,
that is given by Eq.~(\ref{Sglass}). The saddle-point equation that
determines the free energy of the equilibrium states is, from
Eq.~(\ref{Zm1}),
\beq
\label{saddleeq}
\frac{d \Si(\b f)}{d \b f} = 1 \ .
\eeq
From Fig.~\ref{fig:Sif} it is clear that this equation has a solution
$f^* > f_{min}$ for $\f < \f_K = 0.582$. 
Above $\f_K$ Eq.~(\ref{saddleeq}) does not have a solution so the
saddle point is simply $f^* = f_{min}$ and the systems goes in
the glass state. In this sense, the free energy $f_{min}$ of the
lowest states below $\f_K$ can be regarded as the analytic 
continuation of the free energy of the glass, see 
Fig.~\ref{fig:entropy}.
The curves $\Si(\b f)$ in Fig.~\ref{fig:Sif} have been truncated 
arbitrarily at high $\b f$. One should perform a consistency check to 
investigate where the higher free energy states become unstable
(\ie, to compute $f_{max}$). This is not trivial and is left for
future work.

\section{Correlation functions}
\label{sec2:correlations}

The replica approach also allows the study of the pair distribution function 
$\tilde g(r)$ in the glass state. 
In principle a full computation would require the 
evaluation of the corrections proportional to $\sqrt{A}$ in the correlation 
functions of a molecule. 
However these terms will be neglected, as one can argue that they are small, 
and one can start again from the
simple {\it ansatz} (\ref{rrho}), (\ref{gprod}) for the correlation function of the
molecules, in which the information on the shape of the molecule is only encoded
in the function $\r(x)$;
these corrections should be physically more relevant and interesting.

As will be shown in the following, the correlation function of the spheres in the glass is very
similar to the one in the liquid but develops an additional strong peak (that becomes a
$\d$-function at $\f_c$) around $r=D$.  The integral of the latter peak is related to the average
coordination number of the random close packings.

\subsection{Expression of $\tilde g(r)$ in the glass phase}

The following form for the pair distribution function
of the molecular liquid was assumed in Eq.s~(\ref{rrho}) and (\ref{gprod}):
\beq
\label{r2}
\r_2(x,y)=\r(x) g(x,y) \r(y) =
\wh\r^2 \int dX dY \prod_{a=1}^m \r(x_a-X) g(|x_a-y_a|) \r(y_a-Y) \ .
\eeq
The pair correlation $\tilde g(r)$ of a single replica is obtained 
integrating over the coordinates of all the replicas but one:
\beq
\tilde g(|\ul x_1-\ul y_1|) = \wh \r^{-2} \int d\ul x_2\cdots d\ul x_m 
d\ul y_2 \cdots d\ul y_m \r_2(x,y) \ .
\eeq
Using Eq.~(\ref{r2}) one gets, with some simple changes of variable:
\beq
\tilde g(r) = g(r) \int d\ul u d\ul v \r(\ul u) \r(\ul v) F_0(|\ul r+\ul u-\ul v|)^{m-1} \ ,
\eeq
where $F_0(r)$ is defined in Eq.~(\ref{Fp}). The HNC free energy is
optimized by $g(r)=G(r)^{1/m}$, where $G(r)$ is the HNC pair correlation.
Thus the following expression for the pair correlation of a single replica is obtained:
\beq
\label{gvetro}
\begin{split}
&\tilde g(r) = G(r)^{\frac{1}{m}} \int d\ul u \frac{e^{-\frac{u^2}{4A}}}{(\sqrt{4\p A})^d} 
F_0(|\ul r+\ul u|)^{m-1} \ , \\
&F_0(r) =  \int d\ul u \frac{e^{-\frac{u^2}{4A}}}{(\sqrt{4\p A})^d} G(|\ul r+\ul u|)^{\frac{1}{m}} \ .
\end{split}
\eeq
For $m=1$, {\it i.e.} in the liquid phase, this function is trivially equal
to $G(r)$. This is not the case in the glass phase where $m<1$.

\subsection{Small cage expansion of the correlation function}

To expand Eq.~(\ref{gvetro}) for small $A$, note first that, if $r \neq D$,
the function $g(r+u)$ can be expanded in powers of $u$, and the first correction
to $\tilde g(r)$ is of order $A$. Then, as before, one can concentrate only on what happens
around $r=D$.
As already discussed in section~\ref{sec2:smallcageexp}, around $r=D$ one has,
as in Eq.~(\ref{F0sing}), $G(r) \sim Y \th(r-D)$ and
\beq
F_0(r) \sim Y^{\frac{1}{m}} \Th\left(\frac{r-D}{\sqrt{4A}}\right) \ ,
\eeq
and Eq.~(\ref{gvetro}) becomes
\beq
\tilde g(r) = Y \th(r-D) \int d\ul u \frac{e^{-\frac{u^2}{4A}}}{(\sqrt{4\p A})^d} 
\Th\left(\frac{|\ul r+\ul u|-D}{\sqrt{4A}}\right)^{m-1} \ .
\eeq
Applying the same argument used in section~\ref{sec2:smallcageexp} when studying 
the function $F_0(r)$ in dimension $d > 1$, it can be shown that the integration over
the coordinates $u_\m$, $\m \neq 1$, gives a contribution $O(A)$. Then, 
in any dimension $d$:
\beq
\label{gvetro2}
\begin{split}
\tilde g(r) &\sim Y \th(r-D) \int_{-\io}^\io du \frac{e^{-\frac{u^2}{4A}}}{\sqrt{4\p A}}
\Th\left(\frac{r+u-D}{\sqrt{4A}}\right)^{m-1} \\
&= G(r) \left\{ 1 + \int_{-\io}^\io \frac{dt}{\sqrt{\p}} 
e^{-\left(\frac{r-D}{\sqrt{4A}}-t\right)^2}
 \left[ \Th(t)^{m-1} - 1 \right]\right\} \ ,
\end{split}
\eeq
defining the reduced variable $t=\frac{r+u-D}{\sqrt{4A}}$.
The second term in the latter expression is a contribution localized
around $r = D$.

\subsection{Number of contacts}

To compute the average number of contacts, recall that the 
average number of particles in a shell $[r,r+dr]$, if there
is a particle in the origin, is given by
\beq
dn(r) = \O_d r^{d-1} \wh\r \, \tilde g(r) dr \ .
\eeq
Thus the number of contacts can be obtained from the correlation function $\tilde g(r)$.
While the full computation of the correlation function is rather involved, 
here only the second term in Eq.~(\ref{gvetro2}) will be considered;
this term is proportional to a Gaussian with variance $O(\sqrt{A})$ that becomes 
a $\d(|r|-D)$-function in the limit $A \rightarrow 0$.

The value of the number of spheres in contact with the sphere in the
origin is given by
\beq
z = \O_d \wh\r \int_D^{D+O(\sqrt{A})} dr \, r^{d-1} \tilde g(r) \ .
\eeq
The first term in Eq.~(\ref{gvetro2}) gives a contribution $O(\sqrt{A})$
that can be neglected. With the approximation $r \sim D$ and $G(r) \sim Y$ at the
leading order in $\sqrt{A}$ one obtains,
defining the variable $\epsilon = \frac{r-D}{\sqrt{4A}}$,
\beq
z = \O_d D^{d-1} \wh\r Y  \sqrt{4A} \int_0^\io d\epsilon 
\int_{-\io}^\io \frac{dt}{\sqrt{\p}} 
e^{-(\epsilon-t)^2} \left[ \Th(t)^{m-1} - 1 \right] \ .
\eeq
Recalling that
\beq
\frac{1}{\sqrt{\p}}\int_0^\io d\epsilon \, e^{-(\epsilon-t)^2} = \Th(t) \ ,
\eeq
observing that $\int_{-\io}^\io dt \, \big[ \Th(t) - \th(t) \big] = 0$,
and using Eq.~(\ref{Amd}), it follows that
\beq
\label{nD}
z = \Si_d(D) \wh\r Y  \sqrt{4A}
\int_{-\io}^\io dt \, \Th(t) \left[ \Th(t)^{m-1} - 1 \right] 
=  \Si_d(D) \wh\r Y  \sqrt{4A} Q_m = 2 d (1-m) \ .
\eeq
This is the expression of the average number of contacts at the leading
order in $\sqrt{A}$, to be computed at $m=m^*$ in the glass phase.
At $\f=\f_c$, where $m^*=0$, each sphere has on average $2d$ contacts. 
This is exactly what is found in numerical simulations.

Note that this result is independent of the particular expression that has been
chosen for $S(\f)$, $Y(\f)$ and $G(r)$, {\it i.e.} it might hold beyond the 
choice of HNC equations for the molecular liquid provided that
the expression~(\ref{Amd}) for the cage radius is correct.

The condition $z\geq 2d$ is required for the mechanical stability of the 
packings as can be understood by mean of a very simple argument~\cite{Al98}.
Consider the network of forces between the spheres in the packed state.
For Hard Spheres the forces can be considered as independent from the displacements:
indeed, for two spheres in contact an infinitesimal displacement produces a finite
change in the force\footnote{This can be shown for example by considering a potential
$V(r)=r^{-n}$ for $n\to\io$.}. Thus to each pair $\la ij \ra$ of spheres in contact
one can associate a scalar force $F_{ij}=F_{ji}$.
The forces are determined by the linear system
$\sum_{j} F_{ij} (\ul r_i - \ul r_j) = 0$, $\forall i$. 
The total number of forces, if each sphere has $z$
contacts, is $zN/2$, while the number of equations is $N d$. Thus the condition
$z \geq 2d$ is necessary for the system to have a solution.

%%%%%%%%%%%%%%%%%%%%%%%%%%%%%%%%%%%%%%%%%%%%%%%%%%%%%%%%%%%%%%%%%%%%%%%%
%%%%%%%%%%          DISCUSSION         %%%%%%%%%%%%%%%%%%%%%%%%%%%%%%%%%
%%%%%%%%%%%%%%%%%%%%%%%%%%%%%%%%%%%%%%%%%%%%%%%%%%%%%%%%%%%%%%%%%%%%%%%%

\section{Discussion}
\label{sec2:discussion}

The results will now be compared with related ones that appeared in
the literature. The main obstacle for a quantitative comparison is that
the HNC equations are known to yield a not very good description of the 
Hard Sphere liquid at high density~\cite{Hansen}; typically one would obtain the right curves if
one shifts the value of $\f$ of a quantity of order 0.03.  Therefore, only
a {\it qualitative} comparison of the results coming from 
the HNC equations with the results of numerical simulations is possible.
However, note that, although the 
expressions~(\ref{Am}), (\ref{Phim}) for the replicated free energy
have been derived starting from the expression (\ref{HNCfree}) for
the HNC free energy, the final result depends only on the equilibrium
entropy of the liquid $S(\f)$. It is interesting then, for the purpose of 
comparing the results with experiments and numerical simulations, 
to consider a more accurate model for $S(\f)$ in the liquid phase. 
We repeated the calculations of section~\ref{sec2:results}
substituting the Carnahan--Starling (CS) entropy~\cite{Hansen} 
\beq
\begin{split}
&S_{CS}(\f)=-\log\left(\frac{6\f}{\p e}\right)-\frac{4\f-3\f^2}{(1-\f)^2} \ , \\
&Y_{CS}(\f)=\frac{1-\frac{1}{2}\f}{(1-\f)^3} \ .
\end{split}
\eeq
instead of the HNC entropy in Eq.s~(\ref{Phim}), (\ref{Am}).
All the results of section~\ref{sec2:results} are qualitatively reproduced
using the CS entropy, but the latter gives results in better agreement
with the numerical data. However, this procedure is not completely
consistent from a theoretical point of view: one should always keep
in mind that the aim of this work is not to present a quantitative theory,
but only to show that the replica approach yields a reasonable 
qualitative scenario for the glass transition in Hard Sphere systems.

\subsection{Complexity of the liquid and Kauzmann density}

\begin{figure} 
\centering 
\includegraphics[width=.55\textwidth,angle=0]{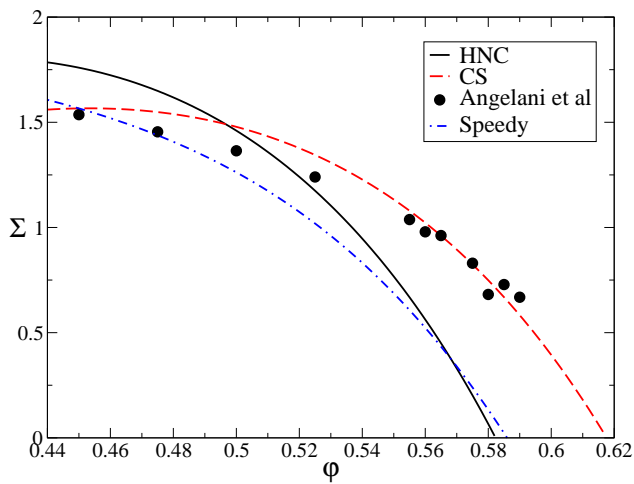}
\caption[Comparison of the results for the equilibrium complexity]
{Equilibrium complexity $\Si(\f)$ as a function of the packing fraction.
The full line is from the HNC equation of state (see Fig.~\ref{fig:Sieq}), 
the dashed line is from the
Carnahan--Starling equation of state. The black dots are from the numerical
computation of Angelani {\it et al.}~\cite{Luca05}, the dot--dashed line
is extrapolated from the numerical data reported by Speedy~\cite{Sp98}.
}
\label{fig:Sicomp} 
\end{figure} 
In Fig.~\ref{fig:Sicomp} the equilibrium complexity $\Si(\f)$,
obtained substituting the HNC and the CS expression for $S(\f)$ and $Y(\f)$ 
in Eq.~(\ref{Sm1}), is reported. The results are compared with recent numerical results
of Angelani {\it et al.}~\cite{Luca05} obtained on a $50:50$ binary mixture 
of spheres (to avoid crystallization) with diameter ratio equal to $1.2$:
the vibrational entropy was estimated using the procedure described 
in~\cite{CMPV99,AFST05} and the complexity was computed as $S(\f)-S_{vib}(\f)$.
A quantitative comparison is difficult here because in the case of a mixture
there can be corrections related to the mixing entropy, $S_{mix} \sim \log 2$.
Nevertheless the data are in good agreement with our results. A detailed
comparison would require the extension of the computation to binary mixtures
following~\cite{CMPV99}.

Another numerical estimate of $\Si(\f)$ was previously reported by 
Speedy~\cite{Sp98}, who
rationalized his numerical data assuming a Gaussian distribution of
states and a particular form for the vibrational entropy inside a
state. The free parameters were then fitted from the liquid equation
of state. The curve obtained by Speedy also agrees with our results.

Both the HNC and the CS estimates of the Kauzmann density ($\f_K=0.582$ and
$\f_K=0.617$ respectively) fall, as it should be,
between the Mode--Coupling dynamical transition that is
$\f_{MCT}\sim 0.56$~\cite{GS91,vMU93}, and the Random Close Packing
density that is estimated in the range $\f=0.64\div 0.67$, see e.g.~\cite{Be83}.

A computation of $\Si(\f)$ based on very similar ideas was presented
in~\cite{CFP98}, where a very similar estimate of $\f_K \sim 0.62$ was
obtained. However in \cite{CFP98} the complexity was found to be
$\Si \sim 0.01$, {\it i.e.} two orders of magnitude smaller than
the one obtained from the numerical simulations.
This negative result is probably due to some technical problem
in the assumptions of~\cite{CFP98}.

\subsection{Equation of state of the glass}

\begin{figure} 
\centering 
\includegraphics[width=.55\textwidth,angle=0]{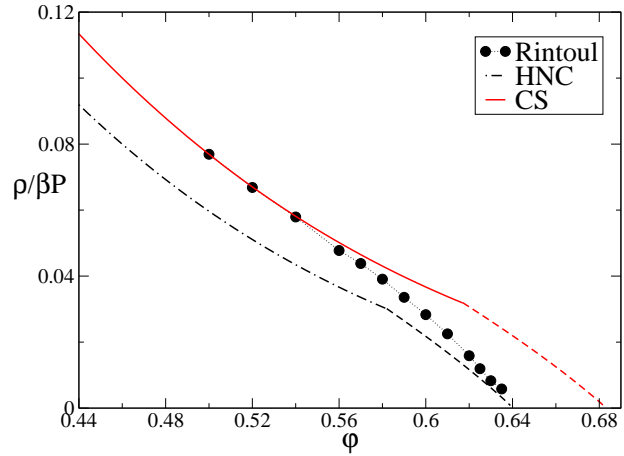}
\caption[Comparison of the equation of state with numerical data]
{Inverse reduced pressure $\frac{\r}{\b P}$ of the Hard Sphere 
liquid as a function of $\f$. The black dots are from the simulation
of Rintoul and Torquato \cite{RT96}. The full line is obtained from
the CS equation of state while the dot--dashed line is from the HNC
equation of state. The dashed parts of the two curves correspond to
the (ideal) glass phase. Note that all the curves are quasi--linear 
functions of $\f$ in the glass phase.
}
\label{fig:Pcomp} 
\end{figure}
In Fig.~\ref{fig:Pcomp} the
numerical data for the pressure of the Hard Sphere liquid 
at high $\f$, obtained by Rintoul and Torquato~\cite{RT96}, are reported.
The data were obtained extrapolating at long times the relaxation
of the pressure as a function of time after an increase of density
starting from an equilibrated configuration at lower density.
Also reported are the curves of the pressure as a function of the
density obtained from the HNC and CS equations, both in the liquid
and in the glass state.

The agreement of the HNC curve with the data is not very good even
in the liquid phase, due to the modest accuracy of the HNC equation of state.
However, its qualitative behavior is similar to
the numerical data, and in particular the quasi--linear behavior of
the inverse reduced pressure in the glass phase found in~\cite{RT96,Sp98},
$\frac{\r}{\b P} \propto \f_c-\f$,
is reproduced by the HNC curve. The HNC pressure of the glass diverges
at $\f_c=0.640$ as discussed in section~\ref{sec2:results}; the latter
is the HNC estimate of the random close packing density.

The CS curve describes well the pressure in the liquid phase~\cite{Hansen}.
Comparing the curve with the data of Rintoul and Torquato, one notices that the glass
transition happens in the numerical simulation at a density $\f_g\sim 0.56$
smaller than the one predicted by the CS 
curve\footnote{The authors of \cite{RT96} interpreted their
data as showing no evidence for a glass transition, the pressure being
a differentiable function of $\f$. However, as recognized in
\cite{RLSB98}, their data are better described by a broken curve
showing a glass transition around $\f_g=0.56$.}, $\f_K=0.617$,
and very close to the Mode--Coupling transition density, $\f_{MCT}\sim 0.56$.
This is not surprising, since the relaxation time grows
fast on approaching the ideal glass transition; at some point it becomes 
larger than the experimental time scale and the liquid falls 
out of equilibrium becoming a {\it real} glass. It is likely that the data of 
Ref.~\cite{RT96} describe the pressure of a real {\it nonequilibrium} glass, while 
the replica computation gives the pressure of the ideal {\it equilibrium} glass,
that cannot be reached experimentally in finite time.

\subsection{Random close packing}

Both the HNC and CS equations predict the existence of a {\it random close packing} density $\f_c$
where the pressure and the value of the radial distribution function $\tilde g(r)$ in $r=D$ diverge.
The HNC estimate is $\f_c=0.640$, in the range of the values ($\f_c=0.64\div 0.67$) reported in the
literature. The CS estimate is $\f_c=0.683$ and it is also a value consistent with numerical
simulations.

The reader should notice that the theoretical value for $\f_c$ is related to the 
{\it ideal} random close packing;
however the states corresponding to this value of $\f_c$ can be reached by local algorithms, like
most of the algorithms that were used in the literature, in a time that should diverge exponentially
with the volume. Some caution should be taken in using the data obtained by numerical simulations.
The question of which is the value of the density that can be obtained in large, but finite amount of time
per particle is very interesting and more relevant from a practical point of view: however we plan
to study it at a later time.

Note that the result for the mean coordination number $z$ 
of section~\ref{sec2:correlations}, that gives $z=6$ at $\f=\f_c$ in $d=3$,
is {\it independent} of the particular form chosen for $S(\f)$,
and thus is valid for both the HNC and CS equations of state.
The value $z=6$ has been reported in many studies
\cite{Be72,Ma74,Po79,Al98,SEGHL02}.

\subsection{Conclusions}

The replica method of~\cite{MP99,Mo95} has been succesfully applied to
the study of the ideal glass transition of Hard Spheres, and 
in general of potentials such that the pair distribution
function $g(r)$ shows discontinuities, starting from
the replicated HNC free energy and expanding it at first
order in the cage radius $\sqrt{A}$.

This result allowed to compute from first principles the 
configurational entropy of the liquid as well as 
the thermodynamic properties of the glass up to the
random close packing density. The computation is based
on the HNC equation of state, that is known to yield a
poor quantitative description of the liquid state at high
density. Nevertheless, it is found that the qualitative scenario
for the ideal glass transition that emerges from the replicated
HNC free energy is very reasonable. In particular, 
a complexity $\Si \sim 1$, a Kauzmann density $\f_K =0.582$, and
a random close packing density $\f_c=0.64$. All these results
compare well with numerical simulations.

Using, on a phenomenological ground, the Carnahan--Starling
equation of state instead of the HNC equation of state as
input for the calculations, it was also possible to compare 
the results with the high--density pressure data 
of Rintoul and Torquato showing that they are indeed compatible
with the observation of a real glass transition.

Moreover, it was found that the mean 
coordination number in the amorphous packed states is $z=2d$
irrespective of the equation of state used for the liquid,
in very good agreement with the result of numerical 
simulations and with theoretical 
arguments~\cite{Be72,Ma74,SEGHL02,Al98}.

It is worth to note that these results do not {\it prove} the 
existence of a glass transition for the Hard Sphere liquid,
as they derive from a particular approximation for the molecular
liquid free energy (the HNC approximation), and, in general, other 
approximation such as the Percus--Yevick are possible~\cite{Hansen}.

%% file: cap3.tex
\chapter[Correlation between fragility and vibrational properties]
{Correlation between fragility of the liquid and the vibrational
properties of its glass}
\label{chap3}

\section{Introduction}

The identification of the microscopic details that, in a given
glass former, determine the temperature dependence of the
viscosity, and thus the value of the fragility, is a long standing
issue in the physics of supercooled liquids and glassy state.
Large numerical and theoretical effort has been devoted to the
attempt to relate the fragility to the specific interparticle
interactions (e.~g. strong glasses are often characterized by
highly directional covalent bonds, while the fragile one have more
or less isotropic interactions).
The phenomenological relevance of the concept of fragility relies on the
correlations that have been found between this index and other
properties of glass-forming liquids. Examples of these
correlations are the specific heat jump at $T_g$ 
(see Eq.~(\ref{frag-DCp}) and \cite{MA01}), 
the degree of stretching in the
non-exponential decay of the correlation functions in the liquid
close to $T_g$ \cite{Ngai}, the visibility of the Boson peak at
the glass transition temperature \cite{Sokolov}, or the temperature
behavior of the shear elastic modulus in the supercooled liquid state
\cite{Dyre}. Recently a strong correlation between fragility of the liquid and
vibrational properties of its glass has been found~\cite{TS}.

\subsection{Fragility and number of states}

Recently, the attention has
been focused on the possible relation existing between
fragility and the properties of the (free) energy landscape, 
more specifically the (free) energy distribution of the minima
and the properties of the basins of attraction of such minima. 
A key point is the validity of 
the Adam-Gibbs relation (\ref{AGrel}):
\begin{equation}
\label{AG}
\t(T) = \t_\infty \exp \left( \frac{\cal E}{T \Sigma(T)} \right) \ .
\end{equation}
By using the
Adam-Gibbs relation, one could expect to relate fragility to the
properties of $\Sigma(T)$, i.e., to the distribution of basins in
the phase space of the system. For example, many authors proposed
that fragile systems should have an higher number of states, \ie
a larger complexity, with respect to strong 
ones~\cite{An95,Sp99,Sa01,DS01}.

However, this possibility is
frustrated by the lack of knowledge on the parameter ${\cal E}$.
Some theories attempting to compute ${\cal E}$,
summarized in section~\ref{sec1:instantons}, appeared only
recently\footnote{In particular, the papers \cite{Fr05,DSW05} appeared
after this work was completed so their results were not known at the
time this calculation was performed.} and in general the theory 
of the Adam--Gibbs relation is still at an early stage
of development.

Unfortunately, even if a model for $\Sigma(T)$ is chosen, so the total number
of states is fixed, one can obtain the whole
range of experimentally observed fragilities by varying 
${\cal E}$: fragility is related to $\Si(T)$ by 
Eq.~(\ref{frag-DCp}), but the value of $T_g$ depends strongly on $\EE$.
More specifically, in \cite{noifrag} it was observed that for
a large class of models for $\Sigma(T)$ - where $\Sigma(T)$ is a
concave function of $T$ that vanishes at a given temperature
$T_K$ and assumes its maximum $\Si_\io$ at high temperature
(``Gaussian-like models'') - the relevant parameter that actually
determines the fragility is
\begin{equation}
\label{Ddef} D=\frac{\cal E}{T_K \Si_\io} \ .
\end{equation}
For example, if $\Si(T)=\Si_\io \left( 1 - \frac{T_K}{T} \right)$ - the form
that is commonly used to fit experimental data, see Eq.~(\ref{ScFIT}) and 
Fig.~\ref{fig1:Sconf} - 
is substituted in Eq.~(\ref{AG}) and the fragility is calculated from
Eq.~(\ref{mAdef}), one gets
\beq
\frac{m_A}{17} = 17 \log 10 \, D^{-1} + 1 \ .
\eeq
Thus, fragility appears to be determined by the ratio between
${\cal E}$ (measured in units of $k_B T_K$) and the total number
of states $\Si_\io/k_B$; it is related to both the distribution
of minima (through $\Si_\io$) and the characteristic of the
transition path between them (through $\cal E$). The relation
between fragility and phase space properties can be even more
complicated, in those cases where the function $\Sigma(T)$ does
not belong to the Gaussian class.

\subsection{Fragility and vibrational properties of the glass}

\begin{figure}[t]
\centering
\includegraphics[width=.45\textwidth,angle=0]{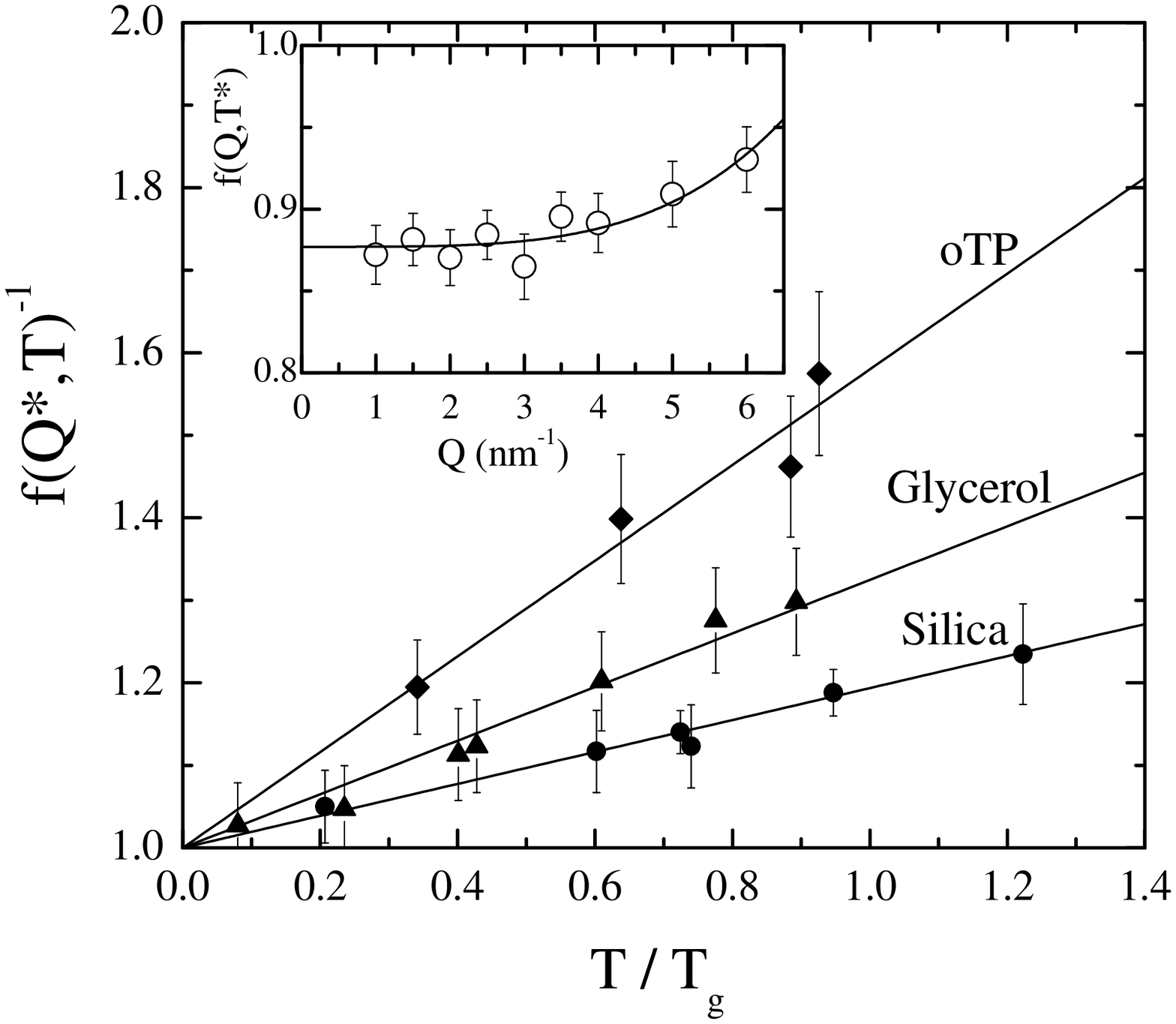}
\includegraphics[width=.45\textwidth,angle=0]{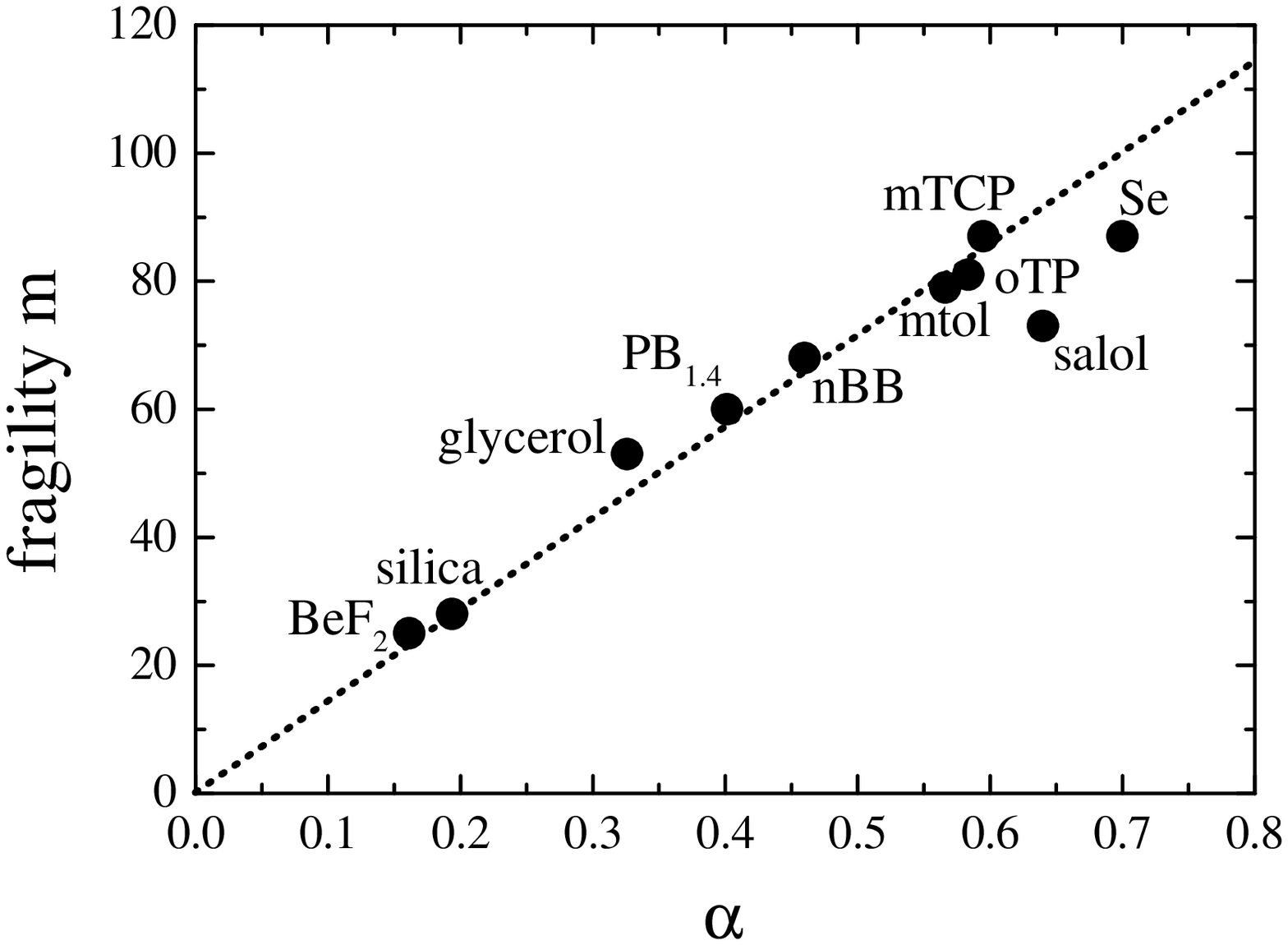}
\vskip-4cm
\caption[Correlation between $\a$ and the fragility from IXS]
{From~\cite{TS}: (Left) The inverse
nonergodicity factor $f(Q^*,T)^{-1}$ for 
$Q^* = 2 \, {\rm nm}^{-1}$ and for three
different substances as a function of $T/T_g$. In the inset, the 
wave vector dependence of $f(Q,T)$ is shown at fixed temperature $T^*$ to 
demonstrate that $f(Q^*,T) \sim \lim_{Q \to 0}f(Q,T)$. 
(Right) Correlation plot of the 
fragility and the index $\a$ defined in Eq.~(\ref{alphaTS}).}
\label{fig3:tullio}
\end{figure}

Recently a strong correlation between fragility
and the vibrational properties of the glass at low temperatures
has been found \cite{TS}. 
The nonergodicity factor $f(k,T)$ defined in Eq.~(\ref{nonergdyn}) was 
measured, by mean of inelastic X-rays scattering, 
in the (nonequilibrium) glass phase after a quench from high 
temperature. In Fig.~\ref{fig3:tullio} the temperature dependence of
$f(k \sim 0,T)$ is shown\footnote{In~\cite{TS} the wave vector was indicated
by $Q$ instead of $k$, so this notation is used in Fig.~\ref{fig3:tullio}}. 
It is found that $f(k\sim 0,T)^{-1}$ is approximately linear in $T/T_g$ and
an index $\a$ is defined as the slope of the curves in Fig.~\ref{fig3:tullio}:
\begin{equation}
\label{alphaTS}
\alpha(T_g) = 
\lim_{k \rightarrow 0} \left. \frac{d [f(k,T)]^{-1}}{d (T/T_g)} \right|_{T=0} \ .
\end{equation}
The index $\a$ has an explicit dependence on $T_g$ 
(as $\frac{d}{d(T/T_g)}=T_g \frac{d}{dT}$). Moreover, it depends on $T_g$ also
because depending on the value of $T_g$ (\ie on the experimental time scale)
different states are selected: in particular, the states that are selected are
the {\it equilibrium} states at $T=T_g$, as exactly at this value of temperature
the system falls out of equilibrium.
The order parameter $f(k,T)$ of a given state
is a measure of the volume of this state in the phase space, and can in principle
depend on the state in which the sistem is frozen below $T_g$.
In particular, in~\cite{TS} an expression of $\a$ in terms
of the harmonic vibrational properties of the states (eigenmodes of
the disordered structure), was derived: and the eigenmodes will depend on the
particular structure in which the system is frozen, that is, on the 
{\it equilibrium structure} at $T=T_g$.

In the right panel of Fig.~\ref{fig3:tullio} it is shown 
that the index $\a$ is strongly correlated with the fragility
index $m_A$. This finding implies the existence of a
relation between three features of the energy landscape: the energy of the
minima, the transition paths between them (that together determine
the fragility) and the Hessian matrix, evaluated at the minima
themselves, that fixes the vibrational properties.

%%%%%%%%%%%%%%%%%%%%%%%%%%%%%%%%%%%%%%%%%%%%%%%%%%
%%%%%%%%%%%%%%%%%%%%%%%%%%%%%%%%%%%%%%%%%%%%%%%%%%

\section{Fragility in mean field $p$-spin models}

As discussed in the first chapter, mean field models such as the 
disordered $p$-spin model provide an useful framework to understanding
many aspects of the glass transition. Using the arguments of sections
\ref{sec1:BBargument} and \ref{sec1:instantons} one can relate the
quantities appearing in the Adam--Gibbs relation, \ie $\Si(T)$ and
$\EE (T)$, to the mean field potential $V(q,T)$ that is expected to
describe short range models at the {\it local} level.
One can then investigate the $p$-spin models as solvable models of ``glass'', 
where the distribution of minima is ``Gaussian-like'' as in real structural
glasses, and both the vibrational properties of the minima and the energy
barrier $\EE(T)$ can be analytically estimated. The $p$-spin models are ``Gaussian-like'', 
in the sense that their complexity - even if the distribution of states is not 
exactly Gaussian - is known to be a concave function of the temperature,
that vanishes at $T_K$ and assumes its maximum at $T_d$, without any inflection point
in between, see~\cite{CS95} and Fig.~\ref{fig1:Scqualit}, yielding a form very
similar to Eq.~(\ref{ScFIT}) for $\Si(T)$.

In this chapter, both the spherical and Ising version of the $p$-spin model
will be investigated, in order to
check whether one can reproduce the correlation between fragility of the
liquid and the vibrational properties of its glass found in
\cite{TS} by studying the geometry of the phase space of these models.
Moreover the question of the existence of a correlation between fragility and 
number of states \cite{Sp99,Sa01,DS01} will be addressed.

The mean field models will be considered as models for the {\it local}
properties of a short range glass, as indicated by the arguments discussed
in sections \ref{sec1:BBargument} and \ref{sec1:instantons}. Then, the existence
of the dynamical transition will be ignored, being an artifact of mean field,
and it will be assumed that it is possible to equilibrate the system below
$T_d$ with a relaxation time following the Adam--Gibbs relation (\ref{AG})
with $\EE$ and $\Si$ determined by the mean field potential $V(q,T)$.
As will be clarified below, the fragility of the models can be varied by
varying the parameter $p$.

\subsection{Definition of the relevant observables}
\label{sec:definitions}

\noindent
It is useful to summarize the definition of all the quantities that will be
computed, that are listed below:
\begin{equation}
\nonumber
\begin{array}{ll}
T_K & \text{Thermodynamical transition temperature} \\
T_g & \text{Glass transition temperature} \\
T_d & \text{Dynamical transition temperature} \\
\Sigma(T_g) & \text{Complexity at $T_g$} \\
m(T_g) & \text{Fragility} \\
\alpha(T_g) & \text{``Volume'' of the equilibrium states at $T_g$} \\
{\cal E}(T_g) & \text{``Barrier height'' at $T_g$}
\end{array}
\end{equation}
Setting $k_B=1$, all the above quantities are either dimensionless or have the
dimension of an energy; in the $p$-spin model -as usual in classical spin
models- a natural energy scale
$J$ appears as the strenght of the couplings between spins. Thus, if one
additionally sets $J=1$, all the quantities become dimensionless.

\subsubsection{Temperatures}

From the two replica potential $V(q,T)$ discussed in section 
\ref{sec1:potentialmethod} the complexity $T\Si(T)=V(q_{min}(T),T)-V(0,T)$ and
the barrier height $\EE(T)=V(q_{max}(T),T)-V(q_{min}(T),T)$ are extracted as 
functions of the temperature.
Then, the thermodynamical transition temperature $T_K$ is defined as the temperature where
the complexity vanishes: $\Sigma(T_K)=0$, 
and the value of $V$ at the secondary minimum becomes equal to zero
(see Fig.~\ref{fig_1}).
The dynamical transition temperature $T_d$ is the temperature at which the metastable
minimum first appears, so it is defined by $\EE(T)=0$.

Then, an Adam--Gibbs like relation in which $\EE(T)$ plays the role of the energy
barrier and $T\Si(T)$ of the configurational entropy is 
considered\footnote{The argument of section \ref{sec1:instantons} predict an
Adam--Gibbs relation where $(T\Si)^2$ enters in the denominator and
$\EE^{\frac{3}{2}}$ in the numerator. However, the exponents will be neglected
as their robustness is still a matter of debate.}.
Starting from the Adam--Gibbs relation
(\ref{AG}) one defines $T_g$ by $\t(T_g)/\t_\infty =$ const, or, equivalently, by
\begin{equation}\label{defTg}
\frac{{\cal E}(T_g)}{T_g \Sigma(T_g)} = {\cal C} \ .
\end{equation} 
The value of the constant ${\cal C}$ determines the value of $T_g$.
It is arbitrary because proportionality factors have always been neglected,
so it will be fixed in order to
obtain reasonable values for the fragility,
$m_A/17 \sim 1 \div 10$, as observed in experiments, see Fig.~\ref{fig3:tullio}. 
It will turn out that the analysis is not strictly dependent on the
value of $T_g$ (and of $\CC$), the behavior of the various quantities at $T_g$
being representative, as will be shown, of a general trend observed
at all temperatures $T\in [T_K,T_d]$ by varying $p$.
Different choices of the constant $\CC$ change only quantitatively
the results, while the qualitative picture stays the same.

\subsubsection{Complexity, barrier heights and fragility}

Given a definition of $T_g$, the complexity at $T_g$ is simply $\Sigma(T_g)$
and the barrier height ${\cal E}(T_g)$: clearly, these two quantities are related
by Eq.~(\ref{defTg}).
Knowing the complexity as a function of the temperature,
the fragility can be defined as in Eq.~(\ref{frag-DCp}). To simplify the notations,
the factor $17$ entering Eq.~(\ref{frag-DCp}) will be neglected in the following 
and the fragility defined as:
\begin{equation}
\label{fragilita}
m(T_g)=1+T_g \frac{\Sigma'(T_g)}{\Sigma(T_g)} \ .
\end{equation}
The latter definition is very useful in a mean field context as - once a definition of
$T_g$ has been chosen - it involves only the complexity,
that is a well-defined quantity in mean field models. It is equivalent to the usual Angell
definition of fragility if $\eta_\infty$ does not depend strongly on the material, 
and the Adam-Gibbs relation is assumed
to be valid~\cite{noifrag}. This definition of fragility has been shown to be correlated to
the fragility defined from the relaxation time using experimental data in~\cite{MA01}.

\subsubsection{Volume of the states}

The index $\alpha$ defined in \cite{TS} can be replaced by other equivalent
- equivalent meaning positively correlated -
definitions. An useful equivalent definition of $\alpha$ is
\begin{equation}
\alpha(T_g) = \lim_{k \rightarrow 0} \Big[ 1 - f(k,T_g) \Big] \ .
\end{equation}
As one can easily check observing Fig.~\ref{fig3:tullio}, this definition is equivalent
to Eq.~\ref{alphaTS} if the curves of $f(k,T)$ as function of $T$ for different materials
do not intersect.

The quantity $f(k,T)$ (in the low-$k$ limit) can be identified in spin models with
the self-overlap of the states as discussed in section \ref{sec1:orderparameter}.
Thus, one can define
\begin{equation}
\label{alphaGP}
\alpha(T_g) = 1 - q(T_g) \ ,
\end{equation}
where $q(T_g)$ is the self-overlap of the equilibrium states at $T_g$, i.e., the value of
$q$ where $V(q,T)$ has the secondary minimum at $T=T_g$ (see Fig.~\ref{fig_1}).

As the self-overlap of the states is related to their volume in phase space (high overlap
corresponding to small states), a small value of $\alpha$ corresponds to
small-volume states, while a big value of $\alpha$ corresponds to large-volume states.
In this sense, $\alpha(T_g)$ will be called ``volume of the equilibrium states at $T_g$''.
Note that a similar identification has been discussed in \cite{TS}: indeed, from
Eq.~(7) of 
\cite{TS}\footnote{Due to a misprint in Eq.(7) of \cite{TS} the power $-1$ has to be disregarded.}
one can see that $\alpha$ is related to the curvatures of the minima
of the potential (in the harmonic approximation), and that small curvatures (large volume)
correspond to large $\alpha$, while high curvatures (small volume) correspond to small $\alpha$.
This is consistent with the equivalence of the definition of $\alpha$ given in \cite{TS}
and the one adopted here.

\subsubsection{Summary of the definitions}

To conclude this section, it is useful to give 
a short summary of all the definition discussed above.
Calling $q_{min}(T)$ the value of $q$ where $V(q,T)$ has the secondary minimum,
and $q_{max}(T)$ the value of $q$ where $V(q,T)$ has a maximum, the definitions are:
\begin{equation}
\nonumber
\begin{array}{lcl}
\Sigma(T) & = & \big[ V(q_{min}(T),T) - V(0,T) \big]/T \\
{\cal E}(T) & = &  V(q_{max}(T),T)-V(q_{min}(T),T) \\
T_K & : & \Sigma(T_K)=0 \\
T_g & : & \frac{{\cal E}(T_g)}{T_g \Sigma(T_g)} = {\cal C} \\
T_d & : & \EE(T_d)=0 \\
m(T_g) & = & 1+T_g \frac{\Sigma'(T_g)}{\Sigma(T_g)} \\
\alpha(T_g) & = & 1- q_{min}(T_g)\\
\end{array}
\end{equation}
The constant ${\cal C}$ has to be chosen in order for the fragility to be in the
experimentally observed range, $m \sim 1\div 10$.

\subsection{Spherical $p$-spin model}
\label{sec:sferico}

\begin{figure}[t]
\includegraphics[width=.5\textwidth,angle=0]{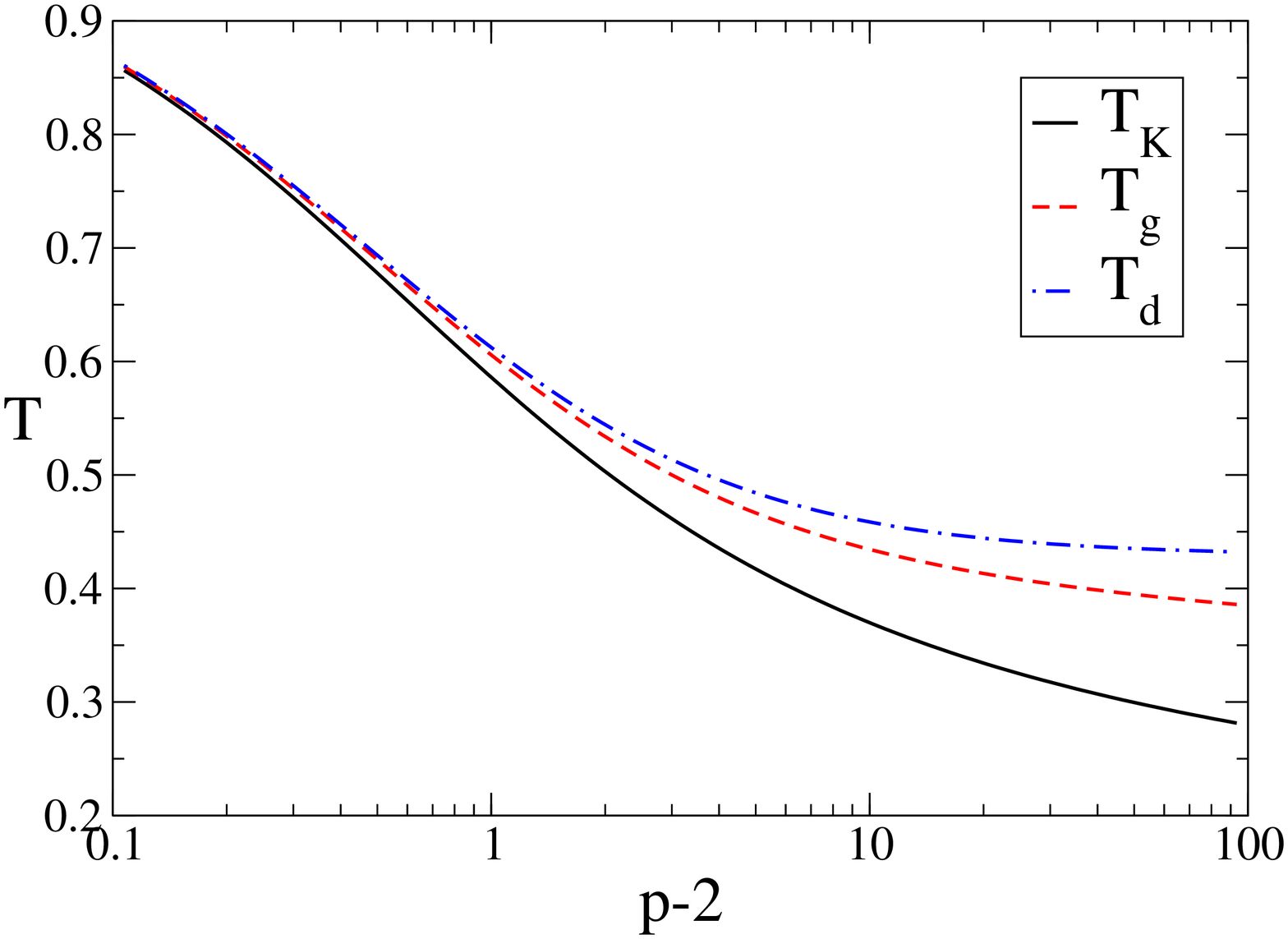}
\includegraphics[width=.5\textwidth,angle=0]{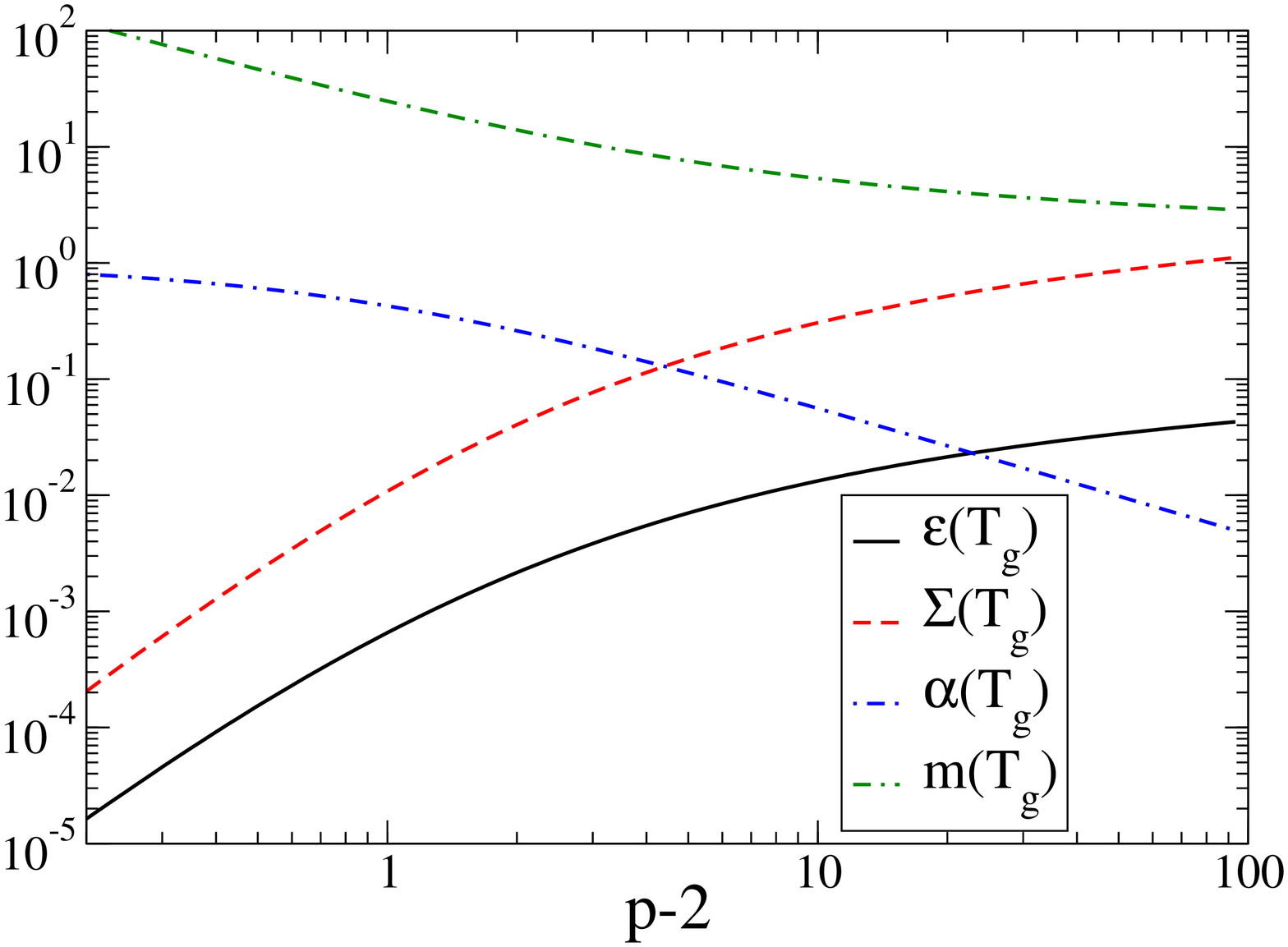}
\caption[Transition temperatures, fragility and volume of the states for the spherical $p$-spin]
{(Left) Thermodynamic transition temperature $T_K$, glass transition temperature $T_g$
and dynamical transition temperature $T_d$, and (right) fragility $m(T_g)$, 
configurational entropy $\Sigma(T_g)$, ``volume'' 
of the equilibrium states $\alpha(T_g)$ and barrier height ${\cal E}(T_g)$ for the $p$-spin 
spherical model as a function of $p-2$.
}
\label{fig_2}
\end{figure}

The full expression for $V(q,T)$ in the $p$-spin spherical model has been computed
in~\cite{FP95,BFP97}.
However, a simplified expression can be used when the value of
$V(q,T)$ {\it on its stationary points} is considered:
\begin{equation}
\label{Vq}
V(q,T) - F(T) = -\frac{\beta}{4} q^p - \frac{T}{2} \log (1-q) - \frac{Tq}{2} \ .
\end{equation}
This function has been shown in section~\ref{sec1:compreplica} to coincide with 
the correct $V(q,T)$ on each stationary point of $V(q,T)$.
If one is interested only in the value of $V(q,T)$ on its stationary points, the use of the
correct $V(q,T)$ calculated in \cite{FP95,BFP97} or
of the one given by Eq.~(\ref{Vq}) gives exactly the same result.

Note that, while the model is defined only for integer $p$, Eq.~(\ref{Vq}) makes sense also
for real $p$; therefore the behavior of the different quantities for any real
$p\geq2$ can be investigated. In particular, the $p\rightarrow2$ limit is interesting 
being related to
a diverging fragility ($T_d\rightarrow T_K$) and to the discontinuous {\sc 1rsb} transition
becoming a continuous one.

\subsubsection{Temperatures}

From Eq.~(\ref{Vq}) one can compute the three temperatures $T_K$, $T_g$ and $T_d$ as
functions of $p$. Their behavior is reported in Fig.~\ref{fig_2}. For $p\sim2$,
the difference between $T_K$ and $T_g$ is very small, therefore the system is very
fragile; for $p\rightarrow\infty$
the Kauzmann temperature approaches zero (as $1/\sqrt{\log p}$),
while the glass transition temperature remains finite.
The system therefore becomes stronger and stronger on increasing $p$.

\begin{figure}[t]
\centering
%\vspace{.05cm}
\includegraphics[width=.6\textwidth,angle=0]{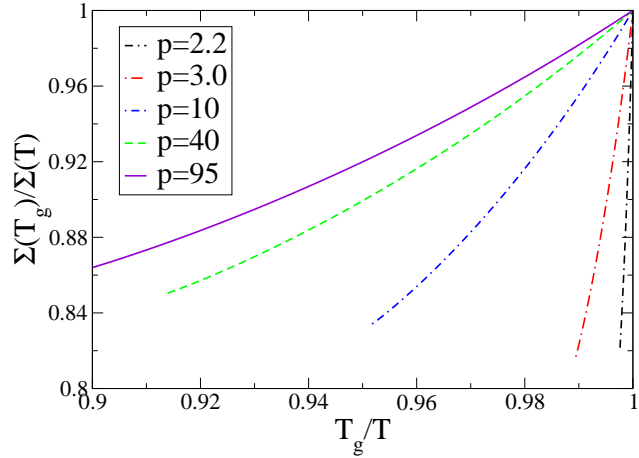}
\caption[Angell plot of the complexity for the spherical $p$-spin]
{Scaled plot of the complexity, $\Sigma(T_g)/\Sigma(T)$, as a function of $T_g/T$ for the $p$-spin
spherical model at different values of $p$. The figure has to be compared with
Fig.~\ref{fig1:Sconf_scaled}; in both figures
fragility is the slope of the curves in $T_g/T=1$.
The system becomes stronger on increasing $p$.}
\label{fig_3}
\end{figure}

\subsubsection{Complexity and fragility}

The same observation can be made more quantitative by considering an ``Angell plot'' for the
complexity~\cite{MA01}:
in Fig.~\ref{fig_3} the complexity $\Sigma(T)$ is plotted as a function of
the temperature, for different values of $p$. The choice of the particular scaling that appears
in Fig.~\ref{fig_3} has been made in order to make a close correspondence with 
Fig.~\ref{fig1:Sconf_scaled}, extracted from \cite{MA01}.
The curves for different values of $p$ are ordered from bottom to top.
The same behavior is observed in glass formers of different fragility. Indeed, the index
of fragility defined in Eq.~(\ref{fragilita}) is exactly one plus the slope in $T_g/T=1$
of the curves in Fig.~\ref{fig_3}:
\begin{equation}
m(T_g) =1+T_g \frac{\Sigma'(T_g)}{\Sigma(T_g)}
= 1 + \left. \frac{d [\Sigma(T_g)/\Sigma(T)]}{d [T_g/T]} \right|_{T=T_g} \ .
\end{equation}
The fragility index $m$ is shown in Fig.~\ref{fig_2} as a function of $p$. It is
a decreasing function of $p$. Its values are in the range observed for experimental system due
to the (arbitrary) choice of the constant ${\cal C}$ appearing in Eq.~(\ref{defTg}), 
${\cal C}=0.1$.
In Fig.~\ref{fig_2} $\Sigma(T_g)$ is also reported as a function of $p$. It is an
increasing function of $p$, that diverges as $\log p$ for $p \rightarrow \infty$:
thus, the number of states in this system is a decreasing function
of the fragility\footnote{A review of some results on the correlation between fragility
and number of states can be found in~\cite{noifrag}.}.

\subsubsection{Barrier heights and volume of the states}

In Fig.~\ref{fig_2} the barrier height ${\cal E}(T_g)$ is also reported as a function of $p$,
together with the index $\alpha(T_g)=1-q(T_g)$ that was called ``volume'' of the equilibrium
states at $T_g$. In this model the states become smaller on increasing $p$,
while the barriers separating them increase.
The correlations between these quantities will be discussed in section~\ref{correlazioni},
where a geometric description of the evolution of the phase space of this model
at different $p$ will be proposed, that relates fragility to geometric properties 
of the phase space.

\subsection{Ising $p$-spin model}
\label{sec:ising}

The Ising $p$-spin model is another popular model for the study 
of the glass transition \cite{KTW87a,KTW87b,Derrida}.
Its Hamiltonian is given by Eq.~(\ref{Hpspin}), where the variables $\sigma_i$ are Ising spins,
$\sigma_i = \pm 1$, and the spherical constraint is absent.
For the Ising $p$-spin model, the two-replica potential $V(q,T)$ is given by
\begin{equation}
V(q,T)-F(T)=\beta \frac{p-1}{4} q^p + \beta \frac{p}{4} q^{p-1}
- \frac{\int {\cal D}z \cosh(\Lambda z) \log \cosh (\Lambda z)}{\int {\cal D}z \cosh(\Lambda z)} \ ,
\end{equation}
where ${\cal D}z = \exp(-z^2/2) \ dz$, and $\Lambda^2 = \beta^2 \frac{p}{2} q^{p-1}$.

Note that the total number of states in the Ising $p$-spin model cannot be greater than $2^N$
(the total number of configurations), and hence $\Sigma(T) \leq \log 2$,
while in the spherical model $\Sigma(T_g)$
diverges as $\log p$ for $p \rightarrow \infty$, as previously discussed.

\subsubsection{Temperatures}

\begin{figure}[t]
\includegraphics[width=.5\textwidth,angle=0]{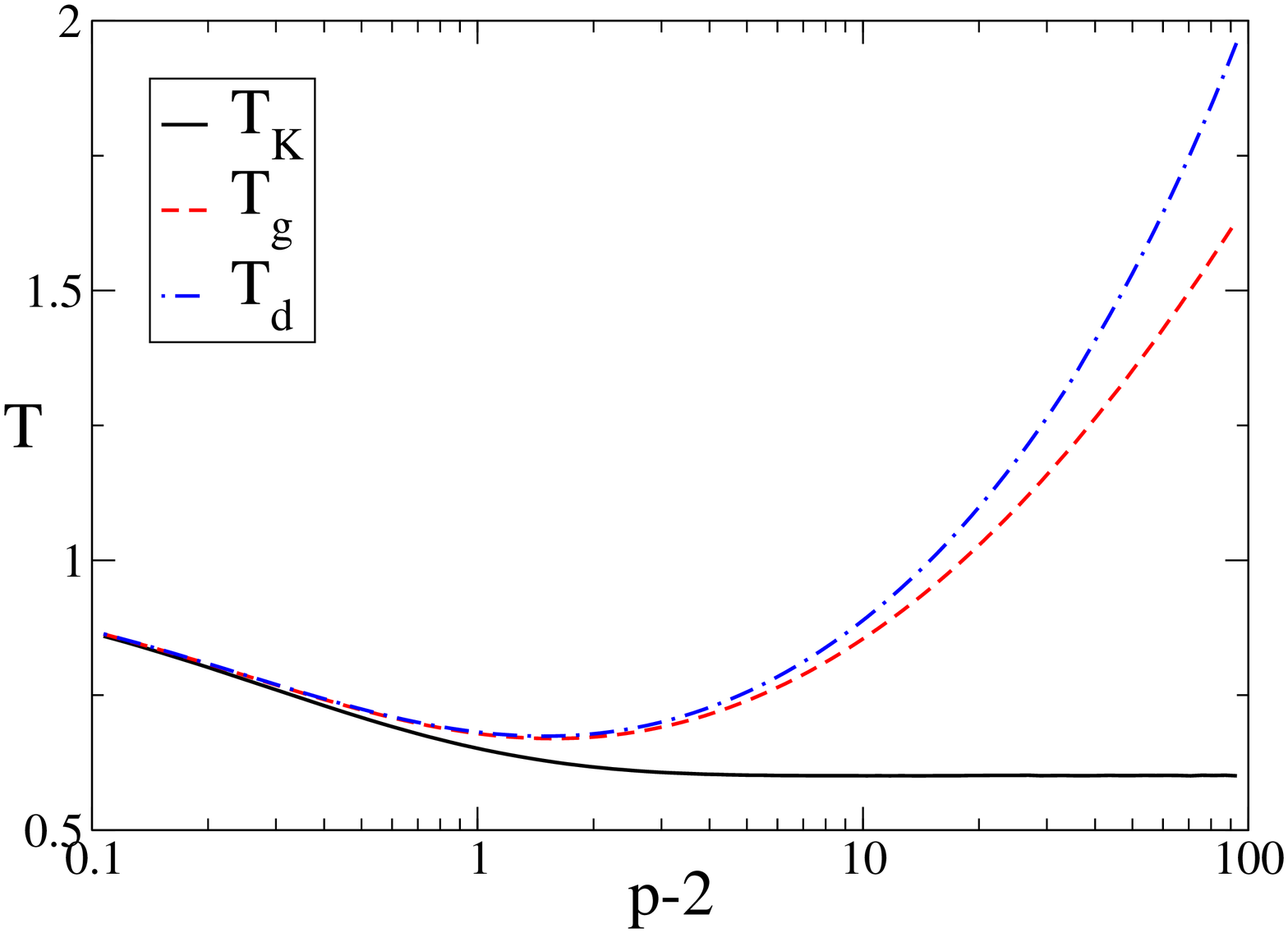}
\includegraphics[width=.5\textwidth,angle=0]{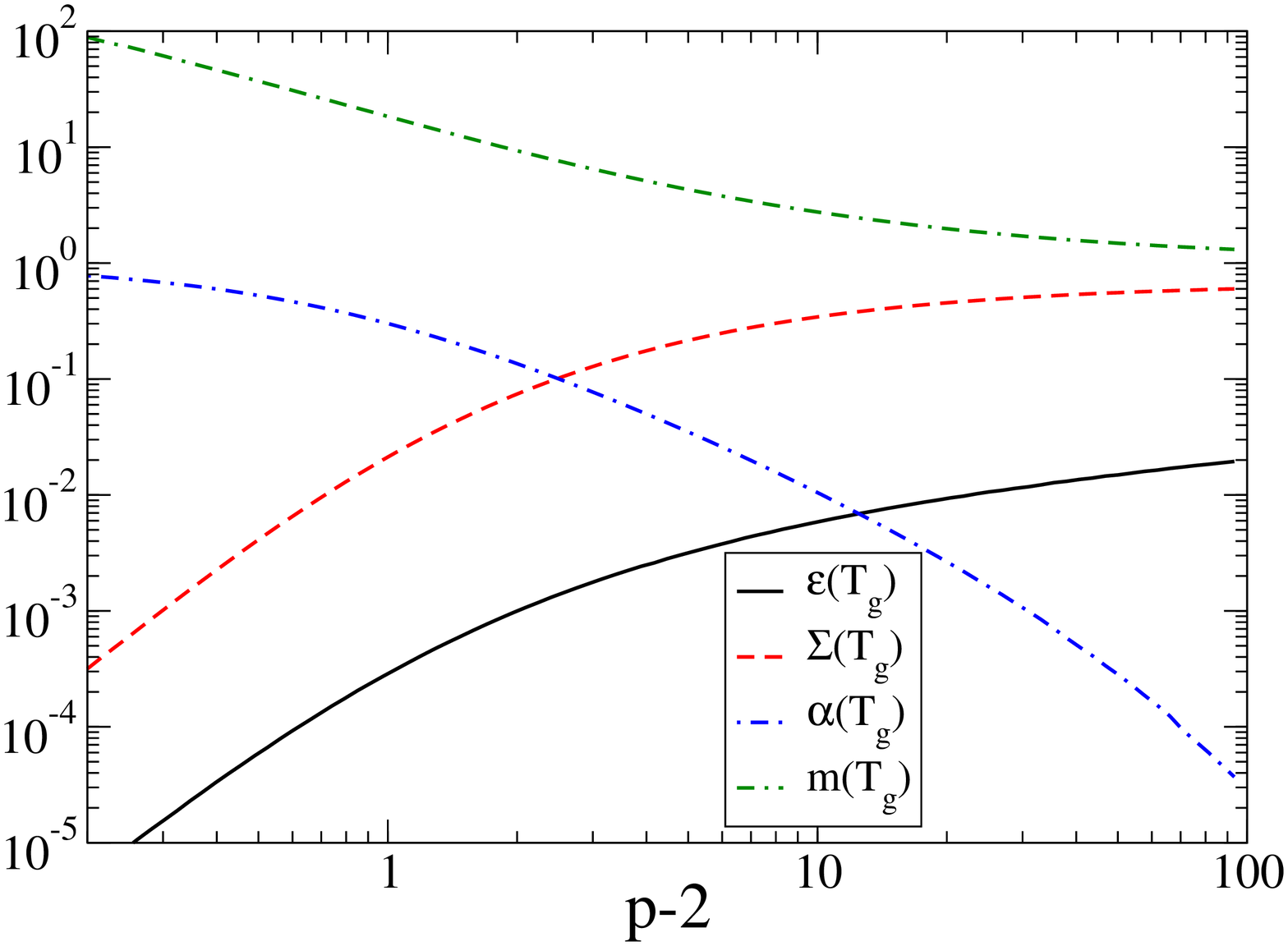}
\caption[Transition temperatures, fragility and volume of the states for the Ising $p$-spin]
{(Left) Thermodynamic transition temperature $T_K$, glass transition temperature $T_g$
and dynamical transition temperature $T_d$ and (right) fragility $m(T_g)$, 
configurational entropy $\Sigma(T_g)$, ``volume'' of the equilibrium
states $\alpha(T_g)$ and barrier height ${\cal E}(T_g)$ for the $p$-spin Ising model as a function
of $p-2$.}
\label{fig_5}
\end{figure}

The first consequence of this difference is observed when studying the transition temperatures
as functions of $p$ (see Fig.~\ref{fig_5}).
Indeed, as in the spherical model, $T_K \sim T_g$ for $p \sim 2$,
and $T_g \gg T_K$ for $p \rightarrow \infty$. But, in this model, $T_K$ tends to a finite value
at large $p$, while $T_g$ and $T_d$ diverge.
This behavior can be understood recalling that for a ``Gaussian-like'' model one has
$T_K \sim 1/\sqrt{\Si_\io}$, $\Si_\io$ being the total number of
states, i.e. the maximum of $\Sigma(T)$ \cite{noifrag}.

\subsubsection{Complexity and geometric properties of the phase space}

The ``Angell plot'' for the complexity of the Ising $p$-spin model looks very similar to the
one of the spherical model, Fig.~\ref{fig_3}, so it is not useful to report it.

Having fixed an appropriate value for the constant ${\cal C}$ in Eq.~(\ref{defTg}) 
(${\cal C}=0.02$, different from the value chosen in the spherical case),
the behavior of the fragility as a function of $p$ is also very similar to the one of the
spherical model. The same behavior is found
for the other quantities under study, as one can deduce from
a comparison of Fig.~\ref{fig_5} and Fig.~\ref{fig_2}, the main difference being the discussed
behavior of $\Sigma(T_g)$ at large $p$.

\subsubsection{Vibrational properties and volume of the states}

Another relevant difference between the spherical and the Ising model is that, in the latter,
harmonic vibrations are not present (the variables being discrete): we have
$q(T) \rightarrow 1$ exponentially for $T \sim 0$, and the definition of $\alpha$
via Eq.~(\ref{alphaTS})
gives $\alpha = 0$ for all $p$. However, the definition given in Eq.~(\ref{alphaGP}) and used in
these calculations gives a reasonable result also in absence of harmonic vibrations.

\section{Correlations between different properties of the phase space}
\label{correlazioni}

In this section the correlations between the quantities under study will be investigated,
trying to relate fragility to the phase space geometry.
The results will be compared with the general consideration of~\cite{noifrag},
and with the experimental results of~\cite{TS}.

\subsection{Fragility and volume of the states}

\begin{figure}[t]
\centering
%\vspace{.05cm}
\includegraphics[width=.6\textwidth,angle=0]{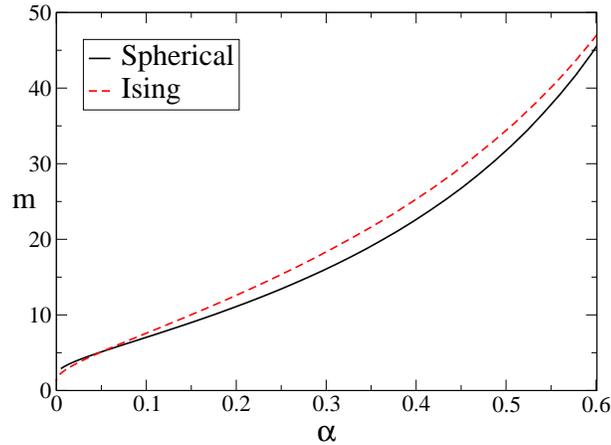}
\caption[Correlation between fragility and $\alpha$ for $p$-spin models]
{Fragility versus $\alpha$ for the spherical and Ising $p$-spin models. 
The curve is very similar for the two models, 
and is consistent with the correlation found in \cite{TS}, see Fig.~\ref{fig3:tullio}.
The linear correlation is reproduced for $\alpha \leq 0.4$.}
\label{fig_7}
\end{figure}

In~\cite{TS} it has been established that fragility is positively correlated with the index
$\alpha$ defined in section~\ref{sec:definitions}.
In other words, {\it fragile systems have large basins while strong systems have small basins}.
In Fig.~\ref{fig_7} the fragility $m$ is plotted as a function of $\alpha$ parametrically in $p$
for the investigated systems. The curve $m(\alpha)$ is very similar for
the two models - remember that the only adjustable parameter is the constant ${\cal C}$
in Eq.~(\ref{defTg}).
By comparison with Fig.~\ref{fig3:tullio}, 
one can conclude that the model has a behavior similar to the one of real systems.
Surprisingly, also the linear correlation between $m$ and $\alpha$ is reproduced for
$\alpha \leq 0.4$.
Thus, mean field $p$-spin models are able to describe the relation between fragility and
the volume of the basins visited around $T_g$ found in~\cite{TS}.

\subsection{Fragility, barrier heights and number of states}

\begin{figure}[t]
\centering
%\vspace{.05cm}
\includegraphics[width=.6\textwidth,angle=0]{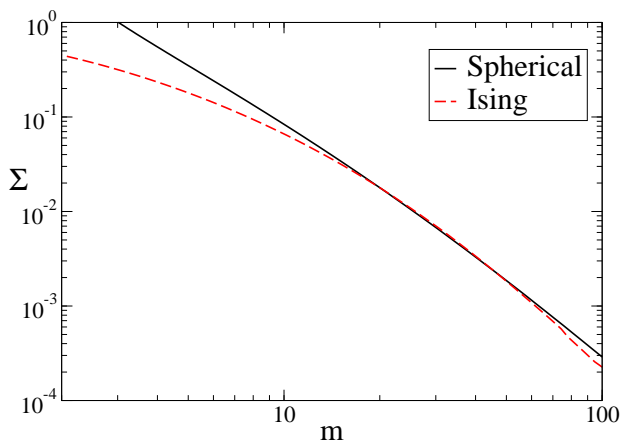}
\caption[Correlation between fragility and number of states in $p$-spin models]
{Total number of states (represented by the complexity at $T_g$) as a function of
the fragility $m$ for the $p$-spin models.}
\label{fig_8}
\end{figure}

It has been conjectured that fragile systems have a larger number of states than strong ones,
even if the total number of states is not an experimentally accessible quantity and numerical
simulations give contradictory results~\cite{Sp99,Sa01,DS01}.
However, in the models considered here the behavior is exactly the opposite. In Fig.~\ref{fig_8}
$\Sigma(T_g)$ is reported as a function of the fragility: the total number of states
is a decreasing function of the fragility.

This point was discussed in detail in \cite{noifrag}, where the possibility
of correlating fragility with the total number of states for general models of $\Sigma(T)$ was
discussed, assuming the validity of the Adam-Gibbs relation, Eq.~(\ref{AG}).
The conclusion was that the knowledge of the distribution of states is not enough to determine
the fragility. Indeed, the relevant
parameter, for a general ``Gaussian-like'' distribution of states, is
\begin{equation}
\label{D}
D = \frac{{\cal E}(T_g)}{T_K \Sigma(T_g)} \ .
\end{equation}
Note that in Eq.~(\ref{D}) ${\cal E}$ has to be evaluated at $T=T_g$ because in the considered models
the barrier height ${\cal E}$ is a $T$-dependent quantity, while in the Adam-Gibbs relation
it is usually assumed to be a constant. However, the Adam-Gibbs relation
has been tested around $T_g$, therefore, to a good approximation, one can fix ${\cal E}$ to be a
constant equal to its $T=T_g$ value.
The parameter $D$ is inversely proportional to the fragility $m$: therefore $m \sim \Sigma/{\cal E}$,
and fragility turns out not to be simply correlated to the total number of states.
If the ``barrier heights'' grow faster than the total number of states,
fragility can be a decreasing function of $\Sigma$:
this is indeed the case in the considered models.
Indeed, from Fig.~\ref{fig_2} and Fig.~\ref{fig_5}, one observes that the barrier height is an
increasing function of $p$.
Using Eq.~(\ref{defTg}), Eq.~(\ref{D}) can be written as
\begin{equation}
D = {\cal C} \frac{T_g}{T_K} \ .
\end{equation}
Therefore, from Fig.~\ref{fig_2} and Fig.~\ref{fig_5}, $D$ is an increasing
function of $p$ that diverges for $p \rightarrow \infty$, as the ratio $T_g/T_K$ increase on
increasing $p$ for both models.
Thus, in the considered models the height of the barriers (in units of $T_K$)
increases faster than the total number of states. This explains why one observes an inverse
correlation between fragility and the total number of states, as discussed above and 
in~\cite{noifrag}.

\subsection{A geometric picture of the phase space}

\begin{figure}[t]
\centering
%\vspace{.05cm}
\includegraphics[width=.6\textwidth,angle=0]{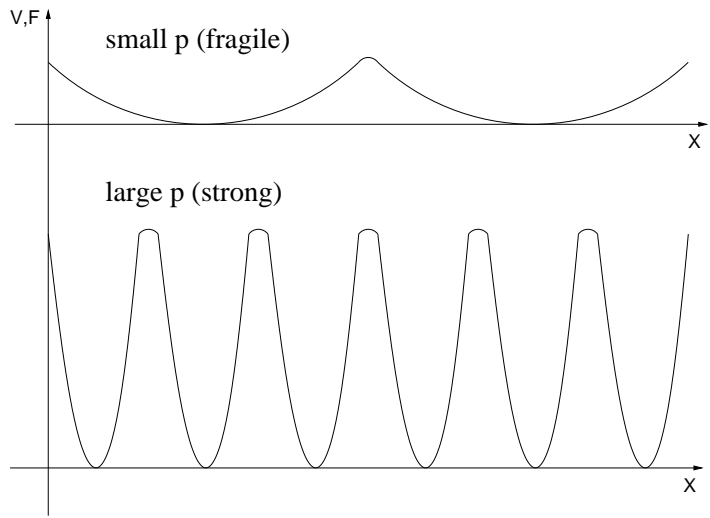}
\caption[Sketch of the evolution with $p$ of the $p$-spin free energy landscape]
{Sketch of the evolution with $p$ of the $p$-spin free energy landscape: at small $p$
there is a small number of states of large volume separated by low barriers; at high $p$
there is a large number of states of small volume separated by high barriers. The height
of the barriers increase faster than the number of states: thus, fragility is a decreasing
function of $p$.}
\label{fig_9}
\end{figure}

The information obtained in the previous sections can be collected in
a geometric picture of the evolution with $p$ of the $p$-spin model free energy landscape.
Indeed, on increasing $p$: \\
{\it i)} The total number of states increases. \\
{\it ii)} The volume of the states decreases ($\alpha$ decreases). \\
{\it iii)} The height of the barriers between states increases. \\
Thus, we get the picture of a landscape where, on increasing $p$, a great number of small
states with very high curvatures and separated by very high barriers appear: a sketch of this
evolution is given in Fig.~\ref{fig_9}.
The behavior of the fragility in this situation is related to the behavior of $\Sigma/{\cal E}$,
the ratio between number of states and height of the barriers between them: in these models,
it turns out that ${\cal E}$ increase faster than $\Sigma$, and the fragility is a decreasing
function of $p$.

This behavior is consistent with the fact that fragility turns out to be
positively correlated with the ``volume'' of the states as measured by $\alpha$. Indeed,
if, on the contrary, the barrier height grew slower than the total number of states
(equivalently, if $m$ would be positively correlated with the total number of states),
there should be also an inverse correlation between $m$ and $\alpha$, in disagreement with what
is experimentally observed.

In the $p\rightarrow 2$ limit, where the fragility becomes infinite, the second
derivative with respect to $q$ of the potential $V(q,T)$ calculated in
$q=0$ and $T=T_K=T_d$ vanishes (see Fig.~\ref{fig_1}) 
and the so-called spin glass susceptibility
diverges at the critical temperature. In other words when the fragility becomes
infinite soft modes appear at the critical temperature supporting the
previously presented physical picture.

Note that the outlined picture is valid for ``Gaussian-like'' models, i.e., models where
the complexity is a concave function of the temperature that vanishes at $T_K$ without any
inflection point. These models seem to describe correctly the distribution of basins in real
systems only for relatively high fragilities. The behavior of the complexity (or configurational
entropy, or excess entropy) as a function of temperature for very strong systems is still an
open problem, see \eg~\cite{SPS01,MBLSSTZ04}; our discussion may not apply to these systems.

The main prediction of this analysis is that the total number of states $\Si_\io$ 
and the Adam-Gibbs barrier ${\cal E}$ should both be decreasing functions of the 
fragility. This prediction disagrees with the statement of \cite{Sp99,Sa01,DS01} that
fragile systems should have an higher number of states with respect to strong ones.
A critical analysis of this works can be found in \cite{noifrag}. Unfortunately,
existing data are not sufficient to strictly test this prediction; the excess entropy
is available only for few experimental systems, and numerical simulations are performed
in a temperature range where the fragility of the investigated systems is approximately
the same. Hopefully this predictions will be tested in the future.

%% file: cap4.tex
\chapter{Nonequilibrium stationary states and the fluctuation theorem}
\label{chap4}

\section{Introduction}

\subsection{A critical review of the ergodic hypothesis}

The statistical mechanics of {\it equilibrium} states can be
constructed starting from the {\it ergodic hypothesis} of 
Boltzmann, see \eg~\cite{Ga99} for a modern review.
An {\it equilibrium} system is, for instance, a system of $N$ classical
particles enclosed in a volume $V$ which interact between themselves 
and with the wall of the container only by {\it conservative} 
forces.
The phase space of the system can be divided in suitable 
cells\footnote{See \cite{Ga99} for a very deep discussion on how these
cells have to be constructed.} $\D$, such that the dynamics can be represented
as a discrete-time map $S \D = \D'$ acting on the cells.
Then, the ergodic hypothesis is the assumption that $S$ is
a {\it one-cycle permutation} of the cells with constant energy
(because the energy is kept constant by the time evolution).
This means that
all the cells are visited sequentially: and if one considers an
observable $F(\D)$, the time average
\beq
\la F \ra = \lim_{\t\to\io} \frac{1}{\t} \sum_{t=0}^{\t-1} F(S^t \D) \ ,
\eeq
starting from {\it any} initial condition $\D$ such that $U(\D)=U$, 
does not depend on the initial condition and is simply given by the flat average 
over the cells with energy $U$,
\beq
\la F \ra = \frac{1}{{\cal N}(U,V)} \sum_{\D : U(\D)=U} F(\D) \ ,
\eeq
where ${\cal N}(U,V)$ is the total number of such cells.
Starting from the ergodic hypothesis Boltzmann was able to prove
that, if one defines the observables $T$ (temperature), $U$ (internal
energy), and $P$ (pressure) in terms of the average kinetic energy,
potential energy, and momentum transfer to the container walls
respectively, a function $S(U,V)$ (entropy) exists such that
\beq\label{TC}
dS = \frac{dU + PdV}{T} \ .
\eeq
The function $S(U,V)$, in the construction of Boltzmann, is 
$S(U,V)=k_B \log \NN(U,V)$.
The ergodic hypothesis says that the statistical measure
describing the equilibrium macroscopic state on the phase space of
the system is simply the {\it flat} measure on the surface of constant
total energy $U$. The {\it ensemble} of these measures, obtained by varying
the values of the external parameters $U$, $V$, is called the
{\it microcanonical ensemble}.
Then one can show that different ensembles of measures can be defined,
and that they are {\it equivalent}, in the sense that they give rise
to the same macroscopic relations between observables and in particular
that Eq.~(\ref{TC}) holds for all of them.

Despite this success, the ergodic hypothesis was criticized, because it
seemed impossible to derive the {\it irreversible} laws of thermodynamics
from the microscopic {\it reversible} equations of motion.
Indeed, if the dynamics $S$ is a one-cycle permutation of the cells,
after a given {\it recurrence time} the system must come back to the
cell in which it was prepared at the initial time.
The observation of Boltzmann was that this recurrence time, for a
macroscopic system of $N \sim 10^{23}$ particles, could be estimated
to be essentially infinite on any reasonable time scale, so that 
reversibility could not be observed for macroscopic systems.

However, this argument seems to make the ergodic hypothesis meaningless:
if the time needed for the system to visit all the cells is 
much larger than any experimentally accessible time scale, the replacement
of the average over the time evolution with the flat average over the
cells is unjustified.
Boltzmann realized this difficulty~\cite{Ga99} and solved it observing
that, for a {\it macroscopic} system, the thermodynamically interesting
observables are essentially {\it constant} on phase space, thanks to
the law of large numbers. Then, even if the system visits only a small
portion of its phase space on the experimental time scale, the replacement
of the time average with a flat average on phase space gives the correct
result.

Thus, the ergodic hypothesis alone is not a justification of the
thermodynamics. However it {\it correctly suggests} that the averages of the
interesting observables can be computed using the flat measure on phase
space, provided that the size of the system is large.
It is even not important to prove that a given physical system
is ergodic, and it is likely that many of the models which are commonly
used to describe physical systems are {\it not} ergodic. 
The ergodic hypothesis is relevant because
it allows to identify the correct measure describing the
macroscopic equilibrium states on the microscopic scale.

\subsection{Nonequilibrium states and the chaotic hypothesis}

One could ask if a similar construction can be repeated in the
case of a stationary {\it nonequilibrium} system.
For instance one can think to a system of $N$ interacting particles
in $d$ dimensions, enclosed in a container of volume $V$, subjected 
to nonconservative forces and kept in a
stationary state by a {\it reversible mechanical thermostat}.
It will be defined by a differential equation $\dot x=X_E(x)$ where 
$x=(\dot{\ul q},\ul q) \in R^{2dN} \equiv M$ ({\it phase space})
and
\beq
\label{INTRO1}
m \ddot{\ul q} = \ul f(\ul q) + \ul g_E (\ul q) - \ul 
\theta_E (\dot{\ul q},\ul q)
\eeq
where $m$ is the mass of the particles, $\ul f(\ul q)$ describes 
the internal (conservative)
forces between the particles (and between the particles and the walls of
the container) and $\ul g_E (\ul q)$ represents 
the nonconservative
``external'' force acting on the system, which is assumed to depend smoothly
on a parameter $E$ (\eg the amplitude of the electric field). Finally, 
$\ul \theta_E (\dot{\ul q},\ul q)$ is a
mechanical force that prevents the system from acquiring energy indefinitely: 
this is why it is
called a {\it mechanical thermostat}.
Systems belonging to this class are frequently 
used as microscopic models to describe
nonequilibrium stationary states induced by the application of a driving force
(temperature or velocity gradients, electric fields, etc.) on a
fluid system in contact with a thermal bath~\cite{EM90,Ru04}. 
The problem of which is the measure describing the stationary states of such a
system is still matter of debate, but important progress have been achieved in
recent times. They will be discussed in the rest of this chapter.

A first difficulty one has to face is that, for nonequilibrium stationary states,
a well-established macroscopic theory similar to equilibrium thermodynamics does not 
exist~\cite{DGM}. From a microscopic point of view, the main difficulty is that
the equation of motion (\ref{INTRO1}) yields a {\it phase space contraction rate}
\beq\label{sxdef}
\s(x) = - {\rm div} X_E(x) \ ,
\eeq
such that $\frac{d}{dt} dx = - \s(x) dx$, which has {\it positive average}~\cite{Ru96}, 
\ie if $S_t x$ is the solution of Eq.~(\ref{INTRO1}) with initial datum $x$, for almost
all $x$ w.r.t. the volume $dx$, one has
\beq\label{s+def}
\s_+ \equiv \la \s(x) \ra = \lim_{\t \to \io} \frac{1}{\t} \int_0^\t dt \, \s(S_t x) > 0 \ .
\eeq
This means that phase space volume is not conserved by the time evolution, so
the invariant measure cannot be the flat measure on phase space, and cannot even admit
a density, \ie be of the form $\m(dx)=\m(x) dx$: it will be concentrated on a set
of zero volume in phase space. Also, this means that the description of the system
in terms of cells $\D$ and the hypothesis that the dynamics is a simple one-cycle
permutation cannot hold for nonequilibrium systems. Still a description in term 
of cells is possible but it does not lead to a satisfactory notion of 
``entropy''~\cite{Ga04}.

A very important step towards the construction of a statistical mechanics of nonequilibrium
stationary states was the explicit construction of the invariant measure for a class
of smooth chaotic dynamical systems, called 
{\it Anosov systems}~\cite{Ga99,Si68,Si72,Si77,Ru78,Ru80,Ga02,GBG04}.
For these systems it was proved that almost all points 
w.r.t. the volume measure\footnote{This is very important as it means that the initial datum
has to be randomly extracted from the flat measure on phase space, 
\ie the system has to be prepared at equilibrium and then
let evolve under the action of the nonconservative forces.}
evolve so that {\it all smooth observables} 
have a well defined average equal to the integral over the invariant measure,
\ie for all smooth $F(x)$
\beq
\la F(x) \ra \equiv \lim_{\t\to\io} \frac{1}{\t} \int_0^\t dt \, F(S_t x) = \int \m_+(dx) F(x) \ .
\eeq
The measure $\m_+(dx)$ is called Sinai--Ruelle--Bowen (SRB) distribution. In particular
this holds for the phase space contraction rate $\s(x)$ and the relation 
$\s_+ \geq 0$, see Eq.~(\ref{s+def}), was proved for these systems~\cite{Ru96}.

Anosov systems are paradigms of chaotic systems: and even if essentially none of the 
physically interesting systems could be proved to be Anosov, a {\it chaotic hypothesis} 
was proposed~\cite{GC95a,GC95b}, which states that nevertheless
one can assume that chaotic motions 
(in the sense of motions with at least one positive Lyapunov exponent) 
exhibit some average properties of truly Anosov systems. In particular, it should
be possible to compute the averages of the physically interesting observables 
using the SRB distribution.

This hypothesis is a natural generalization of the ergodic hypothesis, as the SRB
distribution reduces to the flat distribution in the equilibrium case where $\s_+=0$.
Similarly to the ergodic hypothesis, the chaotic hypothesis simply {\it suggests} that
the invariant measure is the SRB measure, even if the system is not an Anosov system.
But this allows to make some predictions on the macroscopic properties of systems like
the one described by Eq.~(\ref{INTRO1}), similarly to what has been done by Boltzmann
by proving that the ergodic hypothesis implies the validity of Eq.~(\ref{TC}).

\subsection{The fluctuation theorem}

An interesting prediction of the chaotic hypothesis is the validity of a 
relation that concerns the large deviations of the phase space contraction rate,
and that has been named {\it fluctuation relation}.
The validity of this relation for a {\it reversible} Anosov map $S$ was proved 
({\it fluctuation theorem}) in~\cite{GC95a,GC95b}. Reversibility means that there is a
metric--preserving map $I$ on the phase space $M$ such that $IS=S^{-1}I$ and is a key
hypothesis for the validity of the fluctuation relation. For a map $S$ the phase
space contraction rate is defined as
\beq
\s(x)=-\log |\det\partial S(x)| \ ,
\eeq
and it is a smooth function on $M$.
As discussed above the average of $\s(x)$ exists and is $\s_+ \geq 0$.
If $\sigma_+ > 0$, let:
\beq
\label{2}
p(x)=\frac{1}{\tau \sigma_+} \sum_{t=0}^{\tau-1} \sigma(S^t x) \ .
\eeq
The function $p(x)$ has average $\langle\; p\;\rangle = 1$
and distribution $\pi_\tau(dp)$ such that
\beq
\label{3}
\pi_\tau(\{p\in \Delta\})= e^{\tau \max_{p\in\Delta}\zeta_\infty(p)+ o(\t)}\;,
\eeq
where the correction at the exponent is $o(\t)$ w.r.t. $\t$ as $\t\to\io$, and
the function $\z_\io(p)$ is analytic and convex in $p$.
The fluctuation relation then reads:
\beq
\label{FR}
\zeta_\infty(p)=\zeta_\infty(-p)+p \sigma_+ \qquad{\rm for\ all}\ |p|<p^* \ ,
\eeq
where $\infty> p^*\ge1$ is a suitable (model dependent) constant that, in
general, is {\it different} from the maximum over $\tau$ and $x$ of
$p(x)$, and is defined by $\z_\io(p) = -\io$ for $|p|>p^*$.

The interesting fact is that {\it the fluctuation relation has no free parameters}:
thus the simplest and more stringent check of the applicability of the chaotic hypothesis
is a check of the fluctuation relation. Of course even if the
check has a positive result this will not ``prove'' the hypothesis but
it will at least add confidence to it. This test has been performed in a number of
cases\footnote{Historically the fluctuation relation was first discovered numerically 
in~\cite{ECM93}, and this gave the motivation for \cite{GC95a,GC95b}.}
with positive result, by mean of numerical 
simulations~\cite{ECM93,BGG97,BCL98,BPV98,GP99,RS99,ZRA04a,GRS04,GZG05} and 
experiments~\cite{CL98,GGK01,CGHLP03,FM04}.

In the following, after a brief review of the procedure that allows to construct the
SRB measure for Anosov systems, the physical implications of the chaotic hypothesis
will be discussed.

\section{Sinai--Ruelle--Bowen (SRB) measures}

\subsection{Anosov systems}

Anosov systems are paradigms of chaotic systems. They are defined as follows~\cite{GBG04}.
Given a compact, smooth and boundaryless manifold $M$ (phase space), a map
$S \in C^2(M)$ is an {\it Anosov map} if:

\noindent
{\it (1)} For all $x \in M$ the tangent plane to $M$ in $x$, $T_x$, admits a
decomposition $T_x = T^s_x \oplus T^u_x$, such that

\noindent
{\it (2)} the planes $T^{s,u}_x$ vary continuously w.r.t. $x$, \ie the vectors
defining them are continuous functions of $x$;

\noindent
{\it (3)} the angle between $T^s_x$ and $T^u_x$, defined as the minimum angle
between a vector in $T^s_x$ and a vector in $T^u_x$, is not vanishing;

\noindent
{\it (4)} defining $\partial S$ the linearization matrix of $S$ in $x$, \ie
$S(x+\ee v) = S(x) + \ee \ \partial S(x) \cdot v + O(\ee^2)$, $x \in M$, $v\in T_x$,
$\ee \in \RRR$ small, the planes $T_x^{s,u}$ are conserved under $S$, \ie if
$v \in T_x^{s,u}$, then $\partial S(x) \cdot v \in T^{s,u}_{Sx}$;

\noindent
{\it (5)} for all $x \in M$ and for all $v \in T^s_x$ one has 
$|\partial S(x)^k v|_{S^k x} \leq \L^{-k} C |v|_x$, while for
all $v \in T^u_x$ one has 
$|\partial S(x)^{-k} v|_{S^{-k} x} \leq \L^{-k} C |v|_x$, $| \bullet |_x$ being
the norm on $T_x$, for some constants $C,\L > 0$;

\noindent
{\it (6)} there is a point $x$ which has a dense orbit in $M$.

\noindent
The hypotheses above imply that it is possible to identify two families of smooth
manifolds $M^{s,u}$ in $M$, such that $T^{s,u}_x$ are the tangent plane to
$M^{s,u}$ in $x$, and such that points on $M^s$ tend to converge exponentially
while points on $M^u$ tend to diverge exponentially under the action of $S$.

\begin{figure}[t]
\centering
\includegraphics[width=.7\textwidth]{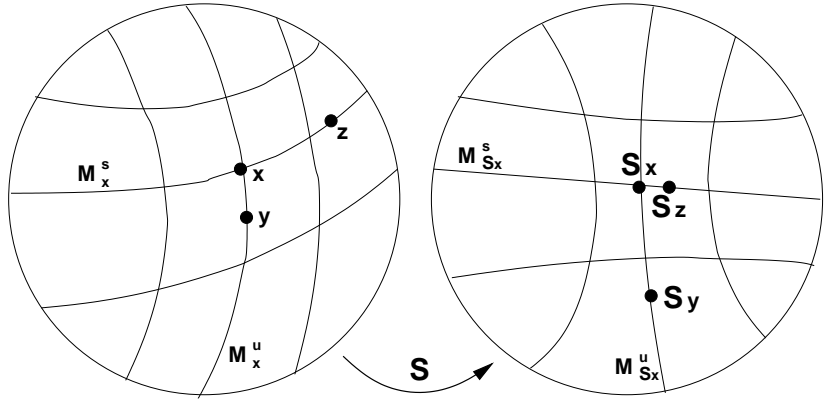}
\caption[Pictorial representation of an Anosov system]{
A pictorial representation of an Anosov system. In the vicinity of a point $x \in M$
it is possible to draw two families of manifolds $M^{s,u}$. The manifolds passing
through $x$ are the stable and unstable manifolds of $x$, $M^{s,u}_x$. 
Under the action of $S$, the manifolds $M^{s,u}_x$ are mapped into the manifolds
$M^{s,u}_{Sx}$ passing through $Sx$. A point $y \in M^u_x$ is mapped into a point
$Sy$ whose distance from $Sx$ is larger by a factor $\sim \L$, while a point 
$z \in M^s_x$
is mapped into a point $Sz$ which is closer to $Sx$ by the same factor $\L$.
}
\label{fig4:anosov}
\end{figure}

\noindent
This means that for each $x \in M$ there is a stable manifold $M^s_x$ such that
for all $y \in M^s_x$ one has $d(S^k x,S^k y) \leq \L^{-k} C d(x,y)$, and
an unstable manifold $M^u_x$ such that
for all $y \in M^u_x$ one has $d(S^{-k} x,S^{-k} y) \leq \L^{-k} C d(x,y)$, where
$d(x,y)$ is the distance on $M$, see Fig.~\ref{fig4:anosov}.

\subsection{Markov partitions}

To each point $x \in M$ one can associate a sequence
$\underline{\varepsilon}=(\varepsilon_i)_{i=-\infty}^\infty$ of 
finitely many digits $\varepsilon=1,\ldots,{\cal K}$ in the following
way: one partitions the phase space in ${\cal K}$ sets $M_1,\cdots,M_\KK$,
such that $\cup_{k=1}^\KK M_k = M$ and $M_i \cap M_j = \emptyset$ for all 
$i\neq j$.
Then one sets $\varepsilon_0 = k_0$ if $x \in M_{k_0}$, $\varepsilon_1 = {k_1}$ 
if $Sx \in M_{k_1}$, $\varepsilon_{-1} = {k_{-1}}$ 
if $S^{-1}x \in M_{k_{-1}}$, and so on, \ie $\varepsilon_i$ is
defined by $S^i x \in M_{\varepsilon_i}$.

It is clear that in such 
a representation the dynamics becomes simply the left
shift, \ie if $\underline{\varepsilon}(x)$ 
represents $x$ then $Sx$ is represented by
the sequence $\underline{\varepsilon}$ shifted to the left by one unit.

To each $x \in M$ a single sequence $\underline{\varepsilon}$ is associated,
apart from that points $x$ such that for some $i$ the point $S^i x$ falls on 
the boundaries of the sets $M_k$. In these cases one has an 
ambiguity: 
but it is possible to show~\cite{GBG04}
that the number of possible sequences that are associated
to a given point $x$ is finite, and that only a set of zero volume has more than
one associated sequence\footnote{This is essentially the same ambiguity that 
one has when writing rational numbers as reals: for example $1$ can also be written 
as $0.\ol 9$. It is clear that the ambiguity can be solved simply by choosing to
write $1$ and not $0.\ol 9$ everywhere.}.

The key observation is that {\it not all sequences are possible}: for example,
the points in the set $M_1$ will evolve under the action of $S$ but in a single 
step they will not reach all the others $M_k$. Thus, the symbol $1$ in the 
sequence $\ul \varepsilon$ can be followed only by the symbols corresponding
to the sets such that $S M_1 \cap M_i \neq \emptyset$.
This can be expressed by a {\it compatibility matrix} 
$T_{ij}$ which is $1$ if $S M_i \cap M_j \neq \emptyset$ and $0$ otherwise.
Only the sequences $\underline \varepsilon$ such that
$\prod_{i=-\io}^\io T_{\varepsilon_i,\varepsilon_{i+1}} \neq 0$ correspond to
points $x \in M$.
But this is not enough: a similar condition is needed also for triples of
symbols, so one must define a matrix $T^{(2)}_{ijk}$ that is $1$ if there is a
point $x\in M_i$ such that $Sx \in M_j$ and $S^2 x \in M_k$ and $0$ otherwise,
and consider only sequences such that 
$\prod_{i=-\io}^\io 
T^{(2)}_{\varepsilon_i,\varepsilon_{i+1},\varepsilon_{i+2}} \neq 0$,
and so on.
It is clear that the full knowledge of the compatibility matrices $T^{(n)}$ 
is completely
equivalent to the full solution of the dynamics (as one will be able to reconstruct
all the trajectories), so it is a very difficult problem even for very simple systems.

A very remarkable consequence of the assumptions defining Anosov systems is,
\cite{Si68,Si72,Si77,GBG04,GC95a,GC95b,Ga95c},
that one can find a partition $M_1,\cdots,M_\KK$ ({\it Markov partition}),
such that the sequences $\ul \varepsilon$ are
subject only to the nearest-neighbors restriction, namely
$\prod_{i=-\io}^\io T_{\varepsilon_i,\varepsilon_{i+1}} \equiv 1$,
as one can prove that $T^{(2)}_{ijk} = T_{ij} T_{jk}$ and so on.
Moreover, one can prove, from assumption {\it (6)} above, that
$(T^N)_{ij}>0$ for some $N>0$
and all $i,j$ ({\it mixing} condition). 
%This means that the phase space of an Anosov system is
%isomorphic to a discrete time Markov process.

\subsection{Observables and the SRB measure}

\subsubsection{Observables}

Smooth observables on phase space can be
represented by {\it short range potentials}: in the case of the
observable $\sigma(x)=-\log |\det \partial S(x)|$ 
this means that there are {\it translationally invariant} functions 
$s(\underline{\varepsilon}_X)$,
where $X$ is an interval $X=(a,\ldots,a+2n+1)$ and
$\underline{\varepsilon}_X=(\varepsilon_{a},
\ldots,\varepsilon_{a+2n+1})$, 
which are {\it exponentially decaying} 
on time scale $\kappa^{-1}$ (\ie 
$|s(\underline{\varepsilon}_X)|< C e^{-\kappa
n}$ for some $C,\kappa>0$), such that 
\beq
\label{FTC2.2}
\sigma(x)= s(\underline{\varepsilon}(x)),\qquad s(\underline{\varepsilon})=
\sum_{X\circ\, 0}
s(\underline{\varepsilon}_X) \ ,
\eeq
where the sum is over the intervals $X$ centered at the origin
(noted by $X\circ 0$). What is important is that the dependence of the
function $s(\ul\varepsilon)$ on the symbols which are far apart from the
origin is exponentially small in the distance from the origin.

Other important smooth observables are the {\it expansion and contraction rates} 
$L_\pm(x)$, defined as the logarithm of the determinant
of the matrix $\partial S(x)$ restricted  
to the unstable (stable) manifold: 
\beq\begin{split}
&L_+ (x)=\log | \det\partial S(x)_{u}| \ , \\
&L_- (x)=-\log| \det\partial S(x)_{s}| \ .
\end{split}\eeq
$L_+(x)$ is the sum of all the positive Lyapunov exponents in $x$, while $L_-$
is minus the sum of all negative Lyapunov exponents\footnote{Note that these are
{\it local} Lyapunov exponents that depend on the metric, at variance to the
usual Lyapunov exponents that are obtained in the limit $t\to\infty$ and do
not depend on the metric used.} in $x$.
They are also expressible via an exponentially decaying potential $\Phi_\pm$:
\beq
\label{expansionrate}
L_\pm(x)=\L_\pm(\underline{\varepsilon}(x)),\qquad
\L_\pm(\underline{\varepsilon})=\sum_{X\circ\,0}
\Phi_\pm(\underline{\varepsilon}_X) \ .
\eeq

\subsubsection{Representation of the volume measure}

A truncated compatible sequence $(\varepsilon_{-T},\cdots,\varepsilon_T)$ 
represents the sets of all points that share the same dynamical history 
in the time interval $[-T,T]$ and have different histories outside. By the
smoothness properties of Anosov systems this set defines a parallelepiped
$\D_T \subset M$. The volume of $\D_T$ is given by~\cite{GBG04}
\beq\label{volumeeps}
{\rm Vol}(\D_T) \propto \exp \left[ B(\ul\varepsilon)
-\sum_{i=0}^{T-1} \L_+ (\TT^i \ul\varepsilon) - \sum_{i=-T}^{-1} \L_-
(\TT^i \ul\varepsilon)
\right] \ ,
\eeq
where $\TT$ is the translation to the left by one unit.
$B(\ul\varepsilon)$ is the sum of boundary terms that decay exponentially 
around $i=0$ and around $i=-T,T$.

The average of a given smooth observable $F$ in stationary state can be
defined extracting an initial datum with respect to the volume measure
and then computing $F(S^t x)$ for $t \to \io$. If the system reaches a
stationary state, the average over the volume measure w.r.t. initial data
is equivalent to the average over the SRB measure describing the stationary
state. Then one has, if $\FF(\ul\varepsilon)$ is the short range function
representing $F$:
\beq
\la F \ra_{srb} = \lim_{t\to\io} \la F(S^t x) \ra_{vol} = \lim_{t\to\io}
\lim_{T\to\io}\frac{ \sum_{\ul\varepsilon}  e^{\left[ B(\ul\varepsilon)
-\sum_{i=0}^{T-1} \L_+ (\TT^i \ul\varepsilon) - \sum_{i=-T}^{-1} \L_-
(\TT^i \ul\varepsilon)\right]} \FF(\TT^t \ul \varepsilon)}
{ \sum_{\ul\varepsilon}  e^{\left[ B(\ul\varepsilon)
-\sum_{i=0}^{T-1} \L_+ (\TT^i \ul\varepsilon) - \sum_{i=-T}^{-1} \L_-
(\TT^i \ul\varepsilon)\right] }} \ .
\eeq

\subsubsection{Representation of the SRB distribution}

The function $ \FF(\TT^t \ul \varepsilon)$, for large $t$,
depends strongly on the symbols around $\varepsilon_t$ and has a very small
dependence on the symbols $\varepsilon_i$ for $i \leq 0$.
And as the functions $B$ and $\L_-$ appearing in the exponent are instead
peaked around the symbols $\varepsilon_i$ for $i \leq 0$, one can write,
defining $\ul\varepsilon_+ = (\varepsilon_i)_{i=0}^\io$ and 
$\ul\varepsilon_- = (\varepsilon_i)_{i=-\io}^{-1}$:
\beq
\la F \ra_{srb} \sim \lim_{t\to\io}
\lim_{T\to\io}\frac{ \sum_{\ul\varepsilon_-}  e^{\left[ B(\ul\varepsilon)
- \sum_{i=-T}^{-1} \L_-(\TT^i \ul\varepsilon) \right]}
 \sum_{\ul\varepsilon_+}  e^{\left[ - \sum_{i=0}^{T-1} \L_+ (\TT^i \ul\varepsilon)
\right]} \FF(\TT^t \ul \varepsilon)}
{  \sum_{\ul\varepsilon_-}  e^{\left[ B(\ul\varepsilon)
- \sum_{i=-T}^{-1} \L_-(\TT^i \ul\varepsilon) \right]}
 \sum_{\ul\varepsilon_+}  e^{\left[ - \sum_{i=0}^{T-1} \L_+ (\TT^i \ul\varepsilon)
\right]}} \ ,
\eeq
so that the terms containing $B$ and $\L_-$ factorize and one finally gets,
with some simple changes of variable in order to center in the origin 
the sequence appearing in the argument of $\FF$:
\beq\label{muSRBAPP}
\la F \ra_{srb} =
\lim_{T\to\io}\frac{
 \sum_{\ul\varepsilon}  e^{\left[ - \sum_{i=-T}^{T-1} \L_+ (\TT^i \ul\varepsilon)
\right]} \FF(\ul \varepsilon)}
{  \sum_{\ul\varepsilon}  e^{\left[ - \sum_{i=-T}^{T-1} \L_+ (\TT^i \ul\varepsilon)
\right]}} \ .
\eeq
The interpretation of the above expression is the following. Given a (large) $T$,
one can consider again the sets $\D_T$ of points which share the same history
in $[-T,T]$. The volume of $\D_T$ was given by Eq.~(\ref{volumeeps}). The total
probability of the set $\D_T$ w.r.t. the SRB distribution describing the stationary
state is instead given by
\beq\label{muSRB}
\mu_+(\D_T) \propto \exp \left[ - \sum_{i=-T}^{T-1} \L_+ (\TT^i \ul\varepsilon)
\right] =  \exp \left[ - \sum_{i=-T}^{T-1} L_+(S^i x) \right] \ ,
\eeq
where the sequence $\ul\varepsilon$ has to be completed 
(rather arbitrarily) to an infinite compatible sequence by continuing 
$\V\e$ to the right with a sequence
$\V\e_{>}$
and to the left with a sequence
$\V\e_{<}$
into $(\V\e_{<},\V\e,\V\e_{>})$ 
so that $\V\e_{<}$ depends only on the symbol $\e_{-T}$ and
$\V\e_{>}$ depends only on the symbol $\e_{T}$: see
\cite{Si68,Ga02,GBG04}. Equivalently, $x \in \D_T$ is the point 
represented by $(\V\e_{<},\V\e,\V\e_{>})$.
Clearly, by sending $T \to\io$ Eq.~(\ref{muSRB}) becomes exact.

\subsubsection{Remarks}

\noindent
{\it (1)}
The SRB distribution, represented as a distribution over
the (compatible) symbolic sequences $\underline{\varepsilon}$, is a
{\it Gibbs distribution} for the short
range potential $\Phi_+=(\Phi_+(\underline{\varepsilon}_X))$ defined in
Eq.~(\ref{expansionrate}), \ie
\beq
\label{15}
\langle F \rangle_{srb}=
\lim_{T\rightarrow\infty} 
\frac{\sum_{\underline{\varepsilon}}e^{-\sum_{X\subset
\Lambda_T}\Phi_+(\underline{\varepsilon}_X)} {\cal F}(\underline{\varepsilon})}
{\sum_{\underline{\varepsilon}}e^{-\sum_{X\subset
\Lambda_T}\Phi_+(\underline{\varepsilon}_X)}}
\eeq
where $\Lambda_T=(-T,\ldots,T)\subset \ZZZ$, the sums extend over
compatible configurations $\underline{\varepsilon}=
(\varepsilon_{-T},\ldots,\varepsilon_T)$ 
(\ie with
$\prod_{i=-T}^{T-1} T_{\varepsilon_i,\varepsilon_{i+1}}=1$), and
${\cal F}(\underline{\varepsilon})$ is an
arbitrary smooth observable defined on phase space regarded as a
function on the symbolic sequences. This representation is equivalent 
to Eq.~(\ref{muSRBAPP}) or (\ref{muSRB}).

\noindent
{\it (2)}
The reduction of the problem of studying the SRB
distribution to that of a Gibbs distribution for a one
dimensional chain with short range interaction (this is the physical
interpretation of Eq.~(\ref{15})) is surprising and 
generates the possibility of studying
more quantitatively at least some of the problems of nonequilibrium
statistical mechanics outside the domain of nonequilibrium
thermodynamics.

\noindent
{\it (3)} The SRB distribution, as expressed by Eq.~(\ref{muSRB}), is
{\it not} absolutely continuous w.r.t. volume, as expected. 
Indeed, Eq.~(\ref{muSRB})
is not proportional to the volume of $\D_T$, which is instead given by
Eq.~(\ref{volumeeps}). The factor relating the two expressions is the
exponential of a sum of Lyapunov exponents that becomes singular in the
limit $T \to \io$.

\noindent
{\it (4)} The SRB measure of a set centered around a point $x$ 
{\it depends on the whole dynamical history} of $x$: indeed, it is
the exponential of the sum of the positive Lyapunov exponents along
the trajectory of $x$ in $[-T,T]$, $T\to\io$. This is very different
to what happens in equilibrium where the measure of a set is simply
its volume, possibly multiplied by some weight, \eg $\exp (-\b V(x))$.
In this sense it is said that 
{\it nonequilibrium ensembles are dynamical}~\cite{GC95a,GC95b}.

\section{The fluctuation relation}

The fluctuation relation is a symmetry property of
the probability distribution function ({\sc pdf}) 
of a quantity $\sigma$ that will be defined below and
coincides with the phase space contraction rate in Anosov systems.

Consider a dynamical system, deterministic or stochastic, and the
space of its trajectories $x(t)$. 
If $\s_+ \equiv \langle \s(t) \rangle > 0$, one can define a variable
$p$ as in Eq.~(\ref{2}):
\beq 
\label{pdef}
\begin{split}
&\s_\t \equiv \int_{-\T/2}^{\T/2} dt \ \sigma(t) \ , \\
&p[x(t)] \equiv \frac{1}{\T \sigma_+} \int_{-\T/2}^{\T/2} dt \ \sigma(t) = 
\frac{\s_\t}{\la \s_\t \ra} \ ,
\end{split}\eeq 
where the angular brackets denote an average over the trajectories $x(t)$ 
weighted with the stationary state measure. The symmetric interval
$[-\t/2,\t/2]$ has been chosen in order to have simpler formulae in the
following.
The {\sc pdf} of $p$ is defined as 
\beq 
\pi_\T(p)={\cal P}\big\{p[x(t)] = p \big\} = \langle 
\, \delta (p[x(t)] - p ) \, \rangle \ ,
\eeq 
and the large deviation function is given by 
\beq 
\label{zetapdef} 
\zeta_\io(p)= \lim_{\T \rightarrow \infty} \T^{-1} \log \big[ \pi_\T(p) / \p_\T(1) \big] \ ,
\eeq  
and is normalized by the condition $\z_\io(1)=0$, where $\la p \ra=1$ is the average value
of $p$, \ie the value of $p$ in which $\z_\io(p)$ assumes its maximum.
The {\it characteristic function} is given by
\beq 
\label{phidef} 
z_\io(\lambda) = - \lim_{\T \rightarrow \infty} \T^{-1}  
\log \langle \exp[-\lambda \sigma_\T ] \rangle \ ,
\eeq 
and is the Legendre transform of $\z_\io(p)$. Indeed, for large $\t$,
\beq
e^{-\T z_\io(\l)} = \langle e^{-\l \s_\T } \rangle = 
\frac{\int dp \, e^{\T [ \z_\io(p) - \l p \s_+ ]}}{\int dp \, e^{\T \z_\io(p)}} \sim
\frac{e^{\T \max_p [ \z_\io(p)-\l p \s_+]}}{e^{\T \z_\io(1)}} \ ,
\eeq
so that, recalling that $\z_\io(1)=0$ by construction,
\beq\label{Legendre1}
z_\io(\l)=-\max_p [ \z_\io(p)-\l p \s_+] \ .
\eeq
The inversion of the Legendre transform yields
\beq\label{Legendre2}
\z_\io(p)=\min_\l [ \l p \s_+ - z_\io(\l) ] \ .
\eeq
The fluctuation relation is given, in terms of $\z_\io(p)$, by Eq.~(\ref{FR}),
and by Eq.~(\ref{Legendre1}), it can be formulated in terms of $z_\io(\l)$ as
$z_\io(\lambda)=z_\io(1-\lambda)$. 

\subsection{Internal symmetries and the fluctuation relation}
\label{sec:LS}

Assume that there exist a map $I$ on the space of trajectories $x(t)$ such 
that $I^2 = 1$ and that the measure ${\cal D} x$ is invariant under $I$,
i.e. ${\cal D} Ix = {\cal D}x$. Then, consider a segment of trajectory
$x(t)$, $t\in [-\t/2,\t/2]$ and define
\beq\label{sigmaLS}
\s_\t = -\log \frac{{\cal P}[Ix(t)]}{{\cal P}[x(t)]} \ ,
\eeq
where ${\cal P}[x(t)]$ is the probability of observing $x(t)$, 
{\it in stationary state}, for
$t\in [-\t/2,\t/2]$ irrespectively of what happens outside the interval
$[-\t/2,\t/2]$. The {\sc pdf} of $\s_\t$ verifies
the fluctuation relation:
\beq\begin{split}
\langle e^{-\l \s_\t} \rangle &= 
\int {\cal D}x \, {\cal P}[x(t)] e^{-\l \s_\t} =
\int {\cal D}x \, {\cal P}[x(t)]^{1-\l} \, {\cal P}[Ix(t)]^\l \\ &=
\int {\cal D}x \, {\cal P}[Ix(t)]^{1-\l} \, {\cal P}[x(t)]^\l =
\langle e^{-(1-\l) \s_\t} \rangle \ .
\end{split}\eeq
Thus, if the limit (\ref{phidef})
exists, it verifies the relation $z_\io(\l) = z_\io(1-\l)$, from which the
fluctuation relation for the {\sc pdf} of $\s_\t$ follows.
This definition of $\s_\t$ was proposed by Lebowitz and Spohn~\cite{LS99},
who showed that the limit $z_\io(\l)$ indeed exists for generic Markov
processes and it is a concave function of $\l$. Moreover they showed that
$\s_\t$ can be identified with the entropy production rate --over the time
interval $\t$-- in stationary state up to boundary terms, 
i.e. terms that do not grow with $\t$, if $I$ is chosen to be the time 
reversal, $Ix(t)=x(-t)$. It turns out that $\s_\t$ can be identified with
the entropy production rate over the time interval $\t$ in many of the physical
interesting cases, as will be discussed below.

\subsection{The fluctuation theorem}
\label{sec4:FTheorem}

The heuristic derivation above can be formulated as a theorem for reversible 
Anosov maps~\cite{GC95a,GC95b}.
In these deterministic systems, the stationary measure over the space of 
trajectories $S^t x$ is simply the SRB measure over the initial 
data $x$, and is given by Eq.~(\ref{muSRB}).
The map $I$ is the time reversal map, which satisfies $I^2=1$ and is metric 
preserving (that means that the measure ${\cal D}x$ is invariant under its action).
Moreover the time reversal $I$ is defined by $IS=S^{-1}I$, so that 
$L_+(S^t Ix) = L_+( I S^{-t} x)=-L_-(S^{-t}x)$, where the last equality
follows from $L_+(Ix)=-L_-(x)$ which holds for reversible systems. 
Then the quantity $\s_\t$ defined in
Eq.~(\ref{sigmaLS}) becomes
\beq\begin{split}
\s_\t &= -\log \frac{\m_+(Ix)}{\m_+(x)} = 
\sum_{t=-\t/2}^{\t/2} L_+(S^t I x) -
\sum_{t=-\t/2}^{\t/2} L_+(S^t x) \\ 
&= -\sum_{t=-\t/2}^{\t/2} L_+(S^t x) + L_-(S^t x) =
- \sum_{t=-\t/2}^{\t/2} \s(S^t x) \ ,
\end{split}\eeq
so that $\s_\t$ can be identified with the phase space contraction rate
$\s(x) = - L_+(x) - L_-(x)$
integrated over the time interval $\t$, as in Eq.~(\ref{2}).
It follows that the validity of the fluctuation relation for Anosov systems 
is a consequence of reversibility and of the structure of the SRB measure, 
Eq.~(\ref{muSRB}).

\section[Entropy production rate]
{Entropy production rate and some consequences of the chaotic hypothesis}
\label{sec4:implications}

In many practical cases~\cite{Ga99,EM90,Ru04,ECM93,BGG97,BCL98,ZRA04a,GZG05,LS99} 
it turns out that the quantity $\s_\t$ defined in
Eq.~(\ref{sigmaLS}) can be interpreted as the total entropy production over
the time interval $\t$, \ie the integral of an {\it entropy production rate}
$\dot s(t)$ which is usually given by the power injected in the system by
the nonconservative forces acting on it divided by some ``temperature'',
which in systems of classical particles is identified with the kinetic 
temperature and in stochastic systems in contact with a thermal bath
is the temperature of the thermal bath.

However this identification is still matter of debate,
see \eg~\cite{Ga99,EM90,Ru04,DGM,Ga04,LS99}.
On a practical ground, there are many nonequilibrium situations in
which it is not clear how to define a ``temperature'': this happens
in strongly nonequilibrium regimes, for instance when the driving
force is very strong or in the glassy cases discussed in the first
chapters. So the definition of entropy production rate as the
injected power divided by the temperature is not always free of
ambiguities.

It is not obvious that 
the problem of defining the notions of ``entropy'' and ``entropy
production rate'' in general nonequilibrium situations can be
solved. In chapter~\ref{chap:sette} some aspects of this problem
will be discussed. For the moment, the entropy production rate
will be identified with Eq.~(\ref{sigmaLS}), or, in the case of
dynamical systems, following the chaotic hypothesis and the discussion
in section~\ref{sec4:FTheorem}, with the
{\it phase space contraction rate} $\s(x)$. Some consequences
of this identification will now be discussed.

\subsection{The fluctuation relation close to equilibrium}
\label{sec4:closetoeq}

\subsubsection{Green-Kubo relations}

It was proved in \cite{Ga96a,Ga96b,GR97} that the fluctuation theorem implies,
in the equilibrium limit ($\s_+ \rightarrow 0$), the Green-Kubo relations for
transport coefficients. This holds {\it if the identification between entropy
production rate and phase space contraction rate is accepted}, at least close
to equilibrium.

Suppose that a constant driving force $E$ is applied to
a system in equilibrium. This generates a corresponding flux $J(t)$ (\eg if $E$
is an electric field $J(t)$ is the electric current) such that, close to
equilibrium, the dissipated power can be written as 
$W(t) = E J(t) + O(E^3)$~\cite{DGM}.  
At first order in the force $E$, the temperature is simply the equilibrium
temperature, and the entropy production rate is~\cite{DGM}: 
\beq
\label{sclosetoeq} 
\s(t) = \frac{W(t)}{T} = \frac{E J(t)}{T} + O(E^3) \ .
\eeq 
The fluctuation relation can
be written as $z_\io(\l)=z_\io(1-\l)$ where $z_\io(\l)$ is defined in
Eq.~(\ref{phidef}) and $\s_\t$ in Eq.~(\ref{pdef}). 
The derivatives of $z_\io(\l)$ are the moments of $\s_\t$,
\ie 
\beq \label{eq:moments}
z^{(k)}_\io \equiv \left. \frac{d^k z_\io}{d\l^k} \right|_{\l=0} =
(-1)^{k-1} \lim_{\t\rightarrow\io} \t^{-1} \langle \s_\t ^k \rangle_c \ , 
\eeq
where $\la A^k \ra_c$ denotes the connected correlation
(\eg $\langle A^2 \rangle_c = \langle A^2 \rangle - \langle A \rangle^2$).
It is possible to show, see Appendix~\ref{app4:closetoeq}, that 
$z^{(1)}_\io = \s_+ \sim E^2$ and that, for $k>1$,
$z^{(k)}_\io \sim E^k$. Then, close to equilibrium ($E \sim 0$)
$z_\io(\l)$ is well approximated by a second order polynomial (corresponding to a
Gaussian {\sc pdf}),
\beq \label{zgauss}
z_\io(\l) = z^{(1)}_\io \l + \frac{z^{(2)}_\io}{2} \l^2 + O(E^3 \l^3) \ ,
\eeq 
and the fluctuation relation, $z_\io(\l)=z_\io(1-\l)$, implies $z^{(2)}_\io = -2z^{(1)}_\io$;
from equation (\ref{eq:moments}), recalling that $\s_\t = \int_{-\t/2}^{\t/2} dt \ \s(t)$ 
and using time-translation invariance, 
\beq
z^{(2)}_\io = -2z^{(1)}_\io
\hspace{10pt} \Rightarrow \hspace{10pt} 
\s_+ = \int_0^\io dt \ \langle \s(t) \s(0) \rangle_c \ . 
\eeq  
Substituting $\s(t)=E J(t)/T$ one obtains 
\beq 
\langle J \rangle_E = 
\frac{E}{T} \int_0^\io dt \  \langle J(t) J(0) \rangle_{E=0} + O(E^2) \ , 
\eeq 
that is to say, the Green-Kubo relation.

In the case where many forces $E_i$ are applied to the system, each of 
them is associated to the corresponding current $J_i$ and the dissipated
power is, at first order in $\ul E$, 
$W(t) = \sum_i E_i J_i(t) = \ul E \, \ul J(t)$.
In the limit $\ul E \rightarrow \ul 0$, an extension of the fluctuation 
theorem to the joint fluctuations of $p$ and of 
$J_i(t) \equiv T \partial_{E_i} \s(t)$ \cite{Ga96a} 
leads then to Green-Kubo's formulas for transport
coefficients:
\beq
\label{GK}
\mu_{ij} \equiv \lim_{\ul E \rightarrow \ul 0} \frac{\langle J_i 
\rangle_{\ul E}}{E_j} = \frac{1}{T}
\int_0^\infty dt \ \langle J_i(t) J_j(0) \rangle_{\ul E=\ul 0} \ ,
\eeq
and to Onsager reciprocity, $\mu_{ij}=\mu_{ji}$ \cite{Ga96a,Ga96b,GR97}.

\subsubsection{Fluxes far from equilibrium}

If the identification between entropy production rate and phase
space contraction rate is accepted also far from equilibrium,
one can define a ``duality'' between fluxes $\ul{{\cal J}}$ and forces 
$\ul E$ using $\sigma(x)$ as a ``Lagrangian'' \cite{Ga04}:
\beq
\label{Jdef}
\ul{{\cal J}}(\ul E,x) = \frac{\partial \sigma(x)}{\partial \ul E} \ .
\eeq
Close to equilibrium one has $\ul{{\cal J}}(t) = \ul J(t)/T$ and the
Green-Kubo relations follow.

\subsubsection{Gaussian distributions}

The only assumption in the derivation of the Green-Kubo relation above
was that in the limit $\s_+ \to 0$ the distribution $\pi_\t(p)$ can be
approximated by a Gaussian, or equivalently Eq.~(\ref{zgauss}),
see Appendix~\ref{app4:closetoeq}.
Thus, if under some particular conditions the distribution of $p$ is observed to be
a Gaussian over the whole accessible range\footnote{The accessible range must include
negative values of $p$, otherwise the fluctuation relation cannot be applied.}, 
one obtains from the fluctuation
relation the same relation $z^{(2)}_\io = -2z^{(1)}_\io$ which is, in this
case, an extension of a Green--Kubo formula, Eq.~(\ref{GK}), to finite forces.

One sees this by considering, for
instance, cases in which $\sigma(x)$ is linear in $E$ (as it will happen in
the cases that will be studied in the following).
Indeed, if $\sigma(t)=E {\cal J}(t)$, one obtains the relation
\beq
\label{GKestesa}
\frac{\langle {\cal J} \rangle_E}{E} = \int_0^\infty dt \
[\langle {\cal J}(t) {\cal J}(0) \rangle_E - \langle {\cal J} \rangle_E^2] \ ,
\eeq
valid, {\it subject to the Gaussian assumption}, 
also for $E\ne0$. If $\s(t)$ is non linear in $E$, Eq.~(\ref{GKestesa})
will assume the appropriate form for the system under investigation.
The leading order in $E$ of the latter relation 
({\it linear response}) is obviously the Green-Kubo formula,
Eq.~(\ref{GK}).

%%%%%%%%%%%%%%%%%%%%%%%%%%%%%%%%%%%%%%%%%%%%%%%%%%%%%%%%%%
%%%%%%%%%%%%%%%%%%%%%%%%%%%%%%%%%%%%%%%%%%%%%%%%%%%%%%%%%%

\subsection{(Dynamical) ensembles equivalence}

Following what is usually done in equilibrium, one can define
a {\it nonequilibrium} ensemble as the collection of probability
distributions (SRB distributions) describing the stationary states 
of a given system, say Eq.~(\ref{INTRO1}), when the parameters 
$N$, $V$, $E$, etc. are varied~\cite{Ga99,Ga04}.

However, as was already remarked above, the SRB distributions
depend explicitly on the dynamics of the system, see Eq.~(\ref{muSRB}), 
hence on the details of the equations of motion, \eg Eq.~(\ref{INTRO1}).
This means that they can in principle depend on the precise form
of the thermostatting force $\ul \th_E(\ul q,\dot{\ul q})$ 
which ensures the existence
of a stationary state by subtracting the energy which is injected
by the nonconservative forces. Then, there is much more freedom in
nonequilibrium to define ensembles: for instance, one can choose
$\ul \th_E(\ul q,\dot{\ul q})$ in order to keep the total energy 
(or kinetic energy)
fixed, or one can simply set 
$\ul \th_E(\ul q,\dot{\ul q})=\nu \dot{\ul q}$, with $\n$ a constant
friction. Also one can
use a {\it stochastic} thermostat, if $\ul \th_E(\ul q,\dot{\ul q})$
is a random variable.
The thermostat could be chosen to act in the bulk of the system or 
only on the boundaries, etc.

All these choices will lead to a different probability distribution 
for the stationary state. For example, the first two define a
{\it reversible} equation, so one can expect that the resulting SRB
distribution will verify the fluctuation relation. But if one sets
$\ul \th_E(\ul q,\dot{\ul q})=\nu \dot{\ul q}$, the resulting
equation {\it is not reversible}, so in principle the resulting SRB
distribution should not satisfy the fluctuation relation.
Moreover, in the case of a stochastic thermostat, the resulting 
probability distribution is expected to admit a density w.r.t. volume
from the general theory of stochastic processes, while the SRB 
distributions describing the other thermostats is concentrated
on a set of zero volume. Nothing could seem more different~\cite{Ga99}.

Nevertheless, it might still be true that, if the size of the system
is large enough, and if one looks only to a small portion of the
system that is far from the boundaries, the resulting statistical
behavior is the same. This is what one should expect on ``physical''
ground, and is similar to what happens in equilibrium where the
canonical and microcanonical ensembles give the same statistical
behavior even if the second is supported on a constant energy surface
which has zero measure w.r.t. the first.
Obviously, the fluctuations of the total energy will be very
different in the canonical and microcanonical ensembles, so one should
keep in mind that the equivalence might hold only if one looks to
a small volume of the total system which is far from the boundaries and
if the details of the system are not probed by the observables under
investigation.

A general theory of ensembles and their equivalence is still lacking so
the statements above are still only conjectures~\cite{Ga99,Ga04,Ga99b,Ru99,Ru00}.
However:

\noindent
{\it (1)} in systems of classical particles like Eq.~(\ref{INTRO1}), some arguments
have been proposed that support
the equivalence of some particular choices of the function $\ul \th_E(\ul q,\dot{\ul q})$
within linear response theory~\cite{EM90,ES93,Ev93} and beyond \cite{Ru00};

\noindent
{\it (2)} in some applications to fluid mechanics the equivalence of the Navier-Stokes
equations to similar, {\it reversible}, equations has been numerically
shown~\cite{Ga99,Ga02,GRS04,SJ93}.

Some numerical results about nonequilibrium ensemble equivalence have been
obtained during this work and were published in~\cite{ZRA04a}. They will not
be reproduced here for reasons of space.

\subsection{Singularities}
\label{sec4:singularities}

The application of the discussion above to concrete cases poses some
problems due to the presence of singularities, \eg the divergence in
the origin of the Lennard--Jones potential. Indeed, the main assumption
defining Anosov systems is smoothness, which is clearly violated by
the Lennard--Jones potential due to the presence of the singularity in
the origin. It turns out that, if the discussion is suitably adapted,
still one can expect the fluctuation relation to hold~\cite{BGGZ05}. 
This is the purpose of this section.

The problems arise if one consider Eq.~(\ref{INTRO1}) in presence of
an unbounded potential $V(\ul q)$ and if $\ul \th(\ul q, \dot{\ul q})$ is
chosen in order to keep the total kinetic energy constant.
In this case the phase space contraction rate 
$\s(\ul q, \dot{\ul q})$ has the form
\beq
\s(\ul q, \dot{\ul q}) = \frac{E J(\ul q,\dot{\ul q})}{T} - \frac{1}{T} \frac{dV}{dt} \ ,
\eeq
\ie it is the sum of the dissipated power (whose precise form in a concrete case will be given
in next chapter) divided by the kinetic temperature and of a term which is the total
derivative of the potential energy divided by $T$.

Even in the conservative case $E=0$, $\s$ is not identically $0$: it has
{\it zero average} but still has fluctuations. Moreover, as the potential
energy is unbounded, the integral $\s_\t$ will contain a term
$[V(-\t/2)-V(\t/2)]/T$ which is unbounded.
The problem is that the fluctuation relation was derived under the assumption
that $\s_\t$ is {\it bounded} (being a smooth function on a compact manifold). 
The ``spurious'' unbounded fluctuations due to the total
derivative term, which has zero average so it does not contribute to the dissipation,
will produce a violation of the fluctuation relation. The purpose of the following 
discussion is to explain the effect of this term and why it should be removed to
obtain a proper definition of phase space contraction rate in singular systems.

\subsubsection{Conservative systems and total derivatives}

The situation described above (for $E=0$) is a particular instance of a system that
is {\it conservative} but does not have an identically vanishing $\s(x)$.

Indeed, Ruelle showed~\cite{Ru96} that in general $\s_+ \geq 0$, and that
if $\s_+ = 0$ then $\s(x)$ has necessarily the form $\s(x)=f(Sx)-f(x)$,
or $\s(x) = \frac{df}{dt}$ in the continuous case, for a suitable function
$f(x)$. Note that this happens also if one considers a system that conserves
the volume, so that $\s(x) \equiv 0$, and changes the metric on phase space
to a new metric $d'x = \exp[-f(x)]dx$. In this case the phase space contraction
rate w.r.t. the new metric is $\s'(x) = \frac{df}{dt}$. Thus total derivatives 
in $\s(x)$ can be eliminated by changing the metric on $M$. This means that
if $\s(x)$ is a total derivative
the system still admits an absolutely continuous SRB distribution w.r.t. 
volume. This will be the general definition of {\it conservative} system that will
be adopted in the following.

If $f(x)$ is bounded, the variable
\beq \label{adef}
a=\frac1\t\sum_{j=-\t/2}^{\t/2-1}
\s(S^jx)\=\frac{f(S^{\frac{\t}2}x)-
f(S^{-\frac{\t}2}x)}\t
\eeq
is also bounded by $\t^{-1}$ and tends {\it uniformly} to $0$. 
In the {\it dissipative} case, $\s_+ > 0$, if $\s(x) = \s_0(x) + \frac{df}{dt}$,
the variable $p$ defined in
Eq.~(\ref{pdef}) will be simply given by $p = p_0 + a/\s_+$, where $p_0$
is obtained substituting in Eq.~(\ref{pdef}) the expression of $\s(x)$
{\it without} the total derivative term. But for large $\t$ the term
$a/\s_+$ tends uniformly to $0$ so one simply has $p=p_0$. This means
that total derivatives of {\it bounded} functions can always be neglected
in the definition of $p$, Eq.~(\ref{pdef}).

\subsubsection{A simple example: Anosov flows}

The latter statement is not true if the function $f(x)$ can become 
infinite in some point $x_0$, as is the case for the Lennard--Jones
potential $V(x)$.

The simplest example (out of many) is provided by the simplest
conservative system which is strictly an Anosov transitive system and
which has therefore an SRB distribution: this is the geodesic flow on
a surface of constant negative curvature \cite{BGM00}.
The phase space $M$ is compact, time reversal is just momentum
reversal and the natural metric, induced by the Lobatchevsky metric
$g(x)$ on the surface, is time reversal invariant: the SRB
distribution is the Liouville distribution and $\s(x)\equiv0$. However
as before one can define a new metric as $g'(x)=e^{-f(x)} g(x)$
where $f(x)$ is a time reversal invariant function
on $M$ whose modulus is very large in a small vicinity of a point $x_0$, 
arbitrarily selected, 
and constant outside a slightly larger vicinity of $x_0$.
The rate of change of phase space volume in the new metric will be
$\sigma'(x)=\frac{df}{dt}$ and
since $f$ is arbitrary one can achieve a value of
$\s'(x)$ as large as wished by fixing suitably the function $f$.

Nevertheless, as long as $f(x)$ is bounded, $a=\frac1\t \int_0^\t
\s'(S_t x) dt= \t^{-1} \big[f(S_\t x)-f(x)\big]\rightarrow_{\t\to\infty}0$ 
(as in the corresponding map case), and the SRB
distribution of $a$ will be a delta function at $0$.
But if $f(x)$ is not bounded
(\eg if it is allowed to become infinite in $x_0$) the distribution of
$a$ can be different from a delta function at $0$ also for
conservative systems, yielding a finite large deviation function 
$\widetilde \z(a) = \lim_{\t\to \io} \frac1\t \log \pi_\t(a)$. 
This will affect the distribution of $p$ also for $\t \to \io$. In this
case it is clear that this is a ``spurious'' effect related to a very
strange choice of the metric on $M$, but realizing this in realistic systems 
can be sometimes difficult.

\subsubsection{The effect of singular boundary terms}

One can show that terms of the form $\big[f(S_\t x)-f(x)\big]$ with
$f(x)$ not bounded can affect the large fluctuations of $\s(x)$.  This
is a valuable and interesting remark brought up for the first time
in \cite{VzC04}.

Consider a system such that
$\s(x)$ has the form $\s(x)=\s_0(x)-\b\frac{dV}{dt}$, $\b=1/T$,
and $\s_0(x)$ is bounded. 
In such cases the non normalized variable $a \equiv p \s_+$, 
see Eq.~(\ref{adef}),
has the form 
\beq
a=\frac1\t\int_0^\t \s(S_tx)dt\=a_0-\frac\b\t (V_f-V_i) \ ,
\eeq
where $V_i,V_f$ are the values of the potential at the initial and
final instants of the time interval of size $\t$ on which $a$ is
defined, and $a_0 \equiv \frac1\t\int_0^\t \s_0(S_tx)dt$.

If the system is chaotic the variables $a_0, V_i,V_f$ can be
regarded as independently distributed and the distribution of $V=V_i$
or $V=V_f$ is essentially $\sim e^{-\b V} dV$, \ie close to the
equilibrium Gibbsian distribution \cite{EM90}, equal to leading
order as $V\to\io$ to $e^{-\b V}$.  Therefore the probability
distribution of the variable $a$ can be computed as
\beq\begin{split}
e^{\t \tilde \z(a)} &= \int_{- p^* \s_+}^{p^*\s_+} da_0 \int_0^\io dV_i 
\int_0^\io dV_f \, 
e^{\t \tilde \z_0(a_0) - \b V_i -\b V_f} \d[\t(a-a_0)-\b V_f + \b V_i]  \\
&= \int_{- p^* \s_+}^{p^*\s_+} da_0 \, e^{\t [ \tilde \z_0(a_0) - |a-a_0| ]} \ ,
\end{split}\eeq
thus
\beq
\widetilde \z(a)=\max_{a_0 \in [-p^* \s_+,p^* \s_+]}
\Big[ \widetilde\z_0(a_0) - |a-a_0| \Big] \ .
\eeq
Defining $a_\mp$ by $\widetilde \z_0 '(a_\mp) = \pm 1$, 
by the strict convexity of $\widetilde \z_0(a_0)$ it follows
\begin{equation}
\widetilde \z(a) = \left\{ 
\begin{array}{ll}
\widetilde \z_0(a_-) - a_- + a \ \ , & a < a_- \ , \\
\widetilde \z_0(a) \ \ , & a \in [a_-,a_+] \ , \\
\widetilde \z_0(a_+) + a_+ - a \ \ , & a > a_+ \ . \\
\end{array} \right.
\end{equation}
Furthermore, if $\widetilde \z_0(a_0)$ verifies the fluctuation relation,
$\widetilde \z_0(a_0) = \widetilde \z_0(-a_0) + a_0$,
by differentiation it follows that $a_-=-\s_+$, where
$\s_+$ is the maximum of $\widetilde\z_0$, \ie the average of $a$.

\begin{figure}[t]
\centering
\includegraphics[width=.65\textwidth]{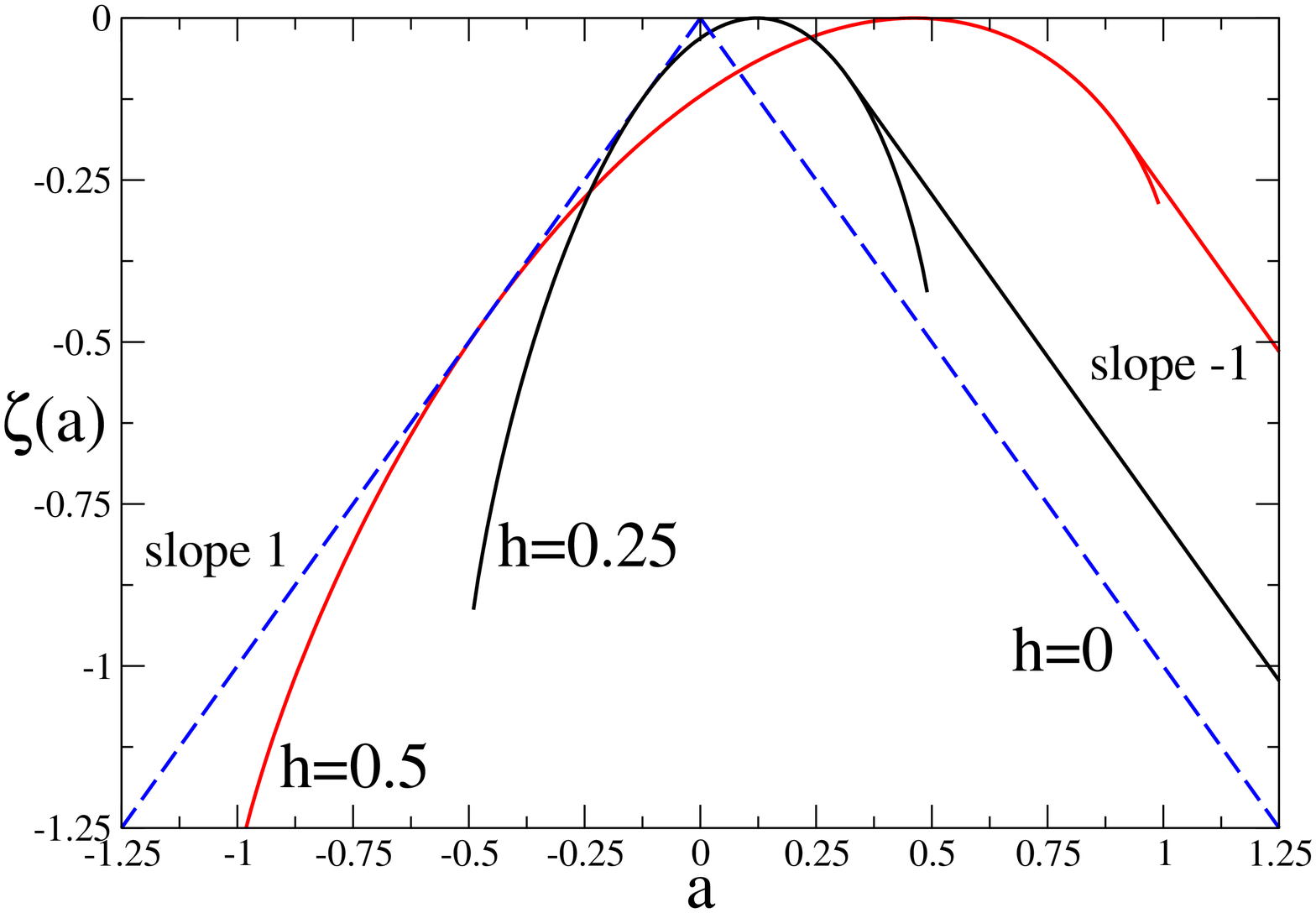}
\caption[An example of the functions $\wt\z(a)$ and $\wt\z_0(a)$]{
An example in a stochastic model. The graph gives
the two functions $\wt\z_0(a)$ and $\wt\z(a)$ for $h=0.,0.25,0.5$.
The average of $a$ is $\la a \ra=\s_+ = 2h \tanh h$,
$a_+ = 2h \tanh 3h$ and $a^* = 2 h$.  The
function $\wt\z(a)$ is obtained from $\wt\z_0(a)$ by continuing it
for $a<a_-=-\s_+$ and $a>a_+$ with straight lines of slope $\pm 1$. 
It does not satisfy the fluctuation relation for $|a|>\la a \ra$. 
As $h\to0$, $\la a \ra \to 0$,
which means that the interval in which the fluctuation relation 
is verified shrinks to $0$.
In this limit $a_+\to 0$, so $\wt \z(a)$ approaches $-|a|$ (dashed lines).
Rephrasing this in terms of
$p=\frac{a}{\la a \ra}$ one obtains that the fluctuation relation 
remains always valid
for $|p|<1$, even as $h\to0$. The three curves for 
$\wt\z_0(a)$ have the same tangent
on left side. The function $\wt\z_0(a)$ is finite {\it only} in the
interval $[-2h,2h]$ and it is $-\infty$ outside it, while the function
$\wt\z(a)$ is finite for all $a$'a and is a straight line outside $[a_-,a_+]$.
}
\label{fig_zeta}
\end{figure}

It follows that, if $\wt\z_0(a_0)$ satisfies the fluctuation relation 
up to $a=p^*\s_+$, then  
$\widetilde \z(a)$ satisfies the fluctuation relation only in the
interval $|a|<|a_-|=\s_+$. Outside this interval $\widetilde\z(a)$
does not satisfy the fluctuation relation and in particular for $a\ge a_+$ it is
$\widetilde\z(a)-\widetilde\z(-a)=const.$, as
already described in \cite{VzC04}.  Translated into the normalized
variables $p_0=a_0/\s_+$ and $p = a/\s_+$, this means that,
even if the large deviation function of $p_0$ satisfies the fluctuation relation up to $p^*>1$, 
the large deviation function of $p$ verifies the fluctuation relation only for
$|p|\leq1$. This effect is due to the presence of the singular
boundary term.  

An example of $\widetilde \z(a)$ is reported in Fig.~\ref{fig_zeta}:
it is a simple stochastic model for the FT (taken from
\cite{BGG97}, see also the extensions in \cite{LS99,Ma99}). The
example is the Ising model without interaction in a field $h$, \ie a
Bernoulli scheme with symbols $\pm$ with probabilities
$p_\pm=\frac{e^{\pm h}}{2\cosh h}$.  Defining
$a_0=\frac1\t\sum_{i=0}^{\t-1}2 h\s_i$, so that 
$\s_+ = \langle a_0 \rangle = 2h \tanh h$,
and setting $x\equiv\frac{1+a_0/(2 h)}2$,
and $s(x)=-x\log x-(1-x) \log (1-x)$, one computes
$\wt\z_0(a_0)=s(x)+\frac12 a_0+ const$ which is {\it not Gaussian} and it is
defined in the interval $[-a^*,a^*]$ with $a^*=2 h$. In
this case the large deviation function $\wt\z_0(a_0)$ satisfies the fluctuation relation for
$|a_0|\le a^*$. If a singular term
$V=-\log (\sum_{i=0}^\io 2^{-i-1}\frac{\s_i+1}2)$ is added to $a_0$,
defining $a=a_0+\b(V_i-V_f)$
(with $\b=\log_2 (1+e^{2h})$ so that the probability distribution of $V$ is 
$\sim e^{-\b V}$ for large $V$),
the resulting $\wt \z(a)$ does not verify the fluctuation relation for
$a > \la a \ra = 2h \tanh h$. In particular, for $h\to 0$, the interval
in which the fluctuation relation is satisfied vanishes.

\subsubsection{How to remove singularities}

From the discussion above it turns out that singular terms which are
proportional to total derivatives of unbounded functions (like the
term $\frac{dV}{dt}$ that appears in the phase space contraction rate
of isokinetic systems) can induce ``undesired'' (or ``unphysical'')
modifications of the large deviation function $\z(p)$.

On heuristic grounds, when dealing with singular systems, one could
follow the prescription that unbounded terms in $\s(x)$ which are
proportional to total derivatives should be {\it subtracted} from the
phase space contraction rate.  If the resulting $\s(x)$ is bounded (as
it is \eg for the isokinetic thermostat) then its distribution should
verify the fluctuation relation for $|p|\leq p^*$, $p^*$ being an intrinsic dynamic
quantity defined by $\lim_{|p|\to p^*} \z(p)=-\io$.
If the singular terms are not subtracted, the fluctuation relation will appear to be
valid only for $|p|\leq 1$ even if $p^* > 1$. 

The heuristic prescription above can be motivated by a careful analysis of
the proof of the fluctuation theorem for Anosov flows.
In the following let us call again $a$ the integral of the total phase space
contraction rate $\s(x)$ (which includes singular terms) and $a_0$ the integral
of the bounded variable $\s_0(x)$ from which singular total derivatives
have been removed.

The fluctuation theorem was proved in \cite{Ge98} for Anosov
  flows. Flows are
  associated with maps via Poincar\'e's sections: excluding singular
  sections which pass through a point of singularity in phase space
  (which is a very natural restriction) the chaotic hypothesis would
  lead to a fluctuation relation for the phase space contraction on a Poincar\'e's
  section and this would lead to a fluctuation relation for the flow by the theory in
  \cite{Ge98} {\it only if the variable $\s(x)$ is bounded}. However
  if the Poincar\'e's section is performed avoiding the singular
  points, like the passage through points in which the potential is
  infinite, then the phase space contraction (taken for
  instance assuming as timing events the instants in which the
  potential energy exceeds some large value) has two contributions:
  the first is from $a_0$ which is bounded (at least in the isokinetic
  thermostats) and the second from the
  potential. The latter however has again the form proportional to the
  difference of the potential energy at the initial timing event and
  at the final event: this {\it vanishes} on the section considered above
  while in general it will have the form of $\frac1\t$ times a {\it bounded
  quantity} (unless the section passes through a singularity of the
  potential). Therefore the distribution of $a_0$ will satisfy the fluctuation relation
  ({\it by the chaotic hypothesis}) for $|a_0|<p^* \s_+$.  By the above
  maximum argument, the distribution of $a$ {\it will also verify, as
  a consequence}, the fluctuation relation but only for $|a|\leq \s_+$ 
(\ie for $|p| \leq 1$).

The (natural) prescription is then to consider only Poincar\'e's
sections which do not pass through a singularity of $\s(x)$. The
integral of $\s(x)$ over a large number of timing events on such
sections is equal to the time integral of $\s_0(x)$ plus a bounded
term which can be neglected. Thus the prescription on the Poincar\'e's
section is equivalent to the heuristic prescription of removing from
$\s(x)$ all the unbounded total derivatives.

It follows that the chaotic hypothesis leads to a prediction on
the outcome of possible numerical simulations in which the isokinetic
thermostat is assumed in particle systems interacting via a
Lennard--Jones potential: the fluctuation relation will hold for all $|a|\leq \s_+$ and,
once the term $\frac{dV}{dt}$ is removed, for all $|a_0| < p^* \s_+$
with $p^* \geq 1$, even for the latter models which admittedly are
quite unphysical as the speed of the particles remains essentially
constant even in ``head on'' collisions in which the potential energy
acquire an infinite value.  The numerical results of the following chapter
\cite{ZRA04a,GZG05} closely agree with this prediction and can be
considered as rather demanding tests of it.

\subsection{Some remarks on the chaotic hypothesis}

\noindent
{\it (1)}
It has to be stressed that the chaotic hypothesis concerns physical
systems: it is very easy to find dynamical systems for which it does
not hold. As it is easy (actually even easier) to find systems in
which the ergodic hypothesis does not hold (\eg harmonic lattices or
black body radiation) but, if suitably interpreted, leads to
physically correct results (the specific heats at high temperature)
or, when it really fails, to new scientific paradigms (like quantum
mechanics from the specific heats at low temperature and Planck's
law).

\noindent
{\it (2)}
Since physical systems are almost always not Anosov systems it is very
likely that probing motions in extreme regimes (\eg when particles go
undisturbed through infinite potential walls, as in the
Lennard--Jones isokinetic models) will make visible the features that
distinguish Anosov systems from non Anosov systems: \eg the isokinetic
thermostats satisfy the fluctuation relation for $|p|\leq 1$ only (if the term
$\frac{dV}{dt}$ is not removed from $\s(x)$). Note that this results
{\it can be derived} from the chaotic hypothesis as discussed above:
this is a quite remarkable fact.

\noindent
{\it (3)} 
If the term $\frac{dV}{dt}$ is removed (as will be done in the following),
the resulting quantity $\s_0(x)$ is bounded and its distribution verifies
the fluctuation relation also for $|p|>1$. This prescription, as discussed above, is equivalent
to the very reasonable prescription that the Poincar\'e's section used for
mapping the flow into a map does not pass through a singularity of $\s(x)$.

\section{Appendix: The large deviation function of $\s(x)$ close to equilibrium}
\label{app4:closetoeq}

Close to equilibrium the entropy production rate has the form 
$\s(t) = E {\cal J}(t)$, as discussed in section~\ref{sec4:closetoeq}.
Thus, from Eq.~(\ref{eq:moments}),
\beq
z^{(k)}_\io = (-1)^{k-1} \lim_{\t \to \io} \t^{-1} \la \s_\t^k \ra_c =
(-1)^{k-1} E^k \lim_{\t \to \io}\t^{-1} \int_{-\t/2}^{\t/2} dt_1 \cdots \int_{-\t/2}^{\t/2} dt_k 
\la {\cal J}(t_1) \cdots {\cal J}(t_k) \ra_c \ .
\eeq
The connected correlations $\la {\cal J}(t_1) \cdots {\cal J}(t_k) \ra_c$ are translationally
invariant due to the stationarity of the SRB distribution and {\it decay exponentially} in the
differences $|t_i - t_j|$ due to the short range property of the SRB potentials.

From stationarity it follows that 
$z^{(1)}_\io = E \la {\cal J} \ra$ and as $\la{\cal J}\ra = 0$ in equilibrium, one has
$\la {\cal J} \ra \sim E$ and $z^{(1)}_\io = \s_+ \sim E^2$.
Using stationarity and the exponential decay of the connected correlations one has,
for $k>1$,
\beq\begin{split}
\lim_{\t \to \io}\t^{-1} \int_{-\t/2}^{\t/2} dt_1 & \cdots \int_{-\t/2}^{\t/2} dt_k 
\la {\cal J}(t_1) \cdots {\cal J}(t_k) \ra_c = \\
=&\Iint dt_1 \cdots \Iint dt_{k-1} \la {\cal J}(0) {\cal J}(t_1) \cdots {\cal J}(t_{k-1}) \ra_c 
\equiv {\cal J}^{(k)}_\io \ ,
\end{split}\eeq
and the ${\cal J}^{(k)}_\io$ are finite for $E\to 0$. This means that the $z_\io^{(k)} \sim E^k$
for $k>1$ and 
\beq
z_\io(\l) = z_\io^{(1)} \l + \frac{z_\io^{(2)}}2 \l^2 + O(E^3 \l^3) \ .
\eeq
Equivalently, from the relation $\z_\io(p) = \min_\l [ \l p \s_+ - z_\io(\l)]$ one can prove
that\footnote{Note that $z^{(2)}_\io$ is negative.}
\beq
\z_\io(p) = \frac{\s_+^2}{2 z_\io^{(2)}} (p-1)^2 - 
\frac{z_\io^{(3)} \s_+^3}{6 \big(z_\io^{(2)}\big)^3} (p-1)^3 + \ldots =
 \frac{\s_+^2}{2 z_\io^{(2)}} (p-1)^2 + O(E^3 (p-1)^3) \ ,
\eeq
\ie that $\z_\io(p)$ is approximated by a
Gaussian in an interval $|p-1| \sim 1/E$ whose size grows for
$E\to 0$, see also \cite{Ga96a}.

%% file: cap5.tex
\chapter{Numerical tests of the Fluctuation Relation}
\label{chap5}

\section{Introduction}

As discussed in section \ref{sec4:closetoeq}, close to equilibrium
the {\sc pdf} of the phase space contraction rate is close to a Gaussian,
and the fluctuation relation is equivalent to the Green-Kubo relations
obtained within linear response theory. Thus, a test of the fluctuation
relation in this context is not really independent from linear response
theory.

A rather stringent test of the chaotic hypothesis would be a check which
{\it cannot be reduced to a kind of Green-Kubo relation}; this requires
at least one of the two following conditions to be satisfied:
\begin{enumerate}
\item the distribution $\pi_\tau(p)$ is distinguishable from a Gaussian, or
\item deviations from the leading order in $E$ in Eq.~(\ref{GKestesa}), 
\ie deviations from the Green-Kubo relation, are observed.
\end{enumerate}
This is very hard to obtain in numerical simulations of 
Eq.~(\ref{INTRO1}) for the following
reasons:
\begin{enumerate}
\item to observe deviations from linearity in Eq.~(\ref{GKestesa}) one has to
apply very large forces $E$, then $\sigma_+$ is very large and 
it becomes very difficult
to observe the negative values of $p(x)$ that are needed to 
compute $\zeta_\infty(-p)$ in Eq.~(\ref{FR});
\item deviations from Gaussianity in $\pi_\tau(p)$ 
are observed only for values of $p$ that
differ significantly (of the order of $2$ times the variance) from $1$ and,
again, it is very difficult
to observe such values of $p$.
\end{enumerate}
Due to the limited computational resources available in the past decade,
all numerical computations that can be found in the literature
on systems described by Eq.~(\ref{INTRO1}) found that the measured
distribution $\pi_\tau(p)$ could not be 
distinguished from a Gaussian distribution
in the interval of $p$ accessible to the numerical experiment 
\cite{ECM93,BGG97,BCL98,ZRA04a}. 

In this chapter a test of the fluctuation relation, 
in a numerical simulation
of a system described by Eq.~(\ref{INTRO1}), for large 
applied force when deviations from linearity can be observed, and
the distribution $\pi_\tau(p)$ is appreciably non-Gaussian, will be presented~\cite{GZG05}.
This has become possible
thanks to the fast increase of computational power in the last decade. However,
it is still very difficult to reach values of $\tau$ which
can be confidently regarded as ``close'' to the asymptotic limit 
$\tau \rightarrow \infty$;
thus to interpret the results a theory of the $o(1)$ 
corrections to the function
$\zeta_\infty(p)$ has to be developed 
in order to extract the limiting function 
$\zeta_\infty(p)$ from the
numerical data. Taking into account the latter finite time corrections, 
the fluctuation relation will be successfully tested for 
non--Gaussian distributions and beyond the
linear response theory.
A similar analysis has been presented in \cite{RM03}, where however
negative values of $p$ were not directly observed.

\section{Finite time corrections to the Fluctuation Relation}
\label{sec7:II}

In the present section
a strategy to study (in principle constructively) 
the $O(1)$ corrections
in the exponent of Eq.~(\ref{3}) will be described. 
The theory holds {\it assuming that the time evolution is hyperbolic} 
so that it can be applied to
physical systems only if the chaotic hypothesis is accepted. For
simplicity only the case of discrete time evolution via a
map $S$ are considered.

\subsection{Finite time corrections to the characteristic function}

The distribution of $p$ at fixed $\tau$ can be studied via its Laplace
transform ({\it characteristic function}) $z_\tau(\lambda)$:
\beq
\label{FTC2.1}
z_\tau(\lambda) = -\frac{1}{\tau} \log 
\langle e^{-\lambda \sum_{j=0}^{\tau-1}\sigma(S^jx)}\rangle \ .
\eeq
From the explicit expression of the SRB measure, see Eq.~(\ref{15}), 
it can computed as 
\beq
\label{4.1}
e^{-\tau z_\tau(\lambda)}=\lim_{T\rightarrow\infty} 
\frac{\sum_{\underline{\varepsilon}}
e^{-\sum_{X\subset \Lambda_T}\Phi_+(\underline{\varepsilon}_X)
-\lambda \,\sum_{X\circ\, [0,\tau-1]}s(\underline{\varepsilon}_X) }}
{\sum_{\underline{\varepsilon}}
e^{-\sum_{X\subset\Lambda_T}\Phi_+(\underline{\varepsilon}_X)}} \ .
\eeq
This means that it is the (limit as $T\rightarrow\infty$ of the) ratio between
the partition functions $Z_T(\Phi_+)$ of a Gibbs distribution in $\Lambda_T$
with potential $\Phi_+$
(the denominator) 
and the partition function $Z_T(\Phi_+,\lambda s)$ with the same potential
{\it modified} in the {\it finite} region $[0,\tau-1]\subset\ZZZ$ by
the addition of a
potential $\lambda s(\underline{\varepsilon}_X)$.

The one dimensional systems are very well understood and the above is a
well studied problem in statistical mechanics, known as {\it a
finite size corrections} calculation.  For instance it can be attacked
by {\it cluster expansion} \cite{GBG04}; this is a technique
to deal with the average of the exponential of a 
spin Hamiltonian which is defined in terms of potentials $\phi$
exponentially decaying
with rate $\k$, such as those appearing in the numerator 
and in the denominator of Eq.~(\ref{4.1}). It allows us to represent
them as:
\beq \label{cluster}\sum_\ul\e e^{-\sum_{X\subset\L_T}\phi(\ul\e_X)}=
e^{-\sum_{X\subset\L_T}\wt\phi_X}\;,\eeq
where $\wt\phi_X$ are new {\it
effective} potentials, explicitly computable in terms of suitable averages 
of products of $\phi(\ul\e_X)$'s, and which can be proven to be still 
exponentially decaying with the diameter of $X$ with a rate $0<\k'\le\k$. 

In particular, a representation like Eq.~(\ref{cluster}) 
allows to rewrite the
partition function in the denominator of Eq.~(\ref{4.1}) as:
\beq
\label{4.2}
Z_T(\Phi_+)=\exp \left[ (2T+1) f_\infty(\Phi_+) - c_\infty(\Phi_+)+ 
O(e^{-\kappa'T}) \right] \ ,
\eeq
and the one in the numerator as
\beq
\label{19}
Z_T(\Phi_+,\lambda s)=\exp \Big[ (2T+1-\tau) f_\infty(\Phi_+)+
\tau f_\infty(\Phi_++\lambda s) 
- c_\infty(\Phi_+)- g_\infty(\lambda)+ O(e^{-\kappa'T}+
e^{-\kappa'\tau}) \Big] \ .
\eeq
Therefore
\beq
\label{20}
z_\tau(\lambda)=
f_\infty(\Phi_+)-f_\infty(\Phi_++\lambda s)+
\frac{g_\infty(\lambda)}{\tau}+O(e^{-\kappa'\tau}) 
\equiv z_\infty(\lambda) 
+\frac{g_\infty(\lambda)}{\tau}+O(e^{-\kappa'\tau}) \ .
\eeq 

The function $z_\infty(\l)$ is
convex in $\l$ and the functions $g_\infty(\lambda)$ and
$z_\tau(\lambda)$ are analytic in $\lambda$ (a consequence of
the $1$-dimensionality and of the short range nature of the SRB
distribution): namely
$g_\infty(\lambda)=g^{(1)}_\infty \lambda
+\frac12g^{(2)}_\infty\lambda^2+\ldots$ and
$z_\tau(\lambda)=z^{(1)}_\tau\lambda
+\frac12z^{(2)}_\tau\lambda^2+\ldots$ and the coefficients of
their expansion in a power series of $\lambda$ can be expressed in
terms of correlation functions of $\sigma(x)$. For instance, from
Eq.~(\ref{FTC2.1}) and using the translational invariance of the SRB
measure, 
\beq\label{zio}
\begin{split}
z^{(1)}_\tau &= \tau^{-1} \langle \sum_{j=0}^{\tau-1}
\sigma(S^jx)\rangle = 
\sigma_+ \ , \\
z^{(2)}_\tau &= \tau^{-1} \left[
\langle \sum_{j=0}^{\tau-1}\sigma(S^jx)\rangle^2-
 \langle  \sum_{j=0}^{\tau-1} \sigma(S^jx) 
\sum_{k=0}^{\tau-1}\sigma(S^kx) \rangle
\right] \\
&=- \sum_{k=-\tau+1}^{\tau-1} \left[1-\frac{|k|}{\tau} \right] 
\langle \sigma(S^kx) \sigma(x) \rangle_{c} \ ,
\end{split}
\eeq
where
$\langle \sigma(S^kx) \sigma(x) \rangle_{c} =
\langle \sigma(S^kx) \sigma(x) \rangle - \sigma_+^2$.
Using Eq.~(\ref{20}), 
$g_\infty(\lambda)=\lim_{\tau \rightarrow \infty} 
\tau [ z_\tau(\lambda)-
z_\infty(\lambda) ]$, and the analyticity of
$z_\tau(\lambda)$, one has
$g^{(j)}_\infty=\lim_{\tau \rightarrow \infty} 
\tau [ z^{(j)}_\tau - z^{(j)}_\infty ]$.
Since the connected correlation function 
$\langle \sigma(S^kx) \sigma(x) \rangle_{c}$ {\it decays exponentially}
for $k \rightarrow \infty$, one obtains
\beq
\label{22}
\begin{split}
&g^{(1)}_\infty = 0 \ , \\
&g^{(2)}_\infty = \sum_{k=-\infty}^\infty |k| 
\langle \sigma(S^kx) \sigma(x) \rangle_{c} \ .
\end{split}
\eeq

\subsection{Finite time corrections to $\zeta_\infty(p)$}
\label{sec7:IIC}

A direct measurement of $z_\tau(\lambda)$ from
the numerical data is difficult. What is really accessible to 
numerical observation are the quantities
$\frac1\tau\log \pi_\tau(\{p\in\Delta\})$ in Eq.~(\ref{3}) because the
measured values of $p$ are used to build an histogram obtained by
dividing the $p$--axis into sufficiently small
bins $\D$ and counting how many values of
$p$ fall in the various bins. 
The size of the bins $\D$ will be chosen to be
$|\D|=O(\e_\t/\t)$, with 
$\e_\t$ a small parameter which will be eventually chosen $\e_\t=o(1)$,
see Appendix~\ref{app7:A} for a discussion of this point.
Let also $p_\D$ be the center of 
the bin $\D$. An application of a local form of central limit theorem,
discussed in Appendix~\ref{app7:A}, shows that
the following asymptotic 
representation of $\pi_\tau(\{p\in\Delta\})$ holds:
\beq\label{24}
\pi_\tau(\{p\in\Delta\})=e^{\t\z_\t(p_\D)}\Big(1+o(1)\Big) \ ,
\eeq 
where $\z_\t(p_\D)$ can be interpolated by an analytic 
function of $p$, satisfying the equation 
\beq\label{23}
\z_\t(p)=-z_\t(\l_p)+\l_p p\s_+-\frac{1}{2\t}
\log\left[\frac{2\p}{\t}\Big(-\frac{z_\t''(\l_p)}{\s_+^2}\Big)\right] \ ,
\eeq
and $\l_p$ is the inverse of $p(\l)=z_\t'(\l)/\s_+$. 
Using the previous equations, the 
lowest order finite time correction to $\z_\io(p)$ around the maximum
can be computed.

First one rewrites $\z_\t(p)$ as $\z_\t(p)=\z_\io(p)+\frac{\g_\io(p)}{\t}+O(\frac{1}
{\t^2})$. 
By the analyticity of $\z_\t(p)$, one can 
write $\z_\io(p),\g_\io(p)$ around $p=1$ in the form:
$\zeta_\infty(p) = \frac12\zeta_\infty^{(2)} (p-1)^2 + 
\frac1{3!} \zeta_\io^{(3)} (p-1)^3 + 
\ldots$ and 
$\gamma_\infty(p) = \gamma_\infty^{(0)} + \gamma_\infty^{(1)} (p-1) +
\ldots$.

Up to terms of order $(p-1)^2$ and 
higher in the series for $\g_\io(p)$, one has:
\beq 
\label{25}
\begin{split}
\z_\t(p)
&=\z_\io(p)+\frac{\g_\io^{(0)}}\t+
\frac{\g_\io^{(1)}}{\t}(p-1)
+O\left(
\frac{(p-1)^2}{\t}\right)+o\left(\frac1\t\right)=\\
&=\z_\io\left(p+
\frac{\g_\io^{(1)}}{\t\z^{(2)}_\io}\right)
+\frac{\g_\io^{(0)}}\t
+O\left(\frac{(p-1)^2}\t\right)+o\left(\frac1\t\right)\;.
\end{split}
\eeq
Thus, the finite time corrections
to $\z_\io(p)$ around its maximum begin with a shift of the maximum 
at 
\beq
\label{26}
p_0 = 1 - \frac{\gamma^{(1)}_\infty}{\tau \zeta^{(2)}_\infty} + 
o\left(\frac1\tau\right) \ .
\eeq
To apply the latter result one needs to compute $\gamma^{(1)}_\infty$ in
terms of observable quantities. And, in order to compute $\gamma^{(1)}_\infty$
one can apply Eq.~(\ref{23}). First of all, note that $\l_p$ is determined
by the condition
\beq \label{27} p\s_+=z'_\t(\l_p)=\s_++z_\t''(0)\l_p+O(\l_p^2) \ ,
\eeq
where Eq.~(\ref{zio}) and Eq.~(\ref{22}) have been used. Then, 
$\l_p=\frac{\s_+(p-1)}{z_\t''(0)}+O\big((p-1)^2\big)$. 
Substituting this result into Eq.~(\ref{23}) and equating the
terms of order $O(\frac{p-1}{\t})$ at both sides one finds:
\beq \label{28}\g_\io^{(1)}=-\frac{1}{2}\frac{z_\io^{(3)}\s_+}{(z_\io^{(2)})^2}\;.
\eeq
The last equation can also be rewritten as:
\beq\label{28a}
\gamma_\io^{(1)} = \frac{\zeta_\io^{(3)}}{2 \zeta_\io^{(2)}} \ .
\eeq
This can be proven recalling that $\z_\io^{(2)}$ and 
$\z_\io^{(3)}$
are derivatives of $\z_\io(p)$ in $p=1$, that 
can be obtained by differentiating w.r.t. $\l$ (two or three times, respectively) 
the definition
$\z_\io\big(z_\io'(\l)/\s_+\big)=-z_\io(\l)+\l z_\io'(\l)$ and computing the 
derivatives in $\l=0$ recalling that $z_\io'(0)/\s_+=1$.
Plugging Eq.~(\ref{28a}) into Eq.~(\ref{26}) one finally gets
\beq \label{wttau}p_0 = 1 - \frac{\z^{(3)}_\infty}{2\tau 
(\zeta^{(2)}_\infty)^2} + 
o\left(\frac1\tau\right) \ ,
\eeq
that is the main result of this section.
The key point is that
the moments $\z_\io^{(2)}$ and $\z_\io^{(3)}$ in Eq.~(\ref{wttau}) are quantities that 
can be measured 
from the numerical data (within an $O(\t^{-1})$ error). 
One then has a verifiable prediction on the expected shift of the maximum
at finite $\t$. The data agree very well with this prediction, 
see Fig.~\ref{fig_7gal_1} and corresponding discussion in section~\ref{sec7:IV} below. 

Substituting Eq.~(\ref{wttau}) in Eq.~(\ref{25}),
one finally finds:
\beq \label{oohh}
\z_\io(p)=
\h_\t(p)+O\Big(\frac{(p-1)^2}{\t}\Big)+o(\t^{-1})\;,\eeq
where $\h_\t(p)$ is defined as
\beq \label{etatau}
\h_\t(p)\equiv -\frac{\g^{(0)}_\io}{\t}+\z_\t\left(p-\frac{\z_\io^{(3)}}{
2\t(\z_\io^{(2)})^2}\right)\;.\eeq
The key point of the above discussion was the validity of 
Eq.~(\ref{24}--\ref{23}); see Appendix~\ref{app7:A} for their derivation. 

\subsection{Remarks}
\label{sec7:IID}

{\it (1)} The shift away from $1$ 
of the maximum of the function $\zeta_\tau(p)$ at
finite $\tau$, expressed by the second term in Eq.~(\ref{etatau}),
is due to the asymmetry of the distribution $\pi_\tau(p)$
around the average value $p=1$; consequently, it is proportional, at
leading order in $\tau^{-1}$, to $\zeta_\io^{(3)}$ which is 
indeed a measure of the asymmetry of $\zeta_\io(p)$ around $p=1$.
This shift would be absent in the case of a symmetric distribution
(\eg a Gaussian) and for this reason it was not observed in 
previous experiments \cite{ECM93,BGG97,BCL98,ZRA04a}.

{\it (2)} The error term in the r.h.s. of Eq.~(\ref{24}) is $o(1)$
w.r.t. $\t$ and it does not affect the computation of 
$\g_\io(p)$. It is then clear that with a calculation similar to that 
performed above, one can get equations for the coefficients $O(\l^k)$ in the 
exponents of Eq.~(\ref{24}); in this way 
one can iteratively construct the whole sequence of coefficients 
$\g_\io^{(k)}$ defining the power series expansion of $\g_\io(p)$.

{\it (3)} In models with continuous time evolution the quantity $\sigma_+$ is
not dimensionless but it has dimensions of inverse time: in such cases
one can imagine that one is still studying a map which maps a system
configuration at a time when some prefixed event happens in the system
(typically a ``collision'') into the next one in which a similar event
takes place. If $\tau_0$ is the average time interval between such events
then $\tau_0\sigma_+$ will play the role played by $\sigma_+$ in the 
discrete time case: it will be the adimensional parameter entering 
the estimates of the error terms. 

Note that the coefficients 
$g_\io^{(k)}$ are of order $\s_+^k$, and their
size is necessarily estimated by the adimensional entropy production to
the $k$--th power. Then, in the continuous time case, the choice 
of $\t_0$ affects the estimates of the remainders, because it
affects the size of the adimensional parameter $\t_0\s_+$; 
and the size of the mixing time (that is connected with the estimated range of 
decay of the potentials, see~\cite{GBG04}). The natural (and physical) 
choice for $\t_0$ is the mixing time. Consistently with this remark,
at the moment of constructing numerically the 
distribution function for the entropy
production rate averaged over a time $\t$,
time intervals of the form
$\t=\t_0 n$, $n\ge 1$, will be considered,
see section~\ref{sec7:IIIC} below. 

\section{Models}
\label{sec7:III}

The model that will be considered in the following is a system of $N$ 
classical particles of equal mass $m$ 
in dimension $d$; they are described by their position $q_i$ and momenta 
$p_i = m \dot q_i$,
$(p_i,q_i) \in R^{2d}, \ i=1, \ldots, N$.
The particles are confined in a cubic box of side $L$ with periodic
boundary conditions.
Each particle is subject to a {\it conservative force}, 
$f_i(\ul q) = - \partial_{q_i} V(\ul q)$, and to a
{\it nonconservative force} $E_i$ that does not depend on
the phase space variables. The force $E_i$ is locally conservative
but not globally such due to periodic boundary conditions.
The {\it mechanical thermostat} is a Gaussian thermostat \cite{EM90},
$\theta_i (\ul p,\ul q) = -\alpha(\ul p, \ul q) \, p_i$, and the
function $\alpha(\ul p, \ul q)$ is defined by the condition that the
total kinetic energy 
$K(\ul p) \equiv \frac1{2m} |\ul p|^2 = \frac1{2m} \sum_i p^2_i$ 
should be a constant
({\it isokinetic ensemble}).
The equations of motion are:
\beq
\label{eqofmotion}
\begin{cases}
&\dot{q}_i = \frac{p_i}{m} \ , \\
&\dot{p}_i = f_i(\ul q) + E_i - \alpha(\ul p,\ul q) \ p_i \ ,
\end{cases}
\eeq
and are a particular instance of Eq.~(\ref{INTRO1}).
From the constraint $\frac{dK}{dt} = 0$ one obtains
\beq
\label{alphadef}
\alpha(\ul p,\ul q)=\frac{\sum_i E_i \ p_i + 
\sum_i f_i(\ul q) \ p_i}{\sum_i p^2_i} \ .
\eeq

\subsection{Entropy production rate}

The total phase space volume contraction rate for this system is given by:
\beq
\begin{split}
&\sigma(\ul p,\ul q)=-\sum_i \left( \frac{\partial \dot{q}_i}{\partial q_i}
+\frac{\partial \dot{p}_i}{\partial p_i} \right)
=  dN\alpha(\ul p,\ul q) + \sum_i \frac{\partial \alpha}{\partial p_i} p_i
= (dN-1) \ \alpha(\ul p,\ul q) \ .
\end{split}
\eeq
Defining the {\it kinetic temperature}, $T \equiv 2 K(\ul p)/(dN-1)$,
\cite{EM90}, the phase space contraction rate can be rewritten
as
\beq\label{36zz}
\sigma(\ul p, \ul q) = 
\frac{\sum_i E_i \ \dot q_i - \dot V}{T} \ .
\eeq
The first term is the power dissipated by 
the external force divided by the kinetic
temperature, and can be identified with the entropy production rate,
see the discussion in section~\ref{sec4:implications} and \cite{EM90,Ga04,ECM93}. 
The second term is the total derivative w.r.t. time of the potential energy
divided by the temperature: following the discussion of 
section~\ref{sec4:singularities} this term will be removed,
and the distribution of the {\it entropy production rate} $\dot s$, 
where $\dot s$ is identified with
$\s$ {\it minus} the total derivative term $-\dot V/T$ in Eq.~(\ref{36zz}), will
be studied:
\beq \label{epr}\dot s(\ul p, \ul q) = 
\frac{\sum_i E_i \ \dot q_i}{T} \ .
\eeq
From now on $\z_\io(p)$ and $\z_\t(p)$ will be the distributions
for the fluctuations of the entropy production rate $\dot s$ averaged
over infinite or finite time, respectively. These will be the 
objects that will be measured and used from now on.

In order to define the {\it current} ${\cal J}(x,E)$, 
it is useful to rewrite $E_i = E \, u_i$, where $u_i$
is a (constant) unit vector that specifies 
the direction of the force acting on the $i$-th particle.
Then, according to Eq.~(\ref{Jdef}),
\beq
\label{Jcolor}
{\cal J}(\ul p, \ul q) = 
\frac{\partial \sigma}{\partial E} = \frac{\sum_i u_i \, \dot q_i}{T} \ .
\eeq
 
\subsection{Discretization of the equations of motion}

To perform the numerical simulation, one has to write the
equations of motion in a discrete form.
One possibility is to use the {\it Verlet algorithm} \cite{AT87};
for Hamiltonian equations of motion ({\it i.e.}, $\ul E=\ul 0$ and $\alpha=0$)
\beq
\label{simpleeqofmotion}
\begin{cases}
&\dot{q}_i = \frac{p_i}{m} \ , \\
&\dot{p}_i = f_i(\ul q) \ ,
\end{cases}
\eeq
the Verlet discretization has the form
\beq
\begin{cases}
&q_i(t+dt) = q_i(t) + \frac{p_i(t)}{m} dt + \frac12 f_i(t) dt^2 \ , \\
&p_i(t+dt) = p_i(t) + \frac12 \big[ f_i(t) + f_i(t+dt) \big] dt \ ,
\end{cases}
\eeq 
where $dt$ is the {\it time step size}. This discretization ensures that
the error is $O(dt^4)$ on the positions $q_i(t)$ in a single time step.
The implementation of this algorithm on a computer is discussed in
detail in \cite{AT87}.

However, this method requires the forces $f_i(t)$ to depend only on the 
positions and not on the velocities: hence, it has to be adapted 
to Eq.s~(\ref{eqofmotion}).
This has been done in the following way.
The discretized equations are written as
\beq
\label{eqdiscrete}
\begin{cases}
q_i(t+dt) &= q_i(t) + \frac{p_i(t)}{m} dt 
+ \frac12 \big[ f_i(t) + E_i - \alpha(t) p_i(t) \big] dt^2 \ , \\
p_i(t+dt) &= p_i(t) + E_i +\frac12 \big[ f_i(t) + f_i(t+dt) 
- \alpha(t) p_i(t) - \alpha(t+dt) p_i(t+dt) \big] dt \ ,
\end{cases}
\eeq 
with the same error as in the standard Verlet discretization.
At time $t$,
the positions $q_i(t)$, the momenta $p_i(t)$,
the forces $f_i(t)$, and the Gaussian multiplier $\alpha(t)$ are
stored in the computer.
Then, the following operations are performed: 
\begin{enumerate}
\item the new positions $q_i(t+dt)$ are calculated using the first equation;
\item using the new positions, the new forces $f_i(t+dt)$ are calculated (the
conservative forces depend only on the positions);
\item the quantity
$\xi_i = p_i(t) +  E_i +\frac12 \big[ f_i(t) + f_i(t+dt) 
- \alpha(t) p_i(t) \big] dt$ is calculated; note that 
$p_i(t+dt)$ can be expressed in terms of the (known) $\x_i$ and the (unknown)
$\a(t+dt)$ as
\beq
\label{DISCRE1}
p_i(t+dt)=\frac{\xi_i}{1 - \alpha(t+dt) dt/2} \ ;
\eeq

\item substituting Eq.~(\ref{DISCRE1}) in the definition of $\alpha(t+dt)$, 
Eq.~(\ref{alphadef}),
a self-consistency equation for $\alpha(t+dt)$ is obtained, whose solution is
\beq
\label{selfconsist}
\begin{split}
&\alpha(t+dt) = \frac{\alpha_0}{1-\alpha_0 dt /2} \ , \\
&\alpha_0 = \frac{ \sum_i E_i \ \xi_i + \sum_i f_i(t+dt) 
\ \xi_i }{\sum_i \xi_i^2} \ ;
\end{split}
\eeq
\item substituting Eq.~(\ref{selfconsist}) in Eq.~(\ref{DISCRE1}) 
$p_i(t+dt)$ can be calculated.
\end{enumerate}
This procedure allows to calculate the new positions, 
momenta, forces, and $\alpha$,
at time $t+dt$ according to Eq.s~(\ref{eqdiscrete}) 
{\it without approximations}, 
defining a map $S$ such that $(\ul p(t+dt),\ul q(t+dt))= S(\ul p(t),\ul q(t))$.

The simulated ({\it discrete}) dynamical system
will be defined by the map $S(\ul p,\ul q)$
and will approximate the differential equations of motion, 
Eq.~(\ref{eqofmotion}), with error $O(dt^4)$ 
for the positions and $O(dt^3)$ for the velocities. \\
The map $S$ satisfies the following properties:
\begin{enumerate}
\item it is {\it reversible}, {\it i.e.} it 
exists a map $I(\ul p, \ul q)$ (simply
defined by $I(\ul p, \ul q)=(-\ul p, \ul q)$)
such that $I S = S^{-1} I$;
\item in the {\it Hamiltonian} case ($\ul E=\ul 0$ and $\alpha=0$, 
Eq.s~(\ref{simpleeqofmotion})) it is {\it volume preserving}.
\end{enumerate}
The first property ensures that {\it assuming the Chaotic Hypothesis} 
the Fluctuation Relation holds for the map $S$.
The second property ensures that at equilibrium the discretization 
algorithm conserves the phase space volume, consistently with the
definition of equilibrium system which has been 
given in section~\ref{sec4:implications}.

\subsection{Details of the simulation}
\label{sec7:IIIC}

In the simulation, the external force has been chosen of the form 
$E_i = E \, u_i$, where
the unit vectors $u_i$ were parallel to the $x$
direction but with different orientation: half of them were oriented in 
the positive
direction, and half in the negative direction, {\it i.e.} $u_i = (-1)^i \wh x$,
in order to keep the center of mass fixed.
Two different systems have been considered, selecting interaction potentials
widely used in numerical simulations (for the purpose
of making easier possible future independent checks and rederivations of the
results):
\begin{enumerate}
\item ({\it model I}) the first investigated system is made by $N=8$ 
particles of equal mass $m$ in $d=2$.
The interaction potential is a sum of pair interactions,
$V(\ul q)= \sum_{i < j} v(|q_i - q_j|)$, 
and the pair interaction is represented by a
WCA potential, {\it i.e.} a 
Lennard-Jones potential truncated at the minimum:
\beq
\nonumber
v(r) =
\begin{cases}
& 4 \epsilon \left[ \left(\frac\sigma r \right)^{12} - 
\left(\frac\sigma r \right)^6 \right] + \epsilon \ ,
\hspace{10pt} r \leq \sqrt[6]{2}\sigma \ ; \\
& 0 \ , \hspace{100pt} r > \sqrt[6]{2}\sigma \ .
\end{cases}
\eeq
The reduced density was $\rho=N\sigma^2/L^2=0.95$
(that determines $L$),
the kinetic temperature was fixed to $T=4\epsilon$
and the {\it time step} to $dt=0.001 t_0$, where 
$t_0 = \sqrt{m\sigma^2/\epsilon}$.
In the following, all the quantities will be 
reported in units of $m$, $\epsilon$
and $\sigma$ ({\it LJ units}). 
This system was already studied in the literature, 
see {\it e.g.} \cite{ECM93,SEC98}.
Different values of the external force $E$ were investigated,
ranging from $E=0$ to $E=25$.
\item ({\it model II})
the second system is a binary mixture of $N$=20 particles (16 of type A and
4 of type B), of equal mass
$m$, in $d=3$, interacting via the same WCA potential of model I; the pair 
potential is
\beq
\nonumber
v_{\alpha\beta}(r) =
\begin{cases}
& 4 \epsilon_{\alpha\beta} \left[ 
\left(\frac{\sigma_{\alpha\beta}} r \right)^{12}
-  \left(\frac{\sigma_{\alpha\beta}} r \right)^{6} \right] 
+ \epsilon_{\alpha\beta} \ , \hspace{21pt} 
r \leq \sqrt[6]{2}\sigma_{\alpha\beta} \ ; \\
& 0 \ , \hspace{150pt} r > \sqrt[6]{2}\sigma_{\alpha\beta} \ ;
\end{cases}
\eeq
$\alpha$ and $\beta$ are indexes that specify the particle species
($\alpha,\beta \in \{A,B\}$).
The parameters entering the potential are the following: 
$\sigma_{AB}=0.8 \sigma_{AA}$;
$\sigma_{BB}=0.88 \sigma_{AA}$;
$\epsilon_{AB}=1.5 \epsilon_{AA}$;
$\epsilon_{BB}=0.5 \epsilon_{AA}$.
Similar potentials have been studied, \cite{KA94,DmSC04},
as models for liquids in the
supercooled regime ({\it i.e.}, below the melting
temperature).
For this system the {\it LJ units} are
$m$, $\epsilon_{AA}$, and $\sigma_{AA}$; the unit of time
is then $t_0 = \sqrt{ m\sigma_{AA}^2/\epsilon_{AA}}$.
The reduced density was $\rho = N \sigma_{AA}^3 / L^3 = 1.2$ 
and the integration step was $dt=0.001 t_0$.
The unit vectors $u_i$ are chosen such that half of the $A$ 
particles and half of the
$B$ particles have positive force in the $x$ direction, 
and the remaining particles 
have negative force in the $x$ direction.
For this system different values of external 
force $E \in [0,10]$ and temperature $T \in [0.5,3]$
were investigated.
\end{enumerate}

For each system and for each chosen value of $T$ and $E$, a very
long trajectory was simulated ($\sim 2 \cdot 10^9 dt$) 
starting from a random initial data; recall that in both
systems $dt=0.001 t_0$, $t_0$ being the natural unit time 
introduced in items (1) and (2) above. After a 
short transient ($\sim 10^3 dt$), still much bigger than the 
decay time $\t_0$ of self-correlations (that 
appears to be $\t_0= 10^2 dt$),
the system reached stationarity, in the sense that the instantaneous
values of observables (\eg potential energy, Lyapunov exponents)
agree with the corresponding asymptotic values within the statistical error
of the asymptotic values themselves.
After this transient ${\cal N}$ values $p_i$, $i=1,\ldots,{\cal N}$, 
of the variable $p(x)$, defined in Eq.~(\ref{pdef}), were recorded,
integrating the entropy production rate, Eq.~(\ref{epr}) 
on adjacent segments of trajectory of length $\tau_0 = 100 dt = 0.1 t_0$.
Note that the length of the time interval over which the entropy
production rate was averaged was chosen to be equal to the mixing time, consistently
with the discussion in Remark~(4) of section~\ref{sec7:IID}.

In conclusion, from each simulation run at fixed $T$ and $E$
${\cal N} \sim 10^7$ values $p_i$ of $p(x)$ are obtained, which are
the starting point of the data analysis.
The value of $\sigma_+$ is estimated by 
averaging the entropy production rate over the whole trajectory.

From a shorter simulation run the Lyapunov exponents
of the map $S$ were also measured 
using the standard algorithm of Benettin {\it et al.} \cite{SEC98,BGS76}.

\subsection{Remarks}

To conclude this section, note that the WCA potential has a discontinuity in
the second derivative. Thus, one should be concerned with the possibility
that the error in the discretization is not $O(dt^4)$ over the $q_i$'s 
on a single time step, as it should be for potentials $V \in C^4$.
To check that this is not the case (or that 
at least this does not affect the results) two independent tests have been made:
\begin{enumerate}
\item a system similar to model I but with a potential $V \in C^4$ was simulated
and (qualitatively) the same results were obtained;
\item model I have been simulated using an 
{\it adaptive step size} algorithm \cite{AT87}; 
this kind of algorithms adapt the step size $dt$ during the simulation
in order to keep constant the difference 
between a single step of size $dt$ and two
steps of size $dt/2$. If the precision of the 
discretization changed at the singular
points of the potential, the time step should change 
abruptly during the simulation,
while a practically constant time step was observed during the simulation.
\end{enumerate}
These checks give evidence of the fact that the (isolated) singularities 
of the potentials do not produce
relevant effects on the observations;
this is probably due to the fact that the set of singular points of the total 
potential energy $V(\ul q)$ has zero measure w.r.t. the SRB measure.

\section{Data analysis}
\label{sec7:IV}

\begin{figure}[t]
%\centering
\includegraphics[width=.50\textwidth,angle=0]{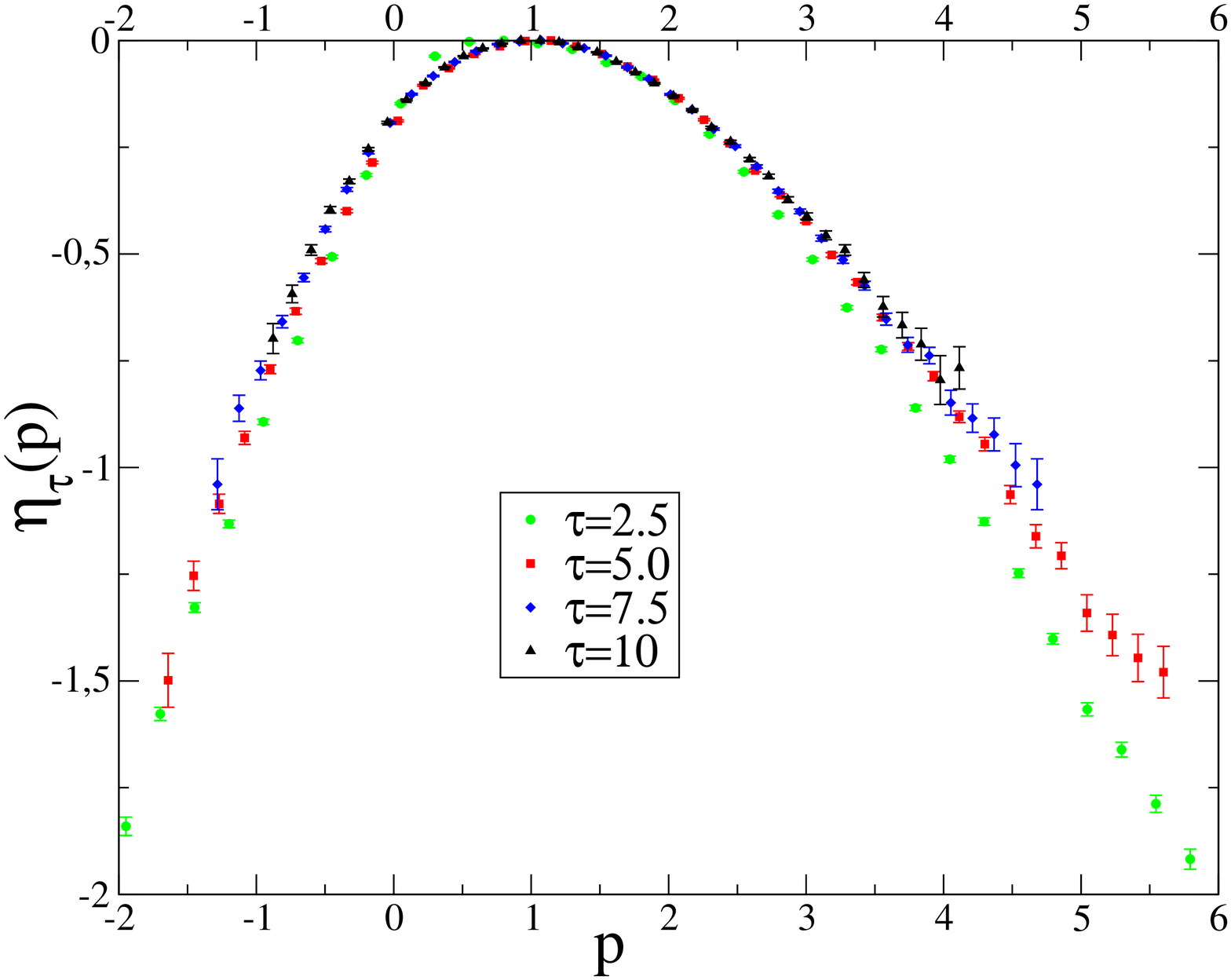}
\includegraphics[width=.50\textwidth,angle=0]{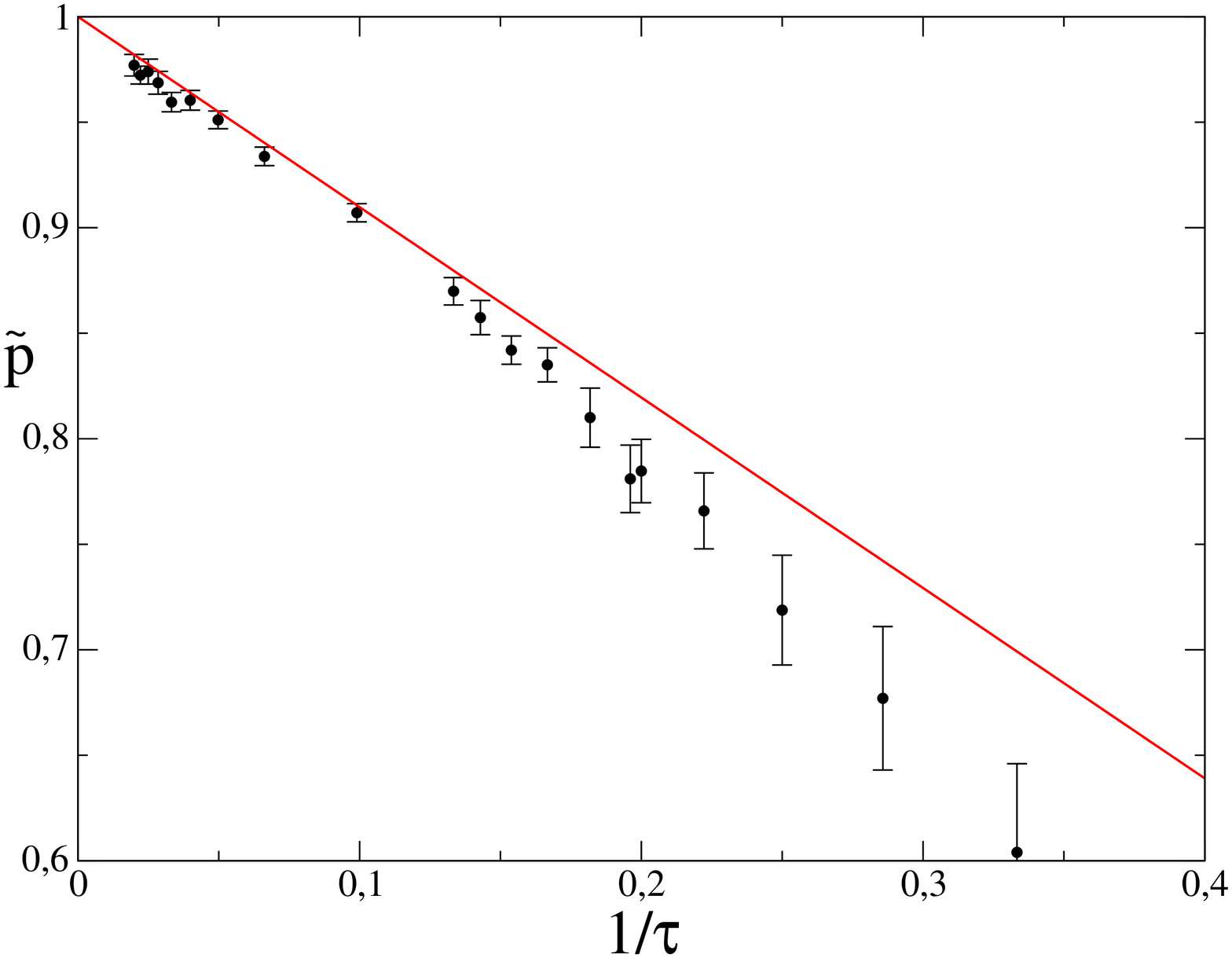}
\caption[The function $\eta_\t(p)$ for Model I at $E=5$]
{Model I at $E=5$: (left) the function 
$\eta_\tau(p)=\zeta_\infty(p)+O((p-1)^2/\tau)$ 
for different values of $\tau$; (right) the maximum 
$\widetilde p_\tau$ of $\zeta_\tau(p)$ 
as a function of $1/\tau$.
The full line is the prediction of Eq.~(\ref{wttau}), 
$\widetilde p = 1 - \zeta_\io^{(3)}/\big[2\tau 
\big(\zeta_\io^{(2)}\big)^2\big]$.}
\label{fig_7gal_1}
\end{figure}

In this section the procedure followed to
analyze the numerical data will be described in detail. 
As an example, the data obtained from the simulation of model I at $E=5$
will be discussed. From the simulation run a set 
${\cal P}_0 = \{ p_i \}_{i=1\ldots{\cal N}}$ of values of the variable $p(x)$ 
is obtained, that correspond to $\tau= \tau_0$ and are 
measured on adjacent segments of trajectory.
As discussed above, $\tau_0 = 0.1 = 100 dt$ is of the order 
of the {\it mixing time}, {\it i.e.} 
the time scale over which the correlation functions 
({\it e.g.} of density fluctuations) decay to zero.

\subsubsection{Probability distribution function}

From the dataset ${\cal P}_0$ the histograms $\pi_\tau(p)$ 
are constructed for different values of $\tau = n \tau_0$ as follows: 
the values of $p(x)$ for $\tau = n\tau_0$ are obtained by averaging 
$n$ subsequent entries of the dataset ${\cal P}_0$; one obtains a new dataset
${\cal P}_n = \{ p^{(n)}_j \}_{j=1\ldots{\cal N}/n}$ such that 
$p^{(n)}_j= n^{-1} \sum_{i=nj+1}^{n(j+1)} p_i$.
Finally, from the dataset ${\cal P}_n$ the histogram of $\pi_{\tau}(p)$ is
constructed for $\tau=n\tau_0$; 
the errors are estimated as the square roots of the number of counts
in each bin.
The function $\zeta_{\tau}(p)$ is then defined as
$\zeta_\t(p)=\t^{-1} \log \pi_\t(p)$.

\subsubsection{Shifting of the maximum}

By fitting the function $\zeta_{\tau}(p)$ in $p\in[-1,3]$ 
with a sixth-order polynomial the position of the maximum $\widetilde p_\tau$ 
is determined within an error that, since $\d p$ is the length of a bin,
is estimated to be $\d p/2$.
Then, the function $\eta_{\tau}(p) = \zeta_\tau(p-1+\widetilde p_\tau)$ 
is constructed, see Eq.~(\ref{etatau}); it
is expected to approximate the limiting function $\zeta_\io(p)$ with error
$O((p-1)^2/\tau)$.
The functions $\eta_\tau(p)$ are reported 
in Fig.~\ref{fig_7gal_1} for different
values of $\tau$. 
A very good convergence for $\tau \gtrsim 5.0 = 50\tau_0$ is observed.

By a fourth-order fit of the so-obtained limiting function $\zeta_\io(p)$ 
around $p=1$ the coefficients $\zeta_\io^{(2)}=-0.287$ and 
$\zeta_\io^{(3)}=0.149$ are extracted in order to test the correctness of 
Eq.~(\ref{wttau}).
In Fig.~\ref{fig_7gal_1} $\widetilde p_\tau$ is reported.
The full line is the prediction of 
Eq.~(\ref{wttau}), that is indeed verified for $\tau \gtrsim 10$.
This result confirms the analysis of section~\ref{sec7:II}.

\begin{figure}[t]
\centering
\includegraphics[width=.55\textwidth,angle=0]{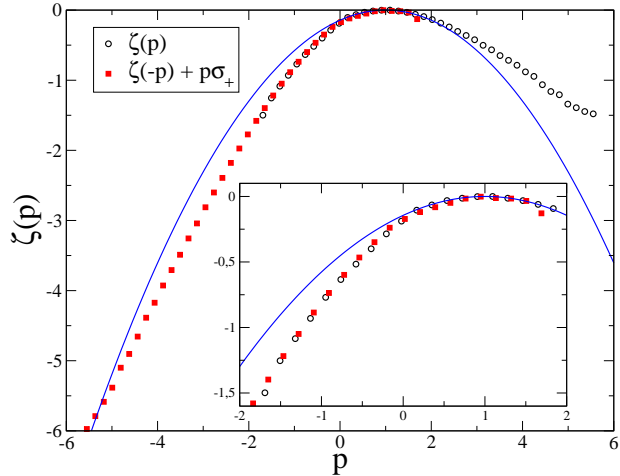}
\caption[Estimate of the function $\z_\io(p)$ for Model I at $E=5$]
{Model I at $E=5$: 
the estimate of the function $\zeta_\io(p)$ (open circles). 
In the same
plot $\zeta_\io(-p) + p\sigma_+$ (filled squares) 
is reported. In the inset, the
interval $p\in[-2,2]$ where the data overlap 
is magnified. The full
line is the Gaussian approximation, 
$\frac12\zeta_\io^{(2)}(p-1)^2$.
The plot shows that the Gaussian is not a good approximation
in the interval $[-2,2]$. The validity of the Fluctuation Relation
in the same interval is shown by the overlap of the open circles and 
filled squares.}
\label{fig_7gal_3}
\end{figure}

\subsubsection{Graphical verification of the fluctuation relation}

From the previous analysis one can conclude that the function $\eta_\tau(p)$ 
for $\tau = 5.0$ provides a good estimate 
of the function $\zeta_\io(p)$ for $p\in[-2,4]$
(see Fig.~\ref{fig_7gal_1});
thus, one can use this function to test the 
fluctuation relation, Eq.~(\ref{FR}),
in this range of $p$. In Fig.~\ref{fig_7gal_3} the estimated functions
$\zeta_\io(p)$ and $\zeta_\io(-p)+p\sigma_+$ are reported. 
An excellent agreement between 
the two functions is observed in the interval $p\in[-2,2]$ where the data 
allows the computation of both $\zeta_\io(p)$ and $\zeta_\io(-p)$.
Note that in this range of $p$ the function $\zeta_\io(p)$ is not Gaussian, see
the inset of Fig.~\ref{fig_7gal_3}.

\subsubsection{Quantitative verification of the fluctuation relation}

The translation of the function $\zeta_\tau(p)$ is crucial to obtain a
correct estimate of the limit $\zeta_\infty(p)$ and to verify the fluctuation
relation.
In this section an attempt to quantify this observation is presented;
as the discussion will
be very technical, the reader who is satisfied with Fig.~\ref{fig_7gal_3} should 
skip to next section.

\begin{figure}[t]
\includegraphics[width=.5\textwidth,angle=0]{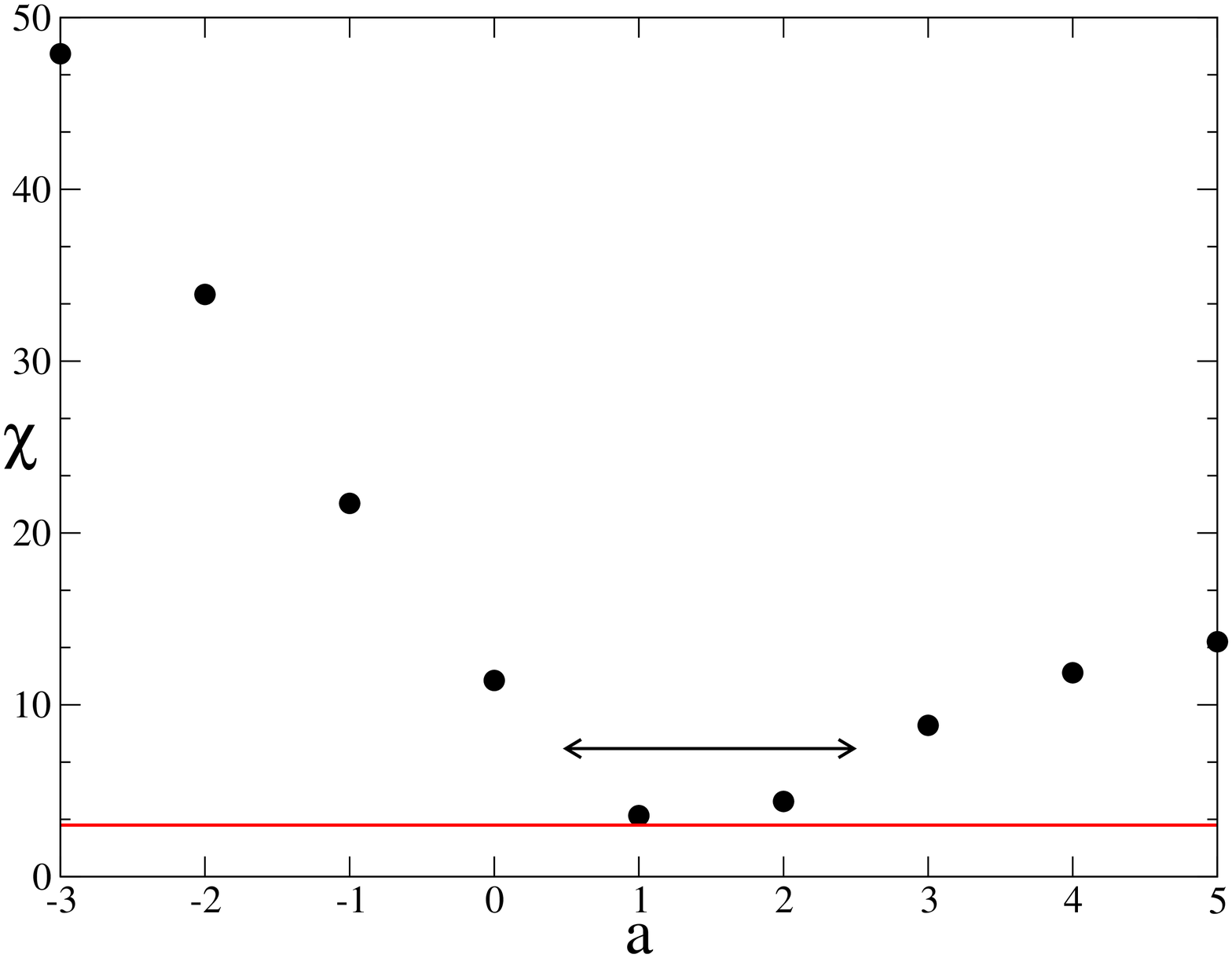}
\includegraphics[width=.5\textwidth,angle=0]{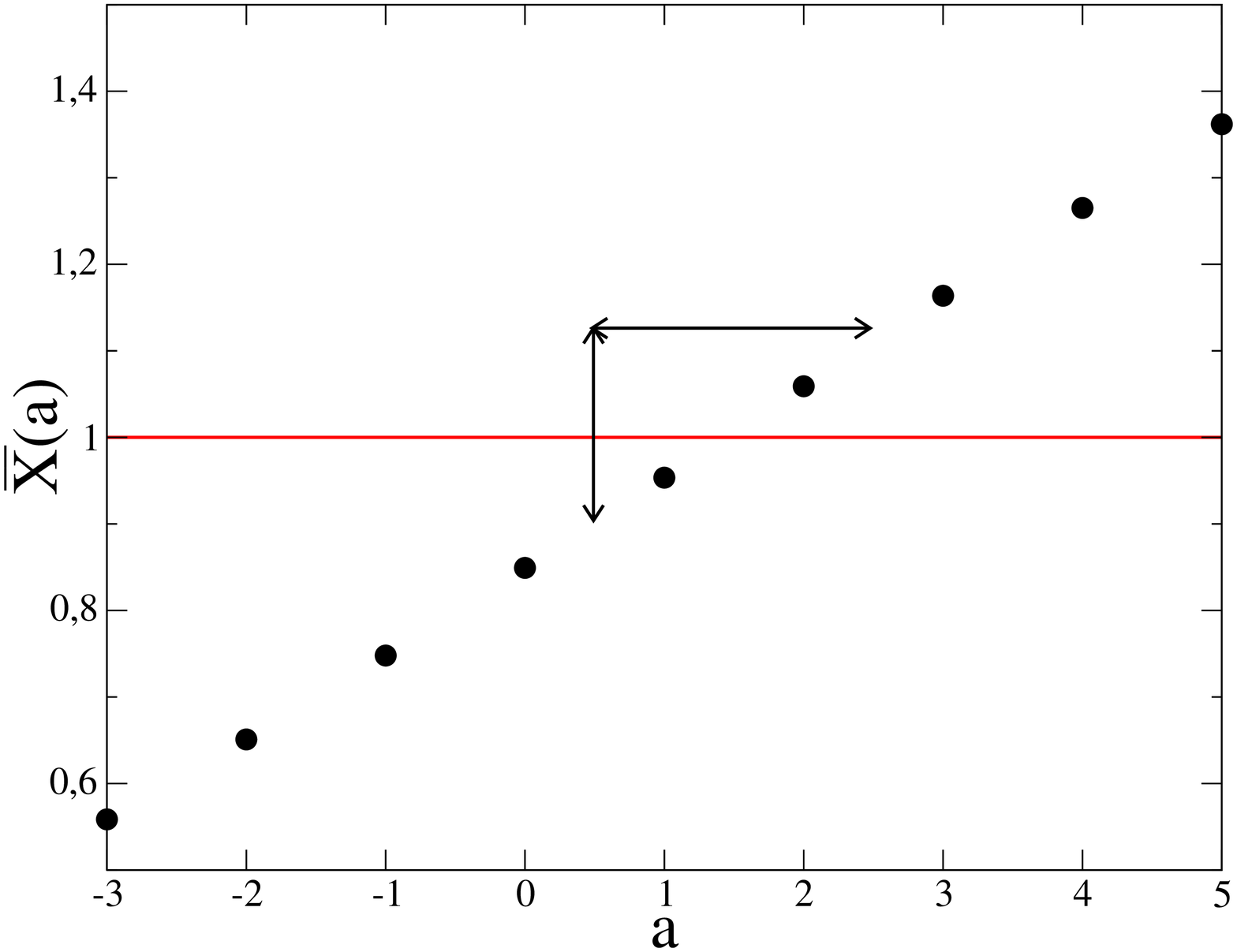}
\caption[Quantitative estimates of the precision of the data analysis]
{Model I at $E=5$: 
(left) the function $\chi(a)$. The full line corresponds to $\chi=3$.
The arrow indicates the interval $\d_0\pm\d p/2$
(note that its length is 2 in units of $a$)
into which the minimum of $\chi$ can be located 
within the accuracy of the histogram.
(right) The function $\overline X(a)$. 
The horizontal arrow marks the interval where 
the minimum of $\chi$ is located, see the left panel.
The vertical arrow indicates the error 
$\delta X$ on the value $X=1$ which is estimated
as $\delta X=2(\overline X(2)-\overline X(1))$. 
The slope of the fluctuation relation without
the translation would have been $\overline X(a=0) \sim 0.85$.
}
\label{fig_7gal_4}
\end{figure}

The histogram $\pi_{n\tau_0}(p)$ derived from the 
dataset ${\cal P}_n$ is constructed
assigning the number of counts $\pi_\alpha$ in the 
$\alpha$-th bin to the middle
of the binning interval, that will be called $p_\alpha$ 
(the latter will be an {\it increasing} function of $\alpha$).
The statistical error 
$\delta \pi_\alpha$ on the number of counts is $\sqrt{\pi_\alpha}$. 
The histograms are constructed
in such a way that if $p_\alpha$ is the center 
of a bin, also $-p_\alpha$ is the center
of a bin\footnote{That is, either $p_\a = (2\a + 1) \d p /2$ or
$p_\a = \a \d p$, where $\d p$ is the size of a bin.}; $\overline\alpha$ is the bin 
such that $p_{\overline\alpha}=-p_\alpha$.
There exists a value $p_m$ such that for $p_\alpha < p_m$ the
number of counts in the bin $\alpha$ is smaller than $m$ ($m=4$ has been chosen
in the present analysis).
Define $p_{\alpha_m}$ the smallest value of $p_\alpha > p_m$. 
Hence, the histogram is characterized by:
\begin{enumerate}
\item a {\it bin size} $\delta p$;
\item
the bin $\alpha_m$ corresponding to the minimum value of $p_\alpha$ such 
that the number of counts in the bin is at least $m$;
\item the total number $M$ of bins such that 
$\alpha \in [\alpha_m,\overline\alpha_m]$; for these values of $p_\alpha$,
both $\pi_\tau(p)$ and $\pi_\tau(-p)$ 
can be computed and they can be used to verify
the fluctuation relation.
\end{enumerate}
The function $\zeta_{\tau}(p)$, derived from the histogram, is specified by
a set of values $(p_\alpha,\zeta_\alpha,\delta \zeta_\alpha)$ 
for each bin $\alpha$,
where $\zeta_\alpha=\tau^{-1} \log \pi_\alpha$ and 
the error $\delta\zeta_\alpha$ has been defined by
\beq
\delta \zeta_\alpha = \frac1\tau \frac{\delta \pi_\alpha}{\pi_\alpha}=
\frac1{\tau \sqrt{\pi_\alpha}} \ .
\eeq
A quantitative verification of Eq.~(\ref{FR}) 
is possible defining the following
$\chi^2$ function:
\beq
\chi^2 \equiv \frac1M \sum_{\alpha=\alpha_m}^{\overline\alpha_m} 
\frac{(\zeta_\alpha - \zeta_{\overline\alpha} - p_\alpha \sigma_+)^2}
{(\delta \zeta_\alpha)^2 + (\delta \zeta_{\overline\alpha})^2} \ .
\eeq
The value of $\chi$ is the average difference between $\zeta_\tau(p)$ and
$\zeta_\tau(-p)+p\sigma_+$ in units of the statistical error.
Translating $p$ of a quantity $a \delta p/2$, $a\in \ZZZ$,
corresponds to shifting the histogram, {\it i.e.} to
consider a new histogram $(p_\alpha + a \delta p/2,\zeta_\alpha,
\delta\zeta_\alpha)$.
This preserves the property that if $p_\alpha$ is the center of a bin, also
$-p_\alpha$ is the center of a bin; let $\overline\alpha(a)$ be the new 
value of $\alpha$
such that $p_{\overline\alpha(a)}+a \delta p/2=
-(p_\alpha + a \delta p/2)$. Also, the number $M_a$ of bins such that
$\alpha(a) \in [\alpha_m, \overline\alpha_m(a)]$ depends on $a$.
Define
\beq
\label{chia}
\chi^2(a) \equiv \frac1{M_a} \sum_{\alpha=\alpha_m}^{\overline\alpha_m(a)} 
\frac{\big(\zeta_\alpha - \zeta_{\overline\alpha(a)} - 
(p_\alpha + a\delta p/2) \sigma_+\big)^2}
{(\delta \zeta_\alpha)^2 + (\delta \zeta_{\overline\alpha(a)})^2} \ .
\eeq
The criterion that will be followed is that 
the fluctuation relation is satisfied if $\chi \leq 3$, which means that
$\zeta_\io(p)$ and $\zeta_\io(-p)+p \sigma_+$
differ, {\it on average}, by less than $3$ times the statistical error 
$\sqrt{ \big( \delta \zeta(p)\big)^2 +\big( \delta \zeta(-p)\big)^2}$.
The function $\chi(a)$ for the case of model I at $E=5$ 
is reported in Fig.~\ref{fig_7gal_4}. 

The minimum of $\chi$
is assumed between $a^*=1$ and $a^*+1=2$ and an upper limit for the value of 
$\chi$ at the minimum is $\chi(1)=3.5$.
The translation that minimizes $\chi$ is estimated as 
$\delta_0 = (a^*+0.5) \delta p/2 = 1.5 \cdot 0.093 = 0.140$, 
and to this estimate an error $\pm\d p/2$ is attributed, 
where $\delta p = 0.186$ is the size of a bin. 
On the other hand, as discussed above,
in order to shift the maximum of $\zeta_\tau(p)$ 
in $p=1$, one has to translate $p$ by a 
quantity $\delta \equiv 1 - \widetilde p = 0.215$. 
The consistency of the analysis
requires that $\delta$ and $\d_0$ coincide within their errors, \ie 
that the intervals $\d\pm\d p/2$ and $\d_0\pm\d p/2$ overlap, or in other
words $|\d-\d_0|<\d p$. In the present case $0.075=|\d-\d_0|<\d p=0.186$,
then $\d$ and $\d_0$ coincide within the errors. This means that
the translation of $p$ 
brings the maximum of $\zeta_\tau(p)$ in
$p=1$ and, {\it at the same time}, minimizes the difference between
$\h_\tau(p)$ and $\h_\tau(-p)+p \sigma_+$, where $\h_\t$ is the
finite time estimate of $\z_\io(p)$. 
The value $\chi(a^*)$ quantifies
this difference and is a first estimate of the precision of the analysis.

Another estimate of the precision of the analysis can be obtained as follows. 
Define a parameter $X$ as the slope of $\zeta_\io(p)-\zeta_\io(-p)$ as a
function of $p\sigma_+$:
\beq
\zeta_\io(p)= \zeta_\io(-p)+Xp\sigma_+ \ .
\eeq
The fluctuation theorem predicts $X=1$, but other values of $X$ are possible 
under different hypothesis, see \cite{Ga04,BGG97,Ga99b,BG97}.
Defining a function $\chi^2(a,X)$ as
\beq
\label{chiaX}
\chi^2(a,X) \equiv \frac1{M_a} 
\sum_{\alpha=\alpha_m}^{\overline\alpha_m(a)} 
\frac{\big(\zeta_\alpha - \zeta_{\overline\alpha(a)} - 
X (p_\alpha+a \delta p /2) \sigma_+\big)^2}
{(\delta \zeta_\alpha)^2 + (\delta \zeta_{\overline\alpha(a)})^2} \ ,
\eeq
for each value of $a$ one can calculate the optimal value of 
$X$, $\overline X(a)$,
by minimizing $\chi^2(a,X)$. The function $\overline X(a)$ is reported in 
Fig.~\ref{fig_7gal_4}. As the shift of the maximum $\delta$ is between $a=1$ and
$a=2$, the slope $X$ is compatible with one. Moreover, as the
natural error on $p$ is the size of a bin $\delta p$, it is reasonable to 
assign to the value $X=1$ a statistical error 
$\delta X= 2 ( \overline X(2)-\overline X(1) ) = 0.22$.
Note again that without the translation of $p$ the optimal slope would be
$X \sim 0.85$, incompatible with Eq.~(\ref{FR}).

\subsubsection{Discussion}

From the present analysis, one can conclude that:
\begin{enumerate}
\item the translation shifting the maximum of 
$\z_\t(p)$ to $p=1$ at the same time
minimizes the difference between $\h_\tau(p)$ and $\h_\tau(-p)+p \sigma_+$, 
where $\h_\t$ is the
finite time estimate of $\z_\io$; this proves the consistency of the
theory of finite time corrections described above;
\item without the translation of $p$ (that corresponds to $a=0$), 
the function $\zeta_\tau(p)$ for $\tau \sim 5.0$
{\it does not satisfy the fluctuation relation}, as $\chi(a=0) = 11$ 
and $\overline X(a=0)=0.85$;
\item the function $\eta_\tau(p)=\zeta_\tau(p-\delta)$ 
satisfies the fluctuation relation
with $\chi \sim 3$ and an error of about $20\%$ on the slope $X$: 
both quantities measure
the accuracy of the data analysis.
\end{enumerate}
Thus, the check of the fluctuation relation relies 
crucially on the translation
of the function $\zeta_\tau(p)$ that has been discussed 
in section~\ref{sec7:II}.
By considering larger values of $\tau$ one could avoid this problem 
(as $\delta \sim \tau^{-1}$); however, 
as one can see from Fig.~\ref{fig_7gal_1}, for $\tau > 5.0$
the negative tails of $\zeta_\tau(p)$ are not accessible to the computational
resources available during this work.
The computation of the finite time corrections 
is mandatory if one aims to test
the fluctuation relation at high values of the external driving force.

\subsubsection{Summary of the data analysis}

To summarize, the procedure
followed to analyze the data of a given simulation run is:
\begin{enumerate}
\item a value of $\tau$ such that $\zeta_\tau(p)$ appear to
be close to the asymptotic
limit $\zeta_\infty(p)$ is determined;
\item the maximum $\widetilde p$ of $\zeta_\tau(p)$ is obtained
by a sixth-order polynomial fit around $p=1$, in an
interval as big as possible compatibly with the request
that the $\c^2$ from the fit is less than $\sim 10$;
\item the histogram is shifted by an integer multiple $a$ of the half 
bin size $\delta p/2$ and
the function $\chi(a)$ is computed according to Eq.~(\ref{chia}). 
The value $a^*$ 
such that the minimum of $\chi(a)$ is
assumed in the interval $[a^*,a^*+1]$ is determined:
the consistency of the analysis requires that
$\delta = 1 - \widetilde p$ and $\d_0=(a^*+0.5)\d p/2$ 
coincide within their errors
(\ie $|\d-\d_0|<\d p$);
\item The value $\chi^* = \min [\chi(a^*),\chi(a^*+1)]$ is an
upper limit for the value of $\chi$ at the minimum.
The number of bins $\min\{M_{a^*},M_{a^*+1}\}$ involved in 
this estimate will be called $M^*$; 
\item the error $\delta X = 2( \overline X(a^*+1) - 
\overline X(a^*))$ is computed.
\end{enumerate}
The relevant quantities $\tau$, $\delta$, $\d_0$, $|\d-\d_0|$, $\d p$, 
$M^*$, $\chi^*$ and $\delta X$ 
for model I are
reported in table~\ref{tab:I} for different values of the external force $E$.

\section{Numerical simulation of model I}
\label{sec5:modelI}

\begin{figure}[t]
\centering
\includegraphics[width=.50\textwidth,angle=0]{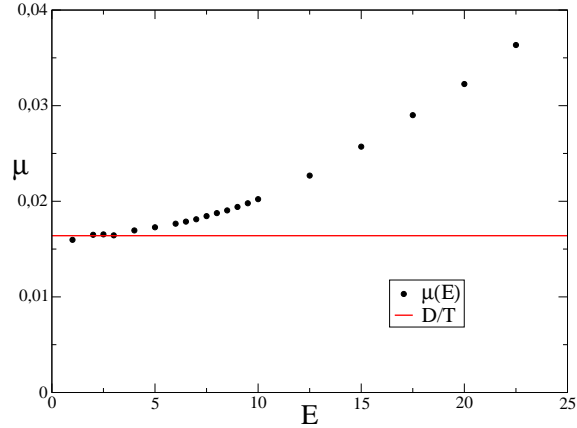}
\caption[Mobility as a function of the driving force for Model I]
{Model I: mobility $\mu$ as a function of the driving force $E$. 
The full line is the
equilibrium diffusion coefficient $D$ divided by the temperature. 
Deviations from the
linear response are observed around $E=5$. 
The error bars are of the order of the
dimension of the symbols. Studying $\m(E)$ 
for values of $E$ bigger than those shown in the figure,
one can verify that
the mobility increases
up to a value $\m_{max}$, reached in correspondence of
$E\sim 45$. For values of $E$ bigger than $E\sim 45$, the mobility
begins to decrease essentially following the limiting curve 
$T {\cal J}_T/(NE)$, where ${\cal J}_T=\sqrt{T(d-1/N)}N/T$
is the maximum allowed value of the current (saturation value).}
\label{fig_7gal_6}
\end{figure}

\begin{table}[b]
\centering
\begin{tabular}{r|ccccccccc}
\hline
$E$ & $\tau$  & $\sigma_+$ & $\delta$ & $\d_0$ & $|\d-\d_0|$ & 
$\d p$ & $M^*$ & $\chi^*$ & $\delta X$ \\
\hline
2.5 & 5.0 & 0.194 & 0.272 & 0.183 & 0.089 & 0.244 & 43 & 2.2 & 0.24 \\
5.0 & 5.0 & 0.810 & 0.215 & 0.139 & 0.076 & 0.187 & 20 & 3.5 & 0.22 \\
7.5 & 4.0 & 1.945 & 0.197 & 0.116 & 0.081 & 0.116 & 18 & 2.8 & 0.18 \\
10.0 & 2.5 & 4.044 & 0.262 & 0.151 & 0.111 & 0.122 & 17 & 4.4 & 0.20 \\
12.5 & 2.5 & 7.090 & 0.257 & 0.137 & 0.120 & 0.111 & 8 & 3.5 & 0.28 \\
\hline
\end{tabular}
\caption{Model I: results of the data analysis 
for some selected values of $E$. 
All the quantities are defined in section~\ref{sec7:IV}. 
For $E > 12.5$ the negative tails of the distribution 
are not accessible to the numerical simulation.}
\label{tab:I}
\end{table}

The numerical data 
obtained from the simulation
of model I (defined in section~\ref{sec7:III}) will now be discussed systematically 
at different values of the driving
force $E$. In Fig.~\ref{fig_7gal_6} the {\it mobility}
$\mu(E) = T \langle {\cal J} \rangle_E / (N E)$, {\it i.e.} 
the l.h.s. of Eq.~(\ref{GKestesa}) times $T/N$, is reported as 
a function of $E$. The current ${\cal J}(\ul p,\ul q)$ has 
been defined in Eq.~(\ref{Jcolor}).
From the Green-Kubo relation, Eq.~(\ref{GK}),
one has~\cite{EM90}
\beq
\lim_{E \rightarrow 0} \mu(E) = \frac{D}{T} \ ,
\eeq
where $D$ is the equilibrium diffusion coefficient,
\beq
D= \lim_{t\rightarrow \infty} \frac{1}{2Nd} \sum_i 
\langle |q_i(t)-q_i(0)|^2 \rangle_{E=0} \ .
\eeq
Deviations from the linear response are observed and 
$\mu(E) \sim D/T + O(E^2)$ above $E=5$.

\begin{figure}[t]
\centering
\includegraphics[width=.50\textwidth,angle=0]{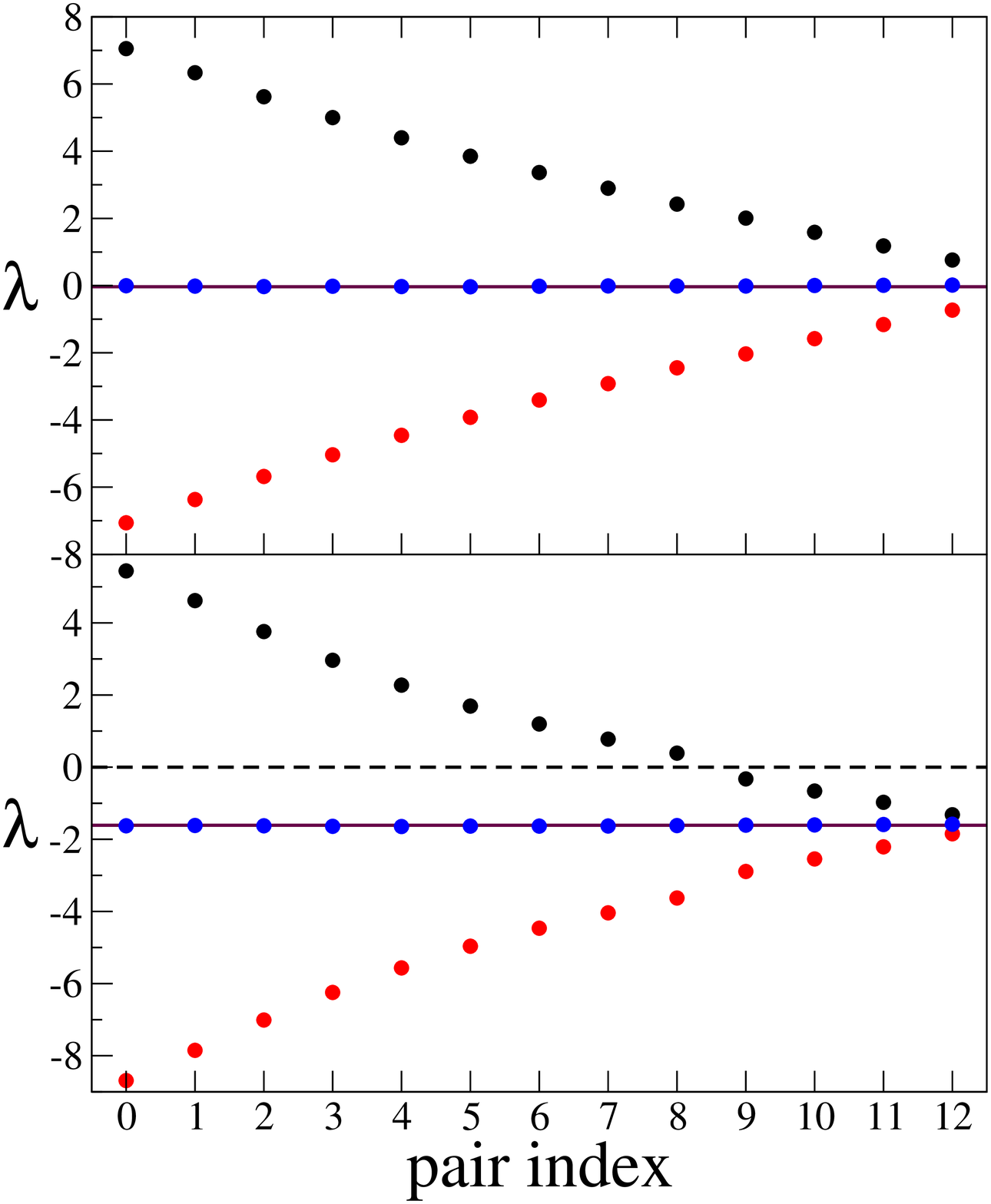}
\caption[Lyapunov exponents for Model I]
{Model I: Lyapunov exponents for $E=5$ (top) and for $E=25$ (bottom).
For each panel, the upper and lower dots are the two paired 
exponents $\lambda^{(+)}_j$
and $\lambda^{(-)}_j$, and the
middle dot is their average $(\lambda^{(+)}_j+\lambda^{(-)}_j)/2$.
The full line is $\sigma_+/2Nd$, the dashed line is at $\lambda=0$.
}
\label{fig_7gal_7}
\end{figure}

In table~\ref{tab:I} the main parameters 
that result from the data analysis
(as discussed in the previous section) are reported
for some selected values of $E$.
The value $|\d-\d_0|$ is always less than $\d p$, consistently with the
discussion above, except for $E=12.5$ where, however, the relative difference 
between the two quantities is small ($\sim 9\%$). 
It can be noted that
$\d$ is systematically bigger than $\d_0$. 
This could be due to the fact that  
the error terms $O((p-1)^2/\t)$ or $o(1/\t)$ that have been discarded
likely produce a systematic shift in $\d$ or in $\d_0$; or that
the velocity of convergence of $\z_\t(p)$ is not the same on the negative or 
on the positive side (because numerically is much more difficult to
observe big negative fluctuations of $\s$ than the positive ones -- 
and the Fluctuation Relation provides a quantitative estimate
of the relative probabilities). 
At the moment, because of the level of precision of the
simulations, it is not possible to investigate this problem in more 
detail, see also Remark (3) in section~\ref{sec7:IID}.
On increasing the value of $E$, one is forced to decrease 
the value of $\tau$ used for
the analysis as, for longer $\tau$, 
the negative tail of the distribution $\zeta_\tau(p)$
becomes unobservable. This can be seen as the number 
$M^*$ of bins used for the computation of
$\chi$ decrease on increasing $E$; 
above $E=12.5$ it is impossible to find a value of
$\tau$ such that $\zeta_\tau(p)$ 
is close to the asymptotic limit and the negative tail
is observable. Thus, the fluctuation relation 
cannot be tested above $E=12.5$ with
the present computational power. However, the fluctuation relation 
has been checked in the region $E > 5$ where deviations from the 
linear response are observed. Moreover,
the estimated distributions $\zeta_\io(p)$ are very similar to the one reported
in Fig.~\ref{fig_7gal_3}: in particular, they are not Gaussian 
in the investigated
interval of $p$ (also for $E < 5$, in the linear response regime).

Finally, in Fig.~\ref{fig_7gal_7} 
the measured Lyapunov exponents of the model for $E=5$ 
and $E=25$ are reported. For this system, the Lyapunov exponents are 
known to be paired \cite{SEC98,SEI98,DM96} like in Hamiltonian systems
and the average of each pair is a constant equal to $\sigma_+/2Nd$.
For $E=5$, each pair is composed of a negative and a positive exponent.
This means that the attractive set is dense in phase space \cite{BGG97,Ga99b}
and the chaotic hypothesis is expected to apply to the system yielding a slope
$X=1$ in the fluctuation relation, as confirmed by the numerical data. The same
happens up to $E \sim 20$. 
Above $E=20$, there is a number $D$ of pairs composed by two negative exponents
(for $E=25$ one has $D=4$, see Fig.~\ref{fig_7gal_7}).
In this situation, the slope $X$ in the fluctuation relation is expected to be
given by $X = 1 - D/Nd$ \cite{Ga99b,BG97}. 
Thus, for $E=25$ one expects $X \sim 0.75$. Unfortunately,
as discussed above, above $E=12.5$ negative fluctuations
of the entropy production are not observed, 
and this prediction could not be tested in
the present simulation.

\section{Numerical simulation of model II}

\begin{figure}[t]
\centering
\includegraphics[width=.50\textwidth,angle=0]{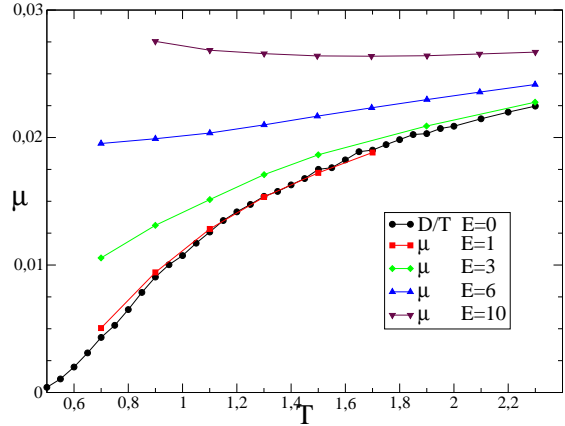}
\caption[Mobility as a function of $T$ and $E$ for Model II]
{Mobility as a function of the temperature $T$ and of the 
driving force $E$ for Model II. The circles correspond to the
equilibrium diffusion coefficient divided by the temperature.
Deviations from the linear response are observed for $E \geq 3$;
they become larger on lowering the temperature, 
as $D \rightarrow 0$.
}
\label{fig_7gal_8}
\end{figure}

\begin{figure}[t]
\centering
\includegraphics[width=.50\textwidth,angle=0]{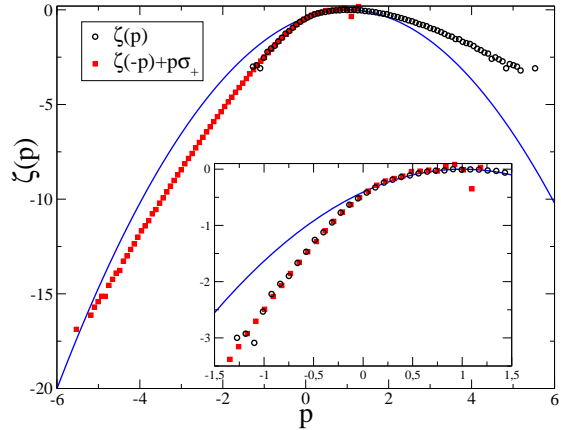}
\caption[The function $\z_\io(p)$ for Model II with $T=1.1$ and $E=3$]
{The estimate of the function $\zeta_\io(p)$ (open circles) for
Model II with $T=1.1$ and $E=3$. In the same
plot $\zeta_\io(-p) + p\sigma_+$ (filled squares) is reported. 
In the inset, the
interval $p\in[-1.5,1.5]$ where the data overlap is magnified. The full
line is the Gaussian approximation, $\zeta_\io(p) = 
\frac12\zeta_\io^{(2)}(p-1)^2$.
The data have been obtained from the histogram of $\pi_\tau(p)$ with $\tau=2.5$
(see table~\ref{tab:II}).
}
\label{fig_7gal_9}
\end{figure}

\begin{table}[h]
\centering
\begin{tabular}{rr|ccccccccc}
\hline
$T$ & $E$ & $\tau$  & $\sigma_+$ & $\delta$ & $\d_0$ & $|\d-\d_0|$ &
$\d p$ & $M^*$ & $\chi^*$ & $\delta X$ \\
\hline
0.9 & 1 & 3.0 & 0.209 & 0.453 & 0.334 & 0.119 & 0.223 & 68 & 1.9 & 0.19 \\
0.9 & 3 & 3.0 & 2.615 & 0.286 & 0.264 & 0.024 & 0.132 & 15 & 1.0 & 0.23 \\
\hline
1.1 & 1 & 4.0 & 0.233 & 0.231 & 0.126 & 0.105 & 0.126 & 79 & 1.7 & 0.24 \\
1.1 & 3 & 2.5 & 2.493 & 0.217 & 0.238 & 0.021 & 0.087 & 30 & 1.0 & 0.12 \\
1.1 & 6 & 1.5 & 13.32 & 0.113 & 0.230 & 0.117 & 0.092 & 7 & 1.1 & 0.21 \\
\hline
1.5 & 1 & 3.0 & 0.230 & 0.179 & 0.140 & 0.039 & 0.140 & 86 & 0.9 & 0.13 \\
1.5 & 3 & 2.5 & 2.227 & 0.145 & 0.123 & 0.022 & 0.082 & 33 & 4.7 & 0.18 \\
1.5 & 6 & 0.5 & 52.14 & 0.074 & 0.130 & 0.056 & 0.052 & 11 & 0.6 & 0.10 \\
\hline
1.7 & 1 & 3.0 & 0.221 & 0.127 & 0.141 & 0.014 & 0.283 & 49 & 1.0 & 0.26 \\
\hline
1.9 & 3 & 2.5 & 1.981 & 0.106 & 0.122 & 0.016 & 0.122 & 26 & 0.8 & 0.12 \\
1.9 & 6 & 0.4 & 43.52 & 0.078 & 0.126 & 0.048 & 0.085 & 14 & 1.7 & 0.11 \\
1.9 & 10 & 0.2 & 139.0 & 0.079 & 0.135 & 0.056 & 0.039 & 7 & 0.8 & 0.10 \\
\hline
2.1 & 6 & 0.4 & 40.48 & 0.074 & 0.110 & 0.036 & 0.110 & 11 & 1.0 & 0.15 \\
\hline
\end{tabular}
\caption{Model II: results of the data analysis 
for some selected values of $T$ and $E$. 
All the quantities are defined in section~\ref{sec7:IV}.
}
\label{tab:II}
\end{table}

Model II differs from model I in the dimension $d=3$, in the larger
number of particles $N=20$, and because it is a binary mixture of
two types of particles.
Binary mixtures are frequently used as models for numerical simulations 
of supercooled liquids as they avoid crystallization also 
at very low temperature on the "physical" time scales (\ie on the time scales
of numerical experiments); for these systems, at low temperature 
deviations from the linear response are observed
also for very low values of the external driving force.

In Fig.~\ref{fig_7gal_8} the equilibrium diffusion coefficient
$D$ (divided by the temperature $T$) and the mobility (for different
values of $E$) are reported as functions of the temperature.
Even though the number of particles is very small, on lowering the 
temperature the systems approaches the supercooled
state and $D$ becomes very small around $T \sim 0.5$.
Slightly above this temperature, \ie around $T=1$, 
strong deviations from the linear response are
observed for $E \geq 3$, where the entropy production $\sigma_+$ is still
close to $0$. Some values of $\sigma_+$ are reported in table~\ref{tab:II}; 
to compare these values with those obtained for model I one should note
that $\sigma_+$ is an {\it extensive} quantity. Thus, the entropy production
{\it per degree of freedom}, $\sigma_+/2Nd$, is much smaller in model II
than in model I. 

In table~\ref{tab:II} the results of the data analysis outlined in 
section~\ref{sec7:IV} are reported. For $E \leq 6$ a very good 
agreement of the data with the predictions of the fluctuation relation 
and with the  theory of finite time 
corrections discussed in section~\ref{sec7:II} is obtained.
For $E=10$ it is very difficult to observe negative fluctuations of $p$
with the available computational power; see {\it e.g.} the result of the 
analysis for $E=10$ and
$T=1.9$, where only $M^*=7$ bins where available and it was mandatory to use
$\tau=0.2$, of the order of the mixing time $\tau_0$.
In Fig.~\ref{fig_7gal_9} the estimated function $\zeta_\io(p)$ obtained
for $T=1.1$ and $E=3$ from the data with $\tau=2.5$ is reported. 
Strong deviations from the Gaussian behavior are observed 
in the accessible range of $p$ (see the
inset of Fig.~\ref{fig_7gal_9}). A similar behavior
of $\z_\io(p)$ is observed in correspondence of all the values
of $E$ and $T$ that were investigated (those listed in Table II):
in particular in all these cases highly non Gaussian behaviors
are observed in the accessible range of $p$. 

The Lyapunov spectrum for this system is very similar to the one reported
in the upper panel of Fig~\ref{fig_7gal_7}. Pairs of two negative exponents were
observed only for $E=10$ at $T \leq 1.3$, where, as in the case of Model I,
$\sigma_+$ is too large to allow for a 
verification of the modified fluctuation relation expected in this case,
see the discussion at the end of section~\ref{sec5:modelI}.

\section{Discussion}
\label{sec5:discussion}

The fluctuation relation has been tested, quite successfully, 
in a numerical simulation of
two models of interacting particles subjected to an external nonconservative
force and to a reversible mechanical thermostat. 
The data satisfy the fluctuation relation with a $\chi \leq 3$
and an accuracy of the order of $20 \%$ also for very large values of the
driving force, where strong deviations from the linear response are observed, 
and where the large deviation function is strongly non-Gaussian.
The comparison of the numerical data with the predictions of the 
fluctuation relation is done by taking into account the 
(lowest order) finite time
corrections to the distribution function for the fluctuations of
the phase space contraction rate. This is crucial: if 
such corrections were not taken into account, the fluctuation relation would
be violated within the precision of the experiment.

In order to compute the finite time corrections, 
an algorithm which allows to reconstruct the asymptotic 
distribution function from measurable quantities at finite time, within
a given precision, has been proposed. The theory of the corrections relies on the 
symbolic representation of the chaotic dynamics, therefore it is 
applicable if one accepts the Chaotic Hypothesis.

The numerical results support the conjecture that the 
{\it chaotic hypothesis} can be applied to these
systems, also very far from equilibrium, and in particular
the fluctuation relation is
satisfied even in regions where its predictions measurably differ 
from those of linear response theory.

The theory of finite time corrections for the analysis of the
numerical data could in principle be of interest for real experimental
settings where non Gaussian fluctuations for the entropy production
rate are observed, see \eg~\cite{CL98,FM04}.

However it should be stressed that in a real experiment there are some
technical differences with respect to the numerical simulation which
could in some cases make inapplicable 
the analysis, namely:
\\ 
{\it (i)} usually the noise in the large deviation function for the
entropy production rate in a real experiment is much bigger than in a
numerical experiment, and it is likely that the translation in 
Eq.~(\ref{etatau}) computed as the ratio $\z^{(3)}/(\z^{(2)})^2$ is not
measurable within an error of some percent;
\\
{\it (ii)} usually in a real experiment the accessible time scales
are naturally much bigger than the microscopic ones so that, if the
negative fluctuations of the entropy production rate are observable at all,
one is automatically in the asymptotic regime, where the finite time
corrections should be negligible;
\\
{\it (iii)} a usual problem in a realistic setting is that there is no
clear connection between the ``natural'' thermodynamic entropy
production rate $\dot s=W/T$ ($W$ is the work of the dissipative
external forces and $T$ is the temperature) and the microscopic phase
space contraction rate, for which a slope $X=1$ in the fluctuation
relation $\z(p)-\z(-p)=X\sigma_+p$ is expected; 
so, often one measures an $X\neq 1$ and correspondingly one {\it
defines} an effective temperature $\Th_{eff}=T/X$ giving a natural
connection between the effective thermodynamic entropy production rate
$\dot s_{eff}=W/\Th_{eff}$ and the phase space contraction rate, see
\cite{Ga04,CL98,FM04} and the following chapters;
in such a situation (where an adjustable parameter $X$ appears) it makes 
no sense to apply this analysis, which is sensible only if one wants to compare
the experimental data with a sharp prediction about the slope $X$ in the
fluctuation relation.

A big problem which is left open is trying to understand 
how the fluctuation relation is modified for values
of the driving force so high that the attractive set is no longer dense in phase
space. It is expected, \cite{BG97},
that in such a case 
$\z_\io(p)-\z_\io(-p)$ is still linear, but the slope is $X\s_+$, with 
$X$ given by the ratio of the dimension of the attractive set and
of that of the whole phase space. An estimate of such quantity can be given 
via the number of negative pairs of exponents in the Lyapunov spectrum 
\cite{Ga99b,BG97}.
Unfortunately negative pairs begin to appear in the Lyapunov spectrum
only for values of the external force so high that no negative fluctuations
are observable anymore.
Hopefully future work will address this point.

\section{Appendix: A Limit Theorem}
\label{app7:A}

In this section Eq.s~(\ref{24}--\ref{23}) will be proved. 
The proof is presented
in the case $p$ is the average 
of independently distributed discrete variables $\s_i^\e$, assuming
values in $\e\ZZZ$, for some small mesh parameter $\e$; then
how this can be applied and adapted to the situation
considered in section~\ref{sec7:II} and subsequent sections 
will be discussed.

Let $\s_i$, $i\in\NNN$, 
be independent continuous random variables with identical 
distributions $\p(d\s_i)$ with positive variance $\d\s^2>0$, supported
on the finite interval $[s_-,s_+]$. Assume that $\p(d\s_i)$
gives positive probability to any finite interval contained in $[s_-,s_+]$.
Let $\p_\l(d\s)$ be the weighted distribution 
$\p_\l(d\s)=e^{-\l\s}\p(d\s)/\int e^{-\l\s}\p(d\s)$ and define
$z_\io(\l)=-\log\int e^{-\l\s}\p(d\s)$ and $\s_+=z_\io'(0)$.
Note that the assumption 
that $\p(d\s_i)$ gives positive probability to an interval of $\s$
in $[s_-,s_+]$ implies that for any finite
$\l$ also $\p_\l(d\s)$ has positive variance $-z_\io''(\l)>0$. 

Also, given $\e>0$ (with the property that $s_+-s_-=N_\e\e$ for some integer 
$N_\e$), 
consider the discretization of 
$\s_i$ on scale $\e$, and call it $\s_i^\e$: $\s_i^\e$ will be a discrete variable 
assuming the values 
$s_k^\e\equiv s_-+(k-\frac{1}{2})\e$, $k=1,\ldots,N_{\e}$, 
with probabilities $\p^\e(s_k^\e)=
\Prob(\s_i^\e=s_k^\e)=\int_{s_k^\e\pm\frac{\e}{2}}
\p(d\s)$. The assumption that $\p(d\s_i)$
gives positive probability to any finite interval contained 
in $[s_-,s_+]$ implies that $\p^\e(s_k^\e)>0$ for any $\e$ and $k$.
Let also $z_\e(\l)=-\log\sum_{k=1}^{N_\e}e^{-\l s_k^\e}\p^\e
(s_k^\e)$ and $\p^\e_\l(s_k^\e)=\p^\e(s_k^\e)e^{-\l s_k^\e+z_\e(\l)}$.
Note that, since $\p^\e(s_k^\e)>0$ for any $k$, for any finite $\l$ one has
$-z_\e''(\l)>0$.

If $p_\t^\e=\frac{1}{\t\s_+}\sum_{i=1}^\t\s_i^\e$ and $\P_\t(\e;I)$ is the 
probability 
that $p_\t^\e$ belongs 
to the finite interval $I$, the following theorem holds.\\
\\
{\bf Theorem:} 
{\it Given a finite interval $I\subset (s_-,s_+)$, let $\s_i^\e$, $\p^\e$ and 
$\P_\t(\e;I)$ be defined as above. Then, for a sufficiently small $\e>0$,
there exists an analytic "rate function" $\widetilde\z_\t(p)$ such that
\beq\label{A1.1}
\lim_{\t\to\io}\frac
{\P_\t(\e;I)} 
{\int_{I} dp e^{\t\wt\z_\t(p)}}
=1\;.\eeq
$\widetilde\z_\t(p)$ is defined by:
\beq\label{A1.2}\begin{split}
&\wt\z_\t(p)+\frac{1}{\t}\log\Big[\frac{\sinh[\e\l_p^\e/(2\s_+)]}
{\e\l_p^\e/(2\s_+)}
\Big]=\z_\t^\e(p)\\
&\z_\t^\e(p)=-z_\e(\l_p^\e)+\l_p^\e p\s_+-\frac{1}{2\t}
\log[\frac{2\p}{\t}\Big(-\frac{z_\e''(\l_p^\e)}{\s_+^2}\Big)]\end{split}
\eeq
and $\l_p^\e$ is the inverse of $p(\l)=z_\e'(\l)/\s_+$. 
The function $\z_\t^\e(p)$ has the 
following property: if $\D\subset I$ is an interval of 
size $\frac{\e}{\t\s_+}$ around
a point $p_\D$, then:
\beq\label{A1.1a}
\lim_{\t\to\io}\frac
{\P_\t(\e;\D)}{|\D|e^{\t\z_\t^\e(p_\D)}}=1\eeq
}
\\
{\bf Proof}
Define the auxiliary variable $q=\frac{1}{\t\s_+}\sum_{i=1}^\t\h_i$,
where $\h_i$ are i.i.d. discrete random variables, with distribution 
$\p_\l^\e(s_k^\e)$. Let $\P_\t^\l(\e;q_0)$ be the probability that 
$q$ assumes the value $q_0\in I$, with $q_0\s_+=s_k^\e/\t$ for some $k\in\NNN$,
and note that
$\P_\t^0(\e;q_0)$ is identical to the probability that $p_\t=q_0$.
By definition $\P_\t^\l(q_0)$ and $\P_\t^0(q_0)$ are related by:
\beq\label{A1.5}
\P_\t^\l(\e;q_0)=\frac{e^{-\l q_0\s_+\t}\P_\t^0(\e;q_0)}{
\big[\sum_k e^{-\l s_k^\e} \p^\e(s_k^\e)\big]^\t} \ .
\eeq
Now, a local form of central limit theorem 
(Gnedenko's theorem, see pag. 211 of \cite{Fi63}) tells that, 
if $q$ is localized near its mean value, that is if 
$|q\s_+-z_\e'(\l)|\le \frac{M\e}{\t}$
for some finite $M$, then $\P_\t^\l(\e;q_0)$ is asymptotically equivalent
to the Gaussian with mean $z_\e'(\l)$ and variance 
$-z_\e''(\l)$, in the sense that
\beq\label{A1.6}\P_\t^\l(q_0)=\frac{\e}{\sqrt{2\p\t(-z_\e''(\l))}}e^{-
\frac{(q_0\s_+-z'_\e(\l))^2}{2(-z_\e''(\l))}\t}(1+o(1))\;,\eeq
for any $q_0$ s.t. $|q\s_+-z_\e'(\l)|\le \frac{M\e}{\t}$
\footnote{\label{central} Note that Gnedenko's Theorem is {\it different}
from the usual central limit theorem, stating instead that for 
$|q\s_+-z_\e'(\l)|\le \frac{C}{\sqrt\t}$ ($C$ big) the sums of 
$\P_\t^\l(\e;q)$ over intervals of amplitude $\frac{1}{\sqrt\t}$ contained in 
$|q\s_+-z_\e'(\l)|\le \frac{C}{\sqrt\t}$
are asymptotically equal to the integrals of the Gaussian 
over the same intervals. 
That is, usual central limit theorem gives informations on the distribution
in a bigger interval around the maximum, but on a rougher scale.}.

So, given $\l_{q_0}^\e$ s.t. $z_\e'(\l_{q_0}^\e)=q_0\s_+$ (such $\l_{q_0}^\e$ 
exists, is unique
and is an analytic function of $q_0$, by the remark that $-z_\e''(\l)>0$ for
any finite $\l$ and
$z_\e(\l)$ is an analytic function of $\l$), using
Eq.~(\ref{A1.6}),
Eq.~(\ref{A1.5}) can be restated as:
\beq\label{A1.7}\P_\t^0(\e;q_0)=\frac{\e}{\sqrt{2\p\t(-z_\e''(\l_{q_0}^\e))}}
e^{\l_{q_0}^\e q_0\s_+\t-z_\e(\l_{q_0}^\e)}(1+o(1)) \ .
\eeq
Now, by the definition of $\z_\t^\e(p)$ in Eq.~(\ref{A1.2}), 
the r.h.s. of the last equation is equal to 
$\frac{\e}{\t\s_+}e^{\t\z_\t^\e(q_0)}(1+o(1))$.
Finally, the statement of the Theorem follows by the remark that 
\beq\label{A1.8}
\frac{\e}{\t\s_+}e^{\t\z_\t^\e(p_0)}=\int_{p_0-\frac{\e}{2\t\s_+}}^{p_0+\frac{\e}
{2\t\s_+}}dp e^{\t\tilde\z_\t(p)}\Big(1+o(1)\Big)\;.\eeq	
In fact the integral in the r.h.s. of the last equation is given by
\beq \label{A1.9}\begin{split}
&e^{\t\tilde\z_\t(p_0)} \int_{p_0-\frac{\e}{2\t\s_+}}^{p_0+\frac{\e}
{2\t\s_+}}dp e^{\t\tilde\z_\t'(p_0)(p-p_0)}\Big(1+O(\frac{\wt\z_\t''(p_0)\e^2}{\t})
\Big)=\\
&=e^{\t\tilde\z_\t(p_0)} \frac{2\sinh[\tilde\z_\t'(p_0)\e/(2\s_+)]}
{\t\tilde\z_\t'(p_0)}
\Big(1+O(\frac{\wt\z_\t''(p_0)\e^2}{\t})\Big)\end{split}\eeq
and in the last expression one has to note that $\wt\z_\t'(p_0)=[\z_\t^\e]'(p_0)+
O(\frac{1}{\t})=\l_{p_0}^\e+O(\frac{1}{\t}).\ $ \qed
\\
\\
A first Remark to be done about the Theorem above is that, in order
to define a ``universal'' rate function in terms of quantities 
depending only on 
$z_\io(\l)$ (instead of quantities depending 
on the ``non universal'' function $z_\e(\l)$, 
which explicitly depends on the discretization step $\e$), it would
desirable to perform (in a sense to be precised) 
the continuum limit $\e\to 0$. To this regard, one can note that 
the only point where in the proof above the fact that 
$\e$ is a constant (\ie is independent of $\t$) has been used 
was in using Gnedenko's Theorem, see \cite{Fi63}. 
However, by a critical analysis of the proof 
of Gnedenko's Theorem, one can realize that it is even possible to 
let $\e=\e_\t$ go to $0$ with $\t$; 
the velocity with which $\e_\t$ is allowed to go to $0$ depends on the 
details of the distribution $\p(d\s)$. 
So one can even study the probability distribution of $p_\t$ on a scale
$\sim \e_\t/\t$: introducing bins $\D_\t$ of size $O(\e_\t/\t)$
and defining $\P_\t(\D_\t)$ to be the probability that $p_\t=\frac{1}{\t\s_+}
\sum_i\s_i$ belongs to the bin $\D_\t$ centered in $p_0$, one can 
repeat the proof above to conclude that
\beq\label{A1.9a}
\lim_{\t\to\io}\frac
{\P_\t(\D_\t)}{|\D_\t|e^{\t\z_\t(p_0)}}=1\eeq
where $\z_\t$ satisfies the equation:
\beq\z_\t(p)=-z_\io(\l_p)+\l_p p\s_+-\frac{1}{2\t}
\log[\frac{2\p}{\t}\Big(-\frac{z_\io''(\l_p)}{\s_+^2}\Big)]\eeq
and $\l_p$ is the inverse of $p(\l)=z_\io'(\l)/\s_+$. \\
 
Another point to be discussed is that in the Theorem above
the $\s_i$ were assumed to be independent. 
This is not the case for the variables $\s(S^i\cdot)$ of section~\ref{sec7:II}. 
However, if, as discussed in Remark (3) of section~\ref{sec7:IID},
the time unit is chosen to be of the order of the mixing time, the variables 
$\s(S^i\cdot)$ have (by construction) a decorrelation time equal to $1$, and the 
analysis of previous theorem can be repeated step by step
in order to construct the probability distribution of 
$p=\frac{1}{\t\s_+}\sum_{i}\s(S^i\cdot)$. The only differences
are that: (1) $\t z_\io(\l)$ should be replaced by $\t z_\t(\l)
=-\log\int 
e^{-\l p\s_+\t}\P_\t(dp)$ throughout the discussion; (2) instead of Gnedenko's
theorem one has to apply a generalization of Gnedenko's to short ranged
Gibbs processes, to be proven via standard cluster expansion
techniques (see for instance \cite{Ga72} for a proof of a generalization
of Gnedenko's theorem to a short ranged Gibbs process
in the context of non critical fluctuations 
of the phase separation line in the 2D Ising model). \\

The conclusion is that, if the bins $\D$ 
in section~\ref{sec7:IIC} are chosen of size $\e_\t/\t$, the probability of the bin $\D$
centered in $p_\D$ is asymptotically given by $\p(p\in\D)\simeq e^{\t\z_\t(p_\D)}$
(in the sense of Eq.~(\ref{24})) and  $\z_\t(p_\D)$ can be interpolated by
an analytic function of $p$ that in fact satisfies Eq.~(\ref{23}).

%% file: cap6.tex
\chapter{Dynamics of glassy systems}
\label{chap6}

\section{Introduction}

From a phenomenological point of view the dynamical behavior of
glassy systems is more relevant than the equilibrium one~\cite{Ku01}.
Indeed, as discussed in chapter~\ref{chap1}, the equilibrium relaxation
time diverges, on approaching $T_K$ from above, following the
VFT law (\ref{VFT}). This divergence is much stronger than the power law
that characterizes second-order phase transitions, and this means
that the relaxation time becomes much larger than any experimentally
accessible time scale at a temperature $T_g$ that is usually much larger
than $T_K$. Indeed, the ratio $\frac{T_g}{T_g-T_K}$, related to {\it fragility},
falls between $1$ and $10$ in molecular glasses\footnote{In numerical simulations 
the accessible time scales are much smaller than in experiments, so the $T_g$ of
numerical simulations is much higher than the experimental one.}, 
see section~\ref{sec1:fragility}.

Below $T_g$ the system cannot be equilibrated anymore, so its equilibrium
properties cannot be investigated and in particular the Kauzmann transition
and the {\it ideal} (equilibrium) glass phase are unobservable.
The system becomes a {\it real} (nonequilibrium) glass whose properties
are not described by the Gibbs distribution. The real glass transition is
a {\it dynamical} phenomenon, so that almost all the experimental data
(apart from the extrapolations discussed in chapter~\ref{chap1}) refer to
dynamical properties of the nonequilibrium glassy state.
The most striking feature of the nonequilibrium glassy state is that 
{\it it is not stationary}: even if the averages of the interesting 
observables (pressure, density, etc.) reach an asymptotic constant value,
the two-time correlation functions $C(t,t')$ depend on the two times
also for $t,t'$ very large (\ie of the order of the experimentally accessible
time scales). This phenomenon is known as {\it aging}, because the properties
of the system depend on its age, \ie on the time elapsed from the initial 
time in which it was prepared.

Thus dynamical theories of glasses are more suitable to be compared with
experimental data. Indeed, many theories have been proposed, either 
phenomenologicals or fundamentals, and many experiments have been performed,
so that a very large literature about the dynamical behavior of glasses
exist, and detailed reviews have been recently 
published, see \eg~\cite{CC05,BCKM98,Cu02,Ku01,Go99,Cu99,CR03,CKP97,CK00}.
Some aspects of the equilibrium dynamics of glasses have been discussed
in chapter~\ref{chap1}. In the following some selected aspects of the
nonequilibrium dynamics will be discussed. Only a particular point of
view, that derives from the exact solution of the dynamics of $p$-spin
models, will be discussed, and the attention will be mainly focused 
on the {\it driven} dynamics, \ie the dynamics in presence of nonconservative
forces, because in next chapter an extension of the fluctuation relation
to driven glasses will be discussed.

\section{Relaxational dynamics of $p$-spin models}

Following the strategy of~\cite{CK93,CK94},
the dynamics of mean field $p$-spin systems can be analytically 
solved~\cite{CS92}. Detailed reviews are in~\cite{Cu02,CC05}.

The Hamiltonian of $p$-spin models is given by Eq.~(\ref{Hpspin}).
If the spherical version is considered, the variable $\s_k$ are
continuous and the simplest possible dynamics is the Langevin
dynamics:
\beq\label{pspinlangevin}
\dot \s_k (t) = -\m(t) \s_k(t) -\frac{\dpr H_p(\s)}{\dpr \s_k(t)} + \h_k(t) + h_k(t) \ ,
\eeq
where $\h_k(t)$ is a Gaussian white noise, with $\la \h_k(t) \ra=0$
and $\la \h_k(t) \h_l(t') \ra = 2 T \d_{kl} \d(t-t')$, $\m(t)$ is a Lagrange
multiplier that is needed to enforce the spherical constraint 
$\sum_k \s_k^2=N$, and $h_k(t)$ is an external field that will be used
only to compute the linear response, see below. 
For Ising systems the Metropolis dynamics can be
considered but the calculations are more complicated.

\subsection{Dynamical generating functional}

The average of a given observable $A$, which is a functional
of the trajectory $\s(t)$, is given by
\beq
\la A \ra = \int {\cal D}\s(t) {\cal P}[\s(t)] A[\s(t)] \ ,
\eeq
where ${\cal P}[\s(t)]$ is the probability distribution on the space
of trajectories induced by the distribution of the noise and by the
equation of motion~(\ref{pspinlangevin}). Examples of interesting observables
are the magnetization $m(t) = N^{-1} \sum_k \la \s_k(t) \ra$ and the correlation
function $C(t,t') = N^{-1} \sum_k \la \s_k(t) \s_k(t') \ra$.
Another interesting observable is the {\it linear response function}, defined by
\beq
R(t,t') = \frac{1}{N}\left. \sum_k \frac{\d \la s_k(t) \ra}{\d h_k(t')}\right|_{h=0} \ .
\eeq

To compute the average of one-time observables $A(t)$, such as the
magnetization, one can introduce a {\it dynamical generating functional}
\beq
Z[J(t)] = \la e^{\int dt J(t) A(t)} \ra = \int {\cal D}\s(t) {\cal P}[\s(t)] \,
 e^{\int dt J(t) A(t)} \ .
\eeq
The average of $A$ and its correlations can be computed as derivatives
of $Z$ w.r.t. to $J(t)$. Note that as long as ${\cal P}[\s(t)]$ is
normalized $Z[J=0]=1$ and one does not need to consider the logarithm of $Z$.

Moreover, one can simply compute $Z[J=0]$ (which is trivially $1$) in order
to see which trajectories $\s(t)$ dominate the generating functional for $N \to \io$.
As the system is mean-field, this procedure allows to write {\it effective} 
equations for a single degree of freedom moving in an environment whose properties
have to be determined self-consistently. If the initial condition is chosen at 
random, one has simply \cite{CC05,Cu02}:
\beq\label{eq6:uno}
1 = \int {\cal D}\s(t) {\cal P}[\s(t)] = 
\int {\cal D}\s(t) {\cal D}\h(t) {\cal P}[\h(t)] 
\d\left[\dot \s_k (t) + \m(t) \s_k(t) +\frac{\dpr H_p(\s)}{\dpr \s_k(t)} - \h_k(t)\right] \ ,
\eeq
and since the distribution of the noise is a Gaussian,
\beq\label{eq6:AAA}
{\cal P}[\h(t)] \propto \exp \left[ -\frac{1}{4T} \sum_k \int dt \, \h_k(t)^2 \right] \ ,
\eeq
one can easily perform the integration over $\h(t)$ in Eq.~(\ref{eq6:uno}) representing
the $\d$-function as
\beq\label{eq6:BBB}
\d\big[ f_k(t) \big] \propto \int \DD \wh\s(t) \, e^{\sum_k \int dt \, i \wh\s_k(t) f_k(t)} \ ,
\eeq
and one finally obtains
\beq\label{eq6:CCC}
1 \propto \int \DD \s(t) \DD \wh\s(t) \, e^{ -T \sum_k \int dt \wh \s_k(t)^2
+ \sum_k \int dt \, i \wh\s_k(t) 
\big[ \dot \s_k (t) + \m(t) \s_k(t) +\frac{\dpr H_p(\s)}{\dpr \s_k(t)} \big]} =
 \int \DD \s(t) \DD \wh\s(t) \, e^{S(\s,\wh\s)} \ .
\eeq
The procedure that leads to Eq.~(\ref{eq6:CCC}) holds for 
a generic Langevin equation. The action $S(\s,\wh\s)$ has a term proportional
to $-\wh\s^2$ whose coefficient is $1/2$ times the variance of the noise,
a term $i\wh\s \dot \s$ and a term $-i\wh\s F(\s)$, where $F(\s)$ is the force
acting on $\s$. 
What is remarkable is that the proportionality factors in Eq.s~(\ref{eq6:AAA}), 
(\ref{eq6:BBB}), (\ref{eq6:CCC}) 
{\it do not depend on the couplings} $J$, \ie on the disorder. This means that
the disorder appears (linearly) only in the exponent of Eq.~(\ref{eq6:CCC}) and one
can now perform the average over $J$ {\it without using replicas}.
Another interesting remark is that the response function $R(t,t')$ is just the
correlation function of $i \wh \s$ and $\s$, 
$R(t,t')=N^{-1} \sum_k \la \s_k(t) i \wh \s_k(t') \ra$ \cite{CC05,Cu02}.

\subsection{The average over the disorder}

Averaging over the disorder in Eq.~(\ref{eq6:CCC}) simply amounts to perform the
Gaussian integral
\beq
\overline{ e^{\sum_k \int dt \, i \wh\s_k(t) \frac{\dpr H_p(\s)}{\dpr \s_k(t)}}} =
\overline{ e^{-\frac{1}{(p-1)!}\int dt \,
\sum_{k_i,\cdots,k_p} J_{k_1,\cdots,k_p} i \wh\s_{k_1}(t) \s_{k_2}(t) \cdots \s_{k_p}(t)}}
\ .
\eeq
The effect of the average over the $J$ is to produce {\it nonlocal} terms,
because in the expression above terms of the form $\exp [J \int dt \, f(t)]$ appear.
After the integration over the (Gaussian) $J$, they give rise to terms of the form
$\exp [\int dt \, f(t) ]^2 = \exp [\int dt dt' f(t) f(t') ]$, which is a nonlocal term.

Without entering into the details of the computation, see \eg~\cite{CC05}, the final
result is
\beq
\overline{ e^{\sum_k \int dt \, i \wh\s_k(t) \frac{\dpr H_p(\s)}{\dpr \s_k(t)}}} =
e^{\frac{N p (p-1)}{4} \int dt dt' R(t,t') R(t',t) C(t,t')^{p-2}} \ .
\eeq
What is important is that the latter expression depend on $\s$, $\wh\s$, only through
$R(t,t')$ and $C(t,t')$, which are {\it macroscopic} observables. This is what usually
happens in mean-field systems.
Substituting this result in Eq.~(\ref{eq6:CCC}),
one can follow the usual procedure of introducing $\d$-functions,
for example $\d[ C(t,t') - N^{-1} \sum_i \s_i(t) \s_i(t') ]$, in order to rewrite
Eq.~(\ref{eq6:CCC}) in the following form, see \cite{CC05} for the details:
\beq\label{eq6:DDD}
\begin{split}
1 &= \int \DD C(t,t') \DD R(t,t') \, 
e^{\frac{N p (p-1)}{4} \int dt dt' R(t,t') R(t',t) C(t,t')^{p-2}} \times \\
&\times \int \DD \s \DD \wh \s \,
e^{ \sum_k \int dt dt' \left[ -\frac{p}{4} C(t,t')^{p-1} \wh\s_k(t) \wh\s_k(t')
-\frac{1}{2} p (p-1) R(t,t') C(t,t')^{p-2} i \wh\s_k(t) \s_k(t') \right]} \times \\
&\hskip25pt \times 
 e^{ -T \sum_k \int dt \wh \s_k(t)^2
+ \sum_k \int dt \, i \wh\s_k(t) 
\big[ \dot \s_k (t) + \m(t) \s_k(t) \big]} \ .
\end{split}\eeq
Comparing the last two lines of the latter expression with Eq.~(\ref{eq6:CCC}),
it turns out that the spins $\s_k$ are now decoupled; and the last expression for
the generating functional could be obtained from the {\it effective equation} for
the dynamics of the spin $\s_k$:
\beq\label{eq6:effective}
\begin{split}
&\dot \s(t) = -\m(t) \s(t) + 
\frac{1}{2} p (p-1) \int dt' R(t,t') C(t,t')^{p-2} \s(t') + \r(t) \ , \\
&\la \r(t) \r(t') \ra = 2T\d(t-t') + \frac{p}{2} C(t,t')^{p-1} \ .
\end{split}\eeq
Thus, the original interacting problem has been mapped into a {\it single-spin}
problem, with a nonlocal force and a noise which is not simply 
$\d$-correlated. Turning back to Eq.~(\ref{eq6:DDD}), the integration over the
functions $C(t,t')$ and $R(t,t')$ can be performed via a saddle-point evaluation.
The saddle point equations are simply\footnote{The saddle point equations cannot
be derived differentiating Eq.~(\ref{eq6:DDD}) because some assumptions have already
been done to obtain it, namely that $P(t,t') \equiv\la \wh\s(t) \wh\s(t') \ra =0$. 
This assumption is related to causality. But one should first differentiate w.r.t.
$P(t,t')$ to obtain the equation $C(t,t') = \la \s(t) \s(t') \ra$, and then set
$P=0$. The equation for $R(t,t')$ is obtained differentiating Eq.~(\ref{eq6:DDD})
w.r.t. $R(t,t')$.}
\beq\begin{split}
&C(t,t') = \la \s(t) \s(t') \ra \ ,\\
&R(t,t') =\left. \frac{\d \la \s(t) \ra}{\d h(t')} \right|_{h=0} = \la \s(t) i\wh \s(t') \ra \ ,
\end{split}\eeq
where the average is now over the dynamics generated by Eq.~(\ref{eq6:effective})
and the field $h(t)$ must be added to Eq.~(\ref{eq6:effective}) to compute
the response.
This means that $C$ and $R$ have to be determined self-consistently as the
correlation and response function for the effective equation (\ref{eq6:effective}).
This procedure makes clear that $C(t,t')$ (and eventually $R(t,t')$) are the dynamical
order parameters, as already discussed in section~\ref{sec1:orderparameter}.
Restricting to causal solutions, such that $R(t,t')=0$ for $t' > t$,
the self-consistency equations can be written starting from 
Eq.~(\ref{eq6:effective}), see \cite{CC05,Cu02}, and are (for $t \geq t'$)
the following:
\beq\label{eq6:selfcons}
 \begin{split}
&\dpr_t C(t,t')=-\m(t) C(t,t') + \frac{1}{2}p(p-1)\int dt'' R(t,t'')C(t,t'')^{p-2}
C(t'',t') + \frac{p}{2} \int dt'' R(t',t'') C(t,t'')^{p-1} \ , \\
&\dpr_t R(t,t')=-\m(t) R(t,t') + \frac{1}{2} p (p-1) \int dt'' R(t,t'') C(t,t'')^{p-2}
R(t'',t') + \d(t-t') \ , \\
&\m(t)= T + \frac{p^2}{2} \int dt' R(t,t') C(t,t')^{p-1} \ . \\
\end{split}\eeq
The equation for $\m(t)$ is obtained from the condition $\frac{d}{dt}C(t,t)=1$
that follows from the spherical constraint.
Note that, {\it once the self-consistency equations are solved}, $C(t,t')$ and
$R(t,t')$ are determined and the dynamics of each spin $\s_k$ is given by
Eq.~(\ref{eq6:effective}), which is a Langevin equation for a single spin moving in an effective
environment defined by $C$ and $R$.

\subsection{The dynamical transition and aging}

As already outlined in section~\ref{sec1:meanfieldgeneral}, Eq.s~(\ref{eq6:selfcons})
admit a stationary solution for $T > T_d$. This means that one can find a solution
such that $\m(t) \equiv \m$, $C(t,t')=C(t-t')$ and $R(t,t')=R(t-t')$. Moreover,
$C$ and $R$ are related by the equilibrium fluctuation-dissipation theorem,
\beq\label{eq6:FirstFDT}
R(t-t') = - \frac{\th(t-t')}{T} \dpr_t C(t-t') \ ,
\eeq
so $C(t)$ is the only independent variable. This means that the system is able to
equilibrate with the thermal bath in a finite time. The relaxation time of the
correlation function $C(t)$, $\t_\a$, grows on lowering the temperature and
diverges as a power law for $T\to T_d^+$. Slightly above $T_d$, the dynamics
is separated in a ``fast'' relaxation, that makes $C(t)$ decrease from $1$ to some
value $q_d < 1$ on a time scale which is $T$-independent, and in a ``slow'' relaxation,
that makes $C(t)$ drop to zero on a scale $\t_\a$ (see the curve for $\ee=0$ in
Fig.~\ref{fig6:corrTa} below), which is strongly $T$ dependent
and diverges at $T_d$ as stated above. 

Below $T_d$ the stationary solution does not exist anymore. One finds a solution
that is not stationary, and has the form
\beq\label{eq6:corraging}
\begin{split}
C(t,t') = C_{f}(t-t') + C_{s}(t,t') \ , \\
R(t,t') = R_{f}(t-t') + R_{s}(t,t') \ ,
\end{split}\eeq
while $\m(t)$ still is asymptotically constant for $t \to \io$. The stationary
part of the correlations still describe the ``fast'' relaxation from $1$ 
to $q_d$, that is very similar to the one observed above $T_d$.
However the ``slow'' relaxation becomes non stationary below $T_d$. In particular,
if one looks to $C(t,t')$ as a function of $\t\equiv t-t'$ for fixed $t'$, it
decays to zero for $\t \gg \t_\a(t')$, where $\t_\a(t')$ is an increasing function
of $t'$ that diverges for $t'\to \io$.

Remember that Eq.s~(\ref{eq6:selfcons}) describe the relaxation of the $p$-spin model
{\it starting from a random initial condition}. This means that, if the system is
prepared in a random configuration and let evolve in contact with a bath at temperature
$T < T_d$, it is not able to equilibrate with the bath. Moreover, if at time $t'$ after
the preparation the system is in a configuration $\s(t')$, the time needed to decorrelate
completely from the configuration $\s(t')$ is an increasing function of $t'$, $\t_\a(t')$,
that diverges for $t' \to \io$. 

\subsection{Interpretation of the dynamics in term of the free energy landscape}

The limit
\beq
q_d(T) = \lim_{t-t' \to \io} \lim_{t'\to \io} \lim_{N\to\io} C(t,t') \ ,
\eeq
is finite for $T < T_d$ and represents the {\it dynamical order parameter}
as discussed in section~\ref{sec1:orderparameter}. 
This means that if one waits a long time $t'$ after the
preparation, the system is no more able to decorrelate completely. It remains {\it trapped}
in a group of configurations $\s$ that have overlap $\gtrsim q_d$ between themselves, \ie
it remains trapped into a {\it metastable state}.

As discussed in section~\ref{sec1:TAP}, for $T_K < T < T_d$, an exponential number
of metastable states can be found in $p$-spin models.
For a given temperature $T < T_d$, their free energies range from $f_{min}$ to
$f_{max}$. Starting from a random initial condition means that the system starts with
a very high free energy. If it is let evolve at $T < T_d$, it will start to descend in
the free energy landscape until it reaches the level $f_{max}$ where metastable states
first appear. 
Indeed, slightly above $f_{max}$ the phase space of the system is still connected,
so the stationary points of the free energy must have some negative eigenvalues
corresponding to unstable directions.
However, the largest negative eigenvalue in stationary points is 
$\l \propto f_{max}-f$~\cite{KPV93,CGP98}.
Thus, the system remains trapped for a long time close to stationary points of
$f$ before it can escape and find states that are closer to $f_{max}$; the latter
have even smaller negative eigenvalues, so the time needed to escape from them
is larger~\cite{Cu02,CGG03b}, and so on. 
This is why the correlations become non stationary and the
system {\it ages} indefinitely.

A confirmation of this scenario is that the value of $q_d(T)$ computed from
the dynamics is equal to the self overlap of the threshold states, \ie the states
with $f=f_{max}$, computed from the TAP equations.
Another confirmation is that, if one studies the dynamics starting from an
equilibrated initial datum\footnote{This is analytically possible but requires 
the introduction of replicas to describe the initial datum. It is impossible
in experiments because the system cannot be equilibrated below $T_d$.} 
at temperature $T < T_d$, one finds
a stationary solution $C(t-t')$ which however do not decay to zero but has a 
finite limit for $t-t' \to \io$. That is, equilibrium configurations below $T_d$
typically belong to a metastable state, so if one starts in one of them, the 
system remains forever trapped into the state. The limit 
$q(T)=\lim_{t-t'\to\io} C(t-t')$ is the self overlap of the equilibrium
states at temperature $T$.

\section{Driven dynamics of $p$-spin models}
\label{sec6:drivenpspin}

\begin{figure}[t]
\centering
\vskip-2cm
\includegraphics[width=15cm]{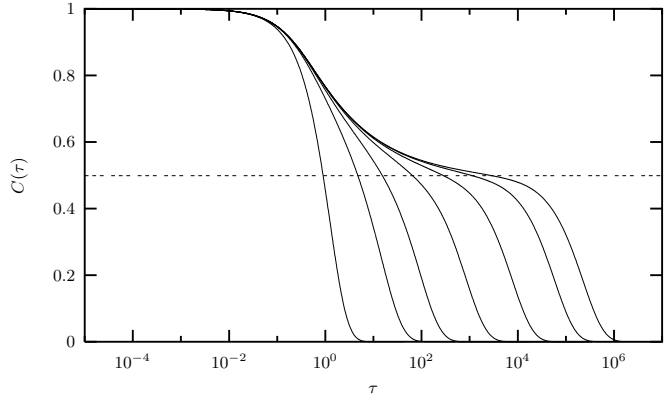}
\vskip-14cm
\caption[Correlation functions for the driven $p$-spin model]
{(From \cite{BBK00})
Correlation function vs. time for Eq.s (\ref{eq6:MCT}), (\ref{eq6:kerneldriven}) 
with $k=p=3$ and $T=0.613>T_d (\sim0.6124)$ for different
driving forces. From left to right: $\epsilon = 5$, 0.333, 0.143, 
0.05, 0.0158, 0.00447 and 0.
The longest plateau corresponds then to the undriven case.}
\label{fig6:corrTa}
\end{figure} 

In presence of a driving force, at the mean-field level, {\it aging} disappears:
the system always reaches a stationary state for $t \to \io$.
This happens because the drive makes the system escape from the trapping
regions around the threshold. Moreover, many interesting phenomena that are
observed when glassy systems are subjected to driving forces, \eg shear forces,
can be reproduced by mean-field models.

\subsection{Dynamical equations for driven systems}

Eq.s~(\ref{eq6:selfcons}), which describe the relaxational dynamics of the
$p$-spin spherical model, are particular instances of a general class of
dynamical equations ({\it mode-coupling} equations) of the schematic form
\beq\label{eq6:MCT}
\begin{split}
&\dpr_t C(t,t')=-\m(t) C(t,t') + \int dt'' \Si(t,t'') C(t'',t') + 
\int dt'' D(t,t'') R(t',t'') \ , \\
&\dpr_t R(t,t')=-\m(t) R(t,t') + \int dt'' \Si(t,t'') R(t'',t') + \d(t-t') \ , \\
&\m(t)= T + \int dt' \big[ D(t,t') R(t,t') + \Si(t,t') C(t,t') \big] \ , \\
\end{split}\eeq
which correspond to an {\it effective} Langevin equation of the form
\beq\label{eq6:MCTeffective}
\begin{split}
&\dot \s(t) = -\m(t) \s(t) +  \int dt' \Si(t,t') \s(t') + \r(t) \ , \\
&\la \r(t) \r(t') \ra = 2T\d(t-t') + D(t,t') \ .
\end{split}\eeq
These equations can be obtained from mean field disordered models as described above
\cite{CC05,BCKM98,Cu02}
or applying resummation schemes to non-mean field ones~\cite{Cu02}. 
They are often used to describe
the dynamics of supercooled liquids and glasses, as they can be derived also
from the memory function formalism with suitable 
approximations~\cite{GS91,Go99,Cu99}. 
In general, the kernels $D(t,t')$ and $\Si(t,t')$ are functionals of the 
correlation and response functions. Within the {\it mode-coupling} scheme,
they become ordinary functions of $C(t,t')$ and $R(t,t')$.
For the $p$-spin spherical model one has $D(C) = \frac{p}{2} C^{p-1}$ and
$\Si(R,C) = \frac{1}{2} p (p-1) R C^{p-2} = R D'(C)$.
In describing realistic systems in finite 
dimension, a wave vector dependence must be introduced in Eq.s~(\ref{eq6:MCT})
and the kernels $D$ and $\Si$ will couple different wave 
vectors\footnote{This is where the name {\it mode-coupling} equations come from.}.

It can be proven that if the forces in the original Langevin equations
describing the interacting system are {\it conservative}, 
so that detailed balance is verified, the relation $\Si(R,C)=R D'(C)$ 
must hold~\cite{BCKM96}. If $\Si(R,C) \neq R D'(C)$, Eq.s~(\ref{eq6:MCT})
describe a driven system in which detailed balance is violated~\cite{CKLP97,BBK00}.

\subsection{A driven $p$-spin model}

\begin{figure}[t]
\centering
\includegraphics[width=.55\textwidth]{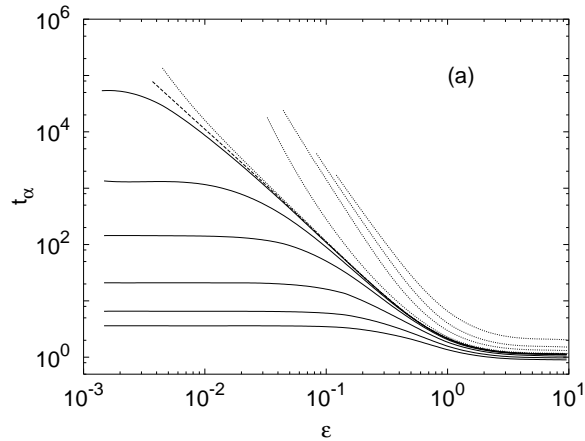}
\caption[Relaxation time for the driven $p$-spin model]
{(From \cite{BBK00})
$\alpha$-relaxation time as a function of drive
for temperatures (from bottom to top) 
 $T=0.9$, 0.8, 0.7,
0.64, 0.62, 0.613, $T_d \simeq 0.61237$, 0.6115, 0.58, 
0.45, 0.3, 0.01.
Full lines are for temperatures above $T_d$, the dashed line is $T=T_d$,
and the dotted lines are for $T<T_d$.}
\label{fig6:relaxdriven}
\end{figure} 

A particular instance of Eq.s~(\ref{eq6:MCT}) with $\Si(R,C) \neq R D'(C)$ was
studied in \cite{CKLP97,BBK00}. It correspond to a driven $p$-spin model,
whose dynamics is defined by Eq.~(\ref{pspinlangevin}), where now $h_j(t)$
represents an {\it external driving force}, which cannot be written as
the derivative of a potential, and is given by
\beq
h_j(t) = \frac{\epsilon}{(k-1)!} 
\sum_{j_1,\cdots,j_{k-1}} \widetilde J_{j,j_1, \cdots, j_{k-1}} \s_{j_1} \cdots \s_{j_{k-1}} \ ,
\eeq
and $\widetilde J$ are independent random Gaussian couplings,
which are also independent from the ones of the
Hamiltonian, and have variance $k!J^2/(2N^{k-1})$. 
They are symmetric in the exchange of two indices $j_l$, but are
{\it not} symmetric in the exchange $j \leftrightarrow j_l$.
These equations
correspond to Eq.s~(\ref{eq6:MCT}) with
\beq \label{eq6:kerneldriven}
\begin{split}
&D = \frac{p}{2} C^{p-1} + \epsilon^2 \frac{k}{2} C^{k-1} \ ,\\
&\Si = \frac{1}{2}p(p-1)R C^{p-2} \ ,
\end{split} \eeq
so the detailed balance condition $\Si(R,C) = R D'(C)$ is violated
by a term proportional to $\epsilon^2$. In \cite{CKLP97,BBK00} it was shown
that these equations admit a stationary solution $C(t-t')$, $R(t-t')$ 
{\it for all temperatures}.
Close to $T_d$ (which is the dynamical transition temperature for $\epsilon=0$)
and for small driving force
the correlation and response function can be decomposed, as in 
Eq.~(\ref{eq6:corraging}), in two parts
\beq\label{eq6:corrstatsep}
\begin{split}
C(t-t') = C_f(t-t')+ C_s(t-t') \ , \\
R(t-t') = R_f(t-t')+ R_s(t-t') \ ,
\end{split} \eeq
corresponding to a ``fast'' and a ``slow'' relaxation which are well
separated around $T_d$ and for $\epsilon \sim 0$, see Fig.~\ref{fig6:corrTa}.
The fast relaxation depends weakly on the driving force and on the 
temperature. The slow relaxation time, in absence of drive, would diverge
for $T\to T_d^+$; in presence of drive, however, it remains finite also
for $T < T_d$. For $T \gtrsim T_d$, a strong dependence on $\epsilon$ 
of the relaxation time $\t_\a$ is observed. Below $T_d$, one has
$\lim_{\epsilon \to 0} \t_\a(\epsilon,T) = \io$, so the relaxation time
diverges (again as a power law) if the driving force is sent to zero.
The relaxation time $\t_\a(\epsilon,T)$ is reported in 
Fig.~\ref{fig6:relaxdriven}.

\begin{figure}[t]
\centering
\vskip-2cm
\includegraphics[width=15cm]{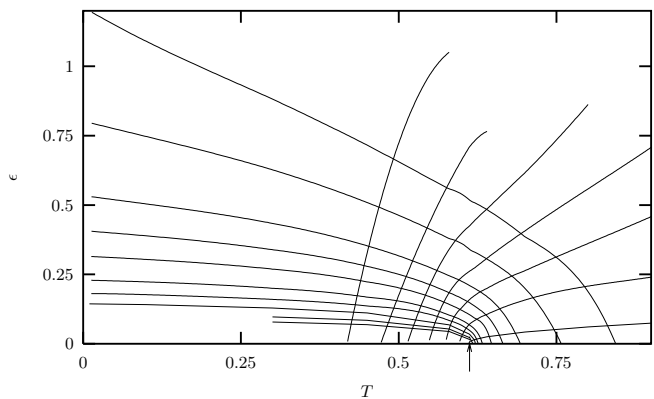}
\vskip-14cm
\caption[``Phase diagram'' of the driven $p$-spin model]
{(From \cite{BBK00})
2D view of the glass transition. Curves bent to the left are
the iso-$t_\alpha$, curves bent to the right are the iso-$X$ (see text).
The critical temperature is indicated by the arrow. Times are
$t_\alpha =5$, 10, 25, 50, ..., 5000 (from top to bottom), and
$X=0.4$, 0.5, 0.6, 0.7, 0.8, 0.9, 0.99 (from left to right).
}
\label{fig6:phasediagram}
\end{figure} 

\subsection{Temperature-drive ``phase diagram''}

If one plots the iso-$\t_\a$ curves in the plane $(\epsilon,T)$,
see Fig.~\ref{fig6:phasediagram}, a temperature-drive ``phase diagram''
can be obtained~\cite{BBK00,LN98}.
At zero drive, the system is in equilibrium above $T_d$ (marked by an
arrow in Fig.~\ref{fig6:phasediagram}) while below $T_d$ it is out
of equilibrium and {\it ages}. The aging dynamics, as discussed above,
happens slightly above the threshold level $f_{max}(T)$ in the free
energy landscape. The system is not able to penetrate below the threshold
because it is trapped by infinite-lived metastable states.

In presence of drive, the system becomes stationary for all temperatures
and driving forces, so it reaches a {\it nonequilibrium stationary state}.
This is because, for $T<T_d$, the system lives
above the threshold {\it also in absence of drive}. Thus, an arbitrary
small drive, that continuously injects a small amount of energy into
the system, is enough to give to the system the freedom to explore the
free energy landscape without being trapped by the metastable states.
This behavior is due to the mean-field nature of the model, that gives
also a dynamical transition $T_d > T_K$ and power-law divergences
of $\t_\a$ on approaching the line $(\epsilon=0, T<T_d)$.
The expected behavior for finite dimensional systems will be discussed
later.

\section{The effective temperature} 
\label{sect:intro-fdt-teff} 
 
To summarize, the analytic solution to the relaxation of mean-field 
glassy models following a quench into their glassy phase (\ie below $T_d$)
demonstrates that their 
relaxation occurs out of equilibrium~\cite{Cu02,CK93,CK94}. 
The reason why these models  
do not reach equilibrium when relaxing from a random initial  
condition is that their equilibration time diverges with  
their size. Thus, when the thermodynamic limit  
is taken at the outset of the calculation, all times considered  
are finite with respect to the equilibration time. These 
systems approach a slow nonequilibrium regime in which one observes 
a breakdown of stationarity, see Eq.s~(\ref{eq6:corraging}).
A small enough driving force is enough to restore stationarity
at all temperatures.

In the aging regime below $T_d$ as well as in presence of the 
driving force a violation of the  
fluctuation-dissipation theorem, that relates spontaneous and  
induced fluctuations in thermal equilibrium, is observed.
The effective equation (\ref{eq6:MCTeffective}) that describes
the dynamics of a single spin in the mean field of the others
is indeed a Langevin equation that describes for instance the
Brownian motion of a particle in an environment defined by
the functions $D$ and $\Si$. If these functions are stationary
and verify the detailed balance condition $\Si(R,C)=R D'(C)$,
and if $R$ and $C$ verify Eq.~(\ref{eq6:FirstFDT}),
they describe an {\it equilibrium} environment, \ie a thermal
bath at a well defined temperature $T$.

If $R$ and $C$ do not verify Eq.~(\ref{eq6:FirstFDT}), and/or
if $\Si(R,C)\neq R D'(C)$, the environment {\it is not at equilibrium},
which means that the spin $\s$ exchanges heat with a reservoir
that does not have a well defined temperature. In the following,
the precise condition for a thermal bath to be in equilibrium,
as well as the definition of an {\it effective temperature} for
a nonequilibrium bath will be discussed, and the results 
applied to Eq.~(\ref{eq6:MCTeffective}).

\subsection{The fluctuation--dissipation theorem}
\label{sec6:FDTgeneralities}

The fluctuation-dissipation theorem states that for systems evolving 
in thermal equilibrium with their equilibrated environment the linear 
response is related to the correlation function of the same observable,
see \eg \cite{CR03} for a recent review.

The linear response $\chi$ of a generic observable $O$ 
measured at time $t$ to a constant infinitesimal perturbation $h$
applied since a previous `waiting-time'\footnote{Experimentally
$t'$ is usually 
the time elapsed from the preparation of the sample, \ie
at time $t_0=0$ the systems is prepared in some way, and then it
is let evolve in contact with the bath. The preparation of the sample
is modeled by the random extraction of the initial data in 
Eq.~(\ref{eq6:MCTeffective}).}
$t'$, and the     
correlation between the same (unperturbed) observable 
measured at $t$ and $t'$, are defined as
\beq\label{eq:C-def}
\begin{split} 
R(t,t') &\equiv \left. \frac{\delta \langle O(t) \rangle}{\delta h(t')} \right|_{h=0}
\ , \\
\chi(t,t') &\equiv  
\int_{t'}^t dt'' \; R(t,t'') = 
\int_{t'}^t dt'' \;  
\left. \frac{\delta \langle O(t) \rangle}{\delta h(t'')} \right|_{h=0} 
\; , \\ 
C(t,t') &\equiv \langle O(t) O(t')\rangle 
\; , 
\end{split}\eeq 
where, for simplicity, it was assumed that the observable $O$ has a vanishing 
average, $\langle O(t)\rangle =0$ for all $t$. 
In the cases that will be discussed in the following the relation between  
these two quantities takes the form 
\begin{equation} 
\lim_{t>t' \gg t_0} \chi(t,t') = \chi[C(t,t')] \ ,
\end{equation} 
in the long waiting-time limit after the initial time $t_0$.
This equation means that the waiting-time and total  
time dependence in $\chi$ enters only  
through the value of the associated correlation between these times. 
This is trivially true in stationary states since $C(t-t')$ and
$\chi(t-t')$ depend only on the time difference $\t=t-t'$, 
so one can invert\footnote{At least in the relaxational regime
(large $t$) the function $C(t)$ is usually a decreasing function
of $t$. Oscillations might be present in $C$ but only at short times.}
the function $C(t)$ and write 
$\chi(C)=\chi(t(C))$.
If the system is in equilibrium, the 
{\it fluctuation-dissipation theorem} 
states 
\begin{equation} 
\label{fCFDT}
\chi(C) = \frac{1}{T} \; [C(0)-C]
\; , 
\end{equation} 
for all times $t\geq t' \geq t_{eq}$,  
where $t_{eq}$ is the ``equilibration time'',
$T$ is the temperature of the thermal bath 
(and the one of the system as well) and  
the Boltzmann constant has been set to one, $k_B=1$. 

An equivalent form of the fluctuation--dissipation theorem is
\begin{equation} \label{eq6:fdt-timedomain}
R(t) = - \frac{\theta(t)}{ T} \frac{dC(t)}{dt} \; , 
\;\;\;\;\;\;\;\;\;\;\;\;\;\; 
\frac{dC(t)}{dt} =  T [ R(-t) - R(t) ] 
\; .
\end{equation} 
The second expression follows from the fact that
$C(t)$ is an even function 
of $t$ from its definition (and $\dot C(t)$ is odd) 
defining also $\theta(0)\equiv 1/2$ (the 
same convention is used in the following).  After Fourier
transforming the second expression becomes 
\begin{eqnarray} 
\omega C(\omega) = 2  T \im R(\o) \ , 
\label{eq:fdt-imaginary} 
\end{eqnarray} 
and the real part of $R(\o)$ is related to $\im R(\o)$ by the 
Kramers-Kr\"onig relation. 

\subsection{A (driven) Brownian particle in a generic environment} 
\label{sec:I} 

To illustrate the basic concepts, that will be useful also in
the following chapter, a simple Langevin equation of the
form (\ref{eq6:MCTeffective}) will be discussed\footnote{
In the following the attention will be focused 
on the {\it stationary} case. Some of the results that will be
described hold also for the nonstationary case if the formulae
are suitably adapted.}~\cite{CK00}.
It describes the random motion of a particle in a  
confining potential in dimension $d$, in contact with a thermal environment, 
and under the effect of a driving external force, and reads 
\begin{equation} 
m \ddot r_\a(t) + \Iint dt' \; g_{\a\b}(t-t') \dot r_\b(t') =  
-\frac{\delta V({\vec r})}{\delta r_\a(t)} +  
\rho_\a(t) 
+ h_\a(t)
\; , 
\;\;\;\;\;\;\;\;\; 
\a=1,\dots, d 
\; . 
\label{eq:linear-t-harm} 
\end{equation} 
${\vec r}=(r_1,\dots,r_d)$ is the position of the particle.  
$m$ is the mass of the particle and $V({\vec r})$ is a 
potential energy. All analytical calculations are  
done in the simple harmonic case, $V({\vec r})=\frac{k}{2}  
\sum_\a r_\a^2$ with $k$ the spring constant of the  
quadratic potential. ${\vec \rho}(t)$ is a Gaussian thermal noise 
with zero average and generic stationary correlation 
\begin{equation} 
\langle \, \rho_\a(t) \rho_\b(t') \, \rangle = 
\delta_{\a\b} \; \nu(t-t') 
\;\;\;\;\;\;\;\;\; 
\a,\b=1,\dots, d 
\; , 
\end{equation} 
with $\nu(t-t')$ a symmetric function of $t-t'$. 
The memory kernel $g_{\a\b}(t-t')$ extends the notion of  
friction to a more generic case. A simple spatial structure,  
$g_{\a\b}(t-t') = \delta_{\a\b} \; g(t-t')$ will be assumed. 
In order to ensure causality $g(t-t')$ is proportional  
to $\theta(t-t')$. The initial time $t_0$ has been taken to $-\infty$. 
${\vec h}(t)$ is a time-dependent field that will be either used to compute 
the linear response or represents the external forcing. 
 
Eq.~(\ref{eq:linear-t-harm}) is analytically solvable in the simple
case in which there are no applied forces and the potential is 
quadratic.
In the following it will be useful to use Fourier  
transforms to solve the linear Langevin equation, with
the conventions 
\begin{equation} 
\rho(t) = \Iint 
\frac{d\omega}{2\pi} \; e^{-i\omega t} \, \rho(\omega) 
\; ,  
\;\;\;\;\;\;\;\;\;\;\;\;\;\; 
\rho(\omega) = \Iint dt \; e^{i\omega t} \, \rho(t) 
\; . 
\end{equation} 
 
In the harmonic Brownian particle problem with no other  
applied external forces the dynamics of 
different spatial components are not coupled. Thus, without loss of  
generality, one can focus on $d=1$.  
In Fourier space, the Langevin equation reads 
\begin{equation} 
-m \omega^2 x(\omega)  -i\omega g(\omega) x(\omega) =  
-k x(\omega) + \rho(\omega) 
\label{eq:linear-omega-harm} 
\end{equation} 
with the noise-noise correlation 
\begin{equation} 
\langle \, \rho(\omega) \rho(\omega') \, \rangle = 
2\pi \delta(\omega+\omega') \nu(\omega) 
\; . 
\end{equation} 
The linear equation (\ref{eq:linear-omega-harm}) is solved by 
\begin{equation} \label{eq17}
x(\omega) = G(\omega) \rho(\omega) \ ,
\;\;\;\;\;\;\;\;\;\;\;\;\; 
G(\omega) \equiv \frac{1}{-m\omega^2 -i \omega g(\omega) + k}  
\;,  
\end{equation} 
and one finds the correlations 
\beq
\begin{split}
& \langle \, x(\omega) x(\omega') \, \rangle = 
G(\omega) G(-\omega) 2\pi \delta(\omega+\omega') \nu(\omega) \ ,
\\ 
& \langle \, x(\omega) \rho(\omega') \, \rangle = 
G(\omega)  
2\pi \delta(\omega+\omega') \nu(\omega)  
\; . 
\label{eq:resp-omega-harm} 
\end{split}
\eeq 
Note that $G(\omega) G(-\omega)=|G(\omega)|^2$; then  
\begin{eqnarray} 
&& \langle \, x(\omega) x(\omega') \, \rangle = 
C(\omega) 2\pi \delta(\omega+\omega') 
\;\;\;\;\;\;\;\;\;\; 
\mbox{with} 
\;\;\;\;\;\;\;\;\;\; 
C(\omega) \equiv |G(\omega)|^2 \nu(\omega)  
\label{eq:corr-omega-harm2} 
\; . 
\end{eqnarray} 
In a problem solved by  
\begin{equation} 
x(t) = \Iint dt' \; G(t-t') [\rho(t')+h(t')] + \mbox{IC} 
\; , 
\end{equation} 
where IC are terms related to the initial conditions, 
the time-dependent linear response is  
\begin{equation} 
R(t-t') \equiv   
\left. \frac{\delta \langle x(t)\rangle}{\delta h(t')}\right|_{h=0} 
= G(t-t')  
\; , 
\end{equation} 
and 
\beq 
R(\o) = \Iint dt \ e^{i\o t} R(t) = G(\o) \ . 
\eeq 
Note that the response function is related to the correlation  
$\langle \, x(t) \r(t') \, \rangle$ by Eq.~(\ref{eq:resp-omega-harm}): 
\begin{equation} 
2\pi \delta(\omega+\omega') R(\omega) \nu(\omega) =  
\langle \, x(\omega) \rho(\omega') \, \rangle  
\; . 
\end{equation} 
 
\subsection{Effective temperature for a generic environment}
\label{sec6:Teff}

Now, one can check under which conditions on the characteristics of the bath 
[$g(t-t')$ and $\nu(t-t')$] the fluctuation-dissipation theorem
(for the particle) holds and, when it does not hold,  
which is the generic form that the relation between the linear response and  
correlation might take in this simple quadratic model. 
Eq.~(\ref{eq17}) 
implies\footnote{This discussion has to be modified
in the $k=0$ limit in which the particle 
does not have a confining potential and diffuses.}:
\begin{equation} 
\mbox{Im} R(\omega) =  \mbox{Im} G(\omega) =  
\omega \; \mbox{Re}g(\omega) \; |G(\omega)|^2 \ ,
\label{eq:fdt-generic-bath0} 
\end{equation} 
and then using equation~(\ref{eq:corr-omega-harm2})
\begin{equation} 
\frac{\omega C(\omega)}{2 \mbox{Im} R(\omega)}= 
\frac{\nu(\omega)}{2\mbox{Re}g(\omega)} 
\label{eq:fdt-generic-bath} 
\; . 
\end{equation} 
The fluctuation-dissipation theorem 
holds only if this ratio is equal to $T$, 
see Eq.~(\ref{eq:fdt-imaginary}). In general, one can define a frequency 
dependent {\it effective temperature} $T_{eff}(\omega)$ as 
\begin{equation} 
 T_{eff}(\omega) \equiv \frac{\nu(\omega)}{2 \mbox{Re}g(\omega)} 
\; . 
\label{eq:Teff-def} 
\end{equation} 
This function is a property of the bath but it can also be expressed 
(in this linear problem)
in terms of measurable quantities, \ie the correlation and response  
of the position of the particle. 
The use of the name {\it effective temperature} has been  
justified within a number of models with slow dynamics and a  
separation of time-scales\footnote{\label{proviso}The definition 
of effective temperature used
here does not have the good thermodynamic properties for all possible
non-equilibrium systems. Even if it is still not completely 
established which are the precise requirements needed to ensure the
thermodynamic nature of this temperature, it seems to be clear that
the underlying system must have a bounded energy density 
and that the relaxing dynamics should be slow 
(a limit of small entropy production, as described in \cite{CK00}).
Some cases where these requirements fail have been discussed
in~\cite{noTeff}.}~\cite{BCKM98,Cu02,CR03,CKP97,CK00}.
 
\subsubsection{Equilibrated environments: the fluctuation--dissipation theorem
and $T_{eff}=T$ } 
 
For any environment such that the right-hand-side in  
Eq.~(\ref{eq:fdt-generic-bath})  
equals $ T$  the fluctuation-dissipation theorem holds. 
In the time domain, this condition reads 
\beq 
\label{FDTbathTD} 
 T g(t) = \theta(t) \nu(t) \; , 
\;\;\;\;\;\;\;\;\;\; 
 \nu(t) =  T [g(t) + g(-t)] = T g(|t|) \ . 
\eeq 
In particular, this is satisfied by a white noise for which 
$\nu(t) = 2 T \gamma \delta(t)$ and  
$g(t) = 2 \gamma \delta(t) \theta(t)$ (remember that $\theta(0) \equiv 1/2$).
This is the noise that appears in Eq.~(\ref{pspinlangevin}). 
The fluctuation-dissipation theorem also holds for any colored noise -- with  
a retarded memory kernel $g$ and noise-noise correlation $\nu$ --  
such that the ratio between Re$g(\omega)$ and $\nu(\omega)$ equals  
$(2 T)^{-1}$.

\subsubsection{Nonequilibrium environments} 
\label{subsec:noneq-baths} 
 
Instead, for any other generic environment, 
the left-hand-side in  
Eq.~(\ref{eq:fdt-generic-bath}) yields a non-trivial and,  
in general model-dependent, result for the effective temperature. 
 
A special case is that of   
an ensemble of $N$ equilibrated baths with different  
characteristic times and at different temperatures. In  
this case, the noise $\vec\r$ in Eq.~(\ref{eq:linear-t-harm}) is 
the sum of $N$ independent noises, 
\beq 
\vec\r = \sum_{i=1}^N \vec\r_i \ , 
\hspace{20pt} 
\langle  \, 
\r_{i\a}(t) \r_{j\b}(t') \, \rangle = \d_{\a\b} \d_{ij}  T_i \nu_i(t-t') \ , 
\eeq 
and the friction kernel is given by 
\beq 
\label{eq:frict-sumN} 
g(t-t')=\sum_{i=1}^N g_i(t-t')  
\ .   
\eeq  
The temperature $T_i$ has been extracted from the definition of 
$\nu_i(t)$ in order to 
simplify several expressions as will be clear in the following. As 
the $\vec\r_i$ are independent Gaussian variables, $\vec\r = 
\sum_i \vec\r_i$ is still a Gaussian variable with zero mean and 
correlation \beq 
\label{eq:corr-sumN} 
\langle \r_\a(t) \r_\b(t') \rangle = \d_{\a\b} \sum_i T_i \nu_i(t-t')
\ .  \eeq Thus, in the Gaussian case {\it the $N$ equilibrated baths
are equivalent to a single nonequilibrium bath} with correlation given
by Eq.~(\ref{eq:corr-sumN}) and friction kernel given by
Eq.~(\ref{eq:frict-sumN}).  In frequency space
\begin{eqnarray} 
g(\omega) =\sum_{i=1}^N g_i(\omega)  
\; ,  
\;\;\;\;\;\;\;\;\;\;\;\;\;\;\;\;\;\; 
\nu(\omega) =\sum_{i=1}^N  T_i \nu_i(\omega) 
\; ,   
\end{eqnarray} 
with  
\begin{eqnarray} 
\label{FDTbathFD} 
\nu_i(\omega)= 2\mbox{Re} g_i(\omega) 
\; , 
\end{eqnarray} 
as each bath is equilibrated at temperature $T_i$. 
The effective temperature is then given by 
\beq 
\label{TeffNbaths} 
T_{eff}(\o) = \frac{\sum_{i=1}^N T_i \nu_i(\o)}{\sum_{i=1}^N
\nu_i(\o)} \ .  
\eeq 
Note that if the functions $\nu_i(\o)$ are chosen
to be peaked around a frequency $\o_i$, choosing suitable values for
$\o_i$ and $T_i$ one can approximate a single nonequilibrium bath with
$N$ baths equilibrated at different temperatures.
 
\subsubsection{Effective temperature in the time domain}

Alternatively one can define the effective temperature in the time domain
from the generalization of Eq.~(\ref{fCFDT}):
\beq\label{eq:Tefftime-def}
\frac{1}{T_{eff}(C)} = -\frac{d\chi}{dC} \ ,
\eeq
or equivalently, recalling that $R(t)=\th(t)\frac{d\chi}{dt}$, 
from Eq.~(\ref{eq6:fdt-timedomain}),
\beq
R(t)=-\frac{\th(t)}{T_{eff}(C(t))} \frac{dC}{dt} \ .
\eeq
It is important to remark that the effective temperature in the frequency
domain, Eq.~(\ref{eq:Teff-def}), is {\it not} equal in 
general to the Fourier transform of the effective temperature 
$T_{eff}(t)\equiv T_{eff}(C(t))$
which is the ratio between the correlation and 
response functions in the time domain.

In the following it will be useful to define the function
\beq 
\label{TeffNbaths2}
T^{-1}(t) \equiv \int_{-\io}^\io \frac{d\o}{2\p} \frac{1}{ T_{eff}(\omega)}  
e^{-i\o t}= \int_{-\io}^\io \frac{d\o}{2\p}\frac{2 \re g(\o)}{\n(\o)} 
e^{-i \o t} \; , 
\eeq 
i.e. the Fourier transform of $1/T_{eff}(\o)$. This function should not
be confused with the inverse of the effective temperature $T_{eff}(t)$.
Indeed, if the bath is in equilibrium at temperature $T$, one has 
$T_{eff}(t) \equiv T $ and $T_{eff}(\o) \equiv T$ while 
$T^{-1}(t) = \delta(t)/T$.

\subsection{Mean field glassy systems}

\begin{figure}[t]
\centering
\vskip-2cm
\includegraphics[width=15cm]{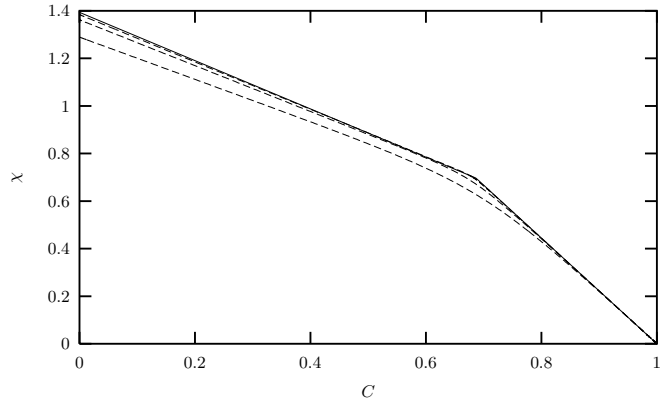}
\vskip-14cm
\caption[Effective temperature for the driven $p$-spin model]
{(From \cite{BBK00})
Integrated response vs. correlation curves for $T=0.45<T_d$.
Full line: asymptotic ($\epsilon = 0$) analytical curve.
Dashed lines (from bottom to top) $\epsilon=0.333$, 0.143, 0.0442.
The breaking point corresponds to the {\it plateau} of $C(t)$.
}
\label{fig6:FDTplot}
\end{figure} 

The relation between spontaneous and induced fluctuations found in 
mean-field glassy models, or, equivalently, the relation between
$R$ and $C$ in the solutions of Eq.~(\ref{eq6:MCT}), 
is surprisingly simple.
Indeed, if the ``fast'' and ``slow'' time scales are well separated
(\ie for $T \leq T_d$ and $\epsilon \sim 0$, where $\t_\a$ is large,
see Fig.~\ref{fig6:relaxdriven}),
so that the decomposition (\ref{eq6:corrstatsep}) holds,
the plot $\chi(C)$ is found to be a broken line:
\begin{equation} \label{twoT}
\chi(C) = \frac{1}{T} \; (1-C) \; \theta(C-q_{d}) +  
\left[\frac{1}{T} \; (1-q_{d}) + \frac{1}{T_{eff}} (q_{d}-C) \right] 
\; \theta(q_{d}-C) 
\; ,
\end{equation} 
see Fig.~\ref{fig6:FDTplot} and \cite{CR03} for a review. 
This broken line has two slopes, $-1/T$ for $C> q_{d}$ 
({\it i.e.}, small $t-t'$), 
and $-1/T_{eff}$ for $C< q_{d}$ ({\it i.e.}, large $t-t'$).
Indeed, the first slope gives the relation between $C_f$ and $R_f$,
while the second gives the relation between $C_s$ and $R_s$:
\beq\begin{split}
&R_f(t)=-\frac{\th(t)}{T} \frac{dC_f}{dt} \ , \\
&R_s(t)=-\frac{\th(t)}{T_{eff}} \frac{dC_s}{dt} \ .
\end{split}\eeq
The breaking-point $q_{d}$, as discussed above, has an interpretation 
as the self-overlap of the states, \ie for $t-t'$ small the system remains
in the same state while for $t-t'$ large it jumps to another 
state\footnote{A similar interpretation is given in term of intra-cage and
cage-rearrangement motions in the relaxation of structural glasses, 
inter-domain and domain wall motion in coarsening systems, etc.
More general forms, with a sequence of  
segments with different slopes appear in mean-field  
glassy models of the Sherrington-Kirkpatrick type, where the structure
of the metastable states is more complicated.}.
Since $T_{eff}$ is found to be 
larger than $T$ the second term violates the fluctuation-dissipation 
theorem.
In order to be consistent with the thermodynamic properties one 
needs to find a single value of $T_{eff}$ in each time
regime as defined by the correlation scales of \cite{CK94}.

In presence of a finite drive $\epsilon > 0$, $\chi(C)$
is a smooth function of $C$. For $\epsilon \to 0$ and $T > T_d$,
the system is in equilibrium and $\chi(C)$ tends to $T^{-1} (1-C)$,
\ie the fluctuation--dissipation relation holds.
For $\epsilon \to 0$ and $T<T_d$, $\t_\a \to \io$
as discussed above, thus the decomposition (\ref{eq6:corrstatsep})
becomes exact and $\chi(C)$ becomes the broken line (\ref{twoT}),
see Fig.~\ref{fig6:FDTplot}. Defining a FDT violation factor
$X \equiv 1/T_{eff}$, one can draw iso-$X$ lines in the $(\epsilon,T)$
plane, see Fig.~\ref{fig6:phasediagram}, and identify a {\it quasi-equilibrium}
region defined by $X\sim 1$ (\eg $X>0.99$). In the latter region the
fluctuation--dissipation relation (FDR) holds: it roughly correspond to the linear
response region.

The same result for $\chi(C)$, Eq.~(\ref{twoT}), is obtained in the
{\it aging} dynamics, \ie for the nonstationary solution of 
Eq.s~(\ref{eq6:selfcons}): for $t,t' \to \io$ the separation
(\ref{eq6:corraging}) holds and $\chi(C)$ is given by Eq.~(\ref{twoT})
as in the driven case.

In the case of relaxing glasses the dynamics occurs out of equilibrium
because below $T_d$ the equilibration time diverges with the size of the
system and falls beyond all accessible time-scales.
These macroscopic systems then evolve out of
equilibrium even if they are in contact with a thermal reservoir,
itself in equilibrium at a given temperature $T$, the white bath in
Eq.~(\ref{pspinlangevin}), due to the interaction between the $N$($\to\io$) 
spins. The effective environment appearing in Eq.~(\ref{eq6:effective})
is self-generated by the system.

In the case of sheared dense liquids, glasses, etc. the systems are driven
into an out of equilibrium stationary regime by the external forces,
so one does not expect the FDR to hold.
However, for $T<T_d$, as the dynamics is very slow and non--stationary at zero 
forcing, it remains slow also for weak forcing, the main effect of a weak 
forcing being that the system becomes stationary. In this situation the
FDR is violated also for small drive, $\epsilon \to 0$, which is a quite
unexpected result.

The suggestive name {\it effective temperature}, 
$T_{eff}$, has been used to parametrize the second slope.  
The justification is that for 
mean-field glassy models --- 
and within all resummation schemes applied to 
realistic ones as well --- $T_{eff}$ 
does indeed behave as a temperature, 
in the sense that it controls heat flows between systems
which are in thermal contact~\cite{Cu02,CR03,CKP97}.

\section{Beyond mean-field}

The results discussed above follow from Eq.s~(\ref{eq6:MCT}), which
describe {\it exactly} the dynamics of mean field systems and 
approximate realistic system as well. For realistic systems 
they can be derived applying suitable resummation and/or approximation 
schemes.

Indeed, the results described above were obtained in numerical
simulations of the slow relaxational dynamics of a number of more 
realistic glassy systems such as Lennard-Jones 
mixtures~\cite{Lennard-FDT},
sheared Lennard-Jones mixtures~\cite{BB00,BB02} and
in a number of other driven low-dimensional models~\cite{lowd-driven}.

\begin{figure}[t]
\centering
\includegraphics[width=.55\textwidth]{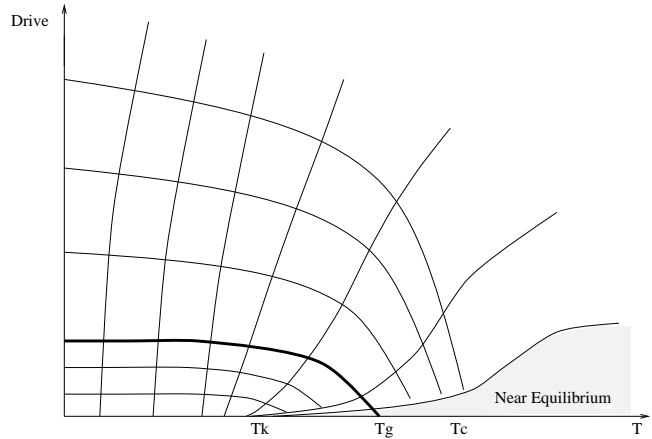}
\caption[Schematic ``phase diagram'' of realistic glassy models]
{(From \cite{BBK00})
Same diagram as in Fig.~\ref{fig6:phasediagram}, taking into account
activated processes. In absence of drive the activated processes restore
the equality of $T_K$ and $T_d$. However below $T_g$ the system cannot
be equilibrated on the experimental time scale. The curve iso-$\t_\a$ 
corresponding to $\t_\a(T_g)=100$s is drawn as a thick line. Inside the
region delimited by this line $\t_\a(\ee,T) > 100$s so the system is not
able to reach stationarity and ages. The region defined by $X > 0.99$ is
again a ``near-equilibrium'' region that roughly correspond to the linear
response region.
}
\label{fig6:realphasediagram}
\end{figure}

However, if one whishes to describe the relaxation of real glasses
in finite dimension, more complicated effects have to be taken into account.
The problem is that, as discussed in section~\ref{sec1:beyondmeanfield},
in finite dimensional models metastable states have a finite time and decay
by {\it activated processes} of barrier crossing. These process are of
{\it nonperturbative} nature so they are missed by any resummation scheme
leading to equations of the class (\ref{eq6:MCT}). 
These processes are relevant also at equilibrium as they are responsible
for the equality of $T_d$ and $T_K$ in finite dimensional systems, while
in mean-field one finds $T_d > T_K$.
Some attempts to describe the {\it equilibrium} properties of such processes
recently appeared in the literature~\cite{Wo97,XW01,LW03,BB04,Fr05,DSW05},
and have been described in section~\ref{sec1:beyondmeanfield}
.
One of the main results coming from these studies is that the equilibrium
dynamics for $T \sim T_g$ is expected to be {\it heterogeneous}: this means
that the relaxation time $\t_\a$ is expected to depend on the region inside
the sample. This is consistent with experimental and numerical results, see
\eg \cite{Ed00} for a review.

Thus one can expect that heterogeneity is relevant also for the 
{\it nonequilibrium} dynamics yielding a {\it local}
effective temperature which may depend on space inside the sample.
This has indeed very recently proposed and found numerically in finite
dimensional spin-glass models and a theory to describe space fluctuations
of $T_{eff}$ has been proposed, see e.g.~\cite{CCCIK03,Cu04}
for a detailed discussion.

On the other hand, it seems that in presence of a driving force heterogeneity
is somehow reduced. For example, the strecthing parameter of the correlation
functions of density fluctuation (which is a measure of the homogeneity of the
sample) is found to be an increasing function of the
driving force in sheared colloidal suspensions, see \eg the discussion
in \cite{DIR05}.

Another important difference follow from the fact that, in finite dimension,
due to activated processes, in absence of drive, 
the system is able to penetrate below the threshold, and is trapped for large
times into metastable states from which it can escape only by jumping over
some barriers. Then, one expects that an infinitesimal drive is not enough to
take the system out of these states. This means that if the drive is small
enough the aging dynamics is recovered, as the system can be trapped by
metastable states as if the drive were not present, 
see~Fig.~\ref{fig6:realphasediagram} and
\cite{BBK00,BB02} for a detailed discussion.

In next chapter, only Langevin equations of the form (\ref{eq6:MCT}) will be
considered.
The extension of the results that will be presented to glassy systems in
finite dimension will require additional work.

%% file: cap7.tex
%\chapter[Extension of the FR to driven glasses]
\chapter{Extension of the fluctuation relation to driven glasses}
\label{chap:sette}

%%%%%%%%%%%%%%%  TEXT  %%%%%%%%%%%%%%%%

\section{Introduction}
\label{sect:intro}

Relatively few generic results for non-equilibrium systems exist.  
Two such results that apply to seemingly very different physical situations
have been discussed in the last chapters.  One is the {\it fluctuation
theorem} that characterizes the fluctuations of the entropy production over
long time-intervals in driven steady states, see chapter~\ref{chap4}.
Another one is the extension
of the fluctuation--dissipation theorem that relates induced and spontaneous
fluctuations in equilibrium to the non-equilibrium slow relaxation of glassy
systems, see chapter~\ref{chap6}.
While the former result has been proven for
reversible hyperbolic dynamical systems~\cite{GC95a,GC95b} and 
for the driven stochastic dynamic evolution of an open system coupled to an
external environment~\cite{LS99,Ku98}, the latter has only been obtained in a
number of solvable mean-field models and numerically in some more realistic
glassy systems~\cite{CK93,CK94,Lennard-FDT,BB00,BB02,lowd-driven}.
As discussed in section~\ref{sect:intro-fdt-teff}, 
the modification of the fluctuation--dissipation 
theorem can be rationalized in terms of the generation of an {\it effective
temperature}~\cite{CKP97,CK00}.  The expected thermodynamic properties of the
effective temperature have been demonstrated in a number of
cases~\cite{CKP97} 
(see however footnote \ref{proviso} in section~\ref{sect:intro-fdt-teff}).

One may naturally wonder whether these two quite generic results
may be included in a common, more generic statement, that reduces
to them in the corresponding limits. The scope of this chapter is 
to discuss this possibility using the very simple working example
discussed in section~\ref{sec:I},
from which one can very easily reach the `driven limit' and the 
`non-equilibrium relaxational' case. This project was pioneered by 
Sellitto~\cite{Se98} who asked the same question some years ago
and tried to give it an answer using a stochastic lattice gas with 
reversible kinetic constraints in diffusive contact with two particle 
reservoirs at different chemical potentials. Other developments 
in similar directions have been proposed and analyzed by 
several authors~\cite{SCM04,CR04,Sasa}. They will be discussed in 
section~\ref{sec:conclusions}.

The fluctuation theorem and the fluctuation--dissipation theorem are
related: indeed, for systems which are able to
equilibrate in the small entropy production limit ($\sigma_+
\rightarrow 0$), the fluctuation theorem implies the Green-Kubo
relations for transport coefficients, that are a particular instance
of the fluctuation--dissipation theorem, see~\cite{Ga96a,Ga96b,Ku98}
and section~\ref{sec4:closetoeq}.
That is, close to equilibrium the fluctuation relation and
the fluctuation--dissipation relation are equivalent.
It is then natural to wonder what is the fate of the fluctuation
theorem if the fluctuation--dissipation is violated even when the
driving force is very small, see Fig.~\ref{fig6:FDTplot}.
One can ask if the fluctuation theorem is modified and, more
precisely, whether the effective temperature enters its
modified version. In particular, this question will arise 
if the limit of large sampling time, $\tau$ in 
Eqs.~(\ref{pdef}) and (\ref{zetapdef}), is taken after the limit of 
large system size. 
The order of the limits is important
because a finite size system will always equilibrate with the thermal bath
in a large enough time $\t$. As the fluctuation theorem concerns the
fluctuations of $\s$ for $\t\to\io$, if one wants to observe {\it nonequilibrium}
effects, the latter limit has to be taken {\it after}
the thermodynamic limit.
 
The idea is to study the relaxational and driven dynamics of the 
simplest system such that the effective temperature is not trivially 
equal to the ambient temperature. For a system coupled to a single 
thermal bath, this happens whenever:
 
{\it (i)} the thermal bath has temperature $T$, but the system is not
able to equilibrate with the bath. This is realized by the glassy
cases discussed above, provided the sampling time is smaller than the
equilibration time; and/or
 
{\it (ii)} the system is very simple (not glassy) but it is set in contact  
with a bath that is not in equilibrium.  One can think of 
two ways of realizing this fact. One is with a single bath 
represented by a thermal noise and a memory friction kernel that do
not verify the fluctuation-dissipation relation \cite{Cu02}.  This 
situation is realized if one considers the diffusion of a Brownian 
particle in a complex medium (\eg a glass, or granular matter) 
\cite{AG04,PM04,Po04}.  In this case the medium, which acts as a 
thermal bath with respect to the Brownian particle, is itself out of 
equilibrium.  Another possibility is to couple the system to a number 
of equilibrated thermal baths with different time-scales and at 
different temperatures~\cite{CK00}. 

These two cases are closely related because, as discussed in 
chapter~\ref{chap6}, at least at the mean-field 
level, the problem of glassy dynamics can be mapped onto the problem of 
a single ``effective'' degree of freedom moving in an out of 
equilibrium environment. Situations {\it (i)} and {\it (ii)} are then 
described by the same kind of equation, 
namely, a Langevin equation for a single degree of freedom coupled to 
a non-equilibrium bath, like Eq.s~(\ref{eq6:MCTeffective}) and
(\ref{eq:linear-t-harm}).

In the following the Langevin equation Eq.~(\ref{eq:linear-t-harm}) will
be considered. Its main characteristic is that the thermal bath,
represented by the functions $g(t-t')$ and $\n(t-t')$, is not at equilibrium.
This equation reproduces many features of the original 
equation~(\ref{eq6:MCTeffective}) in the driven case, where the
functions $\Si(t-t')$ and $D(t-t')$ are stationary.

\section{Entropy production rate for a nonequilibrium bath}

The explicit form of $\s_\t$ for 
the equation of motion (\ref{eq:linear-t-harm}), in the case where
$g_{\a\b}(t)=\d_{\a\b} g(t)$
and $\vec h(t)=\vec h[\vec r(t)]$ is an external nonconservative force
that does not explicitly depend on time
(e.g., in $d=2$, $\vec h = \epsilon (-y,x)$), can be computed following
the procedure outlined in section~\ref{sec:LS}.
Note that the functions $\nu(t)$ and $g(t)$
are such that $\nu(t)=\nu(-t)$
while $g(t)$ is proportional to $\th(t)$, and both decay exponentially
in time. The probability distribution
of the noise $\vec \r(t)$ is
\beq\label{Prho}
{\cal P}[\vec \r(t)] \propto \exp \left[ -\frac{1}{2} \Iint dt dt' \, 
\r_\a(t) \n^{-1}(t-t') \r_\a(t') \right] \ ,
\eeq
where $\n^{-1}(t)$ is the operator inverse of $\n(t)$,
\beq
\n^{-1}(t) = \Iint \frac{d\o}{2\p} \frac{1}{\n(\o)} e^{-i\o t} \ .
\eeq
The probability distribution of $\vec r(t)$ is obtained substituting
$\vec \r(t)$ obtained from Eq.~(\ref{eq:linear-t-harm}) in Eq.~(\ref{Prho}).
Then one has
\beq\label{Pr}
\begin{split}
{\cal P}[\vec r(t)] \propto &\exp \left\{ - \frac{1}{2} \Iint dt dt' \, 
\left[
m \ddot r_\a(t) + \Iint dt'' g(t-t'') \dot r_\a(t'') 
+ \frac{\d V(\vec r)}{\d r_\a(t)} - h_\a[\vec r(t)]
\right] \times \right. \\ &\left. \times \n^{-1}(t-t') 
\left[
m \ddot r_\a(t') + \Iint dt''' g(t'-t''') \dot r_\a(t''') 
+ \frac{\d V(\vec r)}{\d r_\a(t')} - h_\a[\vec r(t')]
\right]
\right\} \ .
\end{split}
\eeq
With some manipulations it is easy to see that
\beq\label{PIr}
\begin{split}
{\cal P}[\vec r(-t)] \propto &\exp \left\{ -\frac{1}{2}  \Iint dt dt' \, 
\left[
m \ddot r_\a(t) - \Iint dt'' g(t''-t) \dot r_\a(t'') 
+ \frac{\d V(\vec r)}{\d r_\a(t)} - h_\a[\vec r(t)]
\right] \times \right. \\ &\left. \times \n^{-1}(t-t') 
\left[
m \ddot r_\a(t') - \Iint dt''' g(t'''-t') \dot r_\a(t''') 
+ \frac{\d V(\vec r)}{\d r_\a(t')} - h_\a[\vec r(t')]
\right]
\right\} \ .
\end{split}
\eeq
To compute $\s_\t$ one should consider the probability of a segment
of trajectory $[-\t/2,\t/2]$ and then send $\t$ to $\io$, neglecting
all boundary terms. As the functions $g(t)$ and $\nu(t)$ have short 
range, the trajectories $\vec r(t)$ decorrelate exponentially fast in
time and up to boundary contributions one can simply truncate the
integrals in ${\cal P}[\vec r(t)]$ in $t,t' \in [-\t/2,\t/2]$ to
obtain the probability of a segment of trajectory for large $\t$.
Equivalently one can consider the integrals
in $(-\io,\io)$ and neglect all the boundary terms: then one obtains the
entropy production $\s_\io$ integrated over the interval $t \in (-\io,\io)$
and one can truncate the
integral in $[-\t/2,\t/2]$ at the end of the computation.

A lot of terms in $\s_\io= -\log {\cal P}[\vec r(-t)] + 
\log {\cal P}[\vec r(t)]$ trivially cancel. 
Before discussing the non-trivial terms, define
$f(t)=g(t)+g(-t)$ and recall that, from Eq.~(\ref{TeffNbaths2})
\beq
T^{-1}(t-t'')=
\Iint \frac{d\o}{2\p} \frac{2\re g(\o)}{\n(\o)} e^{-i\o ( t-t'')}=
\Iint dt' \n^{-1}(t-t') f(t'-t'')
 \ .
\eeq
Note that both $f(t)$ and $T^{-1}(t)$ are even function of $t$, and
if the bath is in equilibrium at temperature $T$, 
$T^{-1}(t-t')= \d(t-t')/T$.

The terms that do not cancel trivially are the following:
\begin{itemize}
\item a ``kinetic'' term of the form
\beq\begin{split}
& \Iint dt dt' \, \left[
m \ddot r_\a(t) \n^{-1}(t-t') \Iint dt'' g(t''-t') \dot r_\a(t'')
+m \ddot r_\a(t) \n^{-1}(t-t') \Iint dt'' g(t'-t'') \dot r_\a(t'')\right] =\\
&= \Iint dt dt' \,
m \ddot r_\a(t) \n^{-1}(t-t') \Iint dt'' f(t'-t'') \dot r_\a(t'')=
 \Iint dt dt' \,
m \ddot r_\a(t) T^{-1}(t-t') \dot r_\a(t') \ .
\end{split}\eeq
If the bath is in equilibrium, this term trivially vanishes as it 
is the integral of the total derivative of the kinetic energy. 
It vanishes also for a nonequilibrium bath:
indeed, integrating by parts first in $t$ and then in $t'$, one has,
recalling that $T^{-1}(t)$ is even and short ranged and up to boundary terms:
\beq\label{zerokin}\begin{split}
&\Iint dt dt' \, \ddot r_\a(t) T^{-1}(t-t') \dot r_\a(t') =
-\Iint dt dt' \, \dot r_\a(t) \frac{d}{dt} T^{-1}(t-t') \dot r_\a(t') = \\
&\Iint dt dt' \, \dot r_\a(t) \frac{d}{dt'} T^{-1}(t-t') \dot r_\a(t') =
-\Iint dt dt' \, \dot r_\a(t) T^{-1}(t-t') \ddot r_\a(t') = 0 \ .
\end{split}\eeq
\item a ``friction'' term of the form
\beq\begin{split}
\frac{1}{2}\Iint dt dt' dt'' dt''' \, & \big[
 \dot r_\a(t'') g(t''-t) \n^{-1}(t-t') 
 g(t'''-t') \dot r_\a(t''') - \\
& \dot r_\a(t'') g(t-t'') \n^{-1}(t-t') 
 g(t'-t''') \dot r_\a(t''')\big] \ .
\end{split}\eeq
This term vanishes because the function
\beq
K(t''-t''')=\int dt dt' \, g(t''-t) \n^{-1}(t-t') 
 g(t'''-t') 
\eeq
is even in its argument as one can easily check.
\item a ``potential'' term of the form
\beq\begin{split}
\s^V_\io &= - \Iint dt dt' dt'' \, 
 f(t-t'') \dot r_\a(t'') 
 \n^{-1}(t-t') 
 \frac{\d V(\vec r)}{\d r_\a(t')} \\ &=
- \Iint dt dt' \, T^{-1}(t-t')
 \dot r_\a(t) 
 \frac{\d V(\vec r)}{\d r_\a(t')} \ .
\end{split}\eeq
This term is related to the work of the conservative forces. If the bath
is in equilibrium, it vanishes being the total derivative of the potential
energy. It vanishes also for an harmonic potential 
$V(\vec r)=\frac{1}{2} k r^2$ because 
$\frac{\d V(\vec r)}{\d r_\a(t)}=k r_\a(t)$ and one can use the same
trick used in Eq.~(\ref{zerokin}).
\item a ``dissipative'' term which is
\beq
\s^{eff}_\io= \Iint dt dt' \, T^{-1}(t-t')
 \dot r_\a(t) 
 h_\a[\vec r(t')] =
\Iint \frac{d\o}{2\p} \frac{-i \o r_\a(\o) h_\a(\o)}{T_{eff}(\o)}
\ .
\eeq
This term is related to the work of the dissipative forces.
If the bath is in equilibrium at temperature $T$, this is exactly the
work of the dissipative forces divided by the temperature of the 
bath. Otherwise, the work done at frequency $\o$ is weighted by the
effective temperature at the same frequency.
\end{itemize}

The expression of the total entropy production over the interval $(-\io,\io)$ is
then
\beq\begin{split}
\s_\io &= \s^V_\io+\st_\io =  \Iint dt dt' \, T^{-1}(t-t')
 \dot r_\a(t) \left[ - \frac{\d V(\vec r)}{\d r_\a(t')} + h_\a[\vec r(t')] \right] \\
& =  \Iint dt dt' \, T^{-1}(t-t')
 \dot r_\a(t) F_\a(t') \ ,
\end{split}\eeq
where $F_\a(t) = h_\a[\vec r(t)]- \frac{\d V(\vec r)}{\d r_\a(t)}$ 
is the total deterministic force acting on the particle at time $t$.
The latter expression can be rewritten as
\beq
\s_\io = \Iint dt \int_{-\io}^t dt' \, T^{-1}(t-t')
\big[ \dot r_\a(t) F_\a(t') + \dot r_\a(t') F_\a(t) \big] \ ,
\eeq
and this leads to identify the entropy production per unit time $\s_t$ with
\beq \label{EPR} \begin{split}
&\s_t = \s^V_t + \st_t = \int_{-\io}^t dt' \, T^{-1}(t-t')
\big[ \dot r_\a(t) F_\a(t') + \dot r_\a(t') F_\a(t) \big] \ , \\
&\s^V_t =- \int_{-\io}^t dt' \, T^{-1}(t-t')
\left[ \dot r_\a(t)\frac{\d V(\vec r)}{\d r_\a(t')}  + 
\dot r_\a(t') \frac{\d V(\vec r)}{\d r_\a(t)} \right] \ , \\
&\st_t = \int_{-\io}^t dt' \, T^{-1}(t-t')
\big[ \dot r_\a(t) h_\a[\vec r(t')] + \dot r_\a(t') h_\a[\vec r(t)] \big] \ .
\end{split} \eeq
If the bath is at equilibrium this expression reduces to
the work done by the nonconservative forces divided by the temperature of the
bath, as expected. Also if the bath is not at equilibrium, but the potential is harmonic,
only the contribution $\st_t$ of the nonconservative force has to be taken into account.
The reason why the work of the conservative forces produces entropy if the bath is
out of equilibrium {\it and} the interaction is nonlinear is that the nonlinear interaction
couples modes of different frequency which are at different temperature, thus producing an
energy flow between these modes; this energy flow is related to the entropy production.

It is also important to remark that boundary terms, that have been neglected in the
calculation above, can have important effects on the large fluctuations of $\s_\t$ 
even for $\t \to \io$~\cite{VzC04}, as discussed in section~\ref{sec4:singularities}.

%%%%%%%%%%%%%%%%%%%%%%%%%%%%%%%%%%%%%%%%%%%%%%%%
%%%%%%%%%%%%%%%%%%%%%%%%%%%%%%%%%%%%%%%%%%%%%%%%

\section{Large deviation function for an harmonic potential}
 
The large deviation function in the harmonic case, 
$V(\vec r) = \frac{1}{2} k r^2$, will now be computed explicitly.
In this case $\s^V_t$ is a total derivative, and only the term $\st_t$ related to the
nonconservative forces is relevant. This term is proportional to the driving force so
it vanishes identically at equilibrium as requested by the empirical prescription of
section~\ref{sec4:singularities}, so ``spurious'' contributions coming from boundary
terms should be absent.
It will be shown that the characteristic function\footnote{From now on the suffix
$\io$ in $z_\io$ will be omitted because in next sections only the asymptotic large
deviations functions will be considered.}
 $z(\l)$ of $\st_t$ exists,
is a convex function of $\l$ and verifies the fluctuation relation $z(\l)=z(1-\l)$.

\subsection{Equilibrium bath}
\label{sec:II} 

As a first illustrative example the case of an equilibrium white bath will be
considered.
The model is a two dimensional harmonic oscillator with potential
energy $V(x,y)=\frac{k}{2}(x^2+y^2)$ coupled to a simple white bath in
equilibrium at temperature $T$, and driven out of equilibrium by the
nonconservative force $\vec h=\epsilon (-y,x)$.  The equations of motion
are 
\beq
\begin{split} 
&m \ddot x_t + \gamma \dot x_t = -k x_t -\epsilon y_t + \xi_t \ , \\ 
&m \ddot y_t + \gamma \dot y_t = -k y_t +\epsilon x_t + \eta_t \ , 
\end{split} 
\eeq 
where $\xi_t$ and $\eta_t$ are independent Gaussian white noises with  
variance 
$\langle \, \xi_t \xi_0 \, \rangle = \langle \, \eta_t \eta_0 \, \rangle = 
2\gamma T \delta(t)$. 
The memory friction kernels $g_{\a\b}(t-s)$ are simply $\delta_{\a\b} g(t-s)= 
2 \delta_{\a\b} \gamma \delta(t-s) \theta(t-s)$ in this case, with  
$\gamma$ the friction coefficient. 
 
Defining the complex variable $a_t=(x_t+iy_t)/\sqrt{2}$ and the noise  
$\r_t=(\xi_t+i\eta_t)/\sqrt{2}$ the equations of motion can be written as 
\beq 
\label{whitemotion} 
m \ddot a_t + \gamma \dot a_t = -\kappa a_t +\rho_t \ , 
\eeq 
where $\kappa = k -i\epsilon$,  
$\langle \rho_t \rho_0 \rangle =  \langle \bar\rho_t \bar\rho_0 \rangle = 0$ and 
$\langle \rho_t \bar\rho_0 \rangle = 2 \gamma T \delta(t)$. 
The complex noise $\rho_t$ has a Gaussian {\sc pdf}: 
\beq 
{\cal P}[\rho_t] \propto \exp \left[ -\frac{1}{2 \gamma T}  
\int_{-\infty}^\infty dt \ \rho_t \bar \rho_t \right] 
\; . 
\label{eq:Gaussian-pdf-rho} 
\eeq 
The energy of the oscillator is $H= m \dot a \dot{\bar a} + k a \bar 
a$, and its time derivative is given by  
\beq  \label{eq7:dHdt}
\frac{dH}{dt} = 2 m \re 
\dot a_t \ddot{\bar a}_t + 2 k \re a_t \dot{\bar a}_t =2 \epsilon \im \dot 
a_t\bar a_t -2 \gamma \dot a_t \dot{\bar a}_t + 2 \re \dot a_t \bar \rho_t = 
 W_t - \widetilde W_t \ ,  
\eeq  
where $W_t = 2 \epsilon 
\im \dot a_t \bar a_t = \epsilon (x_t \dot y_t - y_t \dot x_t)$ is the 
power injected by the driving force and $\widetilde W_t = 2 \gamma 
\dot a_t \dot{\bar a}_t - 2 \re \dot a_t \bar \rho_t$ is the power 
extracted by the thermostat\footnote{Henceforth the sign of the 
powers are chosen such that they have positive average.}.

The entropy production rate (\ref{EPR}) reduces, as expected, to the injected power divided 
by the temperature, $\sigma_t = \b W_t$, where $\b=1/T$. 
One could also consider 
the entropy production of the bath, $\widetilde \sigma_t = \b \widetilde W_t$; it will
be discussed in Appendix~\ref{app:B}. 
 
The average 
value of $\sigma_t = \b W_t$ is in this case given by $\sigma_+ = 2 \epsilon^2 
/(\g k)$.  
To compute the probability distribution function ({\sc pdf}) 
of $\sigma_t$, it is useful to
rewrite it in terms of the complex variable $a_t$:
\beq 
\sigma_\T = \int_{-\T/2}^{\T/2} dt \ \sigma_t = 2 \epsilon 
\beta \; \im \int_{-\T/2}^{\T/2} dt \ \dot a_t \bar a_t \ .  
\eeq 
As already discussed, it is easier to compute the characteristic function 
$z(\lambda)$, Eq.~(\ref{phidef}),
in terms of which the fluctuation relation reads $z(\l)=z(1-\l)$.
To leading order in $\T$ one can neglect all the boundary terms in the 
integrals. 
After integrating by parts,
\beq 
\sigma_\T = 2 \epsilon \beta i \int_{-\T/2}^{\T/2} dt \ a_t \dot{\bar a}_t  
\ ,
\eeq 
and recalling that 
the {\sc pdf} of the noise is given by Eq.~(\ref{eq:Gaussian-pdf-rho})  
one obtains:
\beq 
\label{AAA1} 
\langle \exp[- \lambda \sigma_\T ] \rangle = {\cal N}^{-1} 
\int d{\cal P}[\rho_t] \ \exp \left[ - \frac{2i \epsilon \lambda}{T} 
\int_{-\T/2}^{\T/2} dt \ a_t \dot{\bar a}_t \right]  
\ , 
\eeq 
where the normalization factor ${\cal N} = \int d{\cal P}[\rho_t]$  
is simply given by the numerator calculated at $\lambda =0$. 
 
At the leading order in $\T$ the function $z(\l)$ should not depend 
on the boundary conditions in Eq.~(\ref{AAA1}). 
Thus, one can impose periodic boundary conditions, 
$a(\T/2)=a(-\T/2)$ and $\dot a(\T/2)=\dot a(-\T/2)$, 
and expand $a_t$ in a Fourier series, 
\beq 
\label{AAA2} 
a_t =  
\frac{\Delta \omega}{2 \pi} \sum_{n=-\infty}^{\infty} a_n \ e^{-i \omega_n t}  
\ , 
\eeq 
where $\Delta \omega= 2 \pi /\T$ and $\omega_n = n \Delta \omega$. 
For $\T \rightarrow \io$ 
\beq 
a_t = \int_{-\infty}^\infty \frac{d\omega}{2\pi} \ e^{-i\omega t} a_\omega \ , 
\hspace{30pt} 
a_\omega = \int_{-\infty}^\infty dt \ e^{i\omega t} a_t \ , 
\eeq 
and the equations of motion become 
\beq 
a_\omega = \frac{\rho_\omega}{ - \omega^2 m +\kappa - i\omega \gamma }  
\equiv \frac{\rho_\omega}{D(\o)} \ . 
\eeq 
Note that in the limit $\epsilon=0$ the Green function $G(\epsilon,\o) = 1/D(\o)$  
reduces to the one used 
in section~\ref{sec:I} to compute the violation of the fluctuation-dissipation 
theorem induced by  
a nonequilibrium bath. 
The distribution of the noise is given by 
\beq 
\label{AAA3} 
{\cal P}[\r_\o] = \exp \left[ -\frac{1}{2 \gamma T}  
\int_{-\infty}^\infty \frac{d\omega}{2\pi} \  
\rho_\omega \bar \rho_\omega \right] 
\sim  \exp \left[ -\frac{1}{2 \gamma T}  
\frac{\D\omega}{2\pi}\sum_{n=-\infty}^\infty \rho_n \bar \rho_n \right]  
\ . 
\eeq 
Substituting Eqs.~(\ref{AAA2}) and~(\ref{AAA3}) into Eq.~(\ref{AAA1}) one has
\beq 
\label{AAA4} 
\begin{split}
\langle \exp[- \lambda \sigma_\T ] \rangle &= 
 {\cal N}^{-1} \int d\r_n 
\exp \left[ -  \frac{\Delta \omega}{2\pi}  
\sum_{n=-\infty}^{\infty} \left( 
\frac{|\r_n|^2}{2 \g T} -  
\frac{2 \epsilon \l \o_n |\r_n|^2}{T |D(\o_n)|^2} \right)\right] 
\\ &=  
\prod_{n=-\io}^\io  
\left[ 1 - \frac{4 \g \epsilon \l \o_n}{|D(\o_n)|^2} \right]^{-1}  
\end{split}\eeq 
and using Eq.~(\ref{phidef}) 
\beq 
\label{phi1bagnoFourier} 
z(\lambda) =  \lim_{\T \rightarrow \infty}  
\frac{1}{\T} \sum_{n=-\infty}^{\infty}  
\log \left[ 1 - \frac{4 \g \epsilon \l \o_n}{|D(\o_n)|^2} \right] = 
\int_{-\infty}^\infty \frac{d\omega}{2 \pi} 
\log \left[ 1 - \frac{ 4 \g \epsilon \l \o} 
{ |D(\o)|^2 } \right] \ . 
\eeq 
To show that $z(\lambda)$ verifies $z(\lambda)=z(1-\lambda)$ and  
hence the fluctuation theorem, note that 
\beq 
\label{AAA5} 
z(\lambda)-z(1-\lambda)=\int_{-\infty}^\infty \frac{d\omega}{2 
\pi} \log \left[ \frac{ |D(\o)|^2 - 4 \g \epsilon \l \o} { |D(\o)|^2 - 4 \g 
\epsilon (1-\l) \o } \right]  
\eeq  
and, as $|D(\o)|^2 - 4 \epsilon \g \o = 
|D(-\o)|^2$, the integrand is an odd function of $\o$ and the integral 
vanishes by symmetry.  In Appendix~\ref{app:A} it is shown that the same 
result is obtained if one uses Dirichlet boundary 
conditions (at least for $m=0$, where the computation is feasible); 
this result supports the approximations made when neglecting all the 
boundary terms in the exponential. Moreover, in the case $m=0$ the 
large deviation function $\z(p)$ can be explicitly calculated; 
defining $\t_0 = \g/k$ and $\s_0 = \s_+ \t_0/2 = \epsilon^2/k^2$, one obtains 
\beq 
\label{zetap} 
\z(p)=\t_0^{-1} \left[ 1 + p \s_0 - \sqrt{(1+\s_0)(1+p^2 \s_0)} 
\right] \ .   
\eeq  
Thus, for $\t \rightarrow \infty$ the {\sc pdf} of $p$ has the form 
\beq 
\pi_\t(p) \propto \exp \left[ \frac{\t}{\t_0} f(p,\s_0)\right] \ . 
\eeq 
Note that $\t_0$ is the decay time of the correlation function of $a_t$  
[\ie $\langle a_t \bar a_0 \rangle \propto \exp(-t/\t_0)$] and $\s_0$ is 
the average entropy production over a time $\t_0 / 2$. 
Thus, $\t_0$ is the natural time unit of the problem (as expected); 
remarkably, the function $f$ depends only on 
$\s_0$ and not on the details of the model. 
It would be interesting to see whether the same scaling holds for more  
realistic models. 
 
In summary,
for all driving forces, {\it i.e.} all values of $\epsilon$, the  
fluctuation theorem holds for the entropy production rate (\ref{EPR}).
For a white equilibrium bath this result is a particolar case of the
general theorem derived in~\cite{Ku98}. 
The temperature entering the fluctuation theorem is the  
one of the equilibrated environment with which the system is in contact, 
although it is not in equilibrium with it, when the force is applied.  
 
One can easily check that the fluctuation-dissipation  
relation holds in the absence of the drive (see section~\ref{sec:I})  
but it is strongly violated when the system is taken out of equilibrium 
by the external force. 
 
%%%%%%%%%%%%%%%%%%%%%%%%%%%%%%%%%%%%%%%%%%%%%%%%%%%%%%%%%%%%%%%%%%%%%%%%%%%%%%%%%%%%%%%%% 
%%%%%%%%%%%%%%%%%%%%%%%%%%         SECTION IV      %%%%%%%%%%%%%%%%%%%%%%%%%%%%%%%%%%%%%% 
%%%%%%%%%%%%%%%%%%%%%%%%%%%%%%%%%%%%%%%%%%%%%%%%%%%%%%%%%%%%%%%%%%%%%%%%%%%%%%%%%%%%%%%%% 

\subsection{Non-equilibrium bath} 
\label{sec:IV}
 
The calculation will be now generalized to the case of a generic 
nonequilibrium bath; the equation of motion becomes: 
\beq 
\label{motioncomplex} 
m \ddot a_t + \int_{-\infty}^\infty dt' \ g(t-t') \dot a_{t'} = -\kappa a_t 
+ \rho_t \ ,  
\eeq  
where as before $\kappa=k-i\epsilon$ and $\langle \r_t \bar 
\r_0 \rangle = \nu(t)$. The functions $\nu(t)$ and $g(t)$ are now arbitrary 
(apart from the condition $g(t)=0$ for $t<0$), hence they do not 
satisfy, in general, Eq.~(\ref{FDTbathTD}). 
As discussed in the introduction of this chapter, Eq.~(\ref{motioncomplex}) provides 
a model for the dynamics of a confined Brownian particle in an out of 
equilibrium medium \cite{AG04,PM04,Po04}.  

The dissipated power is given by  
\beq  
\frac{dH}{dt} = 2\epsilon \; \im \dot a_t \bar a_t -2\re \int_{-\infty}^\infty dt' \ g(t-t') 
\dot a_t \dot{\bar a}_{t'} + 2 
\re \dot a_t \bar \rho_t = W_t - \widetilde W_t \; ,  
\eeq  
where as in the previous case $W_t = 2\epsilon \im \dot a_t \bar 
a_t $ is the power injected by the external force and $\widetilde 
W_t = 2\re \int_{-\infty}^\infty dt' \ g(t-t') \dot a_t \dot{\bar 
a}_{t'} - 2 \re \dot a_t \bar \rho_t$ is the power extracted by the 
bath. 

For this harmonic model $\s^V_\t$ is a boundary term and Eq.~(\ref{EPR}) gives
\beq
\label{sigma3} 
\st_\T = -2 \epsilon  \frac{\D\o}{2\p} 
\sum_{n=-\io}^\io \frac{\o_n |a_n|^2}{T_{eff}(\o_n)} = 
2 \epsilon i \int_{-\T/2}^{\T/2} dt  
\int_{-\T/2}^{\T/2} dt' \ T^{-1}(t-t') a_t \dot{\bar a}_{t'} \ .
\eeq 
Note that the last equality holds neglecting boundary terms. 

It is interesting to consider also an ``alternative'' definition of
entropy production rate, which has been often used in the 
literature~\cite{Ga04,GGK01,CGHLP03,FM04}.
It is obtained assuming that the entropy production rate is
proportional to the power injected by the external drive, 
$\su_t = \Th^{-1} W_t$, via a parameter $\Th$ which has the 
dimension of a temperature.
Then, the total entropy production over a 
time $\T$ for Eq.~(\ref{motioncomplex})
is given by (neglecting boundary terms) 
\beq 
\label{sigma1} 
\su_\T =  
\frac{2 \epsilon i}{\Th} \int_{-\T/2}^{\T/2} dt \ a_t \dot{\bar a}_t = 
-2 \epsilon \frac{\D\o}{2\p}\sum_{n=-\io}^\io \frac{\o_n |a_n|^2}{\Th}  
\; . 
\eeq
Usually, in experiments, $\Th$ is a free parameter which is
adjusted in order for the {\sc pdf} of $\su$ to verify the fluctuation 
relation~\cite{GGK01,CGHLP03,FM04}. It is interesting to compute the large
deviations function of $\su_t$ to check if it verifies the fluctuation relation.

The functions $z_\Th(\l)$ and $z_{eff}(\l)$
corresponding to the two entropy production rates defined above will
now be computed. 
The computation is straightforward following the strategy of section~\ref{sec:II}. 
In Fourier space, Eq.~(\ref{motioncomplex}) reads  
\beq  
a_\o=\frac{\r_\o}{-m 
\o^2 + \k - i\o g(\o)}=\frac{\r_\o}{D(\o)} \ .   
\eeq  
The probability 
distribution of $\r_\o$ is  
\beq  
{\cal P}[\r_\o] = \exp \left[ 
- \int_{-\io}^\io \frac{d\o}{2\p} \frac{|\r_\o|^2}{\nu(\o)}\right] \ , 
\eeq  
thus, as in Eqs.~(\ref{AAA3}) and (\ref{AAA4}),  
\beq  \begin{split}
\langle \exp[- 
\lambda \su_\T ] \rangle &= {\cal N}^{-1} \int d\r_n \exp \left[ - 
\frac{\Delta \omega}{2\pi} \sum_{n=-\infty}^{\infty} \left( 
\frac{|\r_n|^2}{\nu(\omega_n)} - \frac{2 \epsilon \l \o_n |\r_n|^2}{\Th 
|D(\o_n)|^2} \right)\right] \\ &=  
\prod_{n=-\io}^\io \left[ 1 - \frac{2 \epsilon 
\l \o_n \nu(\o_n)}{\Th |D(\o_n)|^2} \right]^{-1} \ ,  
\end{split}\eeq  
and  
\beq 
\label{phi_theta}
z_\Th(\l) = \int_{-\io}^\io \frac{d\o}{2\p} \log \left[ 1 - 
\frac{2 \epsilon \l \o \nu(\o)}{\Th |D(\o)|^2} \right] \ .   
\eeq  
The function $z_{eff}(\l)$ is obtained by  
substituting $\Th \rightarrow T_{eff}(\o)$ with the  
latter defined in (\ref{eq:Teff-def}): 
\beq
\label{phi_eff}
z_{eff}(\l) = \int_{-\io}^\io \frac{d\o}{2\p} \log \left[ 1 - 
\frac{4 \epsilon \l \o \re g(\o)}{|D(\o)|^2} \right] \ .   
\eeq  
It is easy to 
prove that $|D(\o)|^2 - 4 \epsilon \o \re g(\o) = |D(-\o)|^2$. Thus, using 
the same trick used in Eq.~(\ref{AAA5}), one can prove that 
$z_{eff}(\l)=z_{eff}(1-\l)$. On the contrary,
in general, it is not possible to find a value of $\Th$ such that
$z_\Th(\l)$ satisfies the fluctuation theorem. It will be shown in the
following that this is possible only in some particular situations:
essentially, when the dynamics of the particle happens on a single
time scale, that corresponds to the experiments cited above.
 
In conclusion, the fluctuation theorem is satisfied  
when the entropy production rate is defined using the power injected by the  
external drive and the temperature of the environment -- that is not  
defined in the case of a nonequilibrium bath -- is replaced by the  
ratio in (\ref{eq:Teff-def}). 

Note that, as discussed in section~\ref{sec6:Teff}, a single nonequilibrium
bath can be equivalently represented by many equilibrated baths at
different temperatures, eventually acting on different time scales.  In
Appendix~\ref{app:III} it will be shown that also in this case the {\sc pdf}
of the entropy production rate $\st$ defined in Eq.~(\ref{sigma3})
verifies the fluctuation theorem, while the {\sc pdf} of $\su$ does
not.

Moreover, in the latter case one can also consider the entropy production of
the baths, defined as the power extracted by each bath divided by the corresponding
temperature. This quantity is of interest if one can clearly identify the different
thermal baths with which the system is in contact: this is not the case in glassy
systems, where the effective temperature is self-generated by the system.
Nevertheless, the study of systems of particles coupled to many baths at different
temperature is of interest in the study of heat conduction.
In Appendix~\ref{app:B} it will be proven that the entropy production rate of the baths
verifies the fluctuation theorem, at least for $|p|\leq 1$, see the discussion in
section~\ref{sec4:singularities} and \cite{VzC04}.

%%%%%%%%%%%%%%%%%%%%%%%%%%%%%%%%%%%%%%%%%%%%%%%%%%%%%%%%%%%%%%%%%%%%%%%%%%%%%%%%%%%%%%%%% 
%%%%%%%%%%%%%%%%%%%%%%%%%%          SECTION VI         %%%%%%%%%%%%%%%%%%%%%%%%%%%%%%%%%% 
%%%%%%%%%%%%%%%%%%%%%%%%%%%%%%%%%%%%%%%%%%%%%%%%%%%%%%%%%%%%%%%%%%%%%%%%%%%%%%%%%%%%%%%%% 
 
\section{Numerical results} 
\label{sec:V}

Some numerical simulations of Eq.~(\ref{motioncomplex}) for a
particular choice of the nonequilibrium bath and in presence of a linear and
nonlinear interaction have been performed.
In the linear case, the numerical results confirm the analytical results
of the previous section. This finding confirms that the boundary terms neglected in
the analytical computation are indeed irrelevant.
In the nonlinear case, it is found that the fluctuation relations holds 
for $\st_t$, as in the linear case.

The simplest non trivial case has been considered, 
where a massless Brownian particle is coupled to two equilibrated 
baths: a white (or fast) bath at temperature $T_f$ and a colored (or 
slow) bath with exponential correlation at temperature $T_s$. This
model has been studied in detail in \cite{CK00} and is relevant for the 
description of glassy dynamics when the time scales of the two baths 
are well separated, as will be discussed in section~\ref{sec:VI}.
In Appendix~\ref{app:III} the general case of an harmonic oscillator
coupled to $N$ colored baths is studied.
The equations of motion are given by 
Eq.~(\ref{motioncomplex}) where $g(t)=g_f(t)+g_s(t)$, $g_f(t)=\g_f \d(t)$ and 
$g_s(t)=\th(t) \frac{\g_s}{\t_s}e^{-\frac{t}{\t_s}}$, or equivalently $g_f(\o)=\g_f$,
$g_s(\o)=\g_s/(1-i\o\t_s)$. The harmonic potential is replaced by a 
generic rotationally invariant 
potential $V(x,y)={\cal V}\left(\frac{x^2+y^2}{2}\right)={\cal V}(|a|^2)$.
The noise is the sum of a fast and a slow component.
Then Eq.~(\ref{motioncomplex}) becomes:  
\beq 
\label{motion2baths} 
\g_f \dot a_t + \frac{\g_s}{\t_s}  
\int_{-\io}^t dt' \ e^{-\frac{t-t'}{\t_s}} \dot a_{t'} = 
- a_t {\cal V}'(|a_t|^2) + i\epsilon a_t + \r^f_{t} + \r^s_{t} 
\ , 
\eeq 
where $\langle \r^f_{t} \r^f_{t'} \rangle = 2 \g_f T_f \delta(t-t')$, 
$\langle \r^s_{t} \r^s_{t'} \rangle = \frac{T_s \g_s}{\t_s} e^{-|t-t'|/\t_s}$ 
and ${\cal V}'(x)$ is the derivative of ${\cal V}(x)$ w.r.t. $x$. 
It is convenient to rewrite this equation in a Markovian form as follows: 
\beq 
\label{riscritte}
\begin{cases} 
&\dot b_t = - \frac{b_t - \y_t}{\t_s} + \frac{\g_s a_t}{\t_s^2} \ , \\ 
&\g_f \dot a_t = - a_t  {\cal V}'(|a_t|^2) + i\epsilon a_t + \r^f_{t} + b_t - \frac{\g_s a_t}{\t_s} \ , 
\end{cases} 
\eeq 
where the auxiliary variable $b_t$ has been introduced and $\y_t$ is a white 
noise with correlation  
$\langle \y_t \bar\y_{t'} \rangle = 2 \g_s T_s \delta(t-t')$. 
The power injected by the external force is, as usual, $W_t=2\epsilon \im \dot a_t \bar a_t$, 
while the power extracted by the two baths can be written as 
$\widetilde W^f_{t}= 2\re \left[ \dot a_t \left( \g_f \dot{\bar a}_t
- \bar \r^f_{t}\right)\right]$  
and 
$\widetilde W^s_{t} = 2 \re \left[ \dot a_t \left( \frac{\g_s}{\t_s} \bar a_t - \bar b_t \right) 
\right]$. 

In the nonlinear case the potential ${\cal V}(|a|^2)=\frac{g}{2} |a|^4$
as been chosen,  
and the results are compared with the ones obtained for the harmonic 
case, ${\cal V}(|a|^2)=k |a|^2$.
The simulation has been performed for $\epsilon=0.5$, 
$T_f=0.6$, $\g_f=1$, $T_s=2$, $\g_s=1$ and $\t_s=1$, setting $k=1$ in the 
linear case and $g=1$ in the nonlinear one.
The system (\ref{riscritte}) is numerically solved via a standard
discretization of the equations with time step $\d t =0.01$; the noises
are extracted using the routine {\sc gasdev} of the C numerical recipes.

It is numerically found that $\s^V_t$ and $\st_t$ are uncorrelated (within
the precision of the numerical data), so their {\sc pdf} can be studied
separately. Unfortunately, the {\sc pdf} of $\s^V_t$ is too noisy to allow
for a verification of the fluctuation relation in the nonlinear case.
This is probably due to the fact that in the linear case $\s^V_t$ reduces
to a boundary term\footnote{This is also observed in the simulation, because
the average of $\s^V_\t$ vanishes and the variance of $\s^V_\t$ does not grow
with $\t$.}; in the nonlinear case it is not a boundary term, but
still it might contain ``spurious'' boundary terms which should be eliminated,
see the discussion in section~\ref{sec4:singularities}. 
Indeed, for the accessible values of $\t$,
it is observed that the variance of $\s^V_\t$ is much larger than its average
(while the FR would predict a variance of the order of $\s^V_+$). This large
variance can be a finite-$\t$ effect due to the presence of a boundary
term whose fluctuations contribute to the fluctuations of $\s^V$ but not to the
average. If this is the case, the FR should hold at least for $|p|<1$ and very
large $\t$: but the values of $\t$ can be so large that the FR is unobservable
in practice, see~\cite{ZRA04a,BGGZ05}. For this reason, in the
following the data for $\s^V$ will not be discussed.
The validity of the FR for $\s^V$ (possibly minus a boundary term) in the nonlinear
case remains an open question that should be addressed by future work.

\subsection{``Effective'' entropy production rate}

%%%%%%%%%%%%%%%%%%%%%%%%%%%%%%%%%%%%%%%%%%%%%%%%%%%%%%%%%%%%%%%%%%%%%%%%%%%%%%%%%%%%%%% 
 
\begin{figure}[t]
\centering 
\includegraphics[width=.75\textwidth,angle=0]{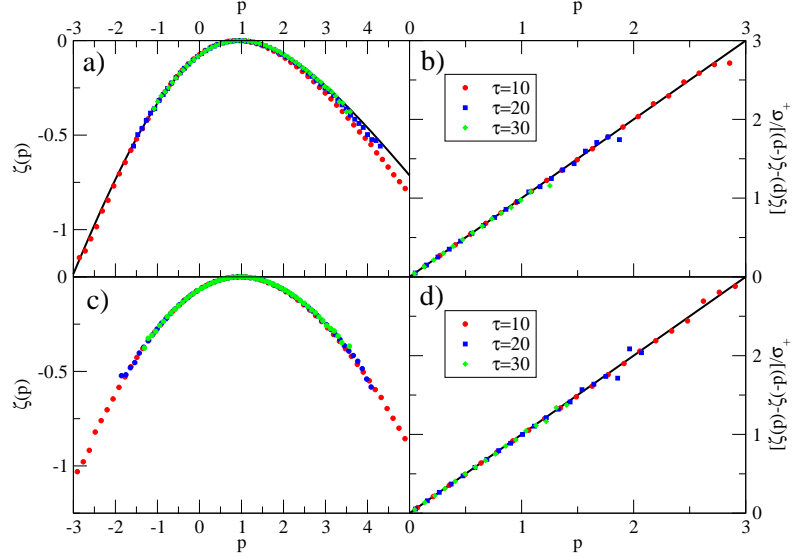} 
\caption[{\sc pdf} of $\st_t$ for the harmonic and anharmonic potentials]
{{\sc pdf} of $\st_t$: a) The large deviation function for the 
harmonic potential at $\T=10,20,30$; the full line is the analytical 
result. b) The function $f(p) \equiv [\z_{eff}(p)-\z_{eff}(-p)]/\st_+$ for the 
harmonic potential: the fluctuation theorem predicts a straight line 
with slope 1, represented by a full line.  c) The large deviation 
function for the quartic potential at $\T=10,20,30$.  d) The function 
$f(p)$ for the quartic potential: also in this case
the fluctuation theorem is well verified.} 
\label{fig:1} 
\end{figure} 
 
%%%%%%%%%%%%%%%%%%%%%%%%%%%%%%%%%%%%%%%%%%%%%%%%%%%%%%%%%%%%%%%%%%%%%%%%%%%%%%%%%%%%%%% 

The effective entropy production rate $\st$, from Eq.~(\ref{EPR}), is given by:
\beq 
\label{sigma3OK} 
\st_\t = \int_{-\io}^t dt' \  T^{-1}(t-t') \, 2\epsilon \im \big[ 
\dot a_t \bar a_{t'} +  \dot a_{t'} \bar a_t \big] \ , 
\eeq 
with $T^{-1}(t)$ the Fourier transform of  
$1/T_{eff}(\o)$. 
From Eq.~(\ref{TeffNbaths}) the latter is given by 
\beq 
\label{TeffV}
\frac{1}{T_{eff}(\o)} =  
\frac{\g_f (1 + \o^2 \t_s^2) + \g_s}{T_f \g_f  (1 + \o^2 \t_s^2) + T_s \g_s} \; . 
\eeq 
Thus 
\beq 
\label{Tstar} 
T^{-1}(t) =  
\frac{1}{T_f} \delta(t) +  
\frac{\g_s}{T_f \g_f\t_s^2}\left(1 - \frac{T_s}{T_f}\right) 
\frac{e^{-\Omega |t|}}{2 \Omega} \ ,  
\;\;\;\;\mbox{with} \;\;\;\; 
\Omega =\frac{1}{\t_s} \sqrt{\frac{T_f \g_f + T_s \g_s}{T_f \g_f}} \ . 
\eeq  
and $T^{-1}(t)$ decays exponentially for large $t$. 
Note that, if the bath is at equilibrium, $T_s=T_f=T$, so one has $T^{-1}(t)=\d(t)/T$ and
$\st_t= 2 \epsilon \im \dot a_t \bar a_t / T = W_t/T$ as expected (recall that by convention
$\int_{-\io}^t dt' \, \d(t-t') = \frac{1}{2} $).

The data for $\st_t$ are shown in Fig.~\ref{fig:1}. 
The large deviation function $\z_{eff}(p)$ is reported in panel a) for the
harmonic and in panel c) for the quartic potential. The average $\st_+$ is
equal to $0.332$ in the harmonic case and to $0.276$ in the quartic case. The
function $\z_{eff}(p)$ converges fast to its asymptotic limit $\T \rightarrow \io$
(note that even the data for $\tau\sim 10$ are in quite good agreement with the 
analytic prediction for the harmonic case).
The fluctuation theorem predicts $f(p) \equiv [\z_{eff}(p)-\z_{eff}(-p)]/\st_+ = p$.
The function $f(p)$ is reported in panel b) for the harmonic and in panel d) for the
quartic potential.
In the harmonic case the numerical data are compatible with the validity of 
the fluctuation theorem, as predicted analytically.
Remarkably, the same happens in the quartic
case where the analytical prediction is no more available.

These results support the conjecture that, if $\s^V_t$ and $\st_t$ are
uncorrelated, the {\sc pdf} of $\st$ verifies the fluctuation
theorem independently of the form of the potential $V(x,y)$.

\subsection{``Classical'' entropy production rate}

It is interesting to investigate numerically also the fluctuations of
the entropy production rate $\su$, defined in section~\ref{sec:IV} as 
\beq 
\su_t=\frac{W_t}{\Th}=\frac{2\epsilon}{\Th}\im \dot a_t \bar
a_t 
\ .  
\eeq 
Rather arbitrarily $\Th=T_f$ was set in the definition of
$\su_t$.  This reflects what is usually done in numerical simulations,
where the dissipated power is divided by the ``kinetic'' temperature,
{\it i.e.} the temperature of the fast degrees of freedom.  Note that
the choice $\Th=T_f$ does not affect the function $\z_\Th(p)$ since
the variable $p$ is normalized, \ie $\z_\Th(p)\equiv \z(p)$
does not depend on $\Th$, see Eq.~(\ref{pdef}), but it changes the
average $\su_+$ that is proportional to $\Th^{-1}$.

%%%%%%%%%%%%%%%%%%%%%%%%%%%%%%%%%%%%%%%%%%%%%%%%%%%%%%%%%%%%%%%%%%%%%%%%%%%%%%%%%%%%%%% 
 
\begin{figure}[t]
\centering 
\includegraphics[width=.75\textwidth,angle=0]{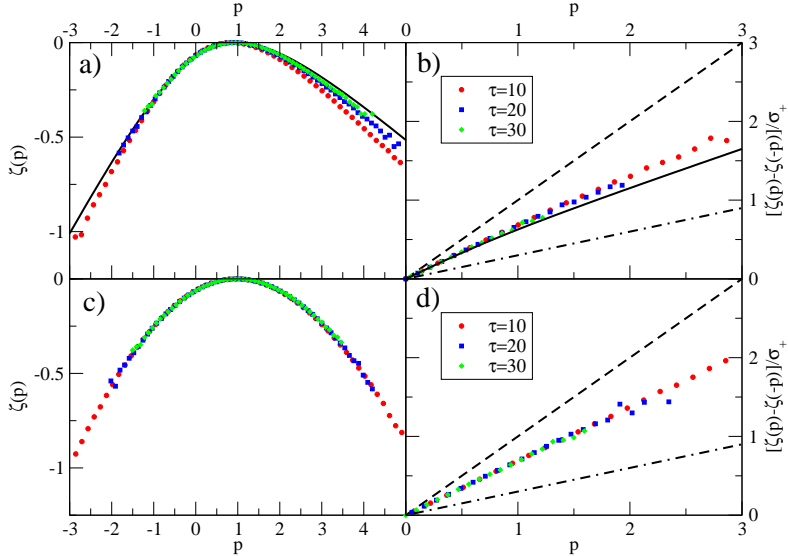} 
\caption[{\sc pdf} of $\su_t$ for the harmonic and anharmonic potentials]
{{\sc pdf} of $\su_t$: a) The large deviation function for the
harmonic potential at $\T=10,20,30$; the full line is the analytical
result.
b) The function $f(p) \equiv [\z(p)-\z(-p)]/\s^{T_f}_+$ for the
harmonic potential. The full line is the analytical prediction, the dashed
line is the prediction of the fluctuation relation, 
the dot-dashed line has slope $T_f/T_s$.  
c) The large deviation function for the quartic potential at $\T=10,20,30$. 
d) The function $f(p)$ for the quartic potential; the dashed line
is the fluctuation theorem, the dot-dashed line has slope $T_f/T_s$.}
\label{fig:2} 
\end{figure} 
 
%%%%%%%%%%%%%%%%%%%%%%%%%%%%%%%%%%%%%%%%%%%%%%%%%%%%%%%%%%%%%%%%%%%%%%%%%%%%%%%%%%%%%%% 
 
The data for $\su_t$ are reported in Fig.~\ref{fig:2}. The harmonic
case is shown in panels a) and b) while the anharmonic case is
presented in panels c) and d). The value $\s^{T_f}_+=0.455$ is obtained
for the harmonic potential and $\s^{T_f}_+=0.366$ for the quartic one. 
The large deviation
function of $\su_t$ agrees very well with the analytical prediction in
the harmonic case but it does not verify the fluctuation
theorem for $\Th=T_f$, as one can clearly see from the right panels in
Fig.~\ref{fig:2}.

Remarkably, in both the harmonic and anharmonic
cases the function $f(p) \equiv [\z(p)-\z(-p)]/\s^{T_f}_+$ 
is approximately linear in $p$ with a slope $X$ such that $1 > X > T_f/T_s$,
{\it i.e.} $\z(p)-\z(-p) \sim  X \; p\s^{T_f}_+$.
If $f(p) \sim X p$, one can tune the value of $\Th$ in order to obtain the fluctuation
relation $\z(p)-\z(-p) = p \su_+$, simply choosing $\Th=\Th_{eff}=T_f/X$, 
thus defining a single ``effective temperature'' $\Th_{eff} \in [T_f,T_s]$.
From the data reported in Fig.~\ref{fig:2} one gets a slope $X \sim 0.66$, that
gives $\Th_{eff} = T_f/X \sim 0.9$.

\begin{figure}
\includegraphics[width=.52\textwidth,angle=0]{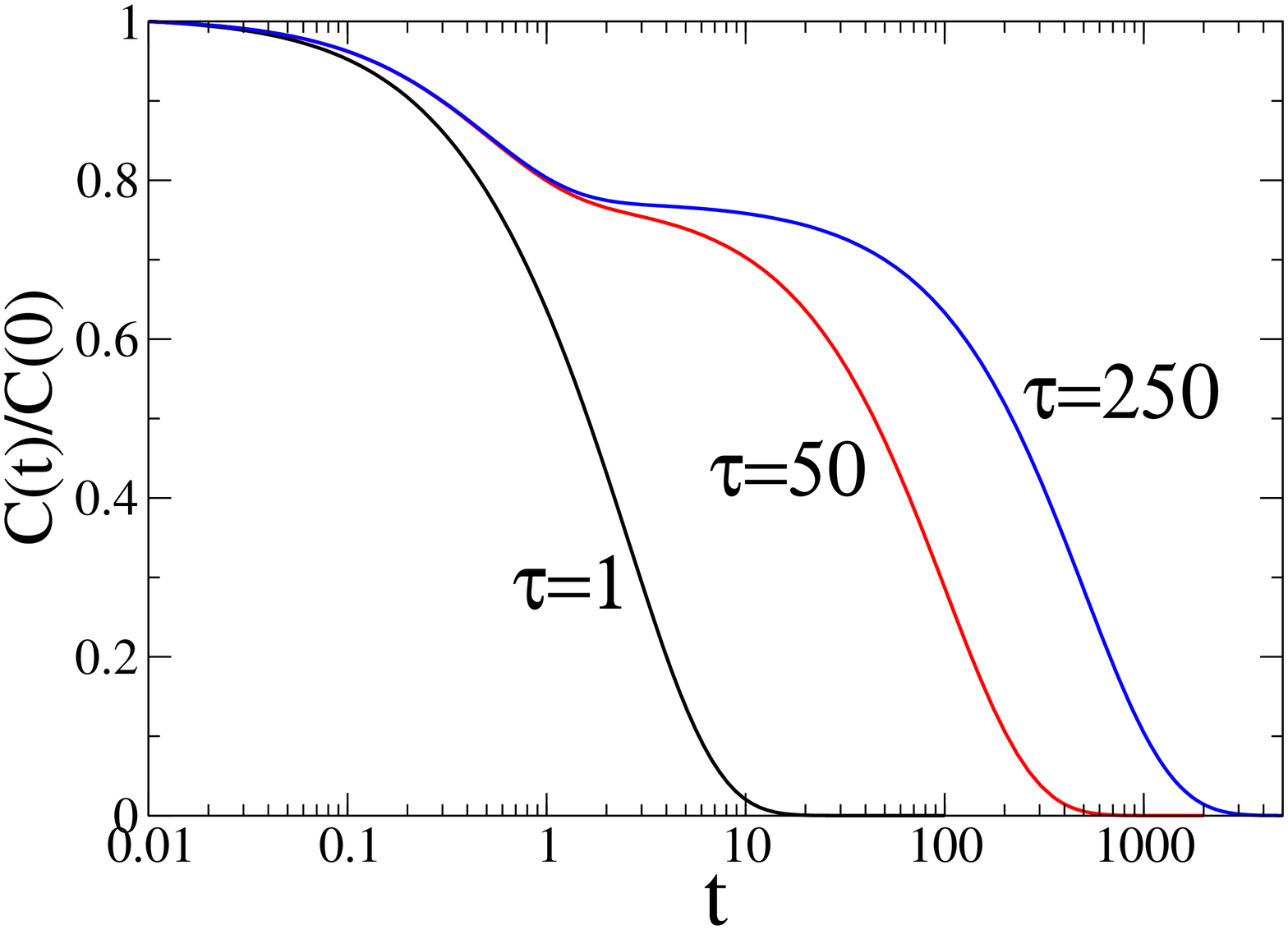} 
\includegraphics[width=.52\textwidth,angle=0]{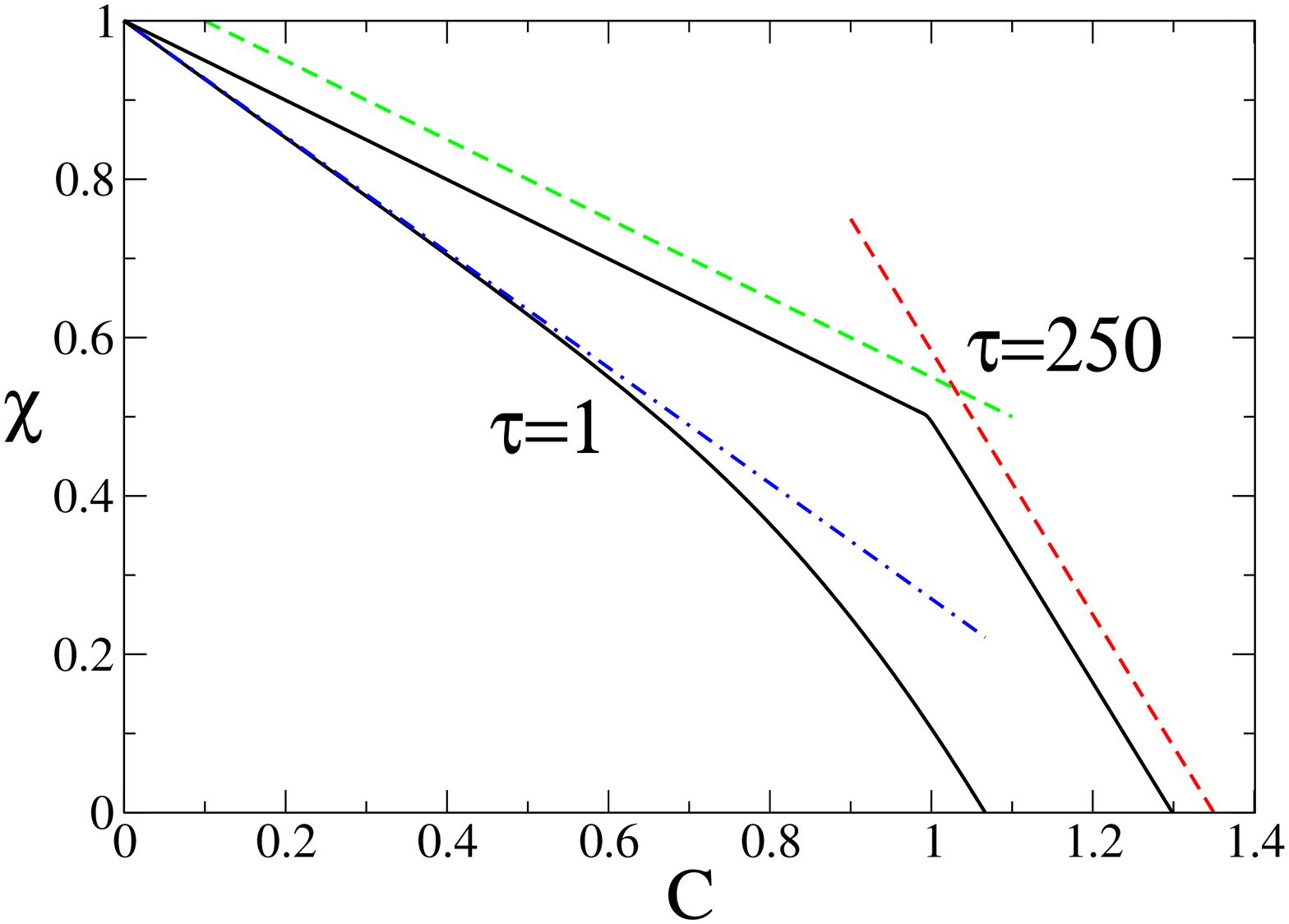}
\caption[Autocorrelation functions and FDT plot for the harmonic oscillator]
{(Left) Normalized autocorrelation functions of $a_t$ for the harmonic oscillator with 
$\epsilon=0.5$, $k=1$, $T_f=0.6$, $T_s=2$, $\g_f=1$, $\g_s = k \t_s$ 
and $\t_s=1,50,250$. (Right)
Parametric plot of the integrated response $\chi(t)$ as a function of the correlation
function $C(t)$ for the same parameters but $\epsilon=0$. The dot-dashed line
has slope $-1/1.37$, the dashed lines have slope $-1/T_s$ and $-1/T_f$.
} 
\label{fig:3b} 
\end{figure}

This behavior reflects the one found in some recent experiments 
\cite{GGK01,CGHLP03,FM04,ZRA04b} in situations where the dynamics of the system 
happens essentially on a single time scale.
This is the case also in the numerical simulation presented here: 
in Fig.~\ref{fig:3b} (left panel) the autocorrelation function 
$C(t) = \re \langle a_t \bar a_0 \rangle$ 
of $a_t$ (computed in Appendix~\ref{app:C}) is reported 
for the harmonic potential. The present simulation refers to the curve with $\t_s=1$,
which clearly decays on a single time scale.

In Fig.~\ref{fig:3b} (right panel) the parametric plot $\chi(C)$ (see
section~\ref{sec:I}) for the same set of parameters,
but $\epsilon=0$, is shown.  The integrated response is given by $\chi(t)=\int_0^t dt'
\, R(t')$ and $R(t)$ is computed in Appendix~\ref{app:C}.
For $\t_s=1$, the function $\chi(C)$ has slope close to $-1/T_f$
at short times (corresponding to $\chi \sim 0$). For longer times, the
slope moves continuously toward $-1/T_{eff}$, with $T_{eff} \sim
1.37$.  This value of $T_{eff}$ is of the order of $\frac{\g_f T_f +
\g_s T_s}{\g_f+\g_s} =1.3$, which means that on time scales of the
order of the (unique) relaxation time the two baths behave like a
single bath equilibrated at intermediate temperature. This would
be {\it exact} if the time scales of the two baths were exactly
equal.

It is worth to note that in this situation one has $T_{eff} \neq
\Th_{eff}$, that is, the effective temperature that one would extract
from the {\it approximate} fluctuation relation of Fig.~\ref{fig:2} is
not coincident with the effective temperature obtained from the
$\chi(C)$ plot of Fig.~\ref{fig:3b}. In particular, $T_f <
\Th_{eff} < T_{eff}$: this relation is consistent with the
results obtained from the numerical simulation of a sheared
Lennard-Jones--like mixture that will be presented below~\cite{ZRA04b}, 
even if the coincidence might be accidental.

%%%%%%%%%%%%%%%%%%%%%%%%%%%%%%%%%%%%%%%%%%%%%%%%%%%%%%%%%%%%%%%%%%%%%%%%%%%%%%%

%%%%%%%%%%%%%%%%%%%%%%%%%%%%%%%%%%%%%%%%%%%%%%%%%%%%%%%%%%%%%%%%%%%%%%%%%%%%%%

\subsection{Summary of the numerical results}

The numerical simulation of the non-linear problem
confirms that the fluctuation theorem is satisfied {\it exactly}
when the entropy production rate $\st$ is defined using the power injected by the  
external drive and the temperature of the environment -- that is not  
defined in the case of a nonequilibrium bath -- is replaced by the  
ratio in (\ref{eq:Teff-def}).

In situations in which the dynamics of the system happens on a single
time scale, a constant effective temperature $\Th_{eff}$ can be
introduced to obtain an {\it approximate} fluctuation relation
defining the entropy production rate as $W_t/\Th_{eff}$.  However,
$\Th_{eff}$ is not necessarily related to the effective temperature
$T_{eff}$ that enters the modified fluctuation--dissipation relation,
and in the systems considered so far \cite{ZRA04b} it seems that
$\Th_{eff}<T_{eff}$.

As will be shown in the following, when the dynamics happens on different, 
well separated, time scales, it is impossible to find a single value 
$\Th_{eff}$ such that $\su_t=W_t/\Th$ verifies the fluctuation relation.

%%%%%%%%%%%%%%%%%%%%%%%%%%%%%%%%%%%%%%%%%%%%%%%%%%%%%%%%%%%%%%%%%%%%%%%%%%%%%%%%%%%%%% 
%%%%%%%%%%%%%%%%%%%%%%%%%%%%%%%       SECTION VI      %%%%%%%%%%%%%%%%%%%%%%%%%%%%%%%% 
%%%%%%%%%%%%%%%%%%%%%%%%%%%%%%%%%%%%%%%%%%%%%%%%%%%%%%%%%%%%%%%%%%%%%%%%%%%%%%%%%%%%%% 
 
\section{Separation of time scales and driven glassy systems} 
\label{sec:VI} 
 
As discussed in chapter~\ref{chap6}, 
in the study of mean-field models for glassy dynamics~\cite{Cu02,CK00} 
and when using resummation techniques within a perturbative approach  
to microscopic glassy models with no disorder, effective equations of motion
of the form of Eq.~(\ref{eq6:MCTeffective}) are obtained.
In the case of a driven mean field system~\cite{CKLP97,BBK00}, the external 
force is also present in Eq.~(\ref{eq6:MCTeffective}) and after a transient 
the system becomes stationary {\it for any temperature}, \ie $\mu(t) 
\equiv \mu$, $\Si(t,t')=\Si(t-t')$, and $D(t,t')=D(t-t')$. The 
functions $D$ and $\Si$ depend on the 
strength $\epsilon$ of the driving force, 
\eg as in Eq.~(\ref{eq6:kerneldriven}), and do not satisfy the detailed
balance condition.
However, it is possible to rewrite Eq.s~(\ref{eq6:kerneldriven}) as
\beq\label{eq6:kernelrivisto}
\begin{split}
&D = \frac{p}{2} C^{p-1} + \epsilon^2 \frac{k}{2} C^{k-1} \equiv D_0 + \ee^2 D_1 \ , \\
&\Si = R D'_0(C) \ ,
\end{split}
\eeq
so that $\Si$ and $D_0$ verify the detailed balance condition.
From the expressions (\ref{eq6:kernelrivisto}), one can rewrite 
Eq.~(\ref{eq6:MCTeffective}), in the following way:
\beq \label{motionglassy-rep}
\begin{split}
&\dot \s(t) = -\mu \s(t) + \int_{-\io}^\io dt' \ \Si(t-t') \s(t') + \r(t) + 
\epsilon h(t) \ , \\
&\la \r(t)\r(t') \ra = 2T\d(t-t') + D_0(t-t') \ , \\
&\la h(t) h(t') \ra = D_1(t-t') \ ,
\end{split}
\eeq 
where $\r(t)$ and $h(t)$ are two uncorrelated Gaussian variables.
Note that $\Si$ and $D_0$ still depend implicitly on $\ee$ as one
has to solve the self-consistency equations for $C$ and $R$ and
substitute the result in $\Si$ and $D_0$.
However, suppose that the equations have been solved and the 
solution for $C_\ee$ and $R_\ee$ has been plugged in $\Si$ and $D_0$.
If the term proportional to $h(t)$ in Eq.~(\ref{motionglassy-rep})
is removed, as $\Si = R D_0'(C)$, Eq.~(\ref{motionglassy-rep})
derives from a Langevin equation where only conservative forces
are present~\cite{BCKM96}. This means that the term corresponding
to the external driving is represented only by $h(t)$, and the
dissipated power is given by $W(t) = \ee h(t) \dot \s(t)$.
Indeed (see Appendix~\ref{app:randomh}),
\beq\label{Wpspin}
\la W \ra = \ee^2 \int_0^{\io} dt \, \dot R(t) D_1(t) 
= \ee^2 \frac{k}{2} \int_0^{\io} dt \, \dot R(t) C^{k-1}(t)
\ ,
\eeq
consistently with the result of \cite{BBK00} where the average
of the injected power was explicitly computed for the driven
spherical $p$-spin. Thus one obtains an equation that is
very similar to Eqs.~(\ref{motioncomplex}) 
and (\ref{motion2baths}), where $\Si(C_\ee,R_\ee)$ and $D_0(C_\ee)$
represent the nonequilibrium bath, and $\ee h(t)$ is the external
driving force: and one can prove that the effective
entropy production rate
\beq\label{EPRhrandom}
\st(t) = \ee \int_{-\io}^t dt' \, T^{-1}(t-t') 
\big[ h(t) \dot \s(t') + h(t')\dot \s(t) \big] \ ,
\eeq
where $T_{eff}(\o) = \frac{2T+D_0(\o)}{2\re [1+\Si(\o)/(i\o)]}$, 
verifies the fluctuation relation, see 
Appendix~\ref{app:randomh}.

As discussed in \cite{BBK00}, for small $\epsilon$ and
$T>T_d$, $R_\ee$ and $C_\ee$ verify the fluctuation--dissipation theorem,
so the same happens for $D_0$ and $\Si$, \ie $T_{eff}(\o) \equiv T$. 
The transport coefficient related to the driving force $\epsilon$ approaches 
a constant value for $\epsilon \rightarrow 0$ (the linear response holds 
close to equilibrium) and the systems behaves like a ``Newtonian 
fluid''. In this situation, the 
system behaves as if coupled to a single equilibrium bath (and the 
fluctuation theorem holds for the entropy production rate
$W_t/T = \epsilon h(t) \dot \s(t)/T$).

Below $T_d$, the fluctuation-dissipation relation is 
violated also in the limit $\epsilon \rightarrow 0$ where it is
a simple broken line, see Fig.~\ref{fig6:FDTplot}, with temperature
$T$ at short time and $T_{eff}$ at long times.
The transport coefficient diverges in this limit: the system is strongly 
nonlinear and behaves as if coupled to two baths acting on different time scales 
and equilibrated at different temperatures, and the correct definition
of entropy production rate is Eq.~(\ref{EPRhrandom}).
In the region $\epsilon \sim 0$ and $T < T_d$, when the two
relaxation scales are well separated, it is possible to separate
the ``fast'' and ``slow'' parts of the equation of motion
({\it adiabatic approximation}). This allows to write all
the relations in a particularly simple way.
 
\subsection{The adiabatic approximation} 
\label{sec:adiabatic}
 
\begin{figure} 
\centering
\includegraphics[width=.55\textwidth,angle=0]{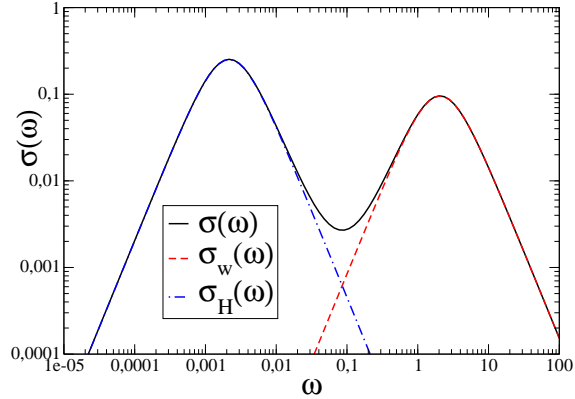} 
\caption[Power spectrum of the entropy production rate for the harmonic oscillator]
{Power spectrum of the entropy production rate (full line) as
a function of the frequency for the harmonic oscillator with $\epsilon=0.5$,
$k=1$, $T_f=0.6$, $T_s=2$, $\g_f=1$, $\g_s = k \t_s$ and $\t_s=250$.
The dot-dashed line is the ``slow'' contribution of $H_t$, the dashed
line is the ``fast'' contribution of $w_t$.  }
\label{fig:2b} 
\end{figure} 

When a simple system is coupled to a complex bath with two (or more)
time scales these are induced into the dynamics of the system.
When the time-scales are well separated, an adiabatic treatment 
is possible in which one separates the dynamic variables in terms 
that evolve in different time-scales (dictated by the baths) and
are otherwise approximately constant. 

In this section an adiabatic approach~\cite{CK00} is used to treat
simple problems coupled to baths that evolve on different scales.  The
motivation for studying this type of problems is that the separation
of time-scales is self-generated in glassy dynamics, as described above.

The {\sc pdf} of $\su_t$ and $\st_t$ will be studied.
The latter satisfies the fluctuation theorem
exactly, and the adiabatic approximation does not spoil this feature. 
The former, instead, does not satisfy the fluctuation theorem in general.
The origin of this difference will be evident in the
adiabatic approximation.

Consider again the Langevin equation (\ref{motion2baths}) with
${\cal V}(|a|^2)=k |a|^2$. In this case, the correlation functions can
be calculated explicitly, see Appendix~\ref{app:C}. In
Fig.~\ref{fig:3b} the autocorrelation function,
$C(t)=\langle a_t \bar a_0 \rangle$, is reported 
for $\epsilon=0.5$, $k=1$, $T_f=0.6$,
$T_s=2$, $\g_f=1$, $\g_s = k \t_s$ and different values of $\t_s$.
Clearly, for $k \t_s = \g_s \gg \g_f$ two very different time scales
--related to the time scales of the two baths-- are present.  
From the plot of Fig.~\ref{fig:3b} one sees that in the
case $k \t_s = 250 \gg \g_f$ the function $\chi(C)$ is a broken line
with slope $-1/T_f$ at large $C$ (short times) and $-1/T_s$ for small
$C$ (large times).

In this situation, the variable $a_t$ can be 
written as the sum of two quasi-independent contributions.
Using the construction introduced
in~\cite{CK00} one can rewrite the equation of motion~(\ref{motion2baths})
as 
\begin{eqnarray}
%\label{adiabatic} 
\left\{ 
\begin{array}{l} 
\g_f \dot a_t = - (k + \g_s/\t_s) a_t + i\epsilon a_t + \r^f_{t} + h_t 
\; , 
\\ 
h_t= - \frac{\g_s}{(\t_s)^2}  
\int_{-\io}^t dt' \ e^{-\frac{t-t'}{\t_s}} a_{t'} + \r^s_{t} 
\; .
\label{eq:second} 
\end{array}
\right. 
\end{eqnarray}
The variable $h_t$ is ``slow''; considering it as a constant in the
first equation, the variable $a_t$ will fluctuate around the equilibrium
position $a_h = h / (k + \g_s/\t_s - i \epsilon) \equiv H$. The latter will
--slowly-- evolve according to the second equation in~(\ref{eq:second}), in
which one can approximate $a_{t'} \sim H_{t'}$. Defining the --fast-- displacement of
$a_t$ w.r.t. $H_t$, $w_t \equiv a_t - H_t$, one obtains the following equations
for $(w_t,H_t)$: \beq
\label{adiabatic2} 
\begin{cases} 
&\g_f \dot w_t = - (k + \g_s/\t_s) w_t + i\epsilon w_t + \r^f_{t} \ , \\ 
&\frac{\g_s}{\t_s}
\int_{-\io}^t dt' \ e^{-\frac{t-t'}{\t_s}} \dot H_{t'} = 
-k H_t +i\epsilon H_t + \r^s_{t} \ . 
\end{cases} 
\eeq 
In this approximation, $a_t = H_t + w_t$ is the sum of two
contributions: $w_t$ is a ``fast'' variable which evolves according to
a Langevin equation with the fast bath only and a renormalized
harmonic constant $k+\g_s/\t_s$, while $H_t$ is a ``slow'' variable
which evolves according to a Langevin equation where the slow bath
only appears. In both equations the driving force $\epsilon$ is present,
thus one expects both $H_t$ and $w_t$ to contribute to the dissipation.
Note that $w_t$ and $H_t$ are completely uncorrelated in this
approximation.
 
\subsection{$\s^V_t$ in the adiabatic approximation}

In the adiabatic approximation, one can argue that
the term $\s^V_t$ in equation~(\ref{EPR})
becomes a boundary term. 
Indeed, the function $T^{-1}(t)$, in the adiabatic approximation, becomes
\beq
T^{-1}(t)=\frac{1}{T_f}\d(t) + T^{-1}_s(t) \ ,
\eeq
where the function $T^{-1}_s(t)$ is ``slow'', see e.g. equation~(\ref{Tstar}).
Inserting this expression in $\s^V_t$, the first term gives a total derivative.
The second term gives
\beq
\int_{-\io}^t dt' \, T^{-1}_s(t-t') 
\left[ \dot r_\a(t) 
\frac{\d V(\vec r)}{\d r_\a(t')} +  \dot r_\a(t') \frac{\d V(\vec r)}{\d r_\a(t)} \right] \ .
\eeq
Due to the convolution with the ``slow'' function $T^{-1}_s(t)$, 
the fast components of $r$ are irrelevant in the integral,
while for the slow ones it is reasonable to replace 
$\dot r_\a(t)$ with $\dot r_\a(t')$ on the scale $\t_s$ over which $T^{-1}_s(t)$ decays.
Thus one obtains again a total derivative times the integral of $T^{-1}_s(t)$ 
which is a finite constant.
Obviously this is not a rigorous proof and should be checked numerically in concrete cases.

\subsection{{\sc pdf} of $\st_t$} 

The entropy production rate defined in
Eqs.~(\ref{sigma3}) and~(\ref{sigma3OK}) can be rewritten in terms of
$H_t$ and $w_t$. Recalling that $T^{-1}(t)$ is defined by Eq.~(\ref{Tstar})
one obtains (the details of the calculation are reported in
Appendix~\ref{app:D}) 
\beq
\label{sigma3adiabatic} 
\st_t \sim 2\epsilon \im \left[ \frac{\dot w_t \bar w_t}{T_f} + \frac{\dot
H_t \bar H_t}{T_s} \right] 
\eeq 
neglecting terms that vanish when
$\st_t$ is integrated over time intervals of the order of $\t_s$. This
is exactly the entropy production expected for two independent
systems.  

\begin{figure} 
\centering 
\includegraphics[width=.55\textwidth,angle=0]{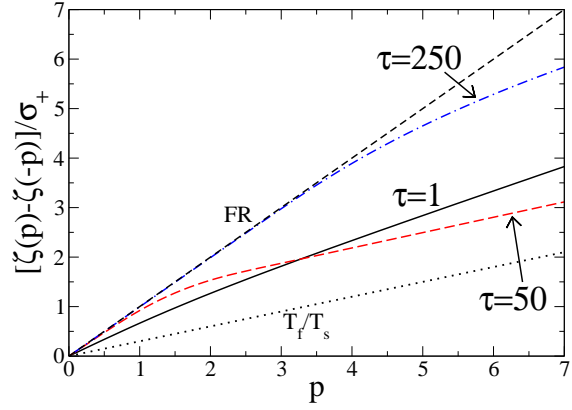} 
\caption[Large deviation function $\z_\Th(p)$ for the harmonic oscillator]
{The function $[\z_\Th(p) - \z_\Th(-p) ] / \su_+$ for the
harmonic oscillator with $\epsilon=0.5$, $k=1$, $\Th=T_f=0.6$, $T_s=2$,
$\g_f=1$, $\g_s = k \t_s$ and $\t_s=1,50,250$. The dashed line is a
straight line with slope 1, the dotted line has slope $T_f/T_s=0.3$.
}
\label{fig:4} 
\end{figure}

To check that this approximation works well, one can introduce a 
``power spectrum'' $\s(\o) d\o$ as the contribution coming from frequencies
$[\o,\o+d\o]$ to the average entropy production rate, 
$\st_+ = \int_0^\io d\o \, \s(\o)$.
From Eq.~(\ref{phi_eff}) one has
\beq
\label{powersp}
\begin{split}
&\st_+ = \left.\frac{d z_{eff}}{d\l}\right|_{\l=0} = 
-\int_{-\io}^{\io} \frac{d\o}{2\p} \frac{ 2 \epsilon\o \n(\o)}{|D(\o)|^2} \\
&\s(\o)=\frac{1}{2\pi} \left[ - \frac{ 2 \epsilon\o \n(\o)}{|D(\o)|^2} + \frac{ 2 \epsilon\o \n(-\o)}{|D(-\o)|^2}
\right] = \frac{\epsilon\o \n(\o)}{\pi} \left[ \frac{1}{|D(-\o)|^2} - \frac{1}{|D(\o)|^2}\right]
\end{split}
\eeq
Substituting the expressions of $\n(\o)$ and of $D(\o)$ appropriate for Eq.~(\ref{motion2baths}) 
one gets the power spectrum $\s(\o)$
as a function of $\o$ which is reported in Fig.~\ref{fig:2b} as a full line.
The contributions of $w_t$ and $H_t$, $\s_w(\o)$ and $\s_H(\o)$, are obtained inserting
in Eq.~(\ref{powersp}) the expression of $\n(\o)$ and $D(\o)$ obtained from the two equations
(\ref{adiabatic2}). They are reported as dashed and dot-dashed lines in Fig.~\ref{fig:2b}.
One can conclude that, for $k \t_s \gg \g_f$, the adiabatic approximation holds and
$\st_t = \s^w_t + \s^H_t$,  
with $\s^w_t = 2\epsilon \im \, \dot w_t \bar w_t / T_f$ 
and $\s^H_t = 2\epsilon \im \, \dot H_t \bar H_t / T_s$, and the two 
contributions are independent. 
Note that the average dissipation
due to $H$ is much larger than the one due to $w$.
Finally, one can write:
\beq 
z_{eff}(\lambda) = z^w(\lambda) + z^H(\lambda) \ , 
\hspace{30pt} z^{w,H}(\l) = -\lim_{\t \rightarrow \infty} \t^{-1} \log
\langle \exp\big[-\l \s^{w,H}\big] \rangle \ .  
\eeq 
Both $z^w(\lambda)$
and $z^H(\lambda)$ verify the fluctuation theorem, as the two equations of
motion (\ref{adiabatic2}) are particular instances of the general case
discussed in section~\ref{sec:IV}. The function $\z_{eff}(p)$ is the
Legendre transform of $z_{eff}(\l)$ and will verify the fluctuation
theorem.

\subsection{{\sc pdf} of $\su_t$}
  
In the same approximation, $\su_t$ is given, for $\Th=T_f$, by 
\beq 
\label{sigma1adiabatic} 
\s^{T_f}_t = 2\epsilon \im \left[ \frac{\dot w_t \bar w_t + 
\dot H_t \bar H_t}{T_f} \right] 
= \s^w_t + \frac{T_s}{T_f}\s^H_t  \ , 
\hspace{20pt} \text{and} \hspace{20pt} z_{T_f}(\l) = z^w(\lambda) +
z^H(\lambda T_s / T_f) \ ; 
\eeq 
the contribution of $H_t$ is weighted with
the ``wrong'' temperature, \ie the temperature of the fast degrees of
freedom. Indeed, as already discussed, $z_\Th(\lambda)$ does not
verify the fluctuation theorem.  The function $f(p) \equiv [\z_\Th(p) -
\z_\Th(-p) ] / \su_+$, obtained from Eq.~(\ref{phi_theta}), is reported in
Fig.~\ref{fig:4}.
As already discussed in section~\ref{sec:V},
when the time scales of the two baths are comparable, $k
\tau_s \sim \g_f$, the two baths act like a single bath at temperature $\Th
\in [T_f,T_s]$ and the function $f(p)$ is approximately linear in $p$ with
slope $X \in [T_f/T_s,1]$.
When the time scales are well separated, $k\t_s \gg \g_f$, the adiabatic
approximation holds; 
in this situation it turns out, from the exact computation of $\z_\Th(p)$,
that the function $f(p)$ has slope
$\sim 1$ for small $p$ and has slope $T_f/T_s$ for large $p$ (see Fig.~\ref{fig:4}).
These results will be compared with numerical simulations of Lennard-Jones
systems below.

\section{Green-Kubo relations} 
\label{sec:Green-Kubo} 
 
The Green-Kubo relations for transport coefficients, that are a 
particular form of the fluctuation--dissipation theorem, follow from
the fluctuation relation, as discussed in section~\ref{sec4:closetoeq}.  
In this section a way to link
the modified fluctuation theorem -- in which the external bath
temperature is replaced by the (frequency dependent) 
effective temperature of the unperturbed system -- to the
modification of the fluctuation--dissipation relation will be
discussed.
 
Note that even out of equilibrium one can define a flux 
${\cal J}_t$ using $\s_t$ as a ``Lagrangian'', see Eq.~(\ref{Jdef})
and \cite{Ga04}: 
\beq 
{\cal J}_t = \frac{\partial \s_t}{\partial E} \ . 
\eeq 
Close to equilibrium $\s_t$ is given by 
Eq.~(\ref{sclosetoeq}) and ${\cal J}_t = J_t/T$. 
If, in the absence of a drive, 
the system has a non trivial effective temperature, 
the entropy production rate should be defined as in Eqs.~(\ref{sigma3})
and (\ref{sigma3OK}).  
Then the flux ${\cal J}_t$ is given by 
\beq 
{\cal J}_t = \frac{\partial \st_t}{\partial \epsilon} =  
4 \, \im \int_{-\io}^t dt' T^{-1}(t-t') \dot a_t \bar a_{t'} = 
2 \int_{-\io}^t dt' T^{-1}(t-t') [\dot y_t x_{t'} - \dot x_t y_{t'}] \ . 
\eeq 
The fluctuation theorem for $\st$ implies then a Green-Kubo 
relation for ${\cal J}_t$: 
\beq 
\label{GKgen} 
\langle {\cal J} \rangle_\epsilon = \epsilon \int_0^\io dt \ \langle {\cal J}_t {\cal J}_0
\rangle_{\epsilon=0} + o (\epsilon^2) \ .  \eeq The physical meaning of the latter
relation becomes clear if one writes the flux ${\cal J}_t$ in the adiabatic
approximation discussed in the previous section; from
Eq.~(\ref{sigma3adiabatic}): 
\beq {\cal J}_t = 2 \im \left[ \frac{\dot w_t
\bar w_t}{T_f} + \frac{\dot H_t \bar H_t}{T_s} \right] = \frac{J^w_t}{T_f} +
\frac{J^H_t}{T_s} \ , \eeq 
and Eq.~(\ref{GKgen}) becomes 
\beq \frac{\langle
J^w \rangle_\epsilon}{T_f} + \frac{\langle J^H \rangle_\epsilon}{T_s} = \frac{\epsilon}{T_f^2}
\int_0^\io dt \ \langle J^w_t J^w_0 \rangle_{\epsilon=0} + \frac{\epsilon}{T_s^2}
\int_0^\io dt \ \langle J^H_t J^H_0 \rangle_{\epsilon=0} + o(\epsilon^2) \ .  
\eeq 
Indeed,
in the adiabatic approximation the Green-Kubo relation holds separately for
$J^w_t$ (with temperature $T_f$) and for $J^H_t$ (with temperature $T_s$).
Eq.~(\ref{GKgen}) encodes the two contributions and holds even when the
adiabatic approximation does not apply and the contributions of the ``fast''
and of the ``slow'' modes is not well separated.
 
Note that the ``classical'' Green-Kubo relation involves the total flux 
$J_t = J^w_t + J^H_t$. For the latter one has, in the adiabatic approximation,
\beq 
\label{GKgenFDR} 
\begin{split} 
\langle J_t \rangle_\epsilon &= 
\langle J^w_t \rangle_\epsilon + \langle J^H_t \rangle_\epsilon = 
\frac{\epsilon}{T_f} \int_0^\io dt \  \langle J^w_t J^w_0 \rangle_{\epsilon=0} +  
\frac{\epsilon}{T_s} \int_0^\io dt \  \langle J^H_t J^H_0 \rangle_{\epsilon=0} \\ 
&=\epsilon \int_0^\io dt \left[ \frac{\langle J^w_t J^w_0 \rangle_{\epsilon=0}}{T_f} +  
\frac{\langle J^H_t J^H_0 \rangle_{\epsilon=0}}{T_s} \right] \sim 
\epsilon \int_0^\io dt \frac{1}{T_{eff}(t)} \langle J_t J_0 \rangle_{\epsilon=0} \ . 
\end{split} 
\eeq 
The latter relation is the generalization of the Green-Kubo
formula that comes from the generalized {\sc FDR} discussed in
section~\ref{sec6:Teff}. It is closely related, but not equivalent, to
Eq.~(\ref{GKgen}).
 
\subsection{The Green-Kubo relation for driven glassy systems} 
 
Equations~(\ref{GKgen}) and (\ref{GKgenFDR}) cannot be applied
straightforwardly to driven glassy systems as for these systems the correlation
function $\langle J_t J_0 \rangle_\epsilon$ is not stationary at $\epsilon=0$ below
$T_d$.  Indeed, the relaxation time of the latter diverges as $\epsilon
\rightarrow 0$ and at some point falls outside the experimentally accessible
range: the system will not be able to reach stationarity on the experimental
time scales and will start to {\it age} indefinitely.

The problem is that in Eq.~(\ref{motionglassy-rep}) 
the functions $\Si(t-t')$ and $D_0(t-t')$, that define the thermal bath,
depend strongly on $\ee$ through the functions
$C$ and $R$ which are determined self-consistently.
However, the Green-Kubo relations above have been obtained sending the 
driving force $\ee \to 0$ {\it keeping the thermal bath fixed}.
This means that in Eq.~(\ref{motionglassy-rep}) one should send the term
$\ee h \to 0$ keeping fixed the functions $\Si$ and $D_0$.
For $\epsilon \sim 0$, the main contribution to the $\epsilon$-dependence of the 
dynamics of $\s(t)$
comes from the $\epsilon$-dependence of $\Si$ and $D_0$, so removing the term
$\ee h$ at fixed $\Si$ and $D_0$ does not
affect too much the correlation function $\langle J_t J_0 \rangle_\epsilon$ if 
$\epsilon$ is small. 
Thus, for small $\epsilon$, one can write the Green-Kubo relations in the form
\beq 
\langle J_t \rangle_\epsilon
\sim \epsilon \int_0^\io dt \frac{1}{T_{eff}(t)} \langle J_t J_0 \rangle_{\epsilon} \ ,
\eeq 
even if the limit $\epsilon \rightarrow 0$ is not well defined. An analogous
relation will be obtained from Eq.~(\ref{GKgen}) (which is equivalent to the
fluctuation theorem in the Gaussian approximation) within the same
approximation.  The latter relations can be tested in numerical simulations as
well as in experiments.

\section{Slow periodic drive and effective temperature}
\label{sec:XI}

A lesson one learns from the previous calculations (see
\eg~Fig.~\ref{fig:2b}) is that the work done at large frequencies is
overwhelmingly larger than that done at very low frequencies --
precisely the one that one wishes to observe in order to detect effective
temperatures. One way out of this is to choose a perturbation that
does little work at high frequencies: a periodically
time-dependent force that {\em derives from a potential} $\cos(\Omega
t) {\tilde V}(r)$, with $1/\Omega$ of the order of timescale of the
slow bath $\tau_s$.  
In the following a one dimensional system will be discussed,
the generalization is straightforward.

Consider a single degree of freedom $r$ moving in a time-independent
potential $V(r)$ and subject to a periodically time-dependent field 
$\cos(\Omega t) {\tilde V}(r)$, and in
contact with a `fast' and a `slow' bath with friction kernel, thermal
noise and temperature $(\rho^f,g_f,T_f)$ and $(\rho^s,g_s,T_s)$,
respectively:
\begin{eqnarray}
m\ddot r(t) + \int_{-\infty}^{t}  dt' \; \big[ g_f(t-t')+g_s(t-t') \big] {\dot r}(t') =
-\frac{\partial V}{\partial r(t)} + \rho^f(t) + \rho^s(t) - \cos(\Omega t) 
\frac{\partial {\tilde V}}{\partial r(t)}
\; ,
\label{uno1} 
\end{eqnarray}
The time scale of the time dependent field $1/\Omega$ is of 
the same order as that of the `slow' bath.
The work in an interval of time $(0,\t)$ done by the time-dependent potential is:
\begin{equation}
W_\t=-\int_0^\t \; \cos(\Omega t') \;
\frac{\partial  {\tilde V}}{\partial r}  \;
\dot r  \; dt' = - {\tilde V}(\t)+{\tilde V}(0) +
\Omega \int_0^\t  \; \sin(\Omega t')  \; {\tilde {V}}  \; dt' \ .
\label{work}
\end{equation}
Only the last term grows with the number of cycles, so for long times one can
neglect the first two.
Now, integrating (\ref{uno1})  by parts, one has:
\begin{eqnarray}
m\ddot r(t)   &=& -\int_{-\infty}^{t}  dt' \;  g_f(t-t') {\dot r}(t') 
-\frac{\partial V}{\partial r(t)}  + \rho^f + h(t) - \hat h(t)
\frac{\partial {\tilde V}}{\partial r(t)}
\label{dos}
\\
h(t) &\equiv& - \int_{-\infty}^{t}  dt' \; g_s(t-t')  r(t')  + \rho^s(t)
\; .
\label{tres} 
\end{eqnarray}
where $\hat h(t) = \cos(\Omega t)$.  In the adiabatic limit when both
the timescales of the slow bath and the period $1/\Omega$
of the potential $\tilde V$ are large, $ h(t)$ and $\hat h(t)$ are quasi-static.
Hence, $r$ has a fast evolution given by Eq.~(\ref{dos}) with $h,\hat
h$ fixed and it reaches a distribution~\cite{CK00}
\begin{equation}
P(r/h,\hat h)= \frac{ 
e^{-\beta_f \left( V+\hat h {\tilde V} + g_f(0) \frac{r^2}{2} - hr \right)}
}
{
\int \; dr \; 
e^{-\beta_f \left( V+\hat h {\tilde V} +  g_f(0) \frac{r^2}{2} - hr \right) }
}
\; .
\label{ll}
\end{equation}
The denominator defines $Z(h,\hat h)$ and 
$F(h,\hat h) \equiv -\beta_f^{-1} \log Z(h,\hat h)$. 
Note that $F(h,\hat h(t))$ is periodically time-dependent
 through $\hat h$.
The approximate evolution of $h$ 
is now given by Eq.~(\ref{tres}) with the replacement of 
$r$ in the friction term by its average $\frac{\partial F(h,\hat h)}{\partial h}$
with respect to the fast evolution:
\begin{equation}
h(t) = \int_{-\infty}^{t} dt' \, g_s(t-t')
\frac{\partial F(h,\hat h)}{\partial h}(t')  + \rho^s(t)
\; .
\label{cinco} 
\end{equation}
Equation (\ref{cinco}) is in fact a generalized Langevin
 equation for a  system coupled to a (slow) bath
of temperature $T_s$. Indeed, it can be shown \cite{CK00} to be equivalent to 
a set of degrees of freedom $y_i$ evolving according to
the ordinary Langevin equation:
\begin{equation}
\left[ m_j \frac{d^2}{dt^2} + \gamma_j \frac{d}{dt} + \Omega_j \right] y_j = \xi_j(t)- 
\frac{\partial F \left( \sum_j A_j y_j \right) }{\partial y_j}
\label{truc}
\end{equation}
with $ \langle \xi_i(t) \xi_j(t') \rangle = 2 T_s \gamma_j \delta_{ij}
\delta(t-t') $, provided that the Fourier transforms $g_s(\omega)$ an
$\nu_s(\omega)$ of friction kernel and noise autocorrelation can be
written as:
\begin{equation}\begin{split}
&g_s(\omega) =
\sum_j \frac{ A^2_j}{m_j (\omega-\omega_j^+)(\omega-\omega_j^-)} \ , \\
&\nu_s(\omega) = 2 T_s \sum_j \frac{ \gamma_j A^2_j}{m^2_j (\omega-\omega_j^+)(\omega-\omega_j^-)
(\omega+\omega_j^+)(\omega+\omega_j^-)} \ ,
\end{split}\end{equation}
where $\omega^\pm_j$ are the roots of $-m_j \omega^2+i\gamma_j\omega  + \Omega_j=0$.

Within the same approximation leading to (\ref{cinco}),  
 the average of $\tilde V(r)$ over a time window $\Delta$ that
 is long compared to the short timescale, but
sufficiently slow that one can consider that $h$ and $\hat h$ are constant is
\begin{equation}
\int_t^{t+\Delta}   \;  \tilde V(r(t')) \; dt' \sim \Delta \int \; dr \;  
P(r/h,\hat h) \; \tilde V(r) =
\Delta \frac{\partial F(h,\hat h)}{\partial \hat h}
\end{equation}
so that one obtains for the work:
\begin{equation}
W_\t \sim  \Omega \int_0^\t \sin(\Omega t')  
\frac{\partial F(h,\hat h(t'))}{\partial \hat h} dt' =
- \int_0^\t \; \frac{\partial F(h,\hat h(t'))}{\partial t'} dt'
\label{work1}
\end{equation} 
which tells that for long time intervals 
the work done by the original time-dependent potential $\tilde V$ is indeed
the same as the work done by the time-dependent effective potential $F$ in (\ref{cinco}).

The fluctuation theorem then holds for the distribution of this work,
with a single temperature $T_s$. One concludes that the distribution of
work due to a slow perturbation satisfies the fluctuation theorem with
only the slow temperature, and can be hence used experimentally to
detect it.

The simplest application of the above general result is obtained
considering $\tilde V(r)=\tilde h r$ and $V(r)=k r^2$. Then, grouping
together the two noises in a single noise with friction $g = g_f +
g_s$ and correlation $\n = T_f \n_f + T_s \n_s$ as described in
section~\ref{sec6:Teff}, Eq.~(\ref{uno1}) becomes simply 
\beq 
m \ddot r(t) +
\int_{-\io}^t dt' \, g(t-t') \dot r(t') = -k r(t) + \r(t) + \tilde h
\cos (\O t) \ .  
\eeq 
This equation describes for instance the motion
of a Brownian particle moving in an out of equilibrium environment and
trapped by an harmonic potential whose center oscillates at frequency
$\O$.
A concrete experimental realization of this setting has been already 
considered in \cite{AG04}: Silica beads of $\sim 2 \m$m diameter were
dispersed in a solution of Laponite (a particular clay of $\sim 30 $nm diameter) 
and water.
The Laponite suspension forms a glass for large enough concentration of
clay and provides the nonequilibrium environment. The Silica beads 
are Brownian particles diffusing in such environment. They can be
trapped by optical tweezing, and the center of the trap can oscillate
with respect to the sample if the latter is oscillated through a
piezoelectric stage. In \cite{AG04} the mobility and diffusion 
of tracer particles were measured obtaining an estimate of $T_{eff}(\O)$.
To check the fluctuation relation one has to
perform a measurement of the work done by the trap on
the tracers.
Indeed, the work dissipated in $(0,\t)$ is linear in $r(t)$ so it should
be possible to measure it simply through the measurement of $r(t)$: 
\beq
\label{workharmonic}
W_\t = \O \tilde h \int_0^\t dt' \sin (\O t') r(t') \ ;
\eeq
note that, as $W_\t$ is linear in $r(t)$, it is a Gaussian variable.
With a simple calculation one finds
\beq
\lim_{\t \rightarrow \io}
\frac{\langle (W_\t - \langle W_\t \rangle)^2 \rangle}{2\langle W_\t \rangle} =  
\frac{\n(\O)}{2 \re g(\O)} = T_{eff}(\O)
\eeq
This means that the (Gaussian) {\sc pdf} of $\st_\t = W_\t / T_{eff}(\O)$ 
satisfies the fluctuation relation. If the two baths are modeled as in 
section~\ref{sec:V}
with $k \g_s = \t_s \gg \g_f$ one has $T_{eff}(\O) = T_s$ for $\O \t_s < 1$,
see Eq.~(\ref{TeffV}).
The measurement of the distribution of the work (\ref{workharmonic}) allows for
the measurement of $T_s$. Note that other experimental settings described by the
same equations should exist.

%%%%%%%%%%%%%%%%%%%%%%%%%%%%%%%%%%%%%%%%%%%%%%%%%%%%%%%%%%%%%%%%%%%%%%%
%%%%%%%%%%%%%%%%%%%%%%   PRE RAPID  %%%%%%%%%%%%%%%%%%%%%%%%%%%%%%%%%%%
%%%%%%%%%%%%%%%%%%%%%%%%%%%%%%%%%%%%%%%%%%%%%%%%%%%%%%%%%%%%%%%%%%%%%%%

\section{Numerical simulation of a binary Lennard--Jones mixture}

It is interesting to test the predictions above in a numerical simulation
of a realistic model for a glassy system, like the ones considered
in~\cite{KA94,DmSC04,Lennard-FDT}.
The predictions obtained from the solution of the dynamics of $p$-spin
models have been succesfully tested in the numerical simulations of these
models. Indeed, on the space-time scales of the numerical simulations
(which are very small) the glass transition is very similar to the
mean-field one, see the discussion in section~\ref{sec1:beyondmeanfield}.
On these scales,
the glass transition $T_g$ reflects the dynamical transition
$T_d$ of $p$-spin models: the numerical results are very well described
by {\it mode-coupling equations} of the form (\ref{eq6:MCT}), see
\eg~\cite{KA94,Go99,Cu99}, and a violation of the fluctuation--dissipation theorem
of the form (\ref{twoT}) is observed~\cite{Lennard-FDT}.
The driven dynamics of these systems has been investigated 
in \cite{BB00,BB02,ARSTZ02} where a uniform velocity gradient $\gamma$ 
was applied on a Lennard-Jones binary mixture, and the results
described in section~\ref{sec6:drivenpspin} were very well 
reproduced by the numerical data.

Thus, it is interesting to see if the generalization of the
fluctuation relation proposed above holds for these systems
below $T_g$ and for $\g \sim 0$.
Note that a possible generalization of the FR, of the form
\beq
\label{FTX}
\zeta_\io(p)-\zeta_\io(-p) = X p \sigma_+ \ ,
\eeq
with $X<1$, was proposed in \cite{BGG97,Ga99b,BG97} in the 
context of chaotic dynamical systems, see the discussion 
in section~\ref{sec5:discussion}.
It has also been proposed to define $\Th_{eff}\equiv T/X$ as the ``temperature''
in nonequilibrium steady states \cite{Ga04}.
However, up to now numerical studies of the FR have been
performed only in the high temperature region ($T \gg T_g$),
where $X=1$.

Eq.~(\ref{FTX}) was shown to hold {\it approximately} if the time
scales of the two baths are not well separated in section~\ref{sec:V}.
The numerical data that will be presented in this section~\cite{ZRA04b} 
show that indeed~(\ref{FTX}) is satisfied by $\z_\io(p)$ below $T_g$.
Unfortunately, {\it (i)} it seems that the proposed connection between
Eq.~(\ref{FTX}) and phase space properties is not confirmed by the
numerical data, and {\it (ii)} a regime in which the time scales are
well separated is not accessible due to limited computational power,
so that the predictions of section~\ref{sec:V} could not be completely
tested. These open points will hopefully be addressed by future works.

\subsection{The model}

The investigated system is a 80:20 binary mixture of 
$N$ particles in $d=3$ of equal mass $m$ interacting via a soft sphere 
potential
\beq
V_{\alpha \beta}(r)=\epsilon_{\alpha \beta} 
\left( \frac{\sigma_{\alpha\beta}}{r} \right)^{12} \ ,
\eeq
$\alpha,\beta \in [A,B]$. It is very similar to 
model II\footnote{Note that for consistency with the definitions
of~\cite{DmSC04} a $4$ is missing.} of
section~\ref{sec7:III}, and the parameters $\epsilon_{\a\b}$
and $\s_{\a\b}$ are the same as in section~\ref{sec7:III}.
This system has been introduced and characterized in equilibrium by 
De Michele {\it et al}~\cite{DmSC04}
as a modification of the standard LJ Kob-Andersen mixture~\cite{KA94} 
that is known to avoid crystallization on very long time scales, 
and hence to be a very good model of glass former;
it has been chosen because the soft sphere potential 
can be cut at very short 
distance ($1.5 \sigma_{AA}$) allowing
the system to be very small\footnote{The minimum size of the system
is determined by the condition that the simulation box is larger than
the range of the potential in order to avoid the interaction of a particle
with its image, see \eg~\cite{AT87}. The WCA potential considered in
section~\ref{sec7:III} has an even shorter range: the soft sphere potential
was chosen in order to make easier the comparison with existing 
results~\cite{DmSC04}.}.

A shear flow is applied to the system
along the $x$ direction with a gradient velocity field along the
$y$ axis. The shear flow was chosen instead of the constant
force of section~\ref{sec7:III} for two reasons. On one hand, it makes
easier a comparison with the existing literature concerning the driven
glassy regime~\cite{BB00,BB02,ARSTZ02}; on the other hand, the shear flow couples
directly with the cooperative structural rearrangemens which are 
responsible for the glassy behavior, while the constant force
used in section~\ref{sec7:III} couples to single-particle diffusion. 

The particles are confined in a cubic box with Lees-Edwards boundary 
conditions and
the molecular dynamics simulation is performed using
SLLOD equations of motion \cite{EM90}:
\beq\label{SLLOD}
\begin{cases}
&\dot{q}_i = \frac{p_i}{m} + \gamma q_{yi} \hat{x} \, , \\
&\dot{p}_i = F_i(q) - \gamma p_{yi} \hat{x} - \alpha(\ul p,\ul q) p_i \, ,
\end{cases}
\eeq
where $F_i(q)=-\partial_{q_i} V(q)$ and 
$\alpha$ is a thermostat which fixed the kinetic temperature $T$, 
as discussed in section~\ref{sec7:III}.
The equation of motion are discretized following the procedure
described in section~\ref{sec7:III}.
The entropy production rate\footnote{Again, total derivatives will
be removed from $\s$, see section~\ref{sec4:singularities}.}
is defined as the
dissipated power $W$ divided by the kinetic temperature $T$:
\beq
\sigma(p,q)=\frac{W(p,q)}{T}=-\frac{\gamma P_{xy}(p,q)}{ T} ,
\eeq
where $P_{xy}(p,q)=\sum_i [ p_{xi} p_{yi} + q_{yi} F_{xi}(q) ]$ is the $xy$
component of the stress tensor \cite{EM90}.
As in section~\ref{sec7:III}, all the quantities are reported in units of
$m$, $\epsilon_{AA}$ and $\sigma_{AA}$.
In these units the integration step is $dt=0.005$.
The density is fixed to $\rho = 1.2$
to compare with~\cite{KA94,DmSC04,Lennard-FDT}.

The main problem is in chosing the size of the system, for the following
reasons:
\begin{enumerate}
\item If $N$ is too small, it is easy for the system to crystallize,
especially in presence of shear. Thus one has to choose $N$ large in 
order to avoid spurious fluctuations due to nucleation of crystals.
\item On the other hand, the fluctuations of $p$ scale as $\exp [-N f(p)]$,
so if $N$ is too large it is impossible to observe negative values of
$p$ which are needed to test Eq.~(\ref{FTX}).
\item One could solve this problem by choosing a very large $N$ ($\sim 1000$, as
in \cite{BB00,BB02}) and looking to the fluctuations of $\s$ in a small volume
inside the sample, following~\cite{GP99,Ga99c}. 
An attempt in this direction was made, but it turned out that it was difficult
to give a good definition of {\it local} entropy production 
rate\footnote{The problem is the following: to observe the phenomenology 
described in section~\ref{sec6:drivenpspin} one has
to apply a shear rate $\g \sim \t_\a(T)^{-1}$~\cite{ARSTZ02}. 
If one consider a volume of linear size $L$, this volume will be deformed
by the dynamical evolution due to the sliding of the different regions in
the sample. The time scale of this deformation is $1/\g$, \ie it is the
same time scale over which one would observe the violation of the 
fluctuation relation. Thus the volume {\it looses its identity} before
any interesting effect is observed. It was not possible to find a clear
way out of this contradiction, so it was preferred to study the {\it global}
entropy production rate to avoid complications and uncontrolled effects
which could alter the slope in Eq.~(\ref{FTX}), as observed \eg in \cite{BCL98}.}.
\end{enumerate}
The value $N$=66, which is large enough to avoid 
crystallization\footnote{The absence of Bragg 
peaks in the dynamic structure factor $S(q)$ was carefully checked.} 
and small enough to allow for the observation of
negative values of $p$, was chosen. However, to reach the asymptotic
regime one has to integrate $\s$ over a time interval $\t \sim 10 \t_\a$,
see the results of chapter~\ref{chap5}. This means that for large $\t_\a$,
\ie for $\g \sim 0$ and $T < T_g$, to obtain a reasonably large number of
values of $p$, enough to observe large deviations, one needs to simulate 
Eq.~(\ref{SLLOD}) for a very large total time. This strongly limits
the values of $\g$ and $T$ which are accessible to investigation, and
in particular completely rules out, for the large system considered here, 
the region where the ``fast'' and ``slow'' time scales are well separated.

It seems that the observation of curves like the one reported in
Fig.~\ref{fig:4} in numerical simulations of glassy systems is a very
difficult task. Probably some difficulties can be avoided considering
\eg the diffusion of a tracer particle in the sample~\cite{BB02}, but this is a
different physical situation that is left for future investigation.

\subsection{Results}

\begin{figure}[t]
\centering
\includegraphics[width=.55\textwidth,angle=0]{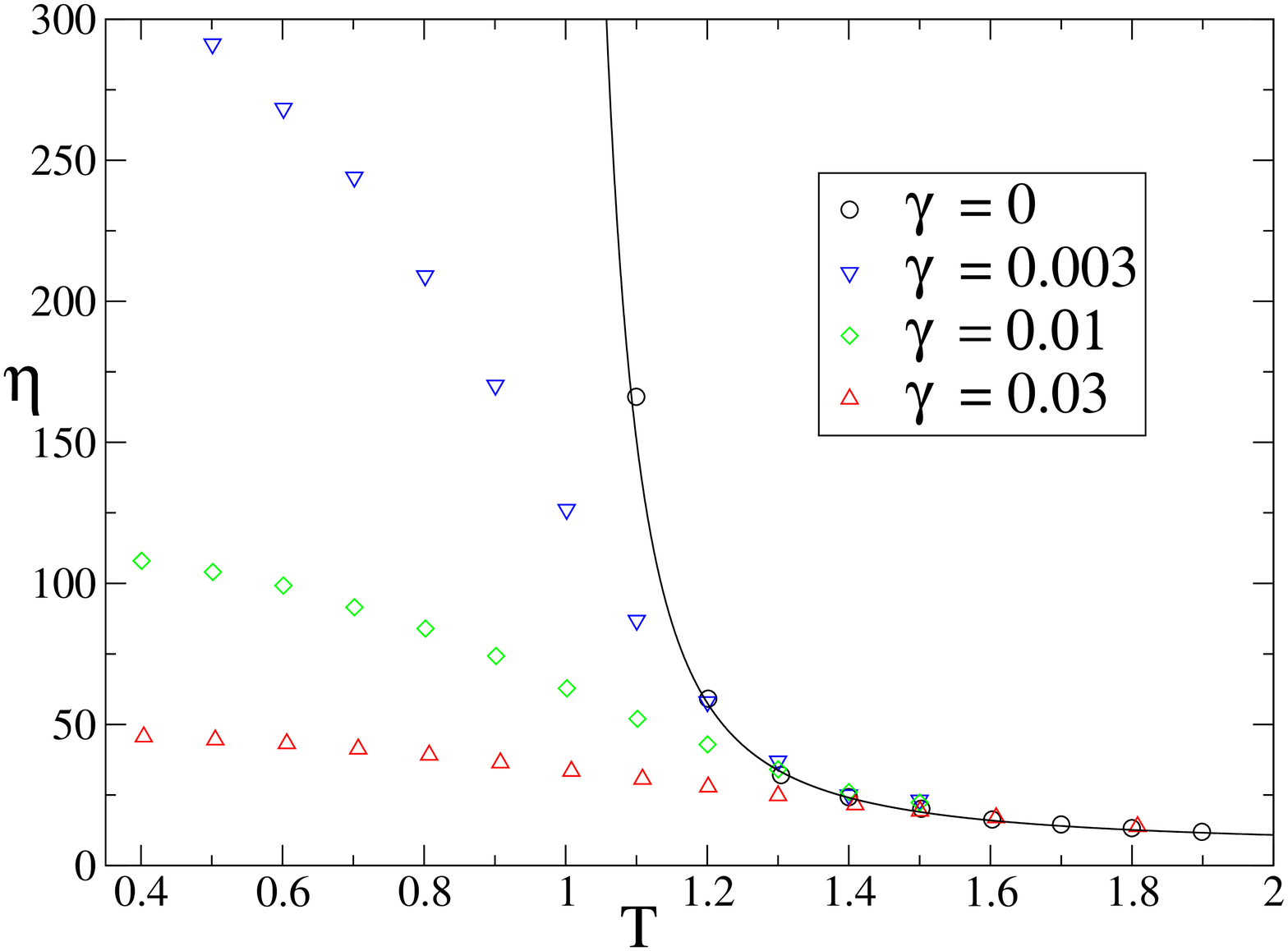}
\caption[Viscosity of the Lennard--Jones binary mixture as a function of $T$]
{Viscosity as a function of temperature for different values of $\gamma$. 
The continuous line is a fit to a Vogel-Tamman-Fulcher law,
$\eta(T)=\eta_\io \exp (\frac{AT_0}{T-T_0})$ with $\eta_\io=5.2$, $A=0.99$, $T_0=0.85$. 
}
\label{fig_7gal_7_1}
\end{figure}

In Fig.~\ref{fig_7gal_7_1} the viscosity $\eta \equiv \langle P_{xy} \rangle /\gamma$
is reported as a function of the temperature $T$ for 
different values of the shear rate $\gamma$.
At $\gamma=0$ the viscosity seems to diverge at a temperature $T_0 \sim 0.85$;
however, the system can be equilibrated only down to $T \sim 1.1$, that provides
an estimate for the glass transition temperature $T_g$.
For $\gamma > 0$ the system becomes stationary and the viscosity is finite 
at all temperatures, even below $T_0$.

Very long simulation runs (up to $2 \cdot 10^9$ time steps) have been performed
to measure the {\sc pdf} of the entropy production rate at different temperatures 
along the line $\gamma=0.03$.
During each run, $p(t)$, given by Eq.~(\ref{pdef}),
has been measured on subsequent time intervals of duration $\tau$. 
From this dataset, the histograms of $\pi_\tau(p)$ and
the large deviation function $\zeta_\tau(p)$ defined in Eq.~(\ref{zetapdef}) are
obtained, following the procedure described in section~\ref{sec7:IV}.

\begin{figure}[t]
\includegraphics[width=.50\textwidth,angle=0]{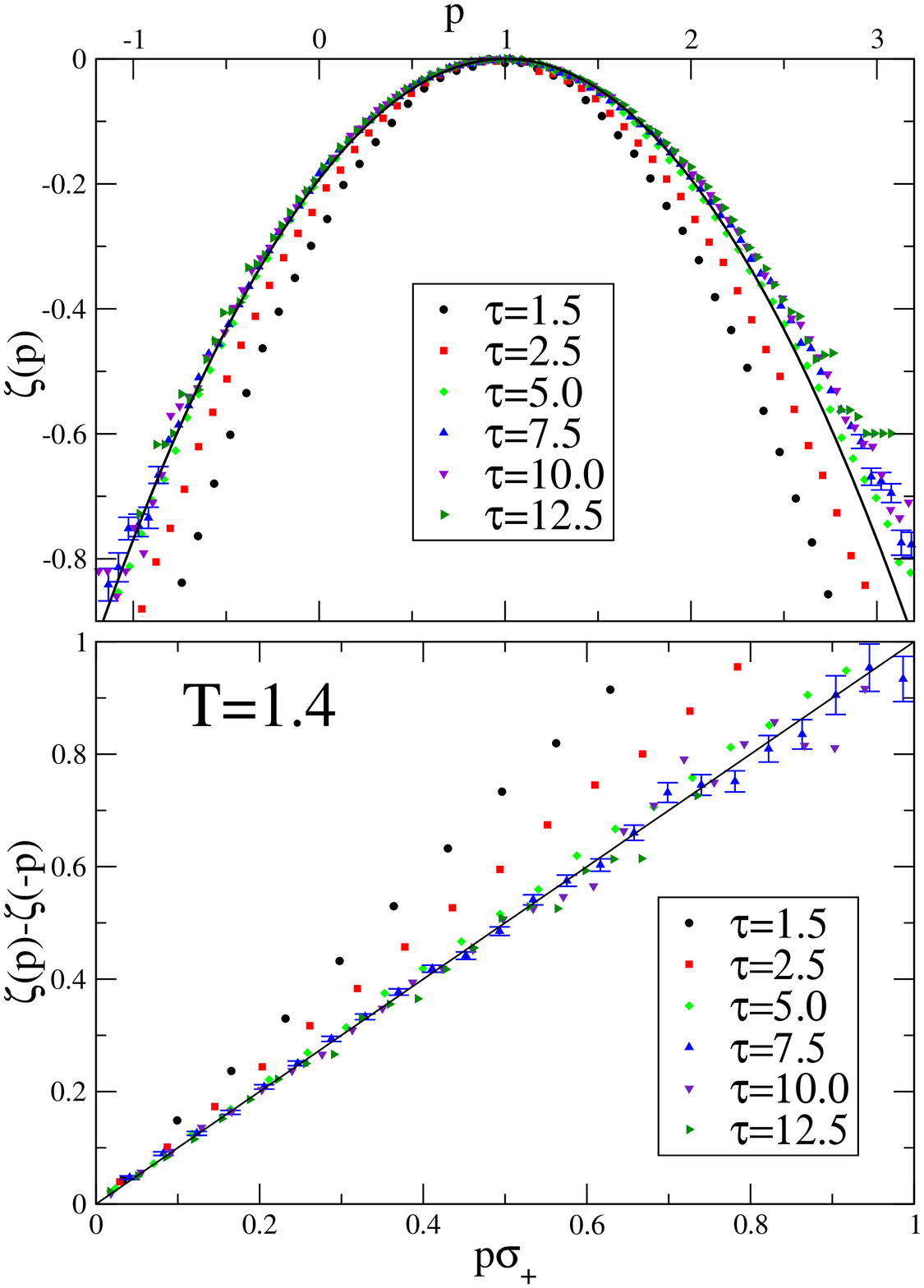}
\includegraphics[width=.50\textwidth,angle=0]{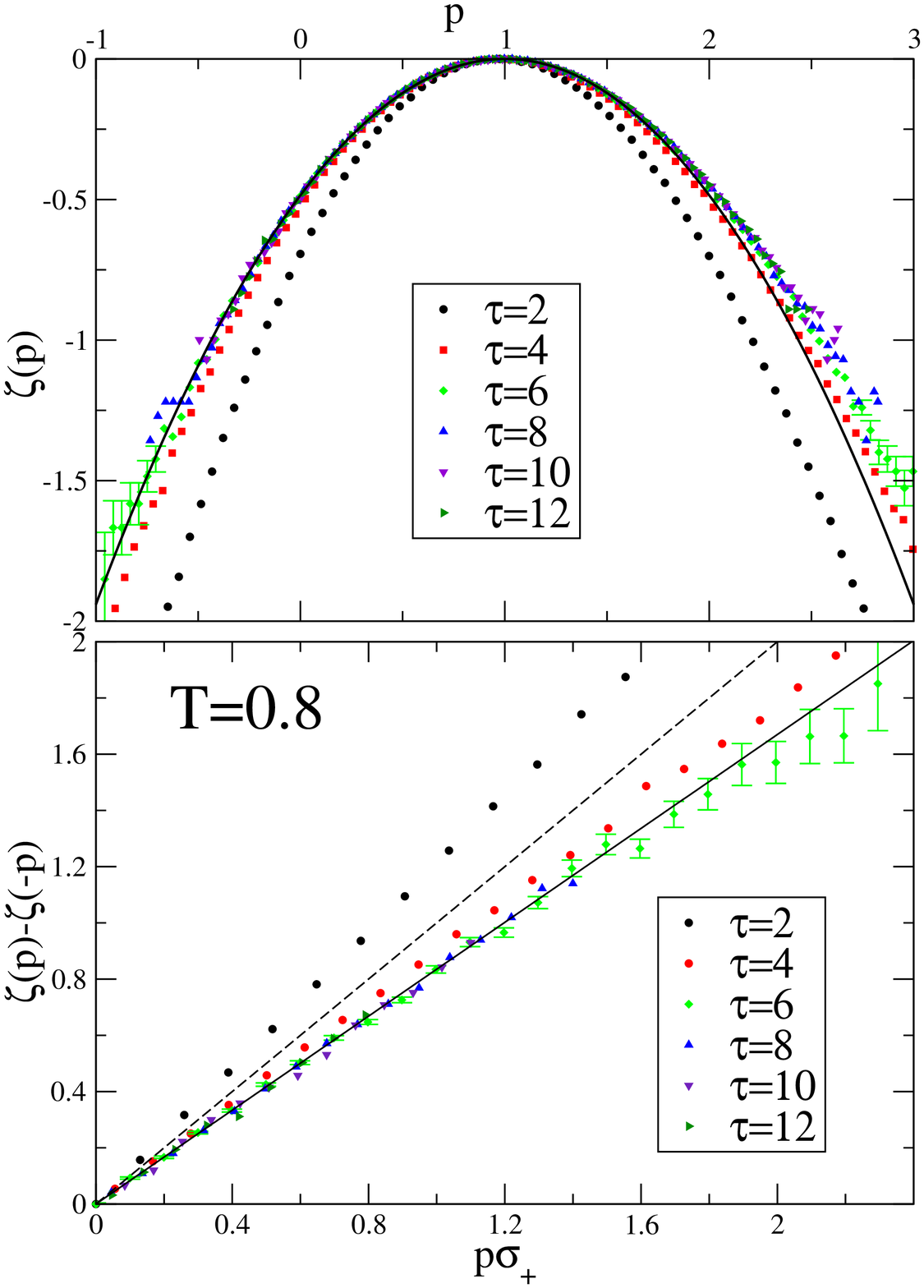}
\caption[The functions $\z(p)$ for the Lennard-Jones system at
$T=1.4 > T_g$ and $T=0.8 < T_g$]
{(Left) Top: the large deviation function 
$\zeta_\tau(p)=\tau^{-1} \log \pi_\tau(p)$ as a function of 
$p$ for different values of $\tau$ at 
$T=1.4 > T_g$ and $\gamma=0.03$.
Error bars are smaller than the symbols except on the tails: they are reported
only for $\tau=7.5$ to avoid confusion.
The line is a Gaussian fit to the data with $\tau > 5$ for $p \in [0,2]$.
Bottom: $\zeta_\tau(p)-\zeta_\tau(-p)$ as a function of $p\sigma_+$.
The FR predicts the plot to be a straight line with slope 1 (full line)
for large $\tau$.
(Right) Same plots for $\gamma=0.03$ and 
$T=0.8 < T_g$. In the lower panel the dashed line has slope
1 while the full line has slope $X = 0.83$.
}
\label{fig_7gal_7_2}
\end{figure}

In the left upper panel of Fig.~\ref{fig_7gal_7_2}, the functions
$\zeta_\tau(p)$ are reported for $\gamma=0.03$ and 
$T=1.4 > T_g$.
The asymptotic function $\zeta_\io(p)$ is obtained for $\tau \gtrsim 5$
and can be described by a simple Gaussian form, 
$\zeta_\io(p) = -(p-1)^2/2\delta^2$, even if small non-Gaussian tails are observed.
Note that this means that the finite time corrections discussed in
section~\ref{sec7:II} are not relevant here.
In the left lower panel of Fig.~\ref{fig_7gal_7_2}
$\zeta_\tau(p)-\zeta_\tau(-p)$ is reported as a function of $p\sigma_+$.
The FR, Eq.~(\ref{FR}), predicts the plot to be a straight line with slope 
1 for large $\tau$; this is indeed the case for $\tau \gtrsim 5$,
consistently with what has been found in the 
literature and in chapter~\ref{chap5}.

In the right upper panel of Fig.~\ref{fig_7gal_7_2}, 
the functions $\zeta_\tau(p)$ for
$\gamma=0.03$ and $T=0.8 < T_g$ are reported. In this case, the asymptotic
regime is reached for $\tau \gtrsim 6$; this value is not so different from
the one obtained in the previous case because the change in viscosity (and hence
in relaxation time) going from $T=1.4$ to $T=0.8$ is very small at this
value of $\gamma$
(see Fig.~\ref{fig_7gal_7_1}).
Also in this case the simple Gaussian form gives a good description 
of the data apart from the small non-Gaussian tails.
In the right lower panel of Fig.~\ref{fig_7gal_7_2},
$\zeta_\tau(p)-\zeta_\tau(-p)$ is reported as a function of $p \sigma_+$.
At variance to what happens for $T > T_g$, in this case the asymptotic
slope reached for $\tau \gtrsim 6$ is smaller than 1;
thus, the FR given by Eq.~(\ref{FR}) has to be 
generalized according to Eq.~(\ref{FTX}).
At this temperature, one has $X = 0.83 \pm 0.05$.

\begin{figure}[t]
\centering
\includegraphics[width=.55\textwidth,angle=0]{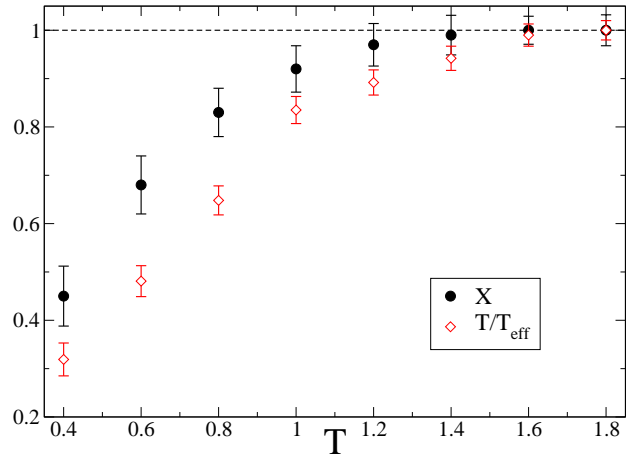}
\caption[FR violation factor $X$ as a function of $T$]
{The violation factor $X=T/\Th_{eff}$ that enters Eq.~(\ref{FTX}) 
(full circles)
and the ratio $T/T_{eff}$ from the generalized FDR (open diamonds)
as a function of the bath temperature $T$ for $\gamma=0.03$.}
\label{fig_7gal_7_4}
\end{figure}

In Fig.~\ref{fig_7gal_7_4}, the violation factor 
$X(T,\gamma=0.03)$ (full circles) is reported 
as a function of the temperature $T$; note that 
$X$ becomes smaller than unity exactly around $T_g \sim 1.1$, {\it i.e.} when the
viscosity starts to diverge strongly (see Fig.~\ref{fig_7gal_7_1}).
Below $T \sim 0.4$, $\sigma_+$ becomes so large that negative
fluctuations of $p$ are extremely rare and the violation factor is no longer
measurable. One can conclude that below $T_g$ the FR does not hold, and the data
are consistent with Eq.~(\ref{FTX}) where the coefficient
$X$ is temperature dependent below $T_g$ and equals $1$ above $T_g$.

Having checked the validity of Eq.~(\ref{FTX}), following \cite{Ga04}
and the analysis of section~\ref{sec:V}, 
one can define a nonequilibrium temperature as $\Th_{eff}=T/X$, such
that defining 
$\st(t) = W(t)/\Th_{eff} = X \sigma(t)$,
the FR for $\st$ is the usual one given by Eq.~(\ref{FR}).

To compare the temperature $\Th_{eff}$ with the effective temperature
$T_{eff}$ that enters the generalized fluctuation--dissipation relation,
one can measure $T_{eff}$ from the relation $T_{eff}=D/\mu$,
where $D$ is the diffusion constant and 
$\mu$ is the mobility of the particles in the considered steady state 
\cite{Lennard-FDT,BB00,BB02}.
This relation generalizes the usual equilibrium FDR $D=\mu T$;
to compute the diffusion constant and the mobility of type-A particles
one can follow the procedure of Di Leonardo {\it et al.}~\cite{Lennard-FDT}.
In Fig.~\ref{fig_7gal_7_4}, together with $X=T/\Th_{eff}$, the ratio $T/T_{eff}$ 
(open diamonds) is reported as a function of the bath temperature $T$.
The two ``effective'' temperatures have a similar qualitative behavior but do not 
coincide, as found in section~\ref{sec:V}, and the relation
$T < \Th_{eff} < T_{eff}$ holds.

\begin{figure}[t]
\centering
\includegraphics[width=.55\textwidth,angle=0]{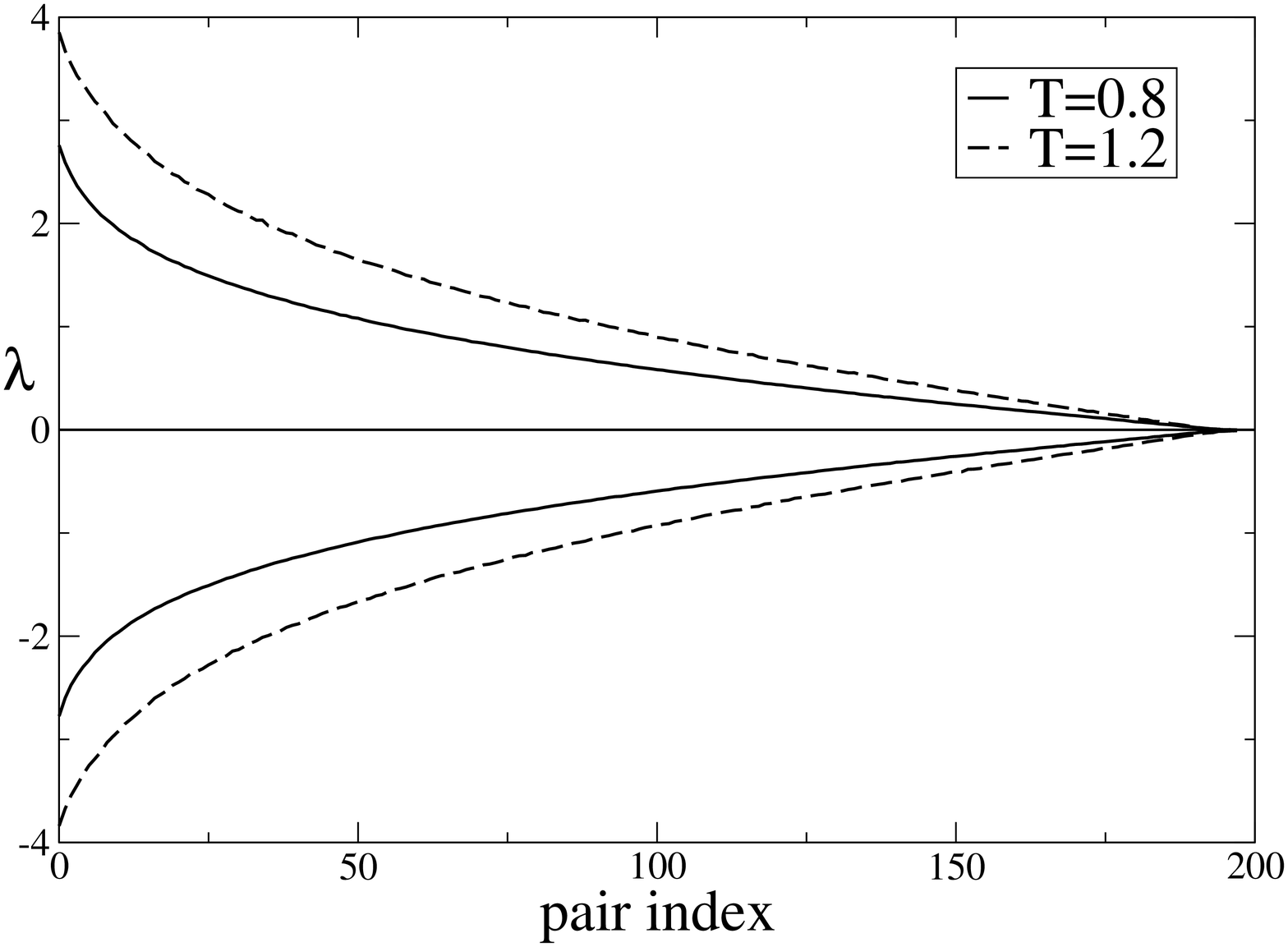}
\caption[Lyapunov exponents at $T=1.2 > T_g$ and $T=0.8 < T_g$]
{Lyapunov exponents for $\gamma=0.03$ and $T=0.8,1.2$. For both temperatures
each pair consists of one positive and one negative exponent.
}
\label{fig_7gal_7_5}
\end{figure}

To test the conjecture of
Bonetto and Gallavotti, see \cite{BG97} and 
section~\ref{sec5:discussion},
the Lyapunov spectra have been computed, see section~\ref{sec5:modelI}.
They
are reported in Fig.~\ref{fig_7gal_7_5}
for $\gamma=0.03$, $T=1.2> T_g$ and $T=0.8<T_g$. 
Unfortunately, no qualitative change in the spectrum 
is observed on crossing $T_g$ and in particular no pairs
of negative exponents are present above and below
$T_g$. Thus, it seems that the theory of \cite{BG97} 
does not apply to the model considered here below $T_g$.
Note however that this theory is developed under the assumption of
a strong chaoticity of the system, while below $T_g$ and for
$\gamma \sim 0$ the dynamics of the system becomes slower and slower.
Thus, the results presented here should not be regarded as 
invalidating the conjecture of
\cite{BG97}, but as indicating that the hypothesis of 
\cite{BG97} (essentially, the requirement of strong chaoticity) 
are not fulfilled by our model below $T_g$. This point requires further
investigation.

%%%%%%%%%%%%%%%%%%%%%%%%%%%%%%%%%%%%%%%%%%%%%%%%%%%%%%%%%%%%%%%%%%%%%%%%%%%%%%%%%%%%%% 
%%%%%%%%%%%%%%%%%%%%%%%%%      CONCLUSIONS      %%%%%%%%%%%%%%%%%%%%%%%%%%%%%%%%%%%%%% 
%%%%%%%%%%%%%%%%%%%%%%%%%%%%%%%%%%%%%%%%%%%%%%%%%%%%%%%%%%%%%%%%%%%%%%%%%%%%%%%%%%%%%% 
 
\section{Discussion} 
\label{sec:conclusions} 

An extension of the fluctuation theorem of Gallavotti and
Cohen to open stochastic systems that are not able to equilibrate with
their environments when relaxing unperturbed has been discussed. 

The simplest example at hand has been used to test several generalized
fluctuation formulas: a Brownian particle in a confining potential
coupled to non-trivial external baths with different time-scales and
temperatures. Independently of the form of the potential energies, due
to the coupling to the complex environment, the particle is not
able to equilibrate. Its relaxational dynamics is characterized by an
{\it effective temperature}, defined via the modification of the
fluctuation-dissipation relation between spontaneous and induced
fluctuations. When no separation of time-scales can be
identified in the bath, the effective temperature is a non-trivial
function of the two times involved. Instead, when the bath evolves in different
time-scales each characterized by a value of a temperature, the
two-time dependent effective temperature is a piece-wise function that
actually takes only these values, each one characterizing the dynamics
of the particle in a regime of times.

The fluctuations of entropy production in a numerical
simulation of a Lennard-Jones like fluid above and below the glass transition
temperature $T_g$ have also been studied, obtaining results that partially
confirm the theoretical analysis. However, many points are still open
and require a much deeper numerical investigation.

Several authors discussed the possibility of introducing the effective
temperature in the fluctuation theorem to extend its domain of
applicability to glassy models driven by external
forces~\cite{Se98,CR04,SCM04,Sasa}. After summarizing the results of this
chapter, it will be discussed
how they compare to the proposals and findings presented here.
  
\subsection{Summary of results}

Different definitions of entropy production rate that are not equivalent when
the {\it effective temperature} is not trivially equal to the ambient
temperature have been discussed. It was found that:

\begin{enumerate} 

\item 
The {\sc pdf} of $\su_t = W_t/\Th$, where $W_t$ is the {\it power 
dissipated by the external force} and $\Th$ is a free parameter with 
the dimensions of a temperature, does not satisfy the fluctuation 
theorem in general.

The large deviation function, $\z_\Th(p)$, still shows some 
interesting features revealing the existence of an effective temperature.
When the bath has, say, two components acting on different time scales
and with different temperatures, the function
$[\z_\Th(p)-\z_\Th(-p)]/\su_+$ may have different slopes corresponding
to these two temperatures, one at small $p$ and the other at large
$p$. The separation of time-scales of the bath 
translates into a separation of 
scales in the function $[\z_\Th(p)-\z_\Th(-p)]/\su_+$.

When the time scales of the baths are not separated, and one observes the 
large deviation function for not too large values of $p$ only, the fluctuation
theorem is verified approximately if $\Th$ is suitably chosen. Note that the
temperature $\Th_{eff}$ defined in this way is not equal to the
effective temperature $T_{eff}$ that enters the modified fluctuation--dissipation
relation in this case.

Instead, when the time-scales are well separated, 
the two scales in the large deviation 
function are clearly visible and a single fitting parameter is not sufficient
to make the fluctuation theorem hold.

\item 
The {\sc pdf} of $\st_t$ defined substituting the {\it frequency dependent}
effective temperature to the constant $\Th$ in the previous definition:
\beq \label{concl-seff}
\s^{eff}_\T= 
 \frac{\D\o}{2\p} 
\sum_{n=-\io}^\io
 \frac{-i \o_n r_\a(\o_n) h_\a(\o_n)}{T_{eff}(\o_n)} =
 \int_{-\T/2}^{\T/2} dt  
\int_{-\T/2}^{\T/2} dt' \,
T^{-1}(t-t')
 \dot r_\a(t) 
 h_\a[\vec r(t')] 
\ ,
\eeq
with $T^{-1}(t)$ the Fourier transform of  
$1/ T_{eff}(\o)$, the effective temperature of the 
relaxing system, see Eq.~(\ref{eq:Teff-def}), 
always verifies the fluctuation theorem, as was shown analytically
for the harmonic potential and numerically for the quartic one.
No requirements on the 
characteristics of the bath are needed. $\st_t$ reduces  
to $\su_t$ when there is only one equilibrated bath. 

\item
The additional term $\s^V_t$ which is obtained from the Lebowitz--Spohn
procedure, see equation~(\ref{EPR}), is not relevant for the applications
discussed in this chapter, as it vanishes identically for harmonic potentials 
and for any potential in the adiabatic approximation. 
However it is relevant when the potential is
nonlinear and the time scales are not separated. Its detailed
investigation is left for future work.

\item
If two time scales are present in the dynamics of a system and if the
applied perturbation is periodic with frequency $\O < 1/\t_s$, $\t_s$
being the largest relaxation time, the {\sc pdf} of the power
dissipated over a (large) number of cycles verifies the fluctuation
relation with temperature $T_s=T_{eff}(\O)$. This is probably the
easiest way of detecting the effective temperature by mean of the
fluctuation relation.

\item
These results should apply to driven glassy systems
as discussed in section~\ref{sec:VI}.
It was shown, in a numerical simulation of a binary Lennard--Jones mixture,
that below $T_g$ the Fluctuation Relation does not
hold for $\su$; the data --obtained in a situation where the time
scales are not well separated-- are consistent with the statements
of item 1. above.
The conjecture of Bonetto and Gallavotti that relates the factor $X$ 
in Eq.~(\ref{FTX}) to properties of the phase space of the considered system
was tested; unfortunately, the
data are not consistent with this conjecture, suggesting that the
violation of the FR is, in the case studied above, 
of different origin than that proposed
in \cite{BG97}. 
This point also requires additional investigation.

\end{enumerate}

Models like the one discussed here have been recently 
investigated \cite{CK00,AG04,PM04,Po04} to describe the dynamics of 
Brownian particles in complex media such as glasses, granular matter, etc. 
Brownian particles are often used as probes in order to study the 
properties of the medium (\eg in Dynamic Light Scattering or Diffusing 
Wave Spectroscopy experiments).  Moreover, confining potentials for 
Brownian particles can be generated using laser beams \cite{Gr03} and 
experiments on the fluctuations of the power dissipated in such 
systems are currently being performed \cite{AG04,WSMSE02}. 

\subsection{Effective temperatures}

It is important to summarize the different definitions of effective temperature
considered above and the relations between them. The effective temperature
in the frequency domain is defined by equation~(\ref{eq:Teff-def}) as a property of the bath
which can also be measured from the ratio between correlation and response functions
in the frequency domain. As discussed above, {\it the same} effective temperature
enters the correct definition of entropy production rate {\it in the frequency domain},
see equation~(\ref{concl-seff}).
Thus, experiments working in the frequency domain should observe the same effective
temperature from the fluctuation--dissipation relation and from the fluctuation relation.

In the time domain the situation is slightly more complicated. On one hand, the effective
temperature obtained from the fluctuation--dissipation relation {\it in the time domain},
defined for example by equation~(\ref{eq:Tefftime-def}),
is {\it not} the Fourier transform of $T_{eff}(\o)$. A convolution with the correlation
function is involved in the relation between $T_{eff}(\o)$ and $T_{eff}(t)$.
On the other hand, the effective temperature $T^{-1}(t)$ entering the entropy production
is exactly the Fourier transform of $1/T_{eff}(\o)$, see again equation~(\ref{concl-seff}).
This can give rise to ambiguities when working in the time domain.

Most of these ambiguities disappear as long as the time scales in the problem are well 
separated. In this case, on each time scale a well defined effective temperature can be
identified, and this temperature enters both the fluctuation--dissipation relation and the
fluctuation relation:
see e.g. the curve for $\t=250$ in Fig.~\ref{fig:3b} and the expression of $\st$ in 
the adiabatic approximation, equation~(\ref{sigma3adiabatic}).
This is essentially related to the validity of the adiabatic approximation discussed in 
section~\ref{sec:adiabatic}.

The difference is relevant when the time scales of the two baths are not well separated,
and a single effective temperature cannot be identified, see the curve for $\t=1$ in
Fig.~\ref{fig:3b}. In this case, it was found that the fluctuation relation holds with
--approximately-- a single effective temperature $\Th_{eff}$ but this temperature 
{\it is not clearly related} to the fluctuation--dissipation temperature 
{\it in the time domain}. This was also observed in numerical simulations on
Lennard--Jones systems.
Still, when moving to the frequency domain, the two effective temperatures should coincide.

Let us remark again that, when applying these results to real glassy systems in finite
dimension, one should take care of the possibility that the effective temperature has
large space fluctuations due to the heterogeneity of the dynamics~\cite{CCCIK03,Cu04}. 
The extension of the results presented here to such a situation is left for future work.

\subsection{Comparison with previous works}

Several proposals to introduce the effective temperature
into extensions of the fluctuation theorem
appeared in the literature. 

Sellitto studied the fluctuations of entropy production in a driven
lattice gas with reversible kinetic constraints~\cite{Se98}. 
When coupling this system to an external particle reservoir 
with chemical potential $\mu$, a dynamic crossover from a 
fluid to a glassy phase is found around $\mu_d$. The glassy
nonequilibrium phase is characterized by a violation of the fluctuation
dissipation theorem in which the parametric relation 
between global integrated response and displacement 
yields a line with slope $\mu_{eff}$~\cite{KPS}.
 
One drives this (possibly already out of equilibrium) 
system by coupling two adjacent layers 
of the three dimensional periodic cube
to particle reservoirs at different
chemical potentials, $\mu_+$ and $\mu_-$. 
The former is allowed to assume values corresponding to
the glassy phase, $\mu_+ > \mu_d$, while $\mu_-$ is always 
below $\mu_d$.
The results of 
the Montecarlo simulation are consistent with a 
generalized form of the fluctuation theorem:
\begin{equation}
\sigma_\t = J_\tau (\mu_{eff}-\mu_-)
\; ,
\label{eq:gen-driven-lattice}
\end{equation} 
where $\sigma_\t$ is the entropy production, $J_\tau$ is the particle
current in the direction of the externally imposed chemical
potential gradient averaged over a time-interval of duration $\tau$;
$\mu_{eff}$ is an effective chemical potential and $\mu_-$ is the
chemical potential of one of the layers.  When the chemical potentials
of the two reservoirs are in the fluid phase, $\mu_{eff}=\mu_+$ and
the usual fluctuation relation holds. Instead, when $\mu_+$ is in the
glassy phase, Sellitto found that Eq.~(\ref{eq:gen-driven-lattice})
holds with $\mu_{eff}$ taking the value appearing in the violation of
fluctuation-dissipation theorem in the aging regime of the {\it
undriven} glassy system at $\mu_+$.

The formula (\ref{eq:gen-driven-lattice}) differs from the ones that
were found to describe the oscillator problem in that in the case
studied here,
when translating from temperature to chemical potential, the 
full time-dependent $\mu(t)$ enters. Strictly,
this improved definition should also apply to the 
lattice gas model. 
However in the case studied by Sellitto the fast dynamics is an ``intra-cage''
dynamics that likely does not contribute to the current. This is a case
in which the perturbation does not produce dissipation at high frequency
so that the difference arising from $\mu(t)\neq \mu_{eff}$ should be tiny 
in this case (see section~\ref{sec:XI}).

More recently, 
Crisanti and Ritort~\cite{CR04} found that the probability
distribution function of the fluctuations of heat exchanges, $Q$,
between an aging random orthogonal model in its `activated regime' (a
long-time regime in which the energy-density decays as a logarithm of
time) and the heat bath is rather well described by a stationary
Gaussian part and a waiting-time dependent exponential tail towards
small values of $Q$.  Assuming that these events are of two types
(`stimulated' and `spontaneous') they proposed to fit the ratio
between the {\sc pdf} of positive and negative `spontaneous' $Q$'s in the
form of a fluctuation theorem, {\it i.e.} to be proportional to
$e^{-2Q/\lambda}$, and relate $\lambda$ to the effective temperature
of the fluctuation-dissipation relation. They found good agreement.
Crisanti, Ritort and Picco
are currently performing simulations to test this hypothesis in
Lennard-Jones mixtures~\cite{Marco}.

Another development is an attempt to generalize the situation considered
by Crooks. He considered a problem that {\it starts from equilibrium in zero
field} and evolves according to some stochastic dynamic rule in the
presence of an arbitrary applied field~\cite{Crooks} and  found that
the ratio between the probability of a trajectory and its
time-reversed one is given by $e^{-\beta \int_0^{t_{max}} dt h(t)
\dot O(t)}$ with $h(t)$ the time-dependent external field that couples
linearly to the observable $O$. For simplicity, one can focus on
$O=\phi$ with $\phi$ a scalar field characterizing the system. In
\cite{SCM04} the extension of this relation to the
non-equilibrium `glassy' case was conjectured. Separating the external
fields $h$ and $\phi$ in their fast and slow
components~\cite{Zannetti}, $h=h_f+h_s$ and $\phi=\phi_f+\phi_s$, one
then proposes that the {\sc pdf}s of the trajectories of the 
slow components satisfy a relation similar to Crooks'
with the temperature replaced by the effective temperature (for a
glassy non-equilibrium system with two correlation
scales~\cite{CK93}).

Finally, it is worth to mention the work of Sasa~\cite{Sasa} where he introduces  
an effective temperature in his definition of entropy production for 
the Kuramoto-Sivashinsky equation.
 
%%%%%%%%%%%%%%%%%%%%%%%%%%%%%%%%%%%%%%%%%%%%%%%%%%%%%%%%%%%%%%%%%%%%%%%%%%%%%%%%%%%%%%%%%%%% 
%%%%%%%%%%%%%%%%%%%%%%%%%%% APPENDIX %%%%%%%%%%%%%%%%%%%%%%%%%%%%%%%%%%%%%%%%%%%%%%%%%%%%%%% 
%%%%%%%%%%%%%%%%%%%%%%%%%%%%%%%%%%%%%%%%%%%%%%%%%%%%%%%%%%%%%%%%%%%%%%%%%%%%%%%%%%%%%%%%%%%% 

\section{Appendix: Dirichlet boundary conditions for the white bath} 
\label{app:A} 
 
A second possibility to calculate the functional integral in Eq.~(\ref{AAA1}) 
is to impose Dirichlet boundary conditions $a(-\T/2)=a(\T/2)=0$. 
However, in this case it is possible to calculate $z(\l)$ only for $m=0$. 
The distribution of $a_t$ is obtained substituting $\r_\o = D(\o) a_\o$ in Eq.~(\ref{AAA3}): 
\beq \begin{split}
{\cal P}[a_t]&= \exp \left[ -\frac{1}{2 \gamma T}  
\int_{-\infty}^\infty \frac{d\omega}{2\pi} \ a_\omega  |D(\o)|^2 \bar a_\omega \right] 
\\ &= 
\exp \left[ -\frac{1}{2 \g T}  
\int_{-\infty}^\infty dt \ a_t   
\left( k^2 + \epsilon^2 - 2 i \epsilon \g \frac{d}{dt} -\g^2 \frac{d^2}{dt^2}\right) 
\bar a_t \right] \ . 
\end{split}\eeq 
From Eq.~(\ref{AAA1}) 
\beq 
\langle \exp[- \lambda \sigma_\T ] \rangle = 
{\cal N}^{-1} \int da_t  
\exp \left[ -\frac{1}{2 \g T}  
\int_{-\infty}^\infty dt \ a_t   
\left( k^2 + \epsilon^2 - 2 i \epsilon \g [1 - 2\lambda \chi_\T(t)]  
\frac{d}{dt} -\g^2 \frac{d^2}{dt^2}\right) 
\bar a_t \right] \ , 
\eeq 
where $\chi_\T(t)$ is the characteristic function of $t\in[-\T/2,\T/2]$. 
At the leading order in $\T$, as the correlation function 
of $a_t$ decays exponentially on a time scale $\t_0=\g k^{-1}$, one can integrate out 
the portion of the trajectory that is outside the interval $[-\T/2,\T/2]$ 
both in the numerator and the denominator, to obtain 
\beq 
\langle \exp[- \lambda \sigma_\T ] \rangle = 
 {\cal N}^{-1} \int da_t  
\exp \left[ -\frac{1}{2 \g T}  
\int_{-\T/2}^{\T/2} dt \ a_t   
\left( k^2 + \epsilon^2 - 2 i \epsilon \g (1 - 2\lambda)  
\frac{d}{dt} -\g^2 \frac{d^2}{dt^2}\right) 
\bar a_t \right] \ . 
\eeq
Then one has to find the eigenvalues of the operator appearing 
in the integral. This corresponds to find the solution of the equation 
\beq 
J \bar a_t =  
\left(k^2 + \epsilon^2 -2i\epsilon \g(1 - 2\lambda)\frac{d}{dt}  -  
\g^2 \frac{d^2}{dt^2}\right) \bar a_t = E \bar a_t 
\eeq 
with boundary conditions $\bar a(\T/2)=\bar a(-\T/2)=0$.  
Note that the operator $J$ is Hermitian, thus the eigenvalues are real; 
they are given by the following expression: 
\beq 
E_n(\lambda)=k^2 + 4\epsilon^2 \lambda (1-\lambda) +\g^2 \frac{\pi^2 n^2}{\T^2} 
\eeq 
with $n = 0,1,\cdots$. 
For each $n$ the integration is performed on one complex variable and 
one gets 
\beq 
\langle \exp[- \lambda \sigma_\T ] \rangle = 
 {\cal N}^{-1} \int da_t  
\exp \left[ -\frac{1}{2 \g T}  
\int_{-\T/2}^{\T/2} dt \ a_t   
J \bar a_t \right] = \prod_{n=0}^\infty \frac{E_n(0)}{E_n(\lambda)} 
\eeq 
recalling that the constant ${\cal N}$ is simply the numerator calculated 
in $\lambda=0$. Finally one obtains, defining $\o=n\pi/\T$, 
\beq 
\label{phi1bagno} 
z(\lambda)= \lim_{\T \rightarrow \infty} \frac{1}{\T}
\sum_{n=0}^{\infty} \log \frac{E_n(\lambda)}{E_n(0)} = \int_0^\infty
\frac{d\o}{\pi} \log \left[ 1 + \frac{ 4\epsilon^2 \lambda (1-\lambda) }{
k^2 + \g^2 \o^2 } \right] \eeq 
The latter expression verifies
obviously the fluctuation theorem. Moreover, in the $m=0$ case
Eq.~(\ref{phi1bagnoFourier}) is equal to Eq.~(\ref{phi1bagno}), as one
can check using suitable changes of variable in the integral.
In this simple
case, $\zeta(p)$ can be computed exactly. Starting from
Eq.~(\ref{phi1bagno}) one has 
\beq z'(\lambda)=\int_0^\infty
\frac{d\o}{\pi} \frac{4 \epsilon^2 (1-2\lambda)} {\g^2 \o^2 + k^2+ 4
\epsilon^2 \lambda(1-\lambda)} = \frac{2\epsilon^2 (1-2\lambda)} {\g
\sqrt{k^2+ 4 \epsilon^2 \lambda(1-\lambda)}} \ , \eeq and, recalling
that $z(0)=0$, \beq z(\lambda)=\int_0^\lambda d\mu \ z'(\mu)
= \g^{-1} \big[ \sqrt{k^2+ 4 \epsilon^2 \lambda(1-\lambda)} - k \big] \
.  \eeq 
The function $\zeta(p)$ is defined by 
\beq \zeta(p) =
\min_\lambda [\lambda p \s_+ - z(\lambda)] = \lambda^* p \s_+ -
z(\lambda^*) \ , 
\eeq 
where $\s_+=2\epsilon^2/(\g k)$ and $\lambda^*$ is
defined by $z'(\lambda^*)= p \s_+$; hence, 
\beq p= \frac{ k
(1-2\lambda^*)}{\sqrt{k^2+ 4 \epsilon^2 \lambda^* (1-\lambda^*)}}
\hskip10pt \Rightarrow \hskip10pt \lambda^* = \frac{1}{2} \left[ 1 - p
\sqrt{\frac{\epsilon^2 + k^2} {\epsilon^2 p^2 + k^2}} \right] \ , 
\eeq 
and finally 
\beq \zeta(p)=\g^{-1} \left\{ k + \frac{\epsilon^2 p}{k} \left[
1 - p \sqrt{\frac{\epsilon^2 + k^2} {\epsilon^2 p^2 + k^2}} \right] - k
\sqrt{\frac{\epsilon^2 + k^2} {\epsilon^2 p^2 + k^2}} \right\} \ .  
\eeq
From the latter expression it is easy to verify that 
\beq
\zeta(p)-\zeta(-p) = \frac{2\epsilon^2 p}{k} = p \s_+ \ , 
\eeq 
as stated
by the FT.  Defining $\t_0 = \g/k$, the relaxation time of the
correlation function of $a_t$, and $\s_0 = \s_+ \t_0/2 = \epsilon^2/k^2$,
the (adimensional) entropy production over a time $\t_0/2$, one obtains
\beq 
\z(p)=\t_0^{-1} \left[ 1 + p \s_0 - \sqrt{(1+\s_0)(1+p^2 \s_0)}
\right] \ .  
\eeq
 
%%%%%%%%%%%%%%%%%%%%%%%%%%%%%%%%%%%%%%%%%%%%%%%%%%%%%%%%%%%%%%%%%%%%%%%%%%%%%%%%%%%%%%%%%%%%%%
%%%%%%%%%%%%%%%%%%%%%%%%%%%%%%%%%%%%%%%%%%%%%%%%%%%%%%%%%%%%%%%%%%%%%%%%%%%%%%%%%%%%%%%%%%%%%%

\section{Appendix: Fluctuation theorem for many equilibrium baths at different temperature}
\label{app:III} 
 
Here the function $z(\l)$ will be computed in the case in which the driven
oscillator is coupled to $N$ colored baths with generic memory functions and
in equilibrium at different temperatures. The violation of the
fluctuation-dissipation theorem for the relaxing particle in such an
environment was discussed in section~\ref{subsec:noneq-baths}.
As discussed there, the equations are mathematically equivalent to the ones
discussed in section~\ref{sec:IV}; thus the strategy as 
well as many details of the calculation are the same as in 
sections~\ref{sec:II} and \ref{sec:IV}.

The equations of motion are  
\beq 
\label{motionNbaths} 
m \ddot a_t +  
\sum_{i=1}^N \Iint ds \ g_i(t-s) \dot a_s  
= -\kappa a_t +  
\sum_{i=1}^N \rho_{it} \ , 
\eeq 
with $\kappa = k - i\epsilon$. The thermal noises satisfy 
\beq 
\label{bathsdef} 
\begin{split} 
&\langle \rho_{it} \rho_{j0} \rangle =  
\langle \bar\rho_{it} \bar\rho_{j0} \rangle = 0 \ ,\\ 
&\langle \rho_{it} \bar\rho_{j0} \rangle =  
\delta_{ij} T_i \nu_i(t) \ . 
\end{split} 
\eeq  
By causality, the functions $g_i(t)$ must vanish for $t <0$.  As 
the baths are in equilibrium at temperature $T_i$, the 
functions $\nu_i(t)$ and $g_i(t)$ are related by 
Eq.~(\ref{FDTbathTD}):  
\beq 
\label{f&g} 
\begin{split} 
& \nu_i(t) = T_i [ g_i(t) + g_i(-t) ] = T_i g_i(|t|) \ , \\ 
& T_i g_i(t) = \theta(t) \nu_i(t) \ . 
\end{split} 
\eeq 
In the frequency domain Eq.~(\ref{motionNbaths}) becomes 
\beq 
a_\omega = \frac{\sum_i \rho_{i\omega}} 
{-m \omega^2 + \kappa - i\omega \sum_i g_i(\omega)} 
\equiv \frac{\sum_i \rho_{i\omega}}{D(\omega)} \ , 
\eeq 
where $D(\omega) = -m \omega^2 + \kappa - i\omega \sum_i g_i(\omega)$. 
 
The dissipated power is given by  
\beq  
\frac{dH}{dt} = 2\epsilon \; \im \dot a_t \bar a_t -2\re 
\sum_i \int_{-\infty}^\infty ds \ g_i(t-s) \dot a_t \dot{\bar a}_s + 2 
\re \sum_i \dot a_t \bar \rho_{it} = W_t - \sum_i \widetilde W_{it} 
\;  ,  
\eeq  
where as in the previous cases $W_t = 2\epsilon \im \dot a_t \bar 
a_t $ is the power injected by the external force and $\widetilde 
W_{it} = 2\re \int_{-\infty}^\infty ds \ g_i(t-s) \dot a_t \dot{\bar 
z}_s - 2 \re \dot a_t \bar \rho_{it}$ is the power extracted by the 
$i$-th bath. 

The first definition of entropy production rate, Eq.~(\ref{sigma1}),
gives (in the following, 
$\frac{\D\o}{2\pi} \sum_{n=-\io}^\io 
\rightarrow \int_{-\io}^\io \frac{d\o}{2\pi}$ as the error is $O(1)$ 
for $\T \rightarrow \io$, see section~\ref{sec:II}): 
\beq 
\su_\T = -\int_{-\infty}^\infty \frac{d\omega}{2\pi}  
\frac{2\epsilon\omega|a_\omega|^2}{\Th} \ . 
\eeq 
Substituting $a_\o = \sum_i \r_{i\o} / D(\o)$, one obtains 
\beq 
\langle \exp[- \lambda \su_\T ] \rangle = {\cal N}^{-1} 
\int d\rho_{i\omega} \  
\exp \left[ - \int_{-\infty}^\infty \frac{d\omega}{2 \pi} 
 \sum_{ij} \rho_{i\omega} A_{ij}^\lambda(\omega) \bar \rho_{j\omega} 
\right] \ , 
\eeq 
where $A^\lambda(\omega)$ is a $N\times N$ {\it real} matrix which elements are 
given by 
\beq 
A_{ij}^\lambda(\omega) = 
\frac{\delta_{ij}}{T_i \nu_i(\omega)} - \frac\lambda{|D(\omega)|^2} 
\frac{2\epsilon\omega}{\Th} \ . 
\eeq 
Then, 
\beq 
z_\Th(\lambda)=\lim_{\T\rightarrow \infty} \T^{-1} \log 
\prod_{n=-\infty}^\infty \frac{\det A^\lambda(\omega_n)} 
{\det A^0(\omega_n)} 
= \int_{-\infty}^\infty \frac{d\omega}{2 \pi} \  
\log \left[ 
\frac{\det A^\lambda(\omega)}{\det A^0(\omega)} \right] \ . 
\eeq 
The determinant of a matrix of the form $A^\l_{ij}=c_i^{-1} \d_{ij} + \l b$ satisfies
the relation
\beq
\frac{\det A^\lambda}{\det A^0} = 1 + \l b \sum_i c_i \; ;
\eeq
one finally obtains 
\beq
\label{phi1Nbaths}
z_\Th(\lambda)= 
 \int_{-\infty}^\infty \frac{d\omega}{2 \pi} \  
\log \left[ 1 - \frac{2 \epsilon \lambda \omega \sum_i \frac{T_i}{\Th} \nu_i(\omega)}  
{|D(\omega)|^2}\right] \ . 
\eeq 
In general, {\it it does not exist a choice of $\Th$ such that $z_\Th(\l)$ 
verifies the fluctuation theorem}, \ie 
$z_\Th(\l) \neq z_\Th(1-\l)$. 
 
For the second definition, given by Eq.~(\ref{sigma3}), the 
computation is identical to the one of the previous section with the 
substitution $\Th \rightarrow T_{eff}(\o)$, where $T_{eff}(\omega)$ is 
given by Eq.~(\ref{TeffNbaths}). The result is then  
\beq 
z_{eff}(\lambda) =  
\int_{-\infty}^\infty \frac{d\omega}{2 \pi} \ \log 
\left[ 1 - \frac{2 \epsilon \lambda \omega \sum_i T_i \nu_i(\omega)} 
{T_{eff}(\omega) |D(\omega)|^2}\right]= \int_{-\infty}^\infty 
\frac{d\omega}{2 \pi} \ \log \left[ 1 - \frac{2 \epsilon \lambda \omega 
\sum_i \nu_i(\omega)} {|D(\omega)|^2} \right] \ .
\eeq  
Observing that 
\beq 
\label{Dsymm} 
\begin{split} 
&D(-\omega)=\overline{D(\omega)} - 2 i\epsilon \ , \\ 
&|D(-\omega)|^2 = |D(\omega)|^2 - 2 \epsilon \omega \sum_i \nu_i(\omega) \ ,
\end{split} 
\eeq 
and using the same trick already used in section~\ref{sec:II}, it is easy 
to show that $z_{eff}(\lambda)=z_{eff}(1-\lambda)$.

%%%%%%%%%%%%%%%%%%%%%%%%%%%%%%%%%%%%%%%%%%%%%%%%%%%%%%%%%%%%%%%%%%%%%%%%%%%%%%%%%
%%%%%%%%%%%%%%%%%%%%%%%%%%%%%%%%%%%%%%%%%%%%%%%%%%%%%%%%%%%%%%%%%%%%%%%%%%%%%%%%%
 
\section{Appendix: Entropy production of the thermal baths}
\label{app:B} 

A different definition of entropy production rate based on the
power {\it extracted by the thermal bath} instead of the one injected by the driving
force will be discussed in this section. 
The two differ by a total derivative if there is only one bath, see Eq.~(\ref{eq7:dHdt}),
so their asymptotic distributions should be identical at least for $|p|\leq 1$,
see \cite{VzC04} and section~\ref{sec4:singularities}.

If there are many baths equilibrated at different temperature, 
the study of the entropy production extracted by each bath
allows to separate the different contributions to the total entropy production
weighting each one with the right temperature, \ie
one can define the entropy production rate of the baths
as
\beq
\label{sigma2} 
\sd_t = \sum_{i=1}^N \frac{\widetilde W_{it}}{T_i} 
\; . 
\eeq  
This quantity takes into account heat exchanges between the 
baths, and its average value does not vanish at $\epsilon = 0$, as will be 
shown in the following. 
 
To compute $z_{baths}(\l)$, one rewrites Eq.~(\ref{sigma2}) as: 
\beq 
\begin{split} 
\sd_\T &=  
\int_{-\T/2}^{\T/2} dt \ \sd_t =  
\int_{-\infty}^\infty \frac{d\omega}{2\pi} \re \sum_i  
\frac2{T_i} \Big[ 
\omega^2 |a_\omega|^2 g_i(\omega) + i \omega a_\omega \bar \rho_{i\omega}  
\Big]\\ 
=&\int_{-\infty}^\infty \frac{d\omega}{2\pi}  
\left[ \frac{\omega^2 \big|\sum_i \rho_{i\omega} \big|^2  
\sum_j \frac{ \nu_j(\omega)}{T_j} }{|D(\omega)|^2}  
+ \sum_{ij} \rho_{i\omega} \bar\rho_{j\omega}  
\left( \frac{i\omega}{D(\omega) T_j} - \frac{i\omega}{\overline{D(\omega)}T_i} 
\right)\right] \ . 
\end{split} 
\eeq 
Defining the functions  
\beq 
\begin{split} 
&p(\omega)=i \omega  \overline{D (\omega)} \ , \\ 
&F(\omega)=\omega^2 \sum_i\frac{\nu_i(\omega)}{T_i} \ , 
\end{split} 
\eeq 
one obtains 
\beq 
\langle \exp[- \lambda \sd_\T ] \rangle = {\cal N}^{-1} 
\int d\rho_{i\omega} \  
\exp \left[ - \int_{-\infty}^\infty \frac{d\omega}{2 \pi} 
 \sum_{ij} \rho_{i\omega} A_{ij}^\lambda(\omega) \bar \rho_{j\omega} 
\right] \ , 
\eeq 
where $A^\lambda(\omega)$ is a $N\times N$ matrix which elements are 
given by 
\beq 
A_{ij}^\lambda(\omega) = \overline{A_{ji}^\lambda(\omega)} = 
\frac{\delta_{ij}}{T_i \nu_i(\omega)} + 
\frac\lambda{|D(\omega)|^2}\left[ F(\omega) + 
\frac{p(\omega)}{T_j}+\frac{\overline{p(\omega)}} 
{T_i} \right] \ . 
\eeq 
Then, 
\beq 
z_{baths}(\lambda)=\lim_{\T\rightarrow \infty} \T^{-1} \log 
\prod_{n=-\infty}^\infty \frac{\det A^\lambda(\omega_n)} 
{\det A^0(\omega_n)} 
= \int_{-\infty}^\infty \frac{d\omega}{2 \pi} \  
\log \left[ 
\frac{\det A^\lambda(\omega)}{\det A^0(\omega)} \right] \ . 
\eeq 
The matrix $A$ has the following form: 
\beq 
A \sim \left( 
\begin{matrix} 
c_i^{-1} + \mu b_{ii} & \cdots & \mu b_{ij} \\ 
\vdots & \ddots & \vdots \\ 
\mu b_{ji} & \cdots & c_j^{-1} + \mu b_{jj} 
\end{matrix} 
\right) \ , 
\eeq 
where $\mu=\lambda/|D(\omega)|^2$, $c_i=T_i \nu_i(\omega)$ and  
$b_{ij}= F(\omega) + \frac{p(\omega)}{T_j}+\frac{\overline{p(\omega)}}{T_i}$. 
Its determinant is an order $N$ polynomial in $\mu$ of the following form: 
\beq 
\label{detexp} 
\frac{\det A^\lambda}{\det A^0} =  
1 + \mu \sum_i  c_i b_{ii} 
+ \mu^2 \sum_{i<j} c_i c_j \left| 
\begin{matrix} 
b_{ii} & b_{ij} \\ 
b_{ji} & b_{jj} 
\end{matrix} \right| 
+ \mu^3 \sum_{i<j<k} c_i c_j c_k \left| 
\begin{matrix} 
b_{ii} & b_{ij} & b_{ik} \\ 
b_{ji} & b_{jj} & b_{jk} \\ 
b_{ki} & b_{kj} & b_{kk} 
\end{matrix} \right| + \cdots \ . 
\eeq 
To compute the coefficients explicitly,
define first $T_{ij}$ by $T_{ij}^{-1}=T_i^{-1}-T_j^{-1}$. 
The coefficient of $\lambda^2$ is given by a sum of determinants of the form 
\beq 
\left| 
\begin{matrix} 
F + \frac{p}{T_i}+\frac{\overline{p}}{T_i} &  
F + \frac{p}{T_i}+\frac{\overline{p}}{T_j} \\ 
F + \frac{p}{T_j}+\frac{\overline{p}}{T_i} &  
F + \frac{p}{T_j}+\frac{\overline{p}}{T_j} 
\end{matrix} \right|= 
\left| 
\begin{matrix} 
F + \frac{p}{T_i}+\frac{\overline{p}}{T_j} &  
\frac{\overline{p}}{T_{ji}} \\ 
F + \frac{p}{T_j}+\frac{\overline{p}}{T_i} &  
\frac{\overline{p}}{T_{ji}} 
\end{matrix} \right|= 
\left| 
\begin{matrix} 
\frac{p}{T_{ij}} &  
0 \\ 
F + \frac{p}{T_j}+\frac{\overline{p}}{T_i} &  
\frac{\overline{p}}{T_{ji}} 
\end{matrix} \right|= 
-\frac{|p|^2}{\big( T_{ij} \big)^2} \ , 
\eeq 
where the first column was first subtracted from the second column,  
and then the second row was subtracted from the first row. 
All the coefficients of the higher powers of $\lambda$ 
vanish. Consider for example the coefficient of $\lambda^3$. It has the form 
\beq \begin{split}
&\left| 
\begin{matrix} 
F + \frac{p}{T_i}+\frac{\overline{p}}{T_i} &  
F + \frac{p}{T_i}+\frac{\overline{p}}{T_j} & 
F + \frac{p}{T_i}+\frac{\overline{p}}{T_k} \\ 
F + \frac{p}{T_j}+\frac{\overline{p}}{T_i} &  
F + \frac{p}{T_j}+\frac{\overline{p}}{T_j} & 
F + \frac{p}{T_j}+\frac{\overline{p}}{T_k} \\ 
F + \frac{p}{T_k}+\frac{\overline{p}}{T_i} &  
F + \frac{p}{T_k}+\frac{\overline{p}}{T_j} & 
F + \frac{p}{T_k}+\frac{\overline{p}}{T_k} 
\end{matrix} \right| = \\
&\left| 
\begin{matrix} 
F + \frac{p}{T_i}+\frac{\overline{p}}{T_i} &  
\frac{\overline{p}}{T_{ji}} & 
\frac{\overline{p}}{T_{ki}} \\ 
F + \frac{p}{T_j}+\frac{\overline{p}}{T_i} &  
\frac{\overline{p}}{T_{ji}} & 
\frac{\overline{p}}{T_{ki}} \\ 
F + \frac{p}{T_k}+\frac{\overline{p}}{T_i} &  
\frac{\overline{p}}{T_{ji}} & 
\frac{\overline{p}}{T_{ki}} 
\end{matrix} \right|= 
\left| 
\begin{matrix} 
\frac{p}{T_{ik}} &  
0 & 
0 \\ 
\frac{p}{T_{jk}} &  
0 & 
0 \\ 
F + \frac{p}{T_k}+\frac{\overline{p}}{T_i} &  
\frac{\overline{p}}{T_{ji}} & 
\frac{\overline{p}}{T_{ki}} 
\end{matrix} \right|=0 \ , 
\end{split}\eeq 
subtracting the first column to the second and third column, and then
the third row to the first and second row. The same argument applies to all the other 
coefficients up to order $N$. Finally, one gets 
\beq 
\begin{split} 
\frac{\det A^\lambda(\omega)}{\det A^0(\omega)} &=  
1 + \frac\lambda{|D(\omega)|^2} \sum_i  T_i \nu_i(\omega)  
\left[ F(\omega) + \frac{p(\omega)}{T_i}+\frac{\overline{p(\omega)}}{T_i} \right] 
- \frac{\lambda^2 |p(\omega)|^2}{|D(\omega)|^4}  
\sum_{i<j} \frac{ T_i T_j \nu_i(\omega) \nu_j(\omega)}{\big( T_{ij} \big)^2} \\ 
&= 1-\frac{2\epsilon\omega\lambda \sum_i \nu_i(\omega)}{|D(\omega)|^2} 
 + \frac{\lambda (1-\lambda)}{|D(\omega)|^2}  
\sum_{i<j} \frac{ T_i T_j \nu_i(\omega) \nu_j(\omega)}{\big( T_{ij} \big)^2} \ , 
\end{split} 
\eeq 
and 
\beq 
z_{baths}(\l)=\int_{-\io}^{\io} \frac{d\o}{2\p} \log \left[ 1-\frac{2\epsilon\omega\lambda  
\sum_i \nu_i(\omega)}{|D(\omega)|^2} 
 + \frac{\lambda (1-\lambda)}{|D(\omega)|^2}  
\sum_{i<j} \frac{ T_i T_j \nu_i(\omega) \nu_j(\omega)}{\big( T_{ij} \big)^2} \right] \ . 
\eeq 
The first term in the logarithm is proportional to $\epsilon$ and is related to the power injected
by the external force, while the second term accounts for heat exchanges between the baths and
does not vanish at $\epsilon=0$.
Finally, observing that 
\beq 
\begin{split} 
&D(-\omega)=\overline{D(\omega)} - 2 i\epsilon \ , \\ 
&|D(-\omega)|^2 = |D(\omega)|^2 - 2 \epsilon \omega \sum_i \nu_i(\omega) \ ,
\end{split} 
\eeq 
and using the same trick already used in section~\ref{sec:II}, it is 
easy to show that $z_{baths}(\lambda)= z_{baths}(1-\lambda)$. 
Thus $\z_{baths}(p)$ should verify the fluctuation relation {\it at least} for $|p|\leq 1$,
if the contribution of boundary terms is not negligible.
This result is of interest for the study of heat conduction and is similar to the one
discussed in \cite{Ga04}.

%%%%%%%%%%%%%%%%%%%%%%%%%%%%%%%%%%%%%%%%%%%%%%%%%%%%%%%%%%%%%%%%%%%%%%%%%%%%%%%%%%
%%%%%%%%%%%%%%%%%%%%%%%%%%%%%%%%%%%%%%%%%%%%%%%%%%%%%%%%%%%%%%%%%%%%%%%%%%%%%%%%%%

\section{Appendix: Fluctuation relation for the spherical $p$-spin model}
\label{app:randomh}

It has been shown in section~\ref{sec:VI} that the dynamics of the mean field
spherical model is described by the following equation:
\beq
\begin{split}
&\dot \s(t) = -\mu \s(t) + \int_{-\io}^\io dt' \ \Si(t-t') \s(t') + \r(t) + 
\epsilon h(t) \ , \\
&\la \r(t)\r(t') \ra = 2T\d(t-t') + D_0(t-t') \ , \\
&\la h(t) h(t') \ra = D_1(t-t') \ ,
\end{split}
\eeq 
where $\r(t)$ and $h(t)$ are two uncorrelated Gaussian variables, and
$\Si$, $D_0$ represent a nonequilibrium bath once the self-consistency
equations for $R$ and $C$ are solved. The term $\ee h(t)$ represent the
external drive.

This equation is a particular instance of the general equation
\beq\label{motionpspineq}
\begin{split}
&m \ddot x(t) + \Iint dt' \, g(t-t') \dot x(t') = -k x(t) + \r(t) + \ee h(t) \ , \\
&\la \r(t) \r(t') \ra = \nu(t-t') \ , \\
&\la h(t) h(t') \ra = \mu(t-t') \ ,
\end{split} \eeq
with $m=0$, $\nu(\o) = 2T + D_0(\o)$ and $g(\o) = 1 + \Si(\o)/(i\o)$.
This gives, in absence of drive, an effective temperature
\beq\label{Teffpppp}
T_{eff}(\o) = \frac{\n(\o)}{2\re g(\o)}=\frac{2T + D_0(\o)}{2\re[1+\Si(\o)/(i\o)]} \ .
\eeq
If $R$ and $C$ are related by the FDR, $R(t) = -\b \th(t) \dot C(t)$, from the relation
$\Si = R D_0'(C)$ follows that $\Si(t) = -\b \th(t) \dot D_0(t)$, \ie
$D_0(\o) = 2T\re[ \Si(\o)/(i\o)]$, see section~\ref{sec6:FDTgeneralities}, so $T_{eff}(\o) \equiv T$
and the bath is in equilibrium, as expected. If $R$ and $C$ do not verify the FDR, 
$T_{eff} \neq T$.

The dissipated power is $W(t) = \ee h(t) \dot x(t)$, and its average is
$\la W(t) \ra = \ee \la h(t) \dot x(t) \ra$. The linear equation (\ref{motionpspineq})
is solved by
\beq
x(t) = \Iint dt' \, R(t-t') \big[ \r(t') + \ee h(t') \big] \ ,
\eeq
so that
\beq
\la W(t) \ra = \ee \la h(t) \dot x(t) \ra = 
\ee \la h(t) \Iint dt' \, \dot R(t-t') \big[ \r(t') + \ee h(t') \big] \ra =
\ee^2 \Iint dt' \, \dot R(t-t') \mu(t-t') \ ,
\eeq
which gives Eq.~(\ref{Wpspin}).

Finally, it is possible to prove that the {\sc pdf} of
\beq
\st(t) = \ee \int_{-\io}^t dt' \, T^{-1}(t-t') \big[ h(t) \dot x(t') + h(t') \dot x(t) \big]
\eeq
verifies the fluctuation relation. Following the strategy of section~\ref{sec:II} and \ref{sec:IV}
one first rewrites
\beq
\st_\t = \ee \frac{\D \o}{2 \pi} \sum_{n=-\io}^{\io} \frac{h_n i \o_n \bar x_n}{T_{eff}(\o_n)} \ ,
\eeq
and then, using $x(\o) = [\r(\o) + \ee h(\o)]/D(\o)$, with $D(\o) = -m \o^2 -i\o g(\o) + k$,
computes\footnote{Some factors are different because now all the quantities are real numbers
instead of complex numbers.}
\beq\begin{split}
\la e^{-\l \st_\t} \ra &= \NN^{-1} \int d\r_n dh_n \exp 
\left[ - \frac{\D \o}{2 \pi} \sum_{n=-\io}^{\io} \left(
\frac{ |\r_n|^2 }{ 2 \nu(\o_n)} + \frac{ |h_n|^2 }{ 2 \m(\o_n)}
+ \l \ee \frac{ h_n i \o_n (\bar \r_n + \ee \bar h_n) }{T_{eff}(\o_n) \overline{ D(\o_n)}}
\right) \right] \\
&= \prod_{n=-\io}^{\io} \left[ 1 + 
2 \ee^2 \l \frac{\m(\o) \o^2 \re g(\o) }{T_{eff}(\o) |D(\o)|^2}
-\ee^2 \l^2 \frac{\m(\o) \n(\o) \o^2 }{[T_{eff}(\o)]^2 |D(\o)|^2} \right]^{-\frac{1}{2}} \ ,
\end{split}
\eeq
so that, substituting Eq.~(\ref{Teffpppp}),
\beq
z_{eff}(\l) = \frac{1}{2} \Iint \frac{d\o}{2\p} \log  \left[ 1 + 
4 \ee^2 \l (1-\l) \frac{\m(\o) \o^2 [\re g(\o)]^2 }{\n(\o) |D(\o)|^2}
\right] \ ,
\eeq
which clearly verifies $z_{eff}(\l) = z_{eff}(1-\l)$.

\section{Appendix: Correlation functions of the harmonic oscillator coupled to two baths} 
\label{app:C} 
 
In the harmonic case, ${\cal V}(|a|^2)=\frac{k}{2} |a|^2$, the correlation function 
of a variable $a_t$, whose time evolution is given by Eq.~(\ref{motion2baths}),
can be computed analytically \cite{CK00}. 
In the frequency domain, Eq.~(\ref{motion2baths}) reads: 
\beq 
a_\omega = \frac{\r^f_{\o} + \r^s_{\o}} 
{ \kappa - i\omega \gamma_f - \frac{i\omega \gamma_s} 
{1-i\omega\tau_s}} \equiv \frac{\r^f_{\o} + \r^s_{\o}}{D(\omega)} \ , 
\eeq 
where $D(\omega) = \kappa - i\omega \gamma_f -  
\frac{i\omega \gamma_s}{1-i\omega\tau_s}$. 
Recalling that 
$\langle \r^f_{\o} \r^f_{\o'} \rangle =  
4\pi \gamma_f T_f \delta(\omega +\omega')$  
and $\langle \r^s_{\o} \r^s_{\o'} \rangle = \frac{4\pi \gamma_s T_s} 
{1+\omega^2\tau_s^2} \delta(\omega + \omega')$, and defining $C(\omega)$ from 
$\langle a_\omega a_\omega' \rangle = 2\pi \delta(\omega + \omega') C(\omega)$ 
one gets 
\beq 
\begin{split} 
&R(\omega)=\frac{da_\omega}{d\r^f_{\omega}}=\frac{1} 
{D(\omega)} \ , \\ 
&C(\omega) = \frac{ 2 \gamma_f T_f + \frac{2 \gamma_s T_s}{1+\omega^2\tau_s^2}} 
{\left| D(\omega) \right|^2} \ . 
\end{split} 
\eeq 
The function 
$(1-\omega \tau_s) D(\omega)$ is a polynomial in $\omega$ and its zeros 
are given by $\omega=- i \gamma_\pm$ where 
\beq 
\gamma_\pm = \frac{1}{2 \gamma_f \tau_s} \left[ (\kappa \tau_s + \gamma_f 
+\gamma_s ) \pm \sqrt{ (\kappa \tau_s + \gamma_f +\gamma_s )^2  
- 4 \kappa \tau_s \gamma_f } \right] \ , 
\eeq 
and $\re \gamma_\pm > 0$. 
The response function is then given by 
\beq 
R(t)=\int_{-\infty}^\infty \frac{d\omega}{2 \pi} e^{-i\omega t}  
\frac{1}{D(\omega)} = \frac{\theta(t)}{\gamma_f \tau_s} 
\left[ \frac{1 - \gamma_+ \tau_s}{\gamma_--\gamma_+} e^{-\gamma_+ t} 
+\frac{1 - \gamma_- \tau_s}{\gamma_+-\gamma_-} e^{-\gamma_- t}\right] \ , 
\eeq 
and the correlation function is given by 
\beq 
C(t)=\frac{1}{(\gamma_f \tau_s)^2} \left[ 
\frac{\gamma_f T_f (1-\gamma_+^2 \tau_s^2) + \gamma_s T_s} 
{( \gamma_- -\gamma_+ ) (\bar\gamma_-+\gamma_+)\re\gamma_+ } 
e^{-\gamma_+ t} +  
\frac{\gamma_f T_f (1-\gamma_-^2 \tau_s^2) + \gamma_s T_s} 
{( \gamma_+ - \gamma_- )(\bar\gamma_+ + \gamma_-) \re\gamma_- }  
e^{-\gamma_- t} \right] \ . 
\eeq 
In the case $\epsilon = 0$, and in the limit $\gamma_f \ll \gamma_s \ll k \tau_s$  
where the time scales of the two baths are well separated, one obtains 
\beq 
\begin{split} 
& \gamma_+ \sim \frac{k \tau_s + \gamma_s}{\gamma_f \tau_s} \ , \\ 
& \gamma_- \sim \frac{1}{\tau_s} \left( 1 - \frac{\gamma_s}{\gamma_s + k\tau_s} \ , 
\right) 
\end{split} 
\eeq 
and 
\beq 
\begin{split} 
& C(t) = \frac{T_s \gamma_s \tau_s}{(k\tau_s + \gamma_s)^2} e^{-t/\tau_s}+ 
\frac{T_f \tau_s}{k\tau_s + \gamma_f} 
e^{- \frac{k \tau_s + \gamma_s}{\gamma_f \tau_s} t} \ , \\ 
& R(t) = \theta(t) \left[ \frac{\gamma_s}{(k\tau_s + \gamma_s)^2} e^{-t/\tau_s}+ 
\frac{1}{\gamma_f} 
e^{- \frac{k \tau_s + \gamma_s}{\gamma_f \tau_s} t} \right] \ . 
\end{split} 
\eeq 
From the latter expressions it is easy to check that one has 
$R(t) \sim - \beta_f \theta(t) \dot C(t)$ for short times ($t\ll\tau_s$) 
and $R(t) \sim - \beta_s \theta(t) \dot C(t)$ for large times ($t\sim\tau_s$). 
The same behavior is found in the limit of small dissipation (small 
$\epsilon$), as one can check  
plotting the exact expression for the functions $R(t)$ and $C(t)$. 

%%%%%%%%%%%%%%%%%%%%%%%%%%%%%%%%%%%%%%%%%%%%%%%%%%%%%%%%%%%%%%%%%%%%%%%%%%%%%%%%%%%%%%%%
%%%%%%%%%%%%%%%%%%%%%%%%%%%%%%%%%%%%%%%%%%%%%%%%%%%%%%%%%%%%%%%%%%%%%%%%%%%%%%%%%%%%%%%%
 
\section{Appendix: The expression of $\st$ in the adiabatic approximation} 
\label{app:D}

Starting from the expression~(\ref{sigma3OK}) for $\st_t$ and from Eq.~(\ref{Tstar}),  
and remembering the convention $\int_{-\io}^t ds \ \delta(t-s) = \theta(0) = 1/2$, one
has: 
\beq 
\int_{-\io}^t dt' \ \frac{\dot a_t \bar a_{t'}}{T^*(t-t')} =  
\frac{\dot a_t \bar a_t}{2T_f} + \frac{\g_s}{2 \Omega T_f \g_f (\t_s)^2}  
\left(1-\frac{T_s}{T_f}\right) \int_{-\io}^t dt' \ e^{-\Omega (t-t')} \dot a_t \bar a_{t'} \ . 
\eeq 
One can substitute $a_t = H_t + w_t$ and neglect all the terms proportional to 
$H_t w_t$: 
indeed, such terms vanish when $\st_t$ is integrated over time intervals of the order 
of $\t_s$, as, on such time scales, $\langle w_t \rangle = 0$ while $H$ is constant. 
The first term gives then 
\beq 
\frac{\dot a_t \bar a_t}{2T_f} = \frac{\dot H_t \bar H_t + \dot w_t \bar w_t}{2T_f} \ . 
\eeq 
In the second term, as $\Omega \sim 1/\t_s$, one approximates  
$\int_{-\io}^t dt' \ e^{-\Omega (t-t')} \bar a_{t'} \sim H_t/\Omega$ to obtain 
\beq 
\frac{\g_s}{2 \Omega^2 T_f \g_f (\t_s)^2}  
\left(1-\frac{T_s}{T_f}\right) \dot H_t \bar H_t \sim  
\frac{1}{2T_s} \left(1-\frac{T_s}{T_f}\right) \dot H_t \bar H_t \ . 
\eeq 
The (imaginary part of the) sum of these two terms times $4 \epsilon$ gives  
Eq.~(\ref{sigma3adiabatic}). 

%%%%%%%%%%%%%%%%%%%%%%%%%%%%%%%%%%%%%%%%%%%%%%%%%%%%%%%%%%%%%%%%%%%%%%%%%%%%%%%%%%%%%%%%%
%%%%%%%%%%%%%%%%%%%%%%%%%%%%%%%%%%%%%%%%%%%%%%%%%%%%%%%%%%%%%%%%%%%%%%%%%%%%%%%%%%%%%%%%%

%% file: ringra.tex
\chapter*{Acknowledgments}
\pagestyle{empty}
\addcontentsline{toc}{chapter}{Acknowledgments}

I had the opportunity to work, during these three years of
Ph.D., under the joint supervision of Giorgio Parisi and Giancarlo
Ruocco: they gave me the possibility to approach the statistical
mechanics of disordered systems from both the theoretical
and experimental side. The interaction with them has been for me
a continuous source of new ideas and of a deeper understanding
of many aspects of physics. I wish to thank them for that and
for their continuous support and encouragement.

Moreover, I benefited from the close collaboration with 
L.~Angelani, F.~Bonetto, and A.~Giuliani,
and with L.~Cugliandolo, J.~Kurchan, and G.~Gallavotti, 
who supervised part of this work. Without them most of the
results that have been presented here would not have been
obtained.

Most of the work was done in the Physics department of
the University of Rome ``La Sapienza'' where the presence of two
research centers (INFM - CRS SMC and INFM - CRS Soft) dedicated
to the physics of disordered systems created a very stimulating
environment, where it is possible to find expert people in
theoretical, computational and experimental physics.
Among them, I wish to thank T.~Castellani, A.~Cavagna, C.~Conti, 
C.~De~Michele, L.~Leuzzi, E.~Marinari, E.~Pontecorvo, F.~Ricci Tersenghi, 
F.~Sciortino, T.~Scopigno, P.~Tartaglia, E.~Zaccarelli, for many useful
discussions that often 
motivated and improved a lot the work presented here.
I wish to especially thank R.~Di Leonardo, who was always a reference
point for discussing experimental results, and with whom I shared many
discussions on ``philosophical'' aspects of the physics of disordered 
systems as well as on practical problems.
I wish also to thank D.~Benedetti, M.~Ortolani, S.~Speziale
for many useful discussions on subjects that were not
directly related to my work: they were very useful to enlarge my 
interests and enter in contact with new interesting subjects 
of research.

During these three years I also spent time in some institutions that
I wish to thank for their hospitality: in particular the 
{\it \'Ecole Normale Sup\'erieure} in Paris, the
{\it Centre de Physique des Houches}, the 
{\it International Center for Theoretical Physics} (ICTP) in Trieste,
the {\it University of La Habana}, Cuba, and the {\it Spinoza Institut} in
Utrecht.

Last but not least, I wish to thank A.~Degasperis who supervised 
my first teaching experience in the best possible way.

\vskip1cm

This work is dedicated to my father Carlo.

\newpage
\pagestyle{empty}

\mbox{}